\documentclass[12pt]{report}
\usepackage[utf8]{inputenc}

\usepackage{graphicx}
\usepackage{cleveref}
\graphicspath{ {images/} }
\usepackage{mathrsfs}
\usepackage{amsmath}
\usepackage{amssymb}
\usepackage{dsfont}
\usepackage{mathtools}
\usepackage[margin=1in]{geometry}
\usepackage[onehalfspacing]{setspace}
\usepackage{cleveref}
\crefname{section}{§}{§§}
\Crefname{section}{§}{§§}
\usepackage{titling}
\usepackage{cite}
\usepackage{subcaption}
\usepackage{enumitem}
\usepackage{gensymb}
\usepackage{bm}

\usepackage{titlesec}
\usepackage{apptools}
\AtAppendix{\titleformat{\chapter}[block]{\LARGE\bfseries}{\appendixname~\thechapter:}{0.333em}{}%
\titlespacing*{\chapter}{0pt}{-20pt}{40pt}}

\begin{document}

\begin{titlepage}
	\centering
	
	{\huge\bfseries On the analogy between black holes and bathtub vortices\par}
	\vspace{0.5cm}
	{\Large \textit{Sam Patrick} \par}
	\par\vspace{1cm}
\includegraphics[width=0.6\textwidth]{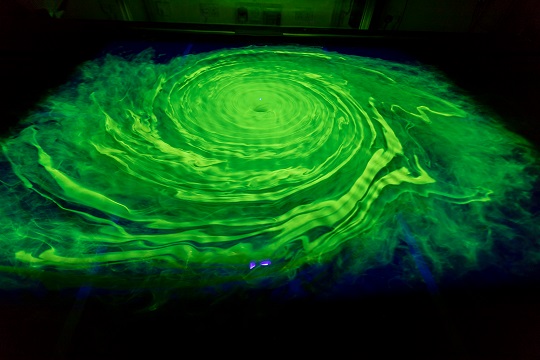}\par\vspace{1cm}
{\scshape School of Mathematical Sciences \par}

    \vspace{2cm}
    
    \includegraphics[width=0.3\textwidth]{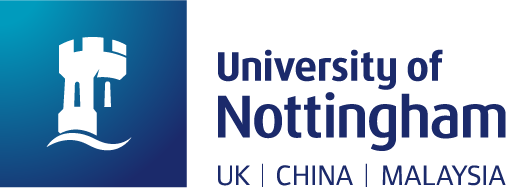}\par\vspace{1cm}
Thesis submitted for the degree of Doctor of Philosophy \\
at the University of Nottingham
\par\vspace{2cm}
	Defended on September 18th 2019 before the jury
\begin{center}
\begin{tabular}{ c c }
 Prof. William G. Unruh & External Examiner \\ 
 Dr. Sven Gnutzmann & Internal Examiner
\end{tabular}
\end{center}

\end{titlepage}

\pagenumbering{gobble}

\clearpage
\thispagestyle{empty}
\phantom{a}
\vfill
\vfill


\begin{abstract}
Analogical thinking is a valuable tool in theoretical physics, since it allows us to take the understanding we have developed in one system and apply it to another. 
In this thesis, we study the analogy between two seemingly unlikely systems:
rotating black holes, elusive cosmic entities that push our theoretical understanding of modern physics to its limits, and bathtub vortices, an occurrence so common that they can be observed on a day-to-day basis in almost any household.
Despite the clear difference between these two systems, we argue that lessons from each can be used to learn something about the other.

We investigate the equivalence between surface wave propagation in shallow water and the propagation of a massless scalar field on an effective spacetime, focussing in particular on the rotating black hole geometry sourced by a rotating draining vortex flow.
Using this analogy, we verify for the first time that three effects  predicted to occur around rotating black holes also occur in a laboratory experiment.
These are \textit{superradiance}, an energy enhancement process whereby waves extract rotational energy from the system, \textit{quasi-normal ringing}, describing the relaxation of the system toward equilibrium, and the \textit{backreaction}, which mediates the exchange of energy between fluctuations and the background they experience.

Previous studies within analogue gravity have focussed on demonstrating that the existence of Hawking radiation does not depend on the details of high frequency dispersion.
Our experimental results indicate that the same can be said for superradiance and quasi-normal ringing, although notable differences do occur when the medium is dispersive.
Using tools originally developed in the context of black hole physics, we propose a new method for flow measurement based on the characteristic ringing frequencies of the vortex, which can be used for a known effective field theory to measure the parameters of the fluid flow, or for known flow parameters to test the effective field theory.
We also study how the characteristic mode spectrum is modified by a purely rotating fluid with vorticity, finding that the system can also support bound state resonances characterised by a much longer lifetime than the usual modes.

Finally, we demonstrate theoretically and experimentally how the backreaction of waves onto the background flow manifests itself in the bathtub system.
This study reveals several interesting features of experimental bathtub vortices, in particular that the presence of waves leads to the removal of fluid mass from the system, leading to a net decrease in the water level.
This dynamical interplay between waves and the background makes it a promising candidate for study in the quantum regime, where the spontaneous emission of waves will also influence the system's evolution.
\end{abstract}


\clearpage
\thispagestyle{empty}
\phantom{a}
\vfill
\vfill

\section*{Acknowledgements}
First of all, I'd like to thank my supervisor, Silke, for the opportunity to work within her group.
Her unfaltering ambition and enthusiasm for the subject has been a source of inspiration for me over the past three years, and the numerous coffees and lunches have also not gone unappreciated!
I will miss our frequent long discussions, which occasionally turned into heated debates, and will remember them all with with fondness.
I am sure we will continue to work together in the future. 
\\
\\
My time during this PhD would not have been the enjoyable experience that it was were it not for my lab mates Th\'eo, Zack, August, Steffen, Antonin, Sebastian and Cisco, as well as our frequent collaborators Harry and Maur\'icio.
I might not miss frantically trying to collect final results until well gone midnight, but I will miss `tea time'.
And August, as Silke once said to me, now it's your time.
Th\'eo and I both wish you good fortune on your journey to search for the quasi-bound states. 
\\
\\
I'd also like to thank my house mates Berry, Will, Aron and John.
At first, I didn't believe anything could live up to my first four years living in Nottingham as an undergraduate but I have to admit, I was wrong.
Thank you all for some truly unforgettable experiences.
\\
\\
To my mother and father for their love and support.
Dad, your ability to find an argument in an empty box undoubtedly prepared me for the task of defending this thesis.
Mum, you have played an equally essential role in helping me tolerate the incessant debates.
I can't thank either of you enough.
\\
\\
To my best friend and partner in crime, Marie-Claire. 
For your patience and continuing encouragement, I am forever grateful.
I could not have finished this thesis without you.
\\
\\
And finally, I'd like to dedicate this thesis to my grandfather, Ernest ``Bernie'' Wenborn, who passed away in the month before my final submission.
You always spoke with pride when talking about how we already have one doctor in the family.
Now we have two.


\section*{Acronyms}

ABHS  (analogue black hole spectroscopy) \\
BS  (bound state) \\
BEC  (Bose-Einstein condensate) \\
BH  (black hole) \\
DBT  (draining bathtub) \\
FCD  (Fast chequerboard demodulation) \\
GR  (general relativity) \\
HR  (Hawking radiation) \\
KG  (Klein-Gordon) \\
MSE  (mean squared error) \\
ODE  (ordinary differential equation) \\
PDE  (partial differential equation) \\
PIV  (particle imaging velocimetry) \\
PV  (potential vorticity) \\
QN  (quasi-normal) \\
QNM  (quasi-normal mode) \\


\tableofcontents
\thispagestyle{empty}

\chapter{Introduction} \label{chap:intro}
\pagenumbering{arabic}
\setcounter{page}{1}
Black holes are arguably the most fascinating objects in the universe.
They are regions of spacetime which become so warped that not even light can escape once it crosses the \textit{horizon} (a limiting surface surrounding the black hole) and everything that does so is destined to collate at the \textit{singularity} (a point at the centre of the hole where the density and curvature become infinite).

The first black hole solution to the Einstein field equations was discovered by Schwarzschild \cite{schwarzschild1916gravitationsfeld} shortly after the publication of General Relativity (GR) in 1915 \cite{einstein1915feldgleichungen}.
At first, the scientific community was dubious as to whether singularities should occur nature, and it wasn't until Chandrasekhar \cite{chandrasekhar1931maximum} and Oppenheimer \cite{oppenheimer1939massive} demonstrated that gravitational collapse could not be halted for stars above a critical mass that black holes began to gain acceptance.
Later, Hawking and Penrose showed that singularities are fundamental in GR \cite{penrose1965gravitational,hawking1973large} which was followed by work on the thermodynamic properties of black holes \cite{bekenstein1973black}, leading eventually to the discovery of Hawking radiation (HR) \cite{hawking1974explosions}.
Since then, evidence for the existence of black holes has accumulated \cite{abbott2016merger,gammie2019first} and they continue to be an active area of research today.
Black holes are central to many conflicts of principle, such as the information paradox and the trans-Planckian problem \cite{giddings1995black,corley1996spectrum}, which challenge the foundations of modern physics.
Deeper insight into these other-worldly objects is therefore essential in developing a complete understanding of the natural world.

By contrast, free-surface vortices or \textit{bathtub vortices} are a familiar phenomenon to anyone who has watched water drain from their kitchen sink or bathtub.
Their occurrence in everyday experience means that they have undoubtedly been recognised for as long as man has regarded bodies of water.
For example, it is common knowledge that water draining from a bathtub (in a very idealised set-up) ought to rotate in opposite directions depending on hemisphere you are in \cite{lugt1983vortex}.
In spite of their widespread recognition, the properties of bathtub vortices (e.g. shape, formation, dynamics) are still not completely understood and have only been addressed in the literature relatively recently \cite{vanden1987free,forbes1995bath,andersen2003anatomy}.
The problem is of both purely academic as well as engineering interest.
Indeed, air entrainment resulting from bathtub type vortices at the outlet of industrial tanks can result in problems with efficiency and the damage of mechanical components (see e.g. \cite{stepanyants2008stationary} and references therein).

In view of such vastly differing contexts, it seems unlikely that bathtub vortices should bear any relation in the slightest to black holes.
The aim of this thesis, however, will be to draw an \textit{analogy} between these two seemingly disparate phenomena and show that we can in fact use each to learn something about the other.


\section{Analogies in physics}

Analogies are a useful tool in science and mathematics, not only for understanding unfamiliar concepts but for developing new theories to describe the natural world.
They are frequently used in the classroom by physics teachers to help students to grasp new ideas \cite{duit1991role,podolefsky2006use}, which is achieved by drawing comparisons between two scenarios which exhibit (at least some) similar aspects.
The key feature of an analogy is that reasoning used in the base domain (which is understood) can be applied to the target domain (which one is trying to understand).
The success of an analogy depends on the extent to which this reasoning can be used to make accurate predictions in the target domain.

An example of this is Rutherfords planetary model of the atom\footnote{In a semiclassical treatment of the Schr\"odinger equation (i.e. in the limit $\hbar\to0$) this approximation (or rather, it's improvement the Rutherford-Bohr model \cite{bohr1913xxxvii}) appears at leading order, hence confirming it's validity as an accurate model \cite{hainz1999centrifugal}.} \cite{rutherford1911lxxix}, which is often used to introduce the concept of electron orbits around a nucleus to physics students \cite{podolefsky2006use}.
In this example, the solar system is the base domain and the atom is target domain.
Another example is the spring ball model of molecules in Chemistry \cite{del2000models}, which well approximates molecular excitations using vibrations of a simplified spring and ball mechanism, despite the fact that the real system is ultimately governed by quantum mechanics \cite{born1927quantentheorie}.

In addition to being a useful tool for learning, there are many historical examples where the use of analogies has led to theoretical developments in physics. 
An example of this is the development of classical electromagnetism, where Maxwell used the fluid equations\footnote{This can be seen easily from the linearised fluid equations which are formally equivalent (upon suitable redefinition of the fields) to the homogeneous Maxwell equation's.} as the ``vehicle of mathematical reasoning'' \cite{maxwell1890scientific}.
It is undeniably curious that such disparate phenomena should be described by the same equations, and this is something that seems to crop up in many different corners of physics.
Another example of this is the formal equivalence of the Korteweg-de Vries equation, describing non-linear surface gravity waves, and the Gross-Pitaevskii equation, which governs atoms in a Bose-Einstein condensate.

Moving on from physical examples, the well-known \textit{Gedankenexperiment} (or thought experiment) is also a type of analogy at its heart.
The premise of the thought experiment is to imagine an idealised version of reality that captures (some of) the features of reality that one is trying to describe.
This type of analogous thinking allows a physicist to make predictions about what he/she expects to happen in the real world based on what occurs in it's idealisation.
Hence, from just these few examples, we can see that analogies are an invaluable tool, that enable us to use our intuition to develop better models of nature.

To differentiate between the natural world and the mathematical models we use to describe it, Unruh uses the terms \textit{territory} and \textit{map} \cite{unruh2018role}.
Analogical thinking is concerned with a scenario in which different territories (different areas of physics) can be covered by the same map (the same theory).
We shall refer to a precise mathematical analogy of this type as meaningful.
The existence of such an analogy is curious: if it is surprising that the overall map (i.e. theoretical physics) should provide a precise description of the terrain at all, then it is astonishing that, in certain cases, the same part of the map can be used to cover seemingly unrelated territories.
Whether a meaningful analogy is indicative of some underlying/universal structure in nature or merely a coincidence arising from our method of description is an on-going debate, and perhaps a question better suited to the philosopher over the physicist.
Nonetheless, it's existence presents an interesting paradigm.
In the words of Unruh \cite{unruh2018role}, any meaningful analogy finds itself subject to the following question:
\begin{quotation}
``\textit{If the map of two regions is the same, how much can we say about the similarity of the territory that the maps describe?}'' - Unruh (2018)
\end{quotation}
Following on from this we can ask, if one territory is poorly understood but has the same map as a second more familiar territory, then can reasoning used to understand the first be applied to make deductions about the second?
If the map were an exact representation of the territory then the answer would indisputably be yes.
However, since the map is only an approximation, the relevant question is whether the difference between reality and description is enough to influence any deductions obtained through analogue reasoning.
It is precisely this question that was encountered in a relatively new field of research called \textit{Analogue Gravity}.




\section{Analogue gravity}

Analogue gravity is a research programme that was pioneered in the 1980s by Unruh \cite{unruh1981experimental} and was later rediscovered by Visser \cite{visser1993lorentzian} in the 1990s.
The story goes (see e.g. \cite{unruh2018role}) that Unruh wanted a simple analogy to describe the idea of a black hole horizon to an audience who were unfamiliar with black hole physics.
Using a waterfall which flows supersonically (faster than sound waves) close to the edge and subsonically (slower than sound waves) upstream\footnote{Fluid flows of this nature are sometimes referred to as trans-sonic.}, he argued that the boundary between these two regions would be an acoustic horizon; a one-way membrane which only lets downstream-directed sound waves pass through, preventing upstream-directed sound waves from escaping once they get too close to the waterfall's edge.
The acoustic horizon for sound waves in this example is analogous to the horizon that encompasses a black hole.

Several years later whilst teaching a fluid mechanics course, Unruh found that linear perturbations to a classical fluid obey the same equation that describes the propagation of massless scalar perturbations on a curved geometry: the Klein-Gordon equation.
The properties of the fluid flow determine an effective curved spacetime seen by the perturbations and thus, by carefully tuning the fluid parameters, one can simulate wave propagation in a range of general relativistic settings, notably in black hole spacetimes.
Unruh realised the importance this might have in understanding the problem of black hole evaporation \cite{unruh1981experimental}.

\section*{The Trans-Planckian problem} 

In the early 1970s, Hawking published his famous result that black holes are not actually black when quantum fields live on the spacetime, but emit a thermal spectrum of electromagnetic radiation \cite{hawking1974explosions}.
The result was both surprising and in some sense expected, since earlier work by Bekenstein suggested that black holes ought to possess thermodynamic properties due to the relation between the black hole area law and entropy, despite the fact that classically a black hole cannot emit \cite{bekenstein1973black}.
Hawking radiation filled in this missing piece of the puzzle.

Soon after Hawking's discovery, people started to realise there was a problem in the derivation.
The out-going flux of particles originates from a region in the past close to the horizon.
Since these particles must climb out of a gravitational well, they are red-shifted, and therefore the thermal radiation at infinity is blue-shifted as it is traced back to its point of origin in the past.
However, these blue-shifted particles have energies which transcend the Planck scale, where the spacetime continuum gives way to quantum mechanics and a full theory of quantum gravity is required.
This issue has become known as the \textit{Trans-Planckian problem}, and can be stated simply as follows:
Do we need to know the details of physics beyond the Planck scale to have faith in Hawking radiation?

\section*{Analogue gravity to the rescue} 

Unruh's idea was to use the analogue gravity formalism to see how the trans-Planckian problem is resolved in the analogue system, where the granularity of the effective geometry at small scales (due to the atomic nature of the medium) is understood.
In other words,
\begin{itemize}
\item We know that the same map can be used to describe low frequency scalar wave propagation in two territories, i.e. trans-sonic fluids and black holes.
\item We also know how to extend this map to cover the high frequency territory in the fluid system.
\item Therefore, in spite our ignorance of the correct map for the high frequency territory of the gravitational system, can we use the fluid to make inferences about the dependence of Hawking radiation on high frequency physics?
\end{itemize} 
The answer turned out to be yes.

Jacobson \cite{jacobson1991ultrashort} pointed out that small scale properties of the medium have the effect of introducing a high frequency cut-off, above which the continuum description ceases to be valid.
In fluids, this cut-off manifests itself as a change in the dispersion relation at high frequencies.
Hence, as the low frequency mode is traced back to the horizon, it is blueshifted into a frequency range where its propagation speed changes: 
if the dispersion relation is superluminal, the mode comes from inside the horizon; 
if it is subluminal, HR originates as high frequency in-going modes which are converted into low frequency out-going modes.
By studying different dispersion relations motivated by analogue systems, both Unruh \cite{unruh1995sonic} and Corley \& Jacobson \cite{corley1996spectrum} were able to show that Hawking radiation is unaffected by the cut-off, at least for low frequencies which dominate the thermal spectrum.
These results are important since they suggest it is correct to assume that Hawking radiation still takes place when the spacetime description is modified at short-distances.
This result was the first triumph of analogue gravity.

\section*{Experiments} 

Analogue gravity was initially proposed with experimental application in mind.
A detection of Hawking radiation in an analogue system would not only give evidence to its existence when the continuum theory breaks down, but could simultaneously test its prevalence over other phenomena which are known to occur in fluid systems, e.g. vorticity, turbulence, dissipation etc.
The same can be said for other effects which are expected to occur around black holes.

Since the inception of analogue gravity, a wide range of base analogue systems have been proposed to model the target gravitational system.
The original system proposed by Unruh and later used by Visser was sound waves (i.e. acoustic perturbations) in a classical fluid \cite{unruh1981experimental,visser1993lorentzian,visser1998acoustic}.
This was extended to surface gravity waves in the early 2000s by Sch\"utzhold \& Unruh \cite{schutzhold2002gravity}.
Other promising candidates for analogue experiments include dilute Bose-Einstein condensates (BECs) \cite{garay2000sonic,garay2001sonic,barcelo2001analogue,barcelo2003towards}, interface waves called ripplons between layers of superfluid helium \cite{volovik2002black,volovik2002effective}, electromagnetic waves in dielectric media \cite{philbin2008fiber,belgiorno2011dielectric} and
slow light in atomic media \cite{leonhardt2002laboratory,unruh2003slow}.
An extensive review of analogue gravity along with a catalogue of models can be found  in \cite{barcelo2011analogue}.

On the HR front, the first analogue gravity experiments to yield results were the surface waves analogues.
Mode conversion was studied as early as 1983 by Badulin \cite{badulin1983laboratory} and evidence of negative norm modes was presented by Rousseaux in 2008 \cite{rousseaux2010horizon,rousseaux2008observation}.
A measurement of the stimulated Hawking effect was finally provided by Weinfurtner et al. in 2011 \cite{weinfurtner2011measurement,weinfurtner2013classical}.
The experiment was performed in an analogue white hole set-up (time-reversed black hole) and it was shown that impingent low frequency modes are converted into high frequency reflected modes. 
By measuring the relative amplitudes of the out-going modes, it was shown that the spectrum was indeed thermal and thus consistent with a detection of stimulated HR.
Spontaneous emission was observed later in a BEC set-up by Steinhauer and collaborators.
The intricate set-up was presented in \cite{lahav2010realization}, evidence for correlation between the out-going radiation and in-going partners was seen in \cite{steinhauer2016entanglement} and thermality of the radiation was demonstrated in \cite{deNova2018thermal}.
Finally, emission of photons in a fibre-optic set-up was reported by Belgiorno et al. in \cite{belgiorno2010hawking}, making this a promising system to further explore spontaneous emission in the future.

More recently, experiments to probe black hole phenomena other than HR have gained increasing amount of momentum on the experimental front.
In particular, surface gravity waves in bathtub vortex set-ups have been used to measure superradiance \cite{torres2017rotational}, analogue black hole ringdown \cite{torres2018application} and the backreaction \cite{goodhew2019backreaction}.
These three experiments will be discussed in Chapters~\ref{chap:super}, \ref{chap:qnm} and~\ref{chap:back} of this thesis.

In addition to the trans-Planckian problem, there are other issues analogue gravity can hope to shed light on.
One aspect that is still not clear is how the backreaction of HR influences the black hole evaporation process at late times.
Some researchers hope to learn how the backreaction can be addressed in general relativity by investigating the evolution of analogue systems where the theory is known at small scales and experiments can (in principle) be performed \cite{balbinot2005quantum,balbinot2006hawking}.
This will be discussed more at the end of Chapter \ref{chap:back}.
Finally, analogue models have also been used to show how dynamical spacetimes can emerge from underlying degrees of freedom \cite{girelli2008gravitational,sindoni2012emergent,belenchia2014emergent}.

\section*{Why analogue gravity experiments?}

Although it may seem like a fairly obvious point, we wish to stress the indisputable value of experiment in developing our understanding of nature.
This value, despite being accepted by an overwhelmingly large faction of the physics community, seems to come into question time and time again following analogue gravity discussions, as evidenced by the number of occasions the question ``Why don't you just simulate it?'' has been asked.
We briefly elucidate on the value of analogue gravity experiments:
\begin{itemize}
\item A theory is only as good as the assumptions that are put into it, and a simulation is only as good as the theory.
If the assumptions neglect certain physical processes which become important in some regime, then the theory will ultimately fail there.
This is the reason why \textit{it is always important to test theory against experiment}.
\item Analogue gravity is in no way a substitute for performing experiments on real gravitational systems.
However, whilst quantum gravity lies beyond our understanding and tests on gravitational systems are not always possible with current technology, analogue systems provide a convenient testing ground to investigate how effects in the classical picture are modified by deviations from the idealised theory.
\item In the analogue system, natural deviations from the idealised set-up (caused by e.g. atomic scale physics, dissipation, turbulence) act as a test of the robustness of phenomena against the details of the theory.
Therefore, if a certain phenomenon predicted by the theory can be observed in an experiment, then the phenomenon rests on more stable ground than the theory it came from.
\end{itemize}
A few more words on this final point.
A theory is usually constructed to explain observed phenomena.
However, in certain circumstances, a theory may also predict the existence of unobserved phenomena.
In such situations, it is important to go out and find these phenomena in the natural world, since the theory is ultimately only an approximation.
Until a phenomenon is observed (and has withstood extensive experimental scrutiny) it cannot be said whether it is (a) an artefact of our description or (b) a reality of the natural world.

To bring the discussion back to analogue reasoning, the observation of a certain phenomenon in one system reinforces our belief that it ought to exist generally in nature, especially if we believe that all physical phenomena are ultimately subject to the same laws (i.e. the laws of nature).
This is generally accepted to be the case since (a) it is commonly believed that there exists a grand unified description (i.e. a Theory of Everything) underpinning all of reality and (b) it is known that disparate areas of physics are subject to the same effective laws in certain limits (e.g. charges in electrostatics and masses in Newtonian gravity).
It is precisely the fact that the laws of nature appear in the same form in different corners of theoretical physics which allows us to use analogies to reinforce our belief in the existence of certain phenomena.

\section{Fluid dynamics}

Fluid dynamics is an effective field theory that emerges when considering the motion of a continuum of \textit{fluid elements} \cite{kundu2008fluid}.
Formally, these are patches which are much smaller than the characteristic length scales of a fluid flow but much larger than the atomic length scale of the medium.
In other words, it is the theory describing a flowing medium once the motions of the individual atoms/molecules are averaged over.
It is one of the oldest disciplines in the physical sciences (dating back to Archimedes in 250 BCE) and has successfully described a multitude of observed phenomena over the past few centuries.
Indeed, it is thought that the second partial differential equation to be written down was the Euler equation (see Eq.~\eqref{euler} in the next chapter) describing incompressible fluid flow \cite{brezis1998partial} (the first was the 1D wave equation).

Historically, the progression of fluid dynamics was quite unlike most areas of physics in the sense that developments often came in the form of phenomenological descriptions inspired by experiment, as opposed to being built on theoretical grounds from the underlying theory\footnote{Rossby puts this nicely in the introductory paragraph to his work on ocean currents \cite{rossby1936dynamics}. Talking about fluid dynamics, he says:
\begin{quotation}
\textit{This science has recognized that the exact character
of the forces controlling the motion of a turbulent fluid is not known and that consequently there is very little  justification for a purely theoretical attack on problems
of a practical character. For this reason fluid mechanics has been forced to develop a research technique all of its own, in which the theory is developed on the basis of experiments and then used to predict the behaviour of fluids in cases which are not accessible to experimentation.} - Rossby (1936)
\end{quotation}
}.
The reason for this is that physical fluid phenomena are often extremely complex, and gross simplifications are usually required to make any analytical headway.
As such, the use of heuristic models is still common in fluid dynamics today.
An example of this that we shall encounter later on is the well known Rankine vortex, describing a forced vortex interior stitched to a free vortex exterior.
Despite the fact that it is not a solution of the governing equations (on top of the fact that the discontinuity between interior and exterior is clearly unphysical) this model has been successful in describing a wide range of experimental observations, from hurricane profiles to vortices near water-tank outlets \cite{smith1995vortex,lautrup2005exotic}.

Although our work in the coming chapters is motivated by black hole physics, our treatment of theoretical problems is more typical fluid dynamics in the sense that we focus on simplified models.
Much of the theory we present will consider the shallow water regime, due to the inherent complexities of dealing with dispersive waves in inhomogeneous media.
On top of this, we will not consider the effects of non-linearities or viscosity on wave propagation.
Despite the fact that all of these will be present to some extent in experiment, we shall see that our simplified theory performs well qualitatively (and quantitatively on occasions).
Improvement to the theory can of course be achieved by including other known physical processes.
In fact, modifications by adding one ingredient at a time to the basic theory can be used to indicate which physical processes are the most important in controlling the effects under consideration.
We will see examples of this in Chapters~\ref{chap:qnm} and \ref{chap:vort}.

We conclude this section with a quote from B\"uhler \cite{buhler2005wave} which neatly captures the approximate nature of fluid dynamics (and indeed theoretical physics in general):
\begin{quotation}
``\textit{... it reminds us that our physical and mathematical categories never fully catch all the facets of the slippery reality we seek to understand – but we try}'' - B\"uhler (2005)
\end{quotation}

\section{Overview}

The analogue black hole set-up based on the bathtub vortex has been well studied in the literature from a theoretical stand-point \cite{basak2003superresonance,basak2003reflection,cardoso2004qnm,berti2004qnm,dolan2012resonances,richartz2015rotating,churilov2018scattering}, but how the analogy performs in an experimental context remains to be seen.
In this thesis we will investigate precisely this, focussing on how the analogy is altered by effects of a purely fluid mechanical origin that naturally arise under experimental conditions.
We have already heard that analogue gravity has played a crucial role in understanding the nature of HR when deviations to the classical theory occur at small scales.
Inspired by this success, we will use analogue gravity to study superradiance, black hole ringdown and backreaction.

In the spirit of maps and territories, we give a brief list of the important destinations to anticipate on our journey.
\begin{itemize}
\item In Chapter~\ref{chap:theory}, we derive from the shallow water equations the map that describes scalar waves in the gravitational scenario, hence demonstrating that analogue reasoning can be applied across the two systems.
\item In Chapter~\ref{chap:super}, we demonstrate how superradiance, an effect expected to occur around black holes, persists in our system far beyond the regime of the gravitational analogy. 
This suggests the robustness of superradiance against modifications to the theory.
\item In Chapter~\ref{chap:qnm}, we use the analogy in reverse to show how tools in gravitational physics (particularly the characteristic modes of black holes) can be used to develop new techniques for experimental fluid dynamics.
This demonstrates that analogue gravity is a two-way street.
\item In Chapter~\ref{chap:vort}, we take a small detour to investigate the effects of vorticity (a natural fluid phenomena) on the characteristic modes of our system, thereby demonstrating how the frequency spectrum can become modified.
\item In Chapter~\ref{chap:back}, we describe how a phenomena which is readily observable our simple fluid experiment is related to an on-going area of research in black hole physics, namely the effects of backreaction.
We hope that this study will motivate future experimental studies, with the ultimate aim of learning how the backreaction can influence the evolution of black holes.
\end{itemize}

\section{Statement of originality}

Below is a description of which parts of this thesis are original, and which parts are my own interpretation of work in the literature.
I also emphasize my particular role within collaborations and whether original work has been published.
\begin{itemize}
\item Chapter~\ref{chap:theory} is my own interpretation of theoretical understanding already present in the literature.

\item Chapters~\ref{chap:super} and \ref{chap:qnm} are based on work in \cite{torres2017rotational} and \cite{torres2018application} respectively. 
The first has been published in \textit{Nature Physics} and the second is available on the \textit{arXiv} as a preprint.
The theory sections, again, are my own interpretation of methods existing in the literature.
The experiments were a collaborative effort between members of our group at the University of Nottingham and are entirely original.
Different aspects of the experiments were led by different members of the team, although all members were involved with and contributed evenly to each aspect of the experiments. 
My specific role was in leading the acquisition and analysis of data obtained for the PIV method.

\item The final part of Chapter~\ref{chap:qnm} is based on \cite{torres2019analogue}, published in \textit{Classical and Quantum Gravity}.
This work expands upon the method developed in \cite{torres2018application} and my particular involvement was performing the numerical simulations.
For this thesis, however, I repeated the full method myself and hence, all of Section~\ref{sec:ABHSmethod} is my own work.

\item Chapter~\ref{chap:vort} is based on the work of myself and collaborators in \cite{patrick2018QBS}, which has been published in \textit{Physical Review Letters}.
However, whilst writing this thesis, I realised this project could be improved and extended and as such, all of Chapter~\ref{chap:vort} is my own unpublished work.

\item Chapter~\ref{chap:back} is based on experiments in \cite{goodhew2019backreaction}, which were performed by myself and a summer student under my supervision.
This work is available as a preprint on the \textit{arXiv}.
The theoretical analysis presented at the start of Chapter~\ref{chap:back} is my own work which I developed after \cite{goodhew2019backreaction} was finalised, and as such is new and unpublished.
\end{itemize} 

\chapter{Theory} \label{chap:theory}
In this chapter, we present a derivation of the equation that governs linear surface gravity wave propagation in a classical fluid.
We show how in shallow water, this equation is analogous to the propagation of scalar perturbations on an effective spacetime geometry, and we show for a draining bathtub fluid flow that this geometry is that of a rotating black hole.
At the end of the chapter, we present some useful tools we will use throughout this thesis to study wave propagation.

\section{The fluid equations}

The equations governing an inviscid, incompressible fluid are,
\begin{equation} \label{euler}
D_t\mathbf{V} + \frac{1}{\rho}\bm\nabla P = \mathbf{g} + \mathbf{f},
\end{equation}
and the equation of mass continuity, which for an incompressible fluid with density $\rho=\mathrm{const}$ reduces to the statement that the velocity field is divergence free, i.e.
\begin{equation} \label{divfree}
\bm\nabla\cdot\mathbf{V}=0.
\end{equation}
These are the incompressible Euler equations, which form a complete set of equations to be solved for the 3 component velocity field $\mathbf{V}$ and the pressure $P$.
We have also defined the material (or convective) derivative $D_t = \partial_t + \mathbf{V}\cdot\bm\nabla$, the acceleration due to gravity $\mathbf{g}$ and an additional external forcing term $\mathbf{f}$ which which we take to be conservative (curl free). 
Significant progress can be made by assuming an irrotational flow. To justify this assumption, the vorticity equation is obtained by taking the curl of Eq.~\eqref{euler},
\begin{equation} \label{vorticity}
D_t\mathbf{\Omega} = (\mathbf{\Omega}\cdot\bm\nabla)\mathbf{V},
\end{equation}
where $\mathbf{\Omega}=\bm\nabla\times\mathbf{V}$ is the vorticity. Hence if a flow is irrotational ($\mathbf{\Omega}=0$) initially, it will remain so for all time. Since $\bm\nabla\times\bm\nabla\Phi=0$ is identically satisfied for  any scalar function $\Phi$, we may express the velocity field in an irrotational flow as,
\begin{equation}
\mathbf{V}=\bm\nabla\Phi,
\end{equation}
thereby reducing the Euler equation in Eq.~\eqref{euler} to Bernoulli's equation
\begin{equation} \label{bernoulli}
\partial_t\Phi + \frac{1}{2}\textbf{V}^2 + \frac{P}{\rho} + gz - V_f = I(t),
\end{equation}
where the forcing potential is given by $\mathbf{f}=\bm\nabla V_f$ and $I$ is a function of time resulting from integration. The problem is thus reduced to solving for two scalars $P$ and $\Phi$, where the latter satisfies the Laplace equation,
\begin{equation} \label{laplace}
\nabla^2\Phi=0.
\end{equation} 
In our analysis, we will be interested in fluids with a free boundary located at $z=H(x,y)$ (called the free surface herein) and perturbations to this boundary (surface waves). We present two derivations of the equations governing the motion of surface waves; the first  is derived from the shallow water equations whereas the second allows for water of arbitrary depth.

\subsection{Shallow water equations}

From here on we adopt coordinates $(\mathbf{x},z)$ where $z$ is the vertical coordinate (in which gravity is directed) and $\mathbf{x}$ spans the horizontal plane (we will use either Cartesian, $\mathbf{x}=(x,y)$ or polar, $\mathbf{x}=(r,\theta)$, coordinates depending on which best suit our needs). $\mathbf{V}$ will be used to refer to the horizontal component and the $z$ component written explicitly.

The shallow water equations simplify the problem in Eqs.~\eqref{bernoulli} and~\eqref{laplace} by integrating out the $z$ dimension. They are obtained when gradients in the fluid quantities occur over a length scale $L$ which is much larger than the height of the fluid $H$, i.e. $L\gg H$. Under this assumption, we can work in the thin layer approximation where $\mathbf{V}$ is not a function of the vertical coordinate $z$\footnote{Of course, the true boundary condition when viscosity is included is $\mathbf{V}(z=0)=0$, and hence $\mathbf{V}$ must be $z$ dependent. However, since we are working in the inviscid theory, the governing equation (i.e. Eq.~\eqref{euler}) is inherently lower order than the full equations (i.e. Navier-Stokes) and this boundary is not specified. This really amounts to assuming that viscosity is confined to a thin layer at the boundary where all of the $z$ dependence in $\mathbf{V}$ is contained.}. Integrating Eq.~\eqref{divfree} in $z$ gives,
\begin{equation} \label{vz_shallow}
V_z = V_z(z=0)-z\bm\nabla\cdot\mathbf{V}.
\end{equation}
In the systems we concern ourselves with, the fluid will move over a flat bed located at $z=0$, with a no penetration boundary condition $V_z(z=0)=0$. Since $z$ is small in our system, the $z$-component of Eq.~\eqref{euler} gives,
\begin{equation} \label{hydrostat}
\mathcal{O}(z) + \frac{1}{\rho}\partial_zP = -g \quad \to \quad P = \rho g (H-z) + \mathcal{O}(z^2),
\end{equation}
where in solving for the pressure we have used the free surface boundary condition $P(z=H)=0$. We have also assumed that $V_f$ is a function only of $(x,y)$. Applying Bernoulli's equation in Eq.~\eqref{bernoulli} at $z=H$ and dropping terms of order $\mathcal{O}(H^2)$ - there is one in $\Phi$ and one in $V_z^2$ - we obtain,
\begin{equation} \label{Hbern}
\partial_t\Phi + \frac{1}{2}\mathbf{V}^2 + gH - V_f = I(t).
\end{equation}
The kinematic condition at the free surface states that a fluid element on the free surface must remain there, i.e. $V_z(z=H)=D_tH$. Hence, Eq.~\eqref{vz_shallow} gives
\begin{equation} \label{Hcont}
\partial_tH+\bm\nabla\cdot(H\mathbf{V})=0.
\end{equation}
Eqs.~\eqref{Hbern} and \eqref{Hcont} are the shallow water equations for an irrotational fluid which will be useful for much of our discussion.

To obtain the wave equation, we perturb the background state to first order,
\begin{equation}
\begin{split}
\Phi & \to\Phi+\epsilon\phi, \\
\mathbf{V} & \to\mathbf{V}+\epsilon\mathbf{v}, \\
H & \to H+\epsilon h,
\end{split}
\end{equation}
where $\epsilon\ll 1$ is a small expansion parameter and lower case quantities represent the perturbations. The dynamics of the perturbations are described by the $\mathcal{O}(\epsilon)$ terms of Eqs.~\eqref{Hbern} and \eqref{Hcont},
\begin{subequations}
\begin{align} 
D_t\phi + gh = 0, \label{shallow_ptb_a} \\
\partial_t h+\bm\nabla\cdot(h\mathbf{V}+ H\mathbf{v})=0, \label{shallow_ptb_b}
\end{align}
\end{subequations}
which combine in the $H\simeq\mathrm{const}$ regime to give the wave equation,
\begin{equation} \label{waveeqn}
D_t^2\phi-c^2\nabla^2\phi=0,
\end{equation}
where $c=\sqrt{gH}$ is the speed of wave propagation. This is the wave equation we will work with for most of this thesis.

It is shown how to deal with free surface gradients for a 2D system, i.e. $(x,z)$, in \cite{unruh2013irrotational,coutant2014undulations} and the 3D system, $(x,y,z)$, is treated for small free surface gradients in \cite{richartz2015rotating}.
The effect of varying $H$ the latter is that perturbations obey the same wave equation as in Eq.~\eqref{waveeqn} but with a spatially varying propagation speed.
This marginally alters the results at the expense of significantly complicating the analysis, hence we choose to work with a flat free surface for most of our theoretical analysis.

\subsection{Dispersion}

Dispersion arises when the $z$-dependence in the horizontal velocity perturbations becomes non-negligible, which results in different wavelengths propagating at different speeds.
The dispersive wave equation can be obtained by solving Laplace's equation and then integrating through the bulk of the fluid up to the free surface.
Bernoulli's equation is evaluated at the free surface as in the previous section.
We assume from the outset that the free surface is approximately flat for simplicity.

The perturbed velocity potential is first expanded in terms of Fourier modes in the horizontal plane,
\begin{equation}
\phi=\iint\phi_k e^{-i\mathbf{k}\cdot\mathbf{x}}dk_xdk_y,
\end{equation}
where the horizontal wavevector is given by $\mathbf{k}=(k_x,k_y)$ with modulus $k=|\mathbf{k}|$. 
Upon insertion into Laplace's equation, we obtain for each $k$ mode,
\begin{equation} \label{phi_zdep}
\partial_z^2\phi_k =  k^2\phi_k \quad \to \quad \phi_k = a_k\cosh(kz),
\end{equation}
where $a_k$ are the Fourier amplitudes and the $\sinh$ solution has been discard by implementing the no-penetration boundary condition at $z=0$.
The aim is to find an expression for the vertical component of the velocity perturbation at the free surface in terms of $\phi$.
Thus, we compute,
\begin{equation}
\partial_z\phi_k = ka_k\sinh(kz) = k\tanh(kz)\phi_k,
\end{equation}
which evaluated in position space at $z=H$ gives,
\begin{equation}
\partial_z\phi|_{z=H} = D_th = -i\nabla\tanh(-iH\nabla)\phi.
\end{equation}
We then combine this with the perturbed Bernoulli equation at the free surface in Eq.~\eqref{shallow_ptb_a} to obtain the dispersive wave equation,
\begin{equation} \label{waveeqndisp}
D_t^2\phi-ig\nabla\tanh(-iH\nabla)\phi=0.
\end{equation}
Note that this reduces to the usual wave equation of Eq.~\eqref{waveeqn} in the long wave limit $kH\ll 1$ (i.e. when the wavelength of perturbations is much greater than the water height).
Also, we can see from Eq.~\eqref{phi_zdep} that the horizontal velocity perturbations are,
\begin{equation}
\mathbf{v}=\iint \mathbf{k}a_k\cosh(kz)e^{-i\mathbf{k}\cdot\mathbf{x}}dk_xdk_y,
\end{equation}
which in the long wave limit is independent of $z$, since the $\cosh$ becomes approximately unity.
This is why dispersion is a result of the $z$-dependence in $\mathbf{v}$.

\subsection{The analogue metric}

Returning to Eq.~\eqref{waveeqn}, this form of the wave equation is completely equivalent to the Klein-Gordon (KG) equation,
\begin{equation} \label{KFG}
\frac{1}{\sqrt{-g}}\partial_{\mu}(\sqrt{-g}g^{\mu\nu}\partial_{\nu}\phi)=0,
\end{equation}
describing the propagation of a massless scalar field $\phi$ on a spacetime with the effective metric $g_{\mu\nu}$ given by,
\begin{equation} \label{EffMetric}
\begin{split}
ds^2 = & \ g_{\mu\nu}dx^{\mu}dx^{\nu} \\
= & \ -(gH-\mathbf{V}^2)dt^2 - 2\mathbf{V}\cdot\mathbf{dx}dt + \mathbf{dx}^2.
\end{split}
\end{equation}
Thus, by tuning the background state $(H,\mathbf{V})$, one is able to simulate an effective spacetime for the waves. As a result, effects experienced by waves in gravitational physics can be mimicked by simulating a suitable geometry. A specific example of such is given now.

\subsection{The Draining Bathtub Vortex} \label{sec:dbt}

The draining bathtub (DBT) vortex has received much attention in the analogue gravity community since it is the simplest fluid system mimicking a rotating black hole \cite{basak2003superresonance,basak2003reflection,cardoso2004qnm,berti2004qnm,dolan2012resonances,richartz2015rotating,churilov2018scattering}. Close to the drain, the flow is fully three dimensional and exact solutions to the fluid equations in Eqs.~\eqref{euler} and~\eqref{divfree} are difficult to obtain. However far from the centre, where $H\simeq\mathrm{const}$, the flow is effectively two-dimensional and irrotational. An ideal DBT is axisymmetric (i.e. $\theta$ independent), hence $\bm\nabla\cdot\mathbf{V}=0$ and $\bm\nabla\times\mathbf{V}=0$ give,
\begin{equation} \label{DBT}
V_r = -\frac{D}{r}, \qquad V_{\theta} = \frac{C}{r},
\end{equation}
where $C>0$ (circulation) and $D>0$ (drain), which for stationary solutions are constant in time. 
Inserting these solutions into Eq.~\eqref{Hbern} we have two choices: 
\begin{enumerate}
\item We can choose a forcing term to balance the flow field. At infinity we have $I=gH_\infty$, hence to maintain a flat free surface, we require $V_f=\mathbf{V}^2/2$.
\item We set the forcing term to zero.
Since the free surface must be approximately flat, Eq.~\eqref{Hbern} tells us the regime of validity of our approximation. Rearranging for $H$ we have,
\begin{equation} \label{height_profile}
H = H_\infty\left(1 - \frac{r_a^2}{r^2}\right), \qquad r_a^2 = \frac{C^2+D^2}{2gH_\infty}.
\end{equation}
(Note that whilst an exact analytic expression for the shape of the free surface is desirable, it is difficult to obtain in practice, even for waterfall type flows in the $(x,z)$-plane \cite{clarke1965waterfall,naghdi1981waterfall}).
\end{enumerate}
In Eq.~\eqref{height_profile}, $r_a$ is the radius at which the free surface intersects the floor of the tank. 
Although this situation arises in many experimental set-ups involving a drain, our solutions are only valid in the regime $r_a/r\ll1$. Specifically, if we treat $H$ as valid to $\mathcal{O}((r_a/r)^3)$, then the continuity equation in Eq.~\eqref{Hcont} is also satisfied to $\mathcal{O}((r_a/r)^3)$. By assumption, $V_\theta$ is exact.

The effective metric corresponding to the flow profiles in Eq.~\eqref{DBT} in Painlev\'e-Gullstrand form is,
\begin{equation} \label{dbtmetric}
ds^2 = -c^2dt^2 + \bigg(dr+\frac{D}{r}dt\bigg)^2 + \bigg(rd\theta-\frac{C}{r}dt\bigg)^2.
\end{equation} 
This metric can be recast into a form reminiscent of Schwarzschild metric by defining Boyer-Lindquist type coordinates $(t',\theta')$ through,
\begin{equation} \label{boyerlind}
dt' = dt + \frac{V_rdr}{c^2-V_r^2}, \qquad d\theta' = d\theta +\frac{V_rV_{\theta}dr}{r(c^2-V_r^2)},
\end{equation}
leading to,
\begin{equation} \label{schwarzform}
ds^2 = -\bigg(1-\frac{r_e^2}{r^2}\bigg)c^2dt'^2 + \left(1-\frac{r_h^2}{r^2}\right)^{-1}dr^2 - 2Cdt'd\theta' + r^2d\theta'^2.
\end{equation}
This effective spacetime exhibits an ergosphere at,
\begin{equation}
r_e=\sqrt{C^2+D^2}/c,
\end{equation} 
and a horizon at, 
\begin{equation}
r_h=D/c.
\end{equation} 
Using the language of the fluid system, the ergosphere is the region where gravity waves cannot propagate against the rotation of the vortex as seen by an observer at infinity.
The horizon is of course the boundary of the region where all waves are dragged into the vortex due to the flow into the drain.

The wave equation pertaining to the metric in Eq.~\eqref{dbtmetric} can be written in a simple form, first by noticing that the background under consideration depends only on the coordinate $r$. Hence we may write the anstaz,
\begin{equation} \label{waveansatz}
\phi(r,\theta,t) = \mathrm{Re}\left[\sum_m\frac{\phi_m(r)}{\sqrt{r}}e^{im\theta-i\omega t}\right],
\end{equation}
where $m$ is the azimuthal number of the mode (number of wavelengths that fit in a circle) and $\omega$ is it's frequency. This is permitted since there is no dependence on $t$ and $\theta$ appearing in the metric of Eq.~\eqref{dbtmetric}, so each $m,\omega$-mode will evolve independently according to the wave equation. The factor $1/\sqrt{r}$ is a matter of convention which is discussed in Appendix \ref{app:polar}. Also, $m\in\mathbb{Z}$ since we require periodic boundary conditions in $\theta$. Finally, we will usually work with perturbations of a single frequency; however perturbations of arbitrary frequency can be obtained by integrating Eq.~\eqref{waveansatz} over $\omega$.

The simple form of the wave equation is obtained by defining a new field $\psi_m$ which is the azimuthal component of $\phi$ in the coordinates defined by Eq.~\eqref{boyerlind} (i.e. in Eq.~\eqref{waveansatz} replace $\phi_m$ with $\psi_m$ and $\theta\to\theta'$, $t\to t'$). The relation between the two fields is,
\begin{equation} \label{fields}
\phi_m = \psi_m \exp\left(-i\int\frac{V_r\tilde{\omega}}{c^2-V_r^2}dr\right),
\end{equation}
where we have defined the intrinsic frequency in the rotating frame of the fluid,
\begin{equation} \label{omtild}
\tilde{\omega} = \omega - \frac{mV_\theta}{r}.
\end{equation}
The reason for this redefinition of field will become apparent when we consider WKB modes in Section~\ref{sec:wkb}. Finally, we define the radial tortoise coordinate,
\begin{equation} \label{tortoise}
dr_* = \frac{cdr}{c^2-V_r^2} \quad \Rightarrow r_* = \frac{r}{c} + \frac{D}{2c^2}\log\left|\frac{r-r_h}{r+r_h}\right|,
\end{equation}
which maps $r\in[r_h,\infty]$ onto $r_*=[-\infty,\infty]$ (as a matter of convention, $r_*$ has the units of time). Hence, the wave equation in Eq.~\eqref{waveeqn} for the velocity profiles in Eq.~\eqref{DBT} can be reduced to an equation for each azimuthal mode in the form of an energy equation,
\begin{equation} \label{modeeqn}
-\partial_{r_*}^2\psi_m + V(r)\psi_m = 0,
\end{equation}
where the effective potential seen by the $m$-mode is,
\begin{equation} \label{potential}
V(r) = -\left(\omega-\frac{mC}{r^2}\right)^2 + \left(c^2-\frac{D^2}{r^2}\right)\left(\frac{m^2-1/4}{r^2}+\frac{5D^2}{4c^2r^4}\right).
\end{equation}
This intuitive description in terms of a mode scattering with an effective potential is extremely useful (and will be exploited extensively in this thesis) as it allows us to draw on tools developed in the semiclassical limit of quantum mechanics \cite{berry1972semiclassical}.

Eq.~\eqref{modeeqn} can be solved asymptotically by noting the limiting forms of the potential, namely $V(r\to\infty)\to-\omega^2$ and $V(r\to r_h)\to-\tilde{\omega}_h^2$, where subscript $h$ denotes a quantity is evaulated at the horizon. In these two limits we have,
\begin{subequations} 
\begin{align}
\psi_m(r\to\infty) = \ & A_m^{\mathrm{in}}e^{-i\omega r/c} + A_m^{\mathrm{out}}e^{+i\omega r/c}, \label{asymp_sols_a} \\
\psi_m(r\to r_h) = \ & A_m^he^{-i\tilde{\omega}_hr_*}, \label{asymp_sols_b}
\end{align}
\end{subequations}
where we have imposed the boundary condition that the mode is in-going on the horizon. 
The potential in Eq.~\eqref{potential} is unfortunately too complicated a function of $r$ to yield analytic solutions for $\psi_m(r)$.
Furthermore, one needs to be careful when constructing a solution in $(r_*,\theta',t')$ coordinates and converting back to $(r,\theta,t)$ since the transformation is singular at $r=r_h$.
This leads to problems if one wants to compute derivatives with respect to $r$ which we discuss in Chapter \ref{chap:back}.
We derive approximate analytic formulae for $\phi_m$ in Section \ref{sec:wkb} of this chapter to highlight some important features of the modes.
Numerical simulations are performed later on in Chapters \ref{chap:super}, \ref{chap:qnm} and \ref{chap:vort} when exact solutions are required.

\subsection{Conserved currents} \label{sec:currents}

Conservation laws are useful since they provide a means of obtaining information about a system in addition to the equations of motion.
They are related via Noether's theorem to the symmetries of a system described by a Lagrangian $\mathcal{L}$.
For example, a system whose dynamics do not depend on time conserves energy, and information about the system may be obtained using the equation for energy conservation without having to know the full solution to the equations of motion.

The Lagrangian describing perturbations to an irrotational, shallow fluid flow is,
\begin{equation} \label{lagrangian}
\mathcal{L} = -\tilde{g}^{\mu\nu}\partial_\mu\phi\partial_\nu\phi^* = (D_t\phi)(D_t\phi^*)-c^2(\bm\nabla\phi)\cdot(\bm\nabla\phi^*),
\end{equation}
where $*$ denotes the complex conjugate (the Lagrangian written in terms of only real valued fields conventionally carries an extra factor of $1/2$ at the front). 
The first equality expresses the Lagrangian in a form which is manifestly covariant using the inverse metric $\tilde{g}^{\mu\nu}=\sqrt{-g}g^{\mu\nu}$ corresponding to Eq.~\eqref{EffMetric}, which is explicitly,
\begin{equation} \label{InvMetric}
\begin{split}
\tilde{g}^{\mu\nu} = \ & \begin{pmatrix}
-1 & -V^i \\
-V^j & c^2\delta^{ij}-V^iV^j
\end{pmatrix} \\
= \ & -\delta^\mu_t\delta^\nu_t +(c^2\delta^{ij}-V^iV^j)\delta^\mu_i\delta^\nu_j - 2V^i\delta^\mu_{(i}\delta^\nu_{t)}.
\end{split}
\end{equation}
The second form of $\mathcal{L}$ expresses the Lagrangian with the space and time derivatives written separately. 
Both of these forms will be useful for our purposes.

The equation of motion for the perturbations is of course the wave equation in Eq.~\eqref{waveeqn}, which is obtained from $\mathcal{L}$ using the field theoretic version of the Euler-Lagrange equations \cite{srednicki2007quantum},
\begin{equation} \label{ELeqns}
\partial_\mu\left(\frac{\partial\mathcal{L}}{\partial(\partial_\mu\phi)}\right) - \frac{\partial\mathcal{L}}{\partial\phi} = 0.
\end{equation}
The application of this to the covariant form of $\mathcal{L}$ directly yields the KG equation in Eq.~\eqref{KFG}, whereas the usual form in Eq.~\eqref{waveeqn} is directly obtained from the form of $\mathcal{L}$ which is split into space and time parts.

A conserved current $j[\phi]$ is a quantity with components $j^{\mu}[\phi]$ satisfying,
\begin{equation} \label{Current}
\partial_{\mu}j^{\mu}[\phi] = 0 \quad \Rightarrow \quad \partial_t\rho[\phi]+\bm\nabla\cdot\mathbf{j}[\phi] = 0.
\end{equation}
The four component object $j=j^\mu\partial_\mu$ is called the 4-current which can be split into a temporal part (the charge $\rho$) and a spatial part (the three component current $\mathbf{j}$). 
Square brackets are use to indicate that a quantity is a functional of $\phi$.

Now, we can derive conservation laws from the Lagrangian as follows. 
Consider an infinitesimal transformation of the field which induces a shift in the Lagrangian,
\begin{equation}
\phi^a\to\phi^a+\delta\phi^a \quad \Rightarrow \quad \mathcal{L}\to\mathcal{L}+\delta\mathcal{L}.
\end{equation}
Noether's theorem \cite{schwartz2014quantum} states that $j$ is a conserved current if the Lagrangian changes by a total derivative $\delta\mathcal{L}=\partial_\mu F^\mu$. 
The components of the current are given by,
\begin{equation}
j^\mu = \frac{\partial\mathcal{L}}{\partial(\partial_\mu\phi_a)}\delta\phi_a - F^\mu,
\end{equation}
where $a=1,2$ with $\phi_1=\phi$ and $\phi_2=\phi^*$. 
We now consider several different conserved quantities which will be useful in our analysis of the bathtub vortex.

\begin{enumerate}

\item \textit{Norm conservation}. We consider first performing a phase rotation on $\phi$. Since the equation of motion is linear in $\phi$, it has an internal symmetry $\phi\to\exp(i\alpha)\phi$ where $\alpha$ is a phase rotation. For infinitesimal $\alpha$, $\phi$ changes by $\delta\phi=i\alpha\phi$ and $\delta\mathcal{L}=0$. The conserved quantities associated with this transformation are the norm $\rho_n[\phi]$ and the norm current $\mathbf{j}_n[\phi]$ given by,
\begin{equation} \label{n_current}
\begin{split}
j^\mu_n[\phi] = & \ i\tilde{g}^{\mu\nu}(\phi^*\partial_\nu\phi-\phi\partial_\nu\phi^*) \\
\qquad \quad \mathrm{or} \\
\rho_n[\phi] = & \ i(\phi^*D_t\phi-\phi D_t\phi^*) \\
\mathbf{j}_n[\phi] = & \ \ i\mathbf{V}(\phi^*D_t\phi-\phi D_t\phi^*) - ic^2(\phi^*\bm\nabla\phi-\phi\bm\nabla\phi^*).
\end{split}
\end{equation}

\item \textit{Angular Momentum}. We will primarily be interested in background states $(H,\mathbf{V})$ which do not depend on $\theta$. Hence, an infinitesimal spatial rotation $\theta\to\theta-\delta\theta$ induces the changes $\phi^a\to\phi^a+\delta\theta\partial_\theta\phi^a$ and $\mathcal{L}\to\mathcal{L}+\delta\theta\partial_\theta\mathcal{L}$, giving rise to the current,
\begin{equation} \label{l_current}
\begin{split}
j^\mu_l[\phi] = & \ \tilde{g}^{\mu\nu}(\partial_\theta\phi^*\partial_\nu\phi+\partial_\theta\phi\partial_\nu\phi^*) \\
\qquad \quad \mathrm{or} \\
\rho_l[\phi] = & \ D_t\phi^*\partial_\theta\phi + D_t\phi\partial_\theta\phi^* \\
j^i_l[\phi] = & \  (V^iD_t\phi^*-c^2\partial^i\phi^*)\partial_\theta\phi + (V^iD_t\phi-c^2\partial^i\phi)\partial_\theta\phi^* - \delta^i_\theta\mathcal{L},
\end{split}
\end{equation}
where have used the index $i=r,\theta$ here since the spatial current has different components ($\delta^\theta_\theta=1$ and $\delta^r_\theta=0$). If we define $\phi\in\mathbb{R}$, then $\rho_l$ is proportional to $hv_\theta$ which in turn is proportional to the angular momentum density (this can be seen by integrating $\mathbf{r}\times\rho\mathbf{V}$ over all space and extracting the wave contribution at quadratic order in $\phi$).

\item \textit{Energy conservation}. Time translation $t\to t-\delta t$ induces a change in the field $\phi^a\to\phi^a+\delta t\partial_t\phi^a$ and the Lagrangian $\mathcal{L}\to\mathcal{L}+\delta t\partial_t\mathcal{L}$. This gives rise to conservation of the energy current which has components,
\begin{equation} \label{e_current}
\begin{split}
j^\mu_e[\phi] = & \ -\tilde{g}^{\mu\nu}(\partial_t\phi^*\partial_\nu\phi+\partial_t\phi\partial_\nu\phi^*) \\
\qquad \quad \mathrm{or} \\
\rho_e[\phi] = & \ \partial_t\phi^*D_t\phi + \partial_t\phi D_t\phi^* - \mathcal{L} \\
\mathbf{j}_e[\phi] = & \ \mathbf{V}(\partial_t\phi^*D_t\phi + \partial_t\phi D_t\phi^*) - c^2(\partial_t\phi^*\bm\nabla\phi + \partial_t\phi\bm\nabla\phi^*).
\end{split}
\end{equation} 
Note that if we define $\partial_t\phi=\dot{\phi}$ and $p=\partial\mathcal{L}/\partial\dot{\phi} = D_t\phi^{*}$ (and similarly for the complex conjugate), $\rho_e=\mathcal{H}$ is the Hamiltonian. 

To make contact with the natural notion of energy in fluid dynamics, we define $\phi\in\mathbb{R}$, as well as $E=\rho_e/2g$ and $\mathbf{I}=\mathbf{j}_e/2g$. 
Using $D_t\phi=-gh$ and $\bm\nabla\phi=\mathbf{v}$, Eq.~\eqref{Current} assumes the form of the energy conservation equation for shallow water waves,
\begin{equation}
\partial_tE + \bm\nabla\cdot\mathbf{I}=0,
\end{equation}
where,
\begin{subequations}
\begin{align}
E = \ & \frac{1}{2}gh^2 + \frac{1}{2}H\mathbf{v}^2 + h\mathbf{V}\cdot\mathbf{v}, \label{energycurrent_a} \\
\mathbf{I} = \ & (h\mathbf{V}+H\mathbf{v})(gh+\mathbf{V}\cdot\mathbf{v}). \label{energycurrent_b}
\end{align}
\end{subequations}
In fluid dynamics, these equations are obtained by contracting the shallow water equations in Eqs.~\eqref{shallow_ptb_a} and~\eqref{shallow_ptb_b} with $(gh,H\mathbf{v}^T)$ where superscript $T$ indicates the transpose.

\end{enumerate}

\noindent Finally, there exists a particularly elegant relation between these three currents in the case of stationary, axisymmetric systems.
Applying $\partial_\theta=im$ and $\partial_t=-i\omega$ to the covariant forms of the $l$ and $e$ currents, we find \mbox{$j_l^\mu = mj_n^\mu$} and \mbox{$j_e^\mu = \omega j_n^\mu$} respectively.
Hence, the ratio of wave energy to angular momentum is fixed by,
\begin{equation}
\frac{j_e}{j_l} = \frac{\omega}{m}.
\end{equation}
We shall see later on that this ratio plays an important role in determining the trajectories of high frequency waves.
Since $\omega$ and $m$ are conserved in a stationary, axisymmetric system, this means the three currents are equivalent.
Applying the divergence theorem to Eq.~\eqref{Current} yields,
\begin{equation} \label{RadCurrCons}
\int^{2\pi}_0 [\mathbf{r}\cdot\mathbf{j}]^{r_2}_{r_1}d\theta = 0,
\end{equation}
which is the conservation of the norm/angular momentum/energy current in the radial direction, and we have used two arbitrary radial locations $r_{1,2}$ as the boundaries of the system.
The conserved radial current in Eq.~\eqref{RadCurrCons} will be relevant in our discussion of superradiant scattering in Chapter~\ref{chap:super}, and we will use the energy current in Eq.~\eqref{energycurrent_b} for our discussion of mass fluxes in Chapter~\ref{chap:back}.

\section{WKB modes} \label{sec:wkb}

The WKB method is a useful approximation scheme that allows us to determine the leading order analytic solutions to the wave equation when exact solutions are not available. 
It was developed originally in the context of quantum mechanics and is now understood to describe solutions to the Schr\"odinger equation in the semi-classical limit, $\hbar\to0$ (see e.g. \cite{berry1972semiclassical}).
The WKB method is routinely applied in a variety of other disciplines, in particular fluid dynamics and black hole physics.

The underlying assumption is that the background - in our case $(H,\mathbf{V})$ - varies over a lengthscale which is large compared to the wavelength of the perturbations. There are many methods of applying the WKB approximation which all turn out to be equivalent. We describe two methods which will prove useful for our purposes:
\begin{enumerate}
\item The first is based on the wave equation in Eq.~\eqref{waveeqn} and is useful when we want to work in the lab frame. This method gives us the dispersion relation on an inhomogeneous background and will be used in Chapter~\ref{chap:super} to obtain scattering coefficients using information about the wave at a finite distance form the origin.
\item The second is based on the mode equation in Eq.~\eqref{modeeqn} and results in a more sophisticated approximation, which reduces to the first solution in the large $m$ limit as we show below. We use this solution in Chapters~\ref{chap:qnm} and \ref{chap:vort} to look for resonant frequencies.
\end{enumerate}

\subsection{Solution to the wave equation} \label{sec:Eik}

The basic WKB assumption is that if a rapidly varying mode propagates over a slowly varying background, then the solution can be approximated as,
\begin{equation} \label{WKB}
\phi^\mathrm{wkb}= A(\xi)e^{iS(\xi)/\epsilon},
\end{equation}
where $A$ and $S$ are the WKB amplitude and phase respectively, $\xi=(t,\mathbf{x})$ are coordinates and $\epsilon$ is a small expansion parameter (different from perturbative parameter in the previous section). The standard procedure is to plug the WKB ansatz in Eq.~\eqref{WKB} into the wave equation in Eq.~\eqref{waveeqn} and rescale derivatives $\partial_{\xi}\to\epsilon\partial_{\xi}$ to express the fact that the background is slowly varying. We then expand $A$ and $S$ in powers of $\epsilon$,
\begin{align*}
A = \sum_n\epsilon^nA_n, \qquad S = \sum_n\epsilon^nS_n,
\end{align*}
where summation is performed over $n=\mathrm{even}$ so as not to double count contributions at each order in $\epsilon$. Defining,
\begin{align*} 
-\omega = \partial_tS_0, \qquad \mathbf{k} = \bm\nabla S_0,
\end{align*}
at $\mathcal{O}(1)$ we have,
\begin{equation} \label{dispersion}
(\omega-\mathbf{V}\cdot\mathbf{k)}^2-c^2\mathbf{k}^2 = 0,
\end{equation}
and at $\mathcal{O}(\epsilon)$,
\begin{equation} \label{WKBamp}
\partial_t(\Omega A_0^2) + \bm\nabla\cdot(\mathbf{v}_g\Omega A_0^2) = 0.
\end{equation}
Eq.~\eqref{dispersion} is the dispersion relation in the shallow water regime. 
Note that this means it is only possible to define a dispersion relation as the first order approximation for small wavelengths.
Eq.~\eqref{WKBamp} describes the adiabatic evolution of the amplitude owing to the non-zero flow field, and we have defined the intrinsic fluid frequency $\Omega$,
\begin{equation}
\Omega = \omega - \mathbf{V}\cdot\mathbf{k},
\end{equation}
and the group velocity,
\begin{equation}
\mathbf{v}_g = \bm\nabla_\mathbf{k}\omega.
\end{equation}
The equation at zeroth order (i.e. Eq.~\eqref{dispersion}) is often called the eikonal (or ray) approximation, and describes the trajectories followed by null particles (i.e. waves with vanishingly small wavelength\footnote{We are neglecting capillarity which introduces corrections to the dispersion relation for small wavelengths.}).
The term WKB is usually reserved for the zeroth order result supplemented by the first order correction, i.e. Eq.~\eqref{WKBamp}. 
Note that the dispersion relation in Eq.~\eqref{dispersion} only becomes exact when $\mathbf{V}=\mathrm{const}$. 
Furthermore, the polar decomposition of the wave vector $\mathbf{k}=(k_r,m/r)$ is also not exact since the $\theta$-component diverges at the origin.
This is discussed further in Appendix~\ref{app:polar}

For axisymmetric, stationary backgrounds, both $m$ and $\omega$ are conserved by the wave equation and the dispersion relation can be reformulated to express the radial component of the wavevector $k_r$ as a function of $r$. Since Eq.~\eqref{dispersion} is quadratic, it has two solutions - indicated by $+$ (out-going) and $-$ (in-going) respectively. These are,
\begin{equation} \label{kr}
k_r^{\pm} = \frac{-V_r\tilde{\omega}}{c^2-V_r^2} \pm \frac{c}{c^2-V_r^2}\sqrt{-V_{\mathrm{geo}}},
\end{equation}
where,
\begin{equation} \label{GeoPotential}
V_\mathrm{geo} = -\tilde{\omega}^2 + \left(c^2-V_r^2\right)\frac{m^2}{r^2},
\end{equation}
is the geodesic potential obtained from the metric in Eq.~\eqref{schwarzform} (this is derived in Appendix~\ref{app:geos}), which is the reason why the eikonal approximation gives the path of the wave. 
In the limit $m\to\infty$, the potential appearing in the mode equation in Eq.~\eqref{modeeqn} reduces to $V_\mathrm{geo}$, hence the eikonal limit can be thought of as the large $m$ limit. Note also that the first term in Eq.~\eqref{kr} is precisely the factor we picked up in Eq.~\eqref{fields} by performing the coordinate transform of Eq.~\eqref{boyerlind}. Hence, the purpose of this coordinate change is to cancel the first term of Eq.~\eqref{kr} and make the in- and out-going wavevectors equal and opposite.

It is worth noting that the $m=0$ mode reduces to,
\begin{equation}
k_{r,m=0}^{\pm} = \frac{-V_r\omega}{c^2-V_r^2} \pm \frac{c\omega}{c^2-V_r^2} = \frac{\omega}{V_r\pm c},
\end{equation}
from which we see the usual red/blueshift behaviour; for a draining flow ($V_r<0$) $k_r^+$ increases toward decreasing $r$ (blueshift) whereas $k_r^-$ decreases (redshift). 

The corresponding values of $\Omega$ are,
\begin{equation}
\Omega^{\pm} = \frac{c^2}{c^2-V_r^2}\left(\tilde{\omega}\mp\frac{V_r}{c}\sqrt{-V_\mathrm{geo}}\right),
\end{equation}
and the evolution of the amplitude $A_0$ can then be obtained from Eq.~\eqref{WKBamp}. In an axisymmetric, stationary geometry, this simplifies to $\partial_r(rv_{g,r}\Omega A_0^2)=0$, where we have defined the radial component of the group velocity,
\begin{equation}
\bm{\mathrm{e}}_r\cdot\mathbf{v}^{\pm}_g = v_{g,r} = V_r + \frac{c^2k^{\pm}_r}{\Omega^{\pm}}.
\end{equation}
We can then easily evaluate the product,
\begin{equation}
\begin{split}
v^{\pm}_{g,r}\Omega^{\pm} = \ & V_r\Omega^{\pm} + c^2k^{\pm}_r \\
= \ & \pm c\sqrt{-V_\mathrm{geo}}.
\end{split}
\end{equation}
Inserting this back into Eq.~\eqref{WKB}, the leading order solution can written as a sum of two WKB modes $\phi_m^\pm$,
\begin{equation} \label{WKBsol1}
\phi^\mathrm{wkb} = \sum_m \begin{pmatrix}
\alpha_m^+, & \alpha_m^-
\end{pmatrix}\cdot\begin{pmatrix}
\phi_m^+ \\ \phi_m^-
\end{pmatrix}
\frac{e^{im\theta-i\omega t}}{\sqrt{r}}, \qquad
\phi_m^\pm = \frac{1}{\sqrt{|(-V_\mathrm{geo})^{1/2}|}}e^{i\int k_r^{\pm}dr},
\end{equation}
where the amplitudes $\alpha_m^{\pm}$ of the separate modes are conserved adiabatically as they propagate. 

\subsection{Solution to the mode equation}

Evaluating the form of the WKB solution from the mode equation is somewhat easier, as most of the hard work has been done by the coordinate transformations. This time, we write the WKB ansatz in the form,
\begin{equation} \label{modeWKB}
\psi_m^\mathrm{wkb} = \alpha_m e^{i\int p(r_*) dr_*},
\end{equation}
where the amplitude $\alpha_m$ is a constant. In the same manner as the previous section, we plug $\psi_m^\mathrm{wkb}$ into the mode equation in Eq.~\eqref{modeeqn} and set $\partial_{r_*}=\epsilon\partial_{r_*'}$ to express that the background is slowly varying. Expanding $p = p_0 + \epsilon p_1 +...$ and collecting orders, we obtain,
\begin{equation}
\begin{split}
\mathcal{O}(1): & \ \ \ p_0^2 + V = 0 \qquad \ \ \to \ p_0 = \pm\sqrt{-V}, \\
\mathcal{O}(\epsilon): & \ \ i\partial_{r_*'}p_0 - 2p_0p_1 = 0 \ \to \ \epsilon p_1 = i\partial_{r_*}\log\sqrt{p_0}.
\end{split}
\end{equation}
Evaluating Eq.~\eqref{modeWKB} to $\mathcal{O}(\epsilon)$, we can write the leading order solution using the WKB modes as a basis,
\begin{equation} \label{WKBsol2}
\phi^\mathrm{wkb} = \sum_m \begin{pmatrix}
\alpha_m^+, & \alpha_m^-
\end{pmatrix}\cdot\begin{pmatrix}
\psi_m^+ \\ \psi_m^-
\end{pmatrix}
\frac{e^{im\theta'-i\omega t'}}{\sqrt{r}}, \qquad
\psi_m^\pm = \frac{1}{\sqrt{|p_0|}}e^{\pm i\int p_0dr_*}.
\end{equation}
To see that this is equivalent in the large $m$ limit to the WKB solution obtained in Eq.~\eqref{WKBsol1}, first convert $\psi_m$ to $\phi_m$ using the coordinate transformations in Eqs.~\eqref{boyerlind} and~\eqref{fields}, then insert the definition of $dr_*$ using Eq.~\eqref{tortoise}. 
We recover the same expression with $V$ replacing the geodesic potential $V_\mathrm{geo}$.
The difference arises because in the previous case we applied the WKB approximation in both spatial dimensions whereas here we have only applied it in $r_*$.

\subsection{Validity}

The WKB approximation is valid in regions where the scale of the perturbations $\lambda$ is much smaller the gradient scale $L$ set by the background, defined by,
\begin{equation}
L^{-1} = \mathrm{max}\left(|\bm\nabla H|/|H|,|\bm\nabla\mathbf{V}|/|\mathbf{V}|\right).
\end{equation}
Combined with the shallow water approximation, this means we are in the regime,
\begin{equation}
H\ll\lambda\ll L.
\end{equation}
Additionally, the solutions in Eqs.~\eqref{WKBsol1} and \eqref{WKBsol2} do not hold close to turning points of the potential where $V\to0$. This is due to the factor $(-V)^{-1/4}$ on the denominator, which causes the solution to diverge. This issue is addressed in the next chapter.

\section{Overview}

In this section, starting from the basic fluid equations we derived an approximate shallow water wave equation on a moving background flow. 
We showed that this is the same equation describing a massless scalar field on an effective spacetime, and specifically that a rotating, draining vortex flow can mimic a rotating black hole geometry.
Note that the effective geometry obtained is not that of the Kerr solution, which is the unique rotating black hole solution in GR.
The main differences are that the effective geometry of our fluid flow has only two spatial dimensions (in contrast to astrophysical black holes which of course have three) and that the fall-off of the gravitational potential in the weak gravity regime of our geometry goes with $1/r^2$, rather than $1/r$ as for astrophysical black holes.
Nonetheless, the existence of an ergosphere and a horizon is sufficient to give rise to the effects we shall be interested in, and therefore the specific details of the effective metric are unimportant.

We then derived some useful results that will serve us well in tackling the forth-coming problems:
\begin{itemize}
\item Exact asymptotic solutions in Eqs.~\eqref{asymp_sols_a} and~\eqref{asymp_sols_b}.
\item Conserved currents in Eqs.~\eqref{n_current}, \eqref{l_current} and \eqref{e_current}.
\item The dispersion relation in Eq.~\eqref{dispersion} and adiabatic amplitude evolution in Eq.~\eqref{WKBamp}.
\item WKB solution in Eq.~\eqref{WKBsol2} using the effective potential of Eq.~\eqref{potential}.
\end{itemize}
Armed with this technology, we are ready to proceed to the next chapter.

\chapter{Superradiance} \label{chap:super}
The term \textit{superradiance} encompasses a broad class of amplification effects in a wide range of systems, from electromagnetic to gravitational \cite{bekenstein1998many}.
It is usually associated with either spontaneous emission in a quantum system or amplification of waves in a classical system.
Our analysis will focus on the latter.
Depending on the context, it may also be referred to as superradiant scattering/amplification, super-resonance or over-reflection.
An excellent review of the topic and it's long history can be found in \cite{brito2015superradiance}.
We give a condensed version here.

Due to it's interdisciplinary nature, different authors cite different works as the original depending on the context.
According to \cite{bekenstein1998many}, the possibility of spontaneous emission due to the superluminal motion of an object through a medium was first realised by Ginzburg and Frank \cite{ginzburg1947doppler}, and this effect is related to Cherenkov radiation \cite{ginzburg1993v}.
The actual name superradiance was coined by Dicke \cite{dicke1954coherence} who studied amplification resulting from coherence in a radiating gas.
Superradiance in rotating systems was discovered in the early 1970's by Zel'dovich, who was looking at scattered electromagnetic radiation from a conducting cylinder \cite{zeldovich1971generation,zeldovich1972amplification}.
Around the same time, Penrose conceived a mechanism in which a particle could extract rotational energy from the ergosphere of a rotating black hole \cite{penrose1971extraction}.
The Penrose process, as it is now known, is the particle analogue of superradiance.
Shortly after, amplification of radiation around black holes was considered \cite{misner1972stability,starobinsky1974waves,starobinsky1974electro}.
As a side note, it was whilst studying superradiance following these developments that Hawking discovered black hole evaporation \cite{hawking2009brief,hawking1975particle}.
Superradiance has a close relation with over-reflection, an effect that occurs in shear flows in fluid mechanics \cite{mckenzie1972reflection,acheson1976overreflexion,kelley2007inertial,fridman2008overreflection}.
More recently superradiance has also been shown to occur around stars \cite{richartz2013eventhorizons}, and a numerical study in \cite{east2014black} showed that the amount of amplification is reduced by backreaction effects.
Finally, following the inception of analogue gravity, the possibility of detecting superradiance in the laboratory has been considered by several authors, \cite{richartz2009generalized,richartz2013dispersive,cardoso2016detecting}, with particular attention paid to the DBT flow outlined in the previous chapter \cite{basak2003superresonance,basak2003reflection,richartz2015rotating}.

Although rotational superradiance has been understood on theoretical grounds for nearly half a century, it had so far evaded detection in the laboratory.
With recent developments in the understanding of analogue models of gravity, this has now become a possibility.
This chapter is structured around our work in \cite{torres2017rotational}, which details the first measurement of superradiance in an analogue rotating black hole.
The experiment was performed in a tub of draining water, which will be described in Section~\ref{sec:detectSR}.

Our theoretical analysis of superradiance will be performed in the shallow water regime following \cite{richartz2015rotating,churilov2018scattering}, since the non-locality of the operator appearing in the dispersive wave equation (see Eq.~\eqref{waveeqndisp}) significantly complicates our lives.
Whilst this approach fails at providing quantitative predictions for scenarios where dispersion comes into play, it can nonetheless provide an qualitative understanding of effects whose existence does not depend on dispersion.
Indeed, superradiance has been shown to occur when dispersion is weak \cite{richartz2013dispersive}.
As we shall see, it is an engineering challenging to devise a system which is both shallow and quickly draining, and as such our experiment was actually performed in the deep water regime.

\section{Wave scattering} \label{wavscat1}

Wave scattering phenomena are ubiquitous to almost all sciences, from biology to physics. 
When an incident wave scatters off of an obstacle, it is partially reflected and partially transmitted. 
Since the scatterer absorbs part of the incident energy via the transmitted part, the reflected wave carries less energy than the incident one. 
However when a wave is superradiantly amplified, the reflected part in fact contains more energy than the incident wave.
Due to energy conservation, the transmitted part must have a negative energy, i.e. an energy that lowers the total energy of the system, and this is how a superradiant mode is able to extract energy from a system.
In this section, we will demonstrate this mechanism mathematically, focussing in particular on the DBT flow outlined in the previous chapter.

\subsection{Scattering coefficients}

A generic wave scattering problem is formulated as follows.
An incident mode is sent into the system with amplitude $A_m^{\mathrm{in}}$.
Part of the mode in transmitted and part is reflected. 
When the reflected mode reaches the point where the in-going mode was sent, its amplitude $A_m^{\mathrm{out}}$ is measured.
In our system the transmitted mode is the one that crosses the horizon with amplitude $A_m^h$. 

The reflection ($\mathcal{R}$) and transmission ($\mathcal{T}$) coefficients are defined by,
\begin{equation} \label{RTcoeffs}
\mathcal{R}(\omega,m) = \frac{|A_m^\mathrm{out}|}{|A_m^\mathrm{in}|}, \qquad \mathcal{T}(\omega,m) = \frac{|A_m^h|}{|A_m^\mathrm{in}|}.
\end{equation}
Without concerning ourselves with the details of the solutions across the entire $r$ range, we can obtain a relation satisfied by these coefficients using the asymptotic solutions in Eqs.~\eqref{asymp_sols_a} and~\eqref{asymp_sols_b}. 
The relation between amplitudes in the two regions is most readily obtained using the Wronskian, which is conserved for solutions of a second order ODE of the form in Eq.~\eqref{modeeqn}. The Wronskian associated with Eq.~\eqref{modeeqn} is defined,
\begin{equation}
W[\psi_m^*,\psi_m] = \psi_m^*\partial_{r_*}\psi_m-\psi_m\partial_{r_*}\psi_m^*,
\end{equation}
where the field $\psi_m$ and it's complex conjugate $\psi_m^*$ are linearly independent solutions. 
This is equivalent to the radial norm current in Eq.~\eqref{n_current} for an axisymmetric, stationary system described by coordinates defined in Eqs.~\eqref{boyerlind} and \eqref{tortoise}.
As discussed at the end of Section~\ref{sec:currents}, this means the Wronskian is also proportional to both the  angular momentum and energy currents in Eqs.~\eqref{l_current} and~\eqref{e_current} respectively.
The Wronksian is evaluated in the two asymptotic regions using Eqs.~\eqref{asymp_sols_a} and~\eqref{asymp_sols_b}, and setting the two equal gives,
\begin{equation} \label{DBT_RadCurr}
-\tilde{\omega}_h|A_m^h|^2 = \omega(|A_m^\mathrm{out}|^2-|A_m^\mathrm{in}|^2),
\end{equation}
where the left/right sides come from the horizon/infinity respectively. 
Rearranging this result gives,
\begin{equation} \label{scattering}
1-\mathcal{R}^2 = \frac{\tilde{\omega}_h}{\omega}\mathcal{T}^2.
\end{equation}
This equation indicates that a mode which satisfies,
\begin{equation} \label{SRcondition1}
\omega < m\Omega_h, \qquad \Omega_h = \frac{V_\theta(r=r_h)}{r_h},
\end{equation}
where $\Omega_h$ is the angular velocity of the background on the horizon, will have a reflected part which is larger than the incident part. This phenomena is known as superradiant scattering (or simply superradiance). A few important points about superradiant modes are noted:
\begin{enumerate} [noitemsep] 
\item The quantity in Eq.~\eqref{DBT_RadCurr} is simply the conservation of the radial current from Eq.~\eqref{RadCurrCons}.
Hence, a superadiant mode has a \textit{positive energy current} (i.e. directed toward $r\to\infty$).
It therefore extracts energy from the system.
\item The norm of a superradiant mode at $r=r_h$ is proportional to $\tilde{\omega}_h$.
Therefore, the extraction of energy from the system is facilitated by a negative energy being carried into it.
Negative energy means something that lowers the overall energy of the system.
\end{enumerate}

\subsection{Prediction at leading order} \label{scatterO1}

In the WKB picture, we can compute the leading order contributions to the scattering coefficients $\mathcal{R}$ and $\mathcal{T}$ using the potential in Eq.~\eqref{potential}.
The WKB solution in Eq.~\eqref{WKBsol2} is valid everywhere except in the vicinity of the turning points $r=r_\mathrm{tp}$ where $V(r_\mathrm{tp})=0$, ultimately due to the factor $(-V)^{1/4}$ in the WKB amplitude which blows up there.
Hence to construct a solution across the full $r$ range, we need a way of relating WKB modes either side of $r_\mathrm{tp}$.


\begin{figure*}[t!]
    \centering
    \begin{subfigure}[t]{0.5\textwidth}
        \centering
        \includegraphics[height=3in]{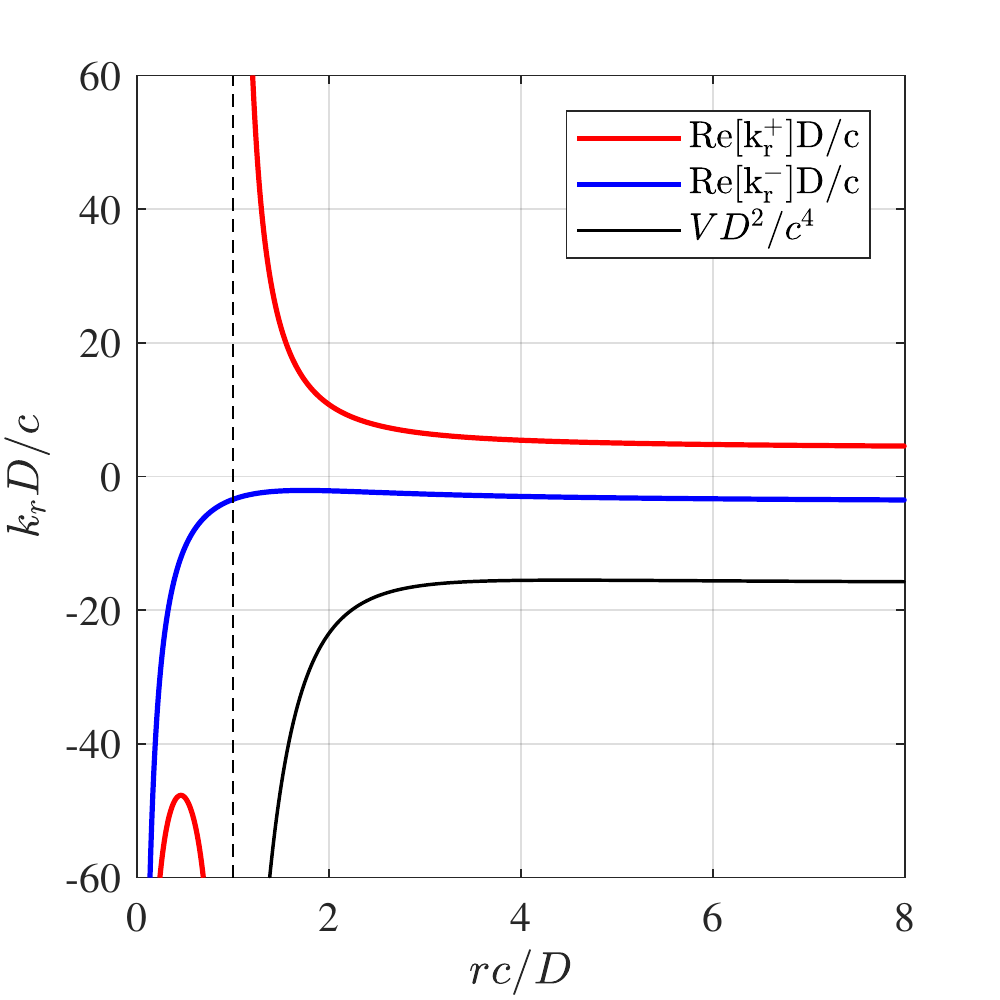}
    \end{subfigure}%
    ~ 
    \begin{subfigure}[t]{0.5\textwidth}
        \centering
        \includegraphics[height=3in]{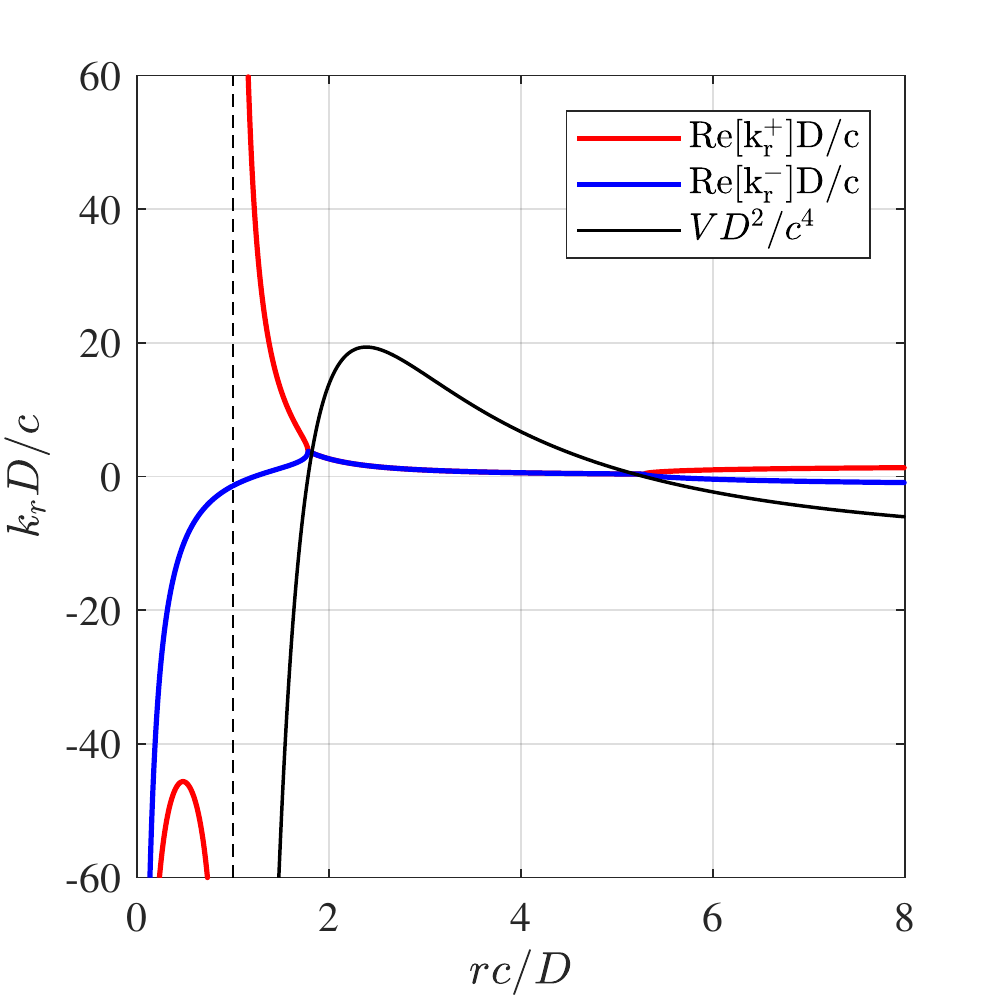}
    \end{subfigure}
    \caption{Real part of the radial wavenumber as a function of radius for $m<0$. The different plots are characterised by the parameters [$\omega D/c^2$, $C/D$, $m$], which for the left hand side are [4, 1, -10] and for the right hand side are [1.5, 1, -10]. All variables are scaled appropriately by $c$ and $D$ such that they are dimensionless. When an in-going mode is sent from far away (i.e. starting on the $k_r^-$ branch at large $r$) there are two possibilities. In the first case (left), the in-going mode is completely absorbed (within the WKB approximation) and does not reflect off the potential barrier. In the second case (right), the in-going mode is partially transmitted and partially reflected through interaction with the potential. This reflection generates the out-going mode on the $k_r^+$ branch. Underneath the potential barrier, the two modes have equal real parts and opposite imaginary parts, which can be seen from Eq.~\eqref{kr}.} \label{fig:kr_negm}
\end{figure*}

\begin{figure*}[t!]
    \centering
    \begin{subfigure}[t]{0.5\textwidth}
        \centering
        \includegraphics[height=3in]{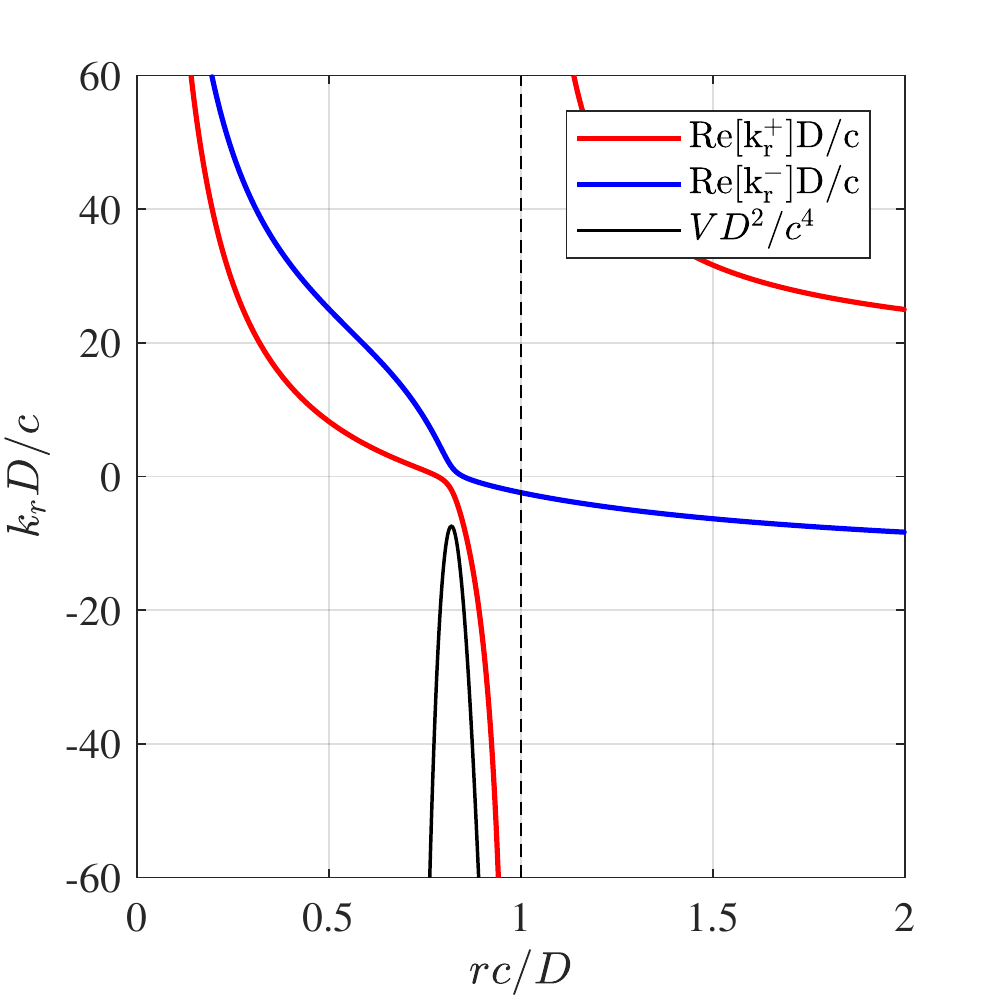}
    \end{subfigure}%
    ~ 
    \begin{subfigure}[t]{0.5\textwidth}
        \centering
        \includegraphics[height=3in]{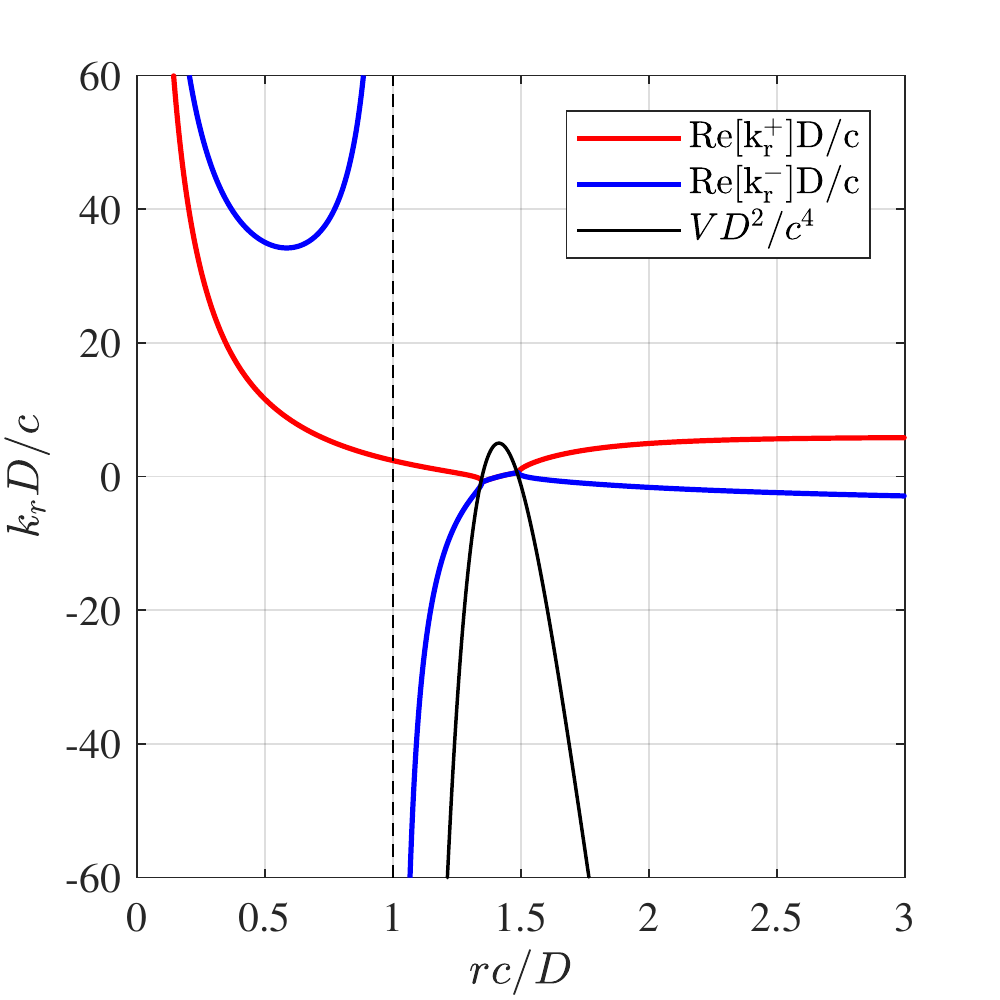}
    \end{subfigure}
    \caption{Real part of the radial wavenumber as a function of radius for $m>0$. The different plots are characterised by the parameters [$\omega D/c^2$, $C/D$, $m$], which for the left hand side are [5, 10, 1] and for the right hand side are [15, 10, 1]. On the left hand side, we have $\tilde{\omega}_h>0$, and the in-going mode is completely absorbed (within the WKB approximation) without generating the out-going mode. If $\tilde{\omega}_h<0$ (right), then the in-going mode interacts with the potential barrier, generating the out-going mode in the process. The mode which is transmitted through the horizon lies on the $k_r^+$ branch, which suggests that out-going mode should have a larger amplitude than the in-going one. This is the origin of rotational superradiance.} \label{fig:kr_posm}
\end{figure*}

To gain some more intuition, consider the plots of $k_r^\pm(r)$ for $m<0$ in Figure~\ref{fig:kr_negm} and $m>0$ in Figure~\ref{fig:kr_posm}.
Starting by sending a mode into the system from far away, i.e. on the $k_r^-$ branch at large $r$,
if there are no turning points (left panels of Figs~\ref{fig:kr_negm} and \ref{fig:kr_posm}) then the in-going mode will be completely absorbed at the horizon (dashed line) without converting to the $k_r^+$ mode\footnote{In reality, even in the absence of turning points the other solution is excited through scattering off the inhomogeneous background. As shown in \cite{coutant2016imprint}, the transition amplitude for such mode conversion is of the form $\sim \exp(-P)$ where $P$ is a function of $|k_r^+-k_r^-|$. Hence the main contribution comes from close to turning points where $k_r^+=k_r^-$, with exponentially small transitions everywhere else. For our purposes, we neglect mode conversion everywhere except in the vicinity of turning points.}.
Since there is no interaction with the background, it's amplitude will vary adiabatically according to Eq.~\eqref{WKBamp}.

If there is a turning point (right panels of Figs~\ref{fig:kr_negm} and \ref{fig:kr_posm}) then the $k_r^-$ mode scatters off the potential, generating the $k_r^+$ mode. 
Only a small proportion of the modes energy will transmit through the barrier and pass through the horizon (we will see in a moment that this proportion is exponentially small).
Since there is an interaction with the background, it is possible to transfer energy between the wave and the system, and non-adiabatic changes in the amplitude must be taken into account.
Indeed, if $\tilde{\omega}_h>0$ then the energy of the transmitted mode is directed toward the horizon (as shown by the left-hand side of Eq.~\eqref{DBT_RadCurr}) and part of the wave is absorbed.
However if $\tilde{\omega}_h<0$, the energy of the transmitted mode is directed outward, allowing for the extraction of energy from the system (superradiant amplification).

\begin{figure} 
\centering
\includegraphics[width=\linewidth]{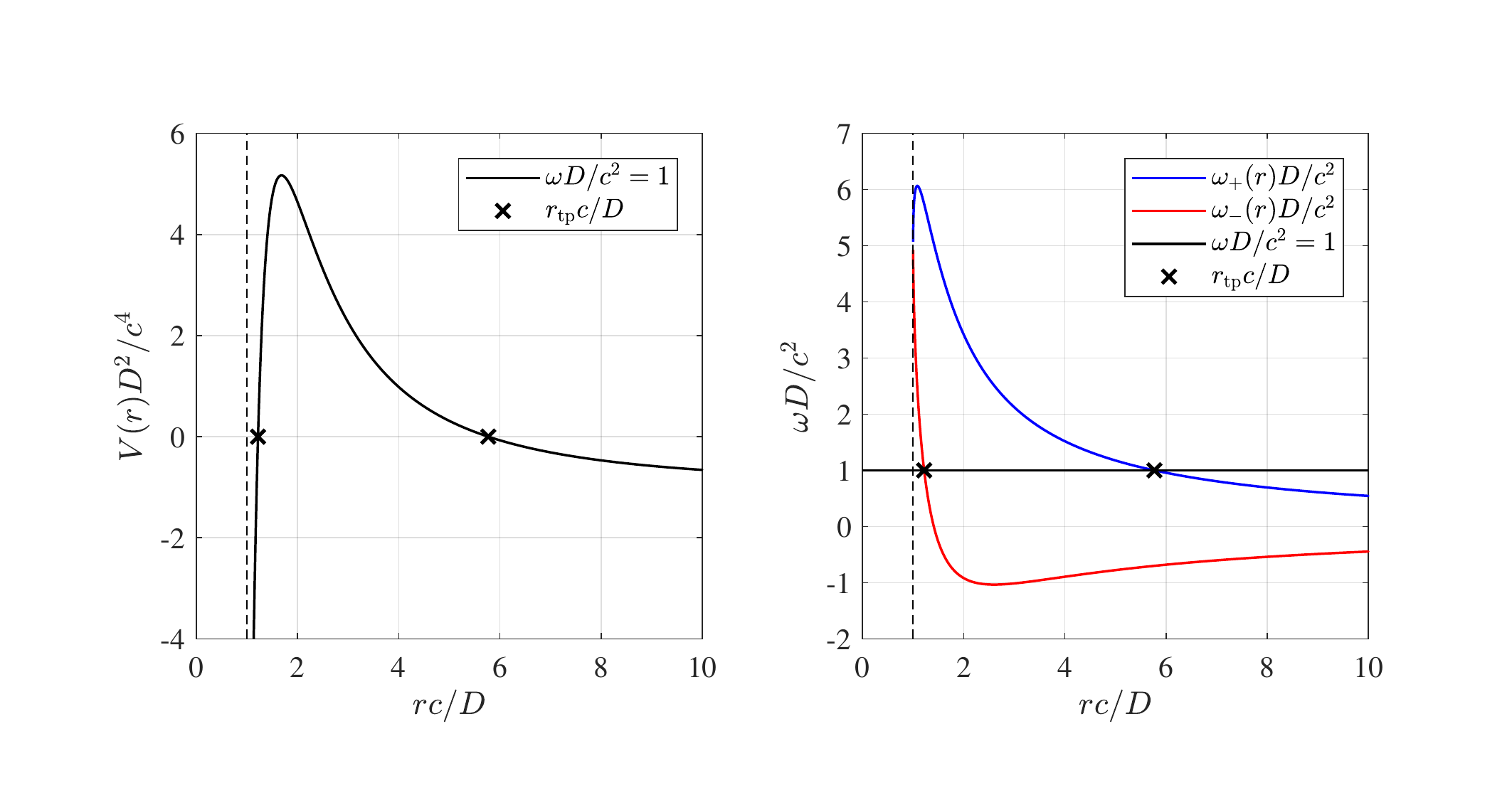}
\caption{Left side: An example of the potential in Eq.~\eqref{potential} for \mbox{$C/D=1$}, \mbox{$m=5$} and the frequency value indicated. 
We have chosen an example where the potential contains two turning points. 
When there are no turning points, the peak either lies below \mbox{$V=0$} or the peak is absent.
Right side: The representation of the potential in Eq.~\eqref{PotEnergyCurves} for the same parameters.
The selected frequency corresponding to $V(r)$ on the left is shown as a black line. 
This representation allows us to see for which values of $\omega$ the potential will have turning points.
} \label{fig:PotentialPlots}
\end{figure}

In the following discussion, we focus on the case where we can approximate the solution everywhere except near two turning points\footnote{As the two turning points become close, a better approximation can be obtained by expanding the potential to quadratic order around the peak and matching to WKB modes either side.
However, since the derivation of the scattering coefficients is more involved (and we will shortly derive exact solutions numerically), we do not pursue the quadratic expansion method here.}.
A useful representation for the potential to find the $r_\mathrm{tp}$'s is,
\begin{equation} \label{PotEnergyCurves}
V = -(\omega-\omega_+)(\omega-\omega_-), \qquad \omega_\pm(r) = \frac{mV_\theta}{r}\pm\sqrt{\frac{m^2-1/4}{r^2}+\frac{5D^2}{4c^2r^4}},
\end{equation}
and the turning points are easily read off as the locations satisfying $\omega=\omega_\pm(r_\mathrm{tp})$ (see for example Figure~\ref{fig:PotentialPlots}).

In the vicinity of $r=r_\mathrm{tp}$, we may expand $V$ to linear order.
Eq.~\eqref{modeeqn} becomes,
\begin{equation} \label{AiryEqn}
-\partial_z^2\psi_m + z\psi_m = 0,
\end{equation}
where $z = (\partial_{r_*}V_\mathrm{tp})^{1/3}(r_*-r_{*\mathrm{tp}})$ and subscript `tp' denotes a quantity is evaluated at the turning point.
We define $z<0$ as the classically allowed region and $z>0$ the classically forbidden region.
Eq.~\eqref{AiryEqn} is the Airy equation whose solutions are the Airy functions $\mathrm{Ai}(z)$ and $\mathrm{Bi}(z)$ \cite{abramowitz1965handbook}, which have the following asymptotics:
\begin{equation} \label{AiryAsym}
\begin{split}
\mathrm{Ai}(z) & \underset{-\infty}\sim \frac{1}{2|z|^{1/4}\sqrt{\pi}}\Big(e^{-i\frac{2}{3}(-z)^{3/2}+i\frac{\pi}{4}}+e^{i\frac{2}{3}(-z)^{3/2}-i\frac{\pi}{4}}\Big), \\
& \underset{+\infty}\sim \frac{e^{-\frac{2}{3}z^{3/2}}}{2|z|^{1/4}\sqrt{\pi}}, \\
\mathrm{Bi}(z) & \underset{-\infty}\sim \frac{i}{2|z|^{1/4}\sqrt{\pi}}\Big(e^{i\frac{2}{3}(-z)^{3/2}-i\frac{\pi}{4}}-e^{-i\frac{2}{3}(-z)^{3/2}+i\frac{\pi}{4}}\Big), \\
& \underset{+\infty}\sim \frac{e^{\frac{2}{3}z^{3/2}}}{
|z|^{1/4}\sqrt{\pi}}.\\
\end{split}
\end{equation}
Close to $r=r_\mathrm{tp}$, the WKB modes in Eq.~\eqref{WKBsol2} have the limiting form,
\begin{equation}
\begin{split}
\psi^{\rightarrow}_m\simeq\frac{e^{-i\frac{2}{3}(-z)^{3/2}}}{2|z|^{1/4}\sqrt{\pi}}, \qquad \psi^{\leftarrow}_m\simeq\frac{e^{i\frac{2}{3}(-z)^{3/2}}}{2|z|^{1/4}\sqrt{\pi}}, \\
\psi^{\uparrow}_m\simeq\frac{e^{\frac{2}{3}z^{3/2}}}{2|z|^{1/4}\sqrt{\pi}}, \qquad \psi^{\downarrow}_m\simeq\frac{e^{-\frac{2}{3}z^{3/2}}}{2|z|^{1/4}\sqrt{\pi}},
\end{split}
\end{equation}
where $\rightarrow$ ($\leftarrow$) denotes the left (right) moving mode for $z<0$ and $\uparrow$ ($\downarrow$) denotes the growing (decaying) mode for $z>0$.
The globally defined solution is a sum of these modes,
\begin{equation}
\begin{split}
\psi_m(z<0) = & \alpha_m^{\rightarrow}\psi^{\rightarrow}_m(z)+\alpha_m^{\leftarrow}\psi^{\leftarrow}_m(z), \\
\psi_m(z>0) = &  \alpha_m^{\downarrow}\psi^{\downarrow}_m(z)+\alpha_m^{\uparrow}\psi^{\uparrow}_m(z).
\end{split}
\end{equation}
and the amplitudes are related to one another using the solution in Eq.~\eqref{AiryAsym}.
This results in a transfer matrix $T$ defined by,
\begin{equation} \label{1tp_cf}
\begin{pmatrix}
\alpha_m^{\rightarrow}\\\alpha_m^{\leftarrow}
\end{pmatrix}
= T\cdot \begin{pmatrix}
\alpha_m^{\downarrow}\\\alpha_m^{\uparrow}
\end{pmatrix}, \qquad T = e^{i\frac{\pi}{4}}\begin{pmatrix}
1 & -i/2\\
-i & 1/2
\end{pmatrix}.
\end{equation}
This argument can be applied directly to the the turning point closest to the horizon (which we will call $r_{\mathrm{tp},1}$) due to the way we defined $z$.
A similar argument can be applied to $r_{\mathrm{tp},2}$, which mirrors the scenario just considered, by inverting $z$. 
The matrix that converts decaying modes to oscillatory modes there is the complex conjugate $T^*$.
Between the two turning points, the modes will incur a phase shift which is captured by the matrix,
\begin{equation} 
J = \begin{pmatrix}
0 & e^{S} \\ e^{-S} & 0
\end{pmatrix}, \qquad S = \int^{r_{*\mathrm{tp},2}}_{r_{*\mathrm{tp},1}}|p_0(r_*)|dr_*.
\end{equation}
Note that $J$ is antidiagonal since the growing and decaying modes switch positions when $z$ is inverted, i.e. a decaying mode on one side is a growing mode on the other side since the coordinate direction is reversed. 

Finally, the effect of the adiabatic change in amplitude between two locations contributes an extra factor, which is the ratio of $p_0^{1/2}$ at those locations.
We then set the amplitude of the in-going mode to $1$ and the reflected mode has amplitude $\mathcal{R}$.
Let the modes on the horizon have amplitudes $\mathcal{A}_{1,2}$.
These amplitudes satisfy,
\begin{equation} \label{ScatterRelation}
\begin{pmatrix}
\mathcal{A}_1 \\ \mathcal{A}_2
\end{pmatrix} = \left|\frac{\omega}{\tilde{\omega}_h}\right|^{\frac{1}{2}} T J (T^*)^{-1} \begin{pmatrix}
\mathcal{R} \\ 1
\end{pmatrix},
\end{equation}
where we have inserted the asymptotic form of $p_0$  using the potential in Eq.~\eqref{potential}.
For $\tilde{\omega}_h>0$, we have \mbox{$\mathcal{A}_1=0$} and \mbox{$\mathcal{A}_2=\mathcal{T}$}, whereas the reverse is true for \mbox{$\tilde{\omega}_h<0$} (this can be seen by comparing the right panels of Figures~\ref{fig:kr_negm} and~\ref{fig:kr_posm}). 
Note that the solution breaks down close to $\tilde{\omega}_h=0$ since the horizon lies on a turning point.

\begin{figure} 
\centering
\includegraphics[width=0.5\linewidth]{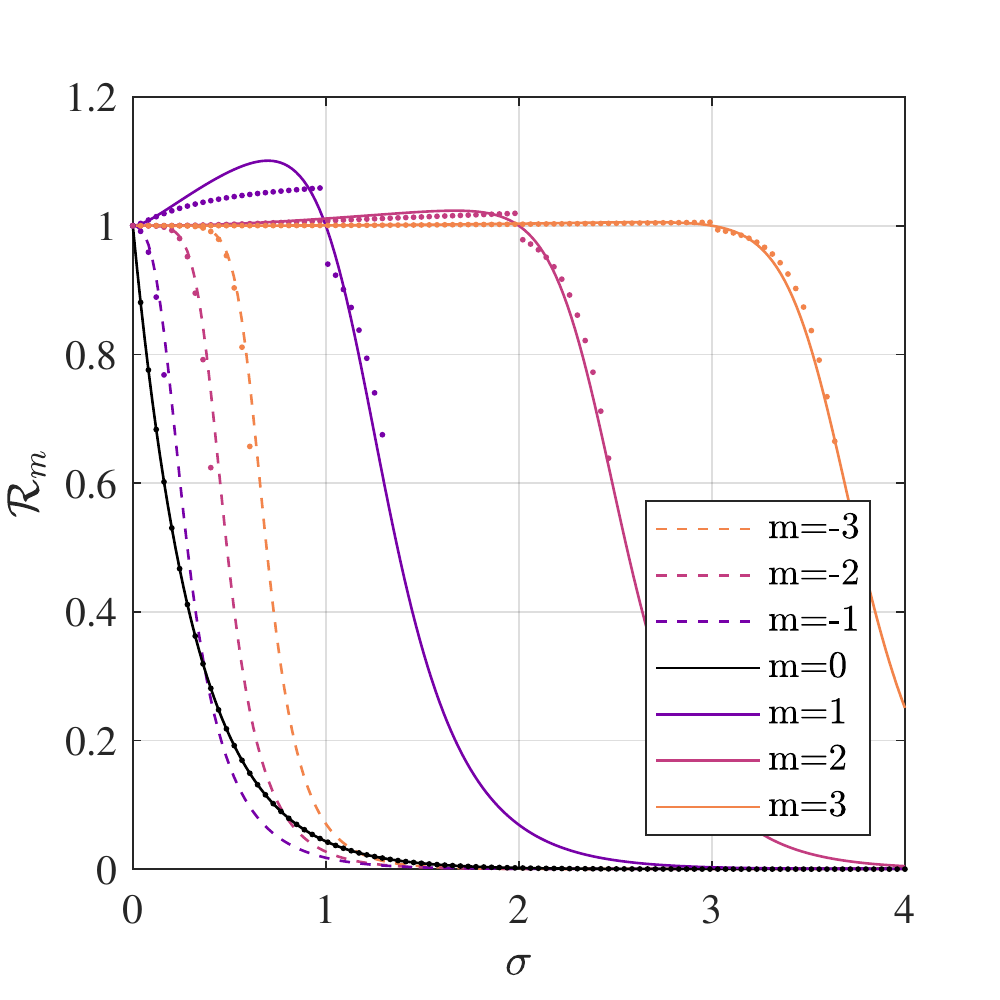}
\caption{Reflection coefficient spectrum for $B=1$ using the WKB approximation (points) and numerically computed values (lines).
The points for $m=0$ are the exact analytical expression \mbox{$\mathcal{R}_0=e^{-\pi\sigma}$} which is derived in \cite{churilov2018scattering}.
Qualitative agreement is seen between approximate and exact solutions.
Note that although WKB is formally only valid for large $m$, 
the agreement is in fact relatively good for all $m$.
Modes with $m\leq0$ are partially absorbed whereas $m>0$ modes are amplified in the range $\sigma<mB$.
Less amplification is seen for higher $m$ modes.
} \label{fig:ReflB1}
\end{figure}

Solving Eq.~\eqref{ScatterRelation} for $\mathcal{R}$ and $\mathcal{T}$ gives,
\begin{equation} \label{ReflWKB}
\mathcal{R} = e^{-i\pi/2}\left(\frac{1-e^{-2S}/4}{1+e^{-2S}/4}\right)^{\mathrm{sgn}(\tilde{\omega}_h)},
\end{equation}
for the reflection coefficient and,
\begin{equation} \label{TransWKB}
\mathcal{T} = \left|\frac{\omega}{\tilde{\omega}_h}\right|^{\frac{1}{2}}e^{-S}~\times \begin{cases}
\left(1+e^{-2S}/4\right)^{-1} \ \ \qquad \tilde{\omega}_h>0 \\
e^{-\frac{i\pi}{2}}\left(1-e^{-2S}/4\right)^{-1} \ \ \tilde{\omega}_h<0
\end{cases},
\end{equation}
for the transmission coefficient. 
It is simple to verify that these expression satisfy Eq.~\eqref{scattering}.
These approximations reveal that $\mathcal{T}$ is exponentially small when there are turning points, with the size of this smallness determined by the area of the region under the square root the potential barrier between the two turning points.
They also reveal that more amplification is achieved the smaller this area is.
We compare this approximation for $\mathcal{R}$ to numerically computed values in Figures~\ref{fig:ReflB1} and \ref{fig:ReflB5}.

\subsection{Numerical prediction} \label{sec:Rnums}

To obtain an exact spectrum from the reflection coefficient, we perform a numerical simulation following~\cite{churilov2018scattering}.
First, we define the dimensionless quantities,
\begin{equation} \label{adim}
\sigma = \frac{\omega D}{c^2}, \qquad B = \frac{C}{D}, \qquad x = \frac{rc}{D}, \qquad \tau = \frac{tc^2}{D},
\end{equation}
then write the wave equation of Eq.~\eqref{waveeqn} in a form suitable for numerical solution,
\begin{equation} \label{waveeqnFx}
\begin{split}
x^2(x^2-1)f'' + & \ \left[1+x^2-2i\left(\sigma x^2-mB\right)\right]xf' \\
& + \left[(\sigma x^2-mB)^2-2imB-m^2x^2\right]f = 0,
\end{split}
\end{equation}
where $'=\partial_x$ and we have written the perturbation for a particular \mbox{$(m,\sigma)$} mode using the ansatz \mbox{$\phi = \mathrm{Re}[f(x)\exp(im\theta-i\sigma\tau)]$}.
Noticing the appearance of powers of $x^2$ in this equation, we define a new variable $y=x^2$ which leads to,
\begin{equation} \label{waveeqnFy}
\begin{split}
y^2(y-1)\partial_y^2f+ & \ \left[y-i(\sigma y-mB)\right]y\partial_yf \\
& + \frac{1}{4}\left[(\sigma y-mB)^2-2imB-m^2y\right]f = 0.
\end{split}
\end{equation}

\begin{figure} 
\centering
\includegraphics[width=\linewidth]{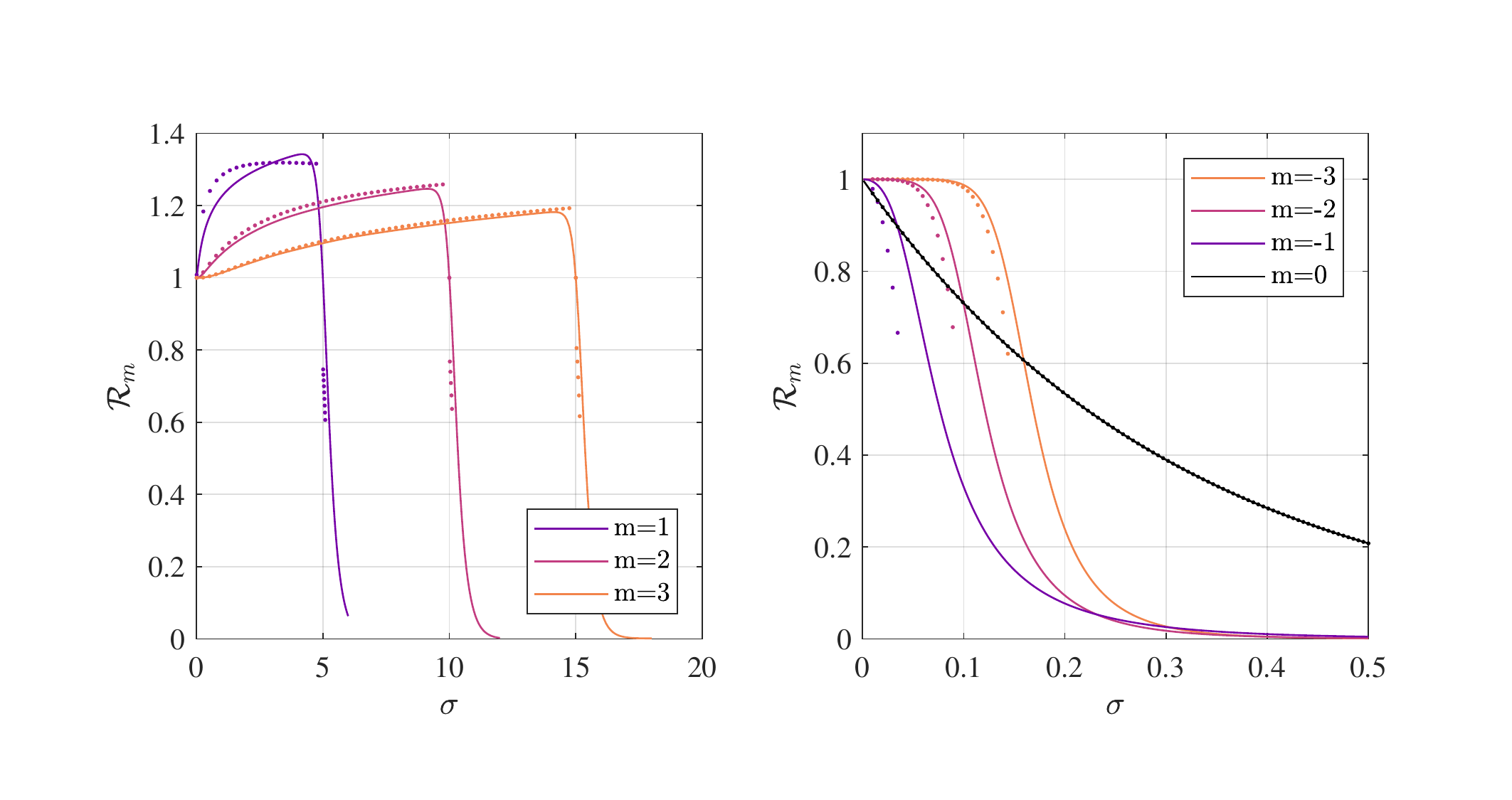}
\caption{Reflection coefficient spectrum for $B=5$ using the WKB approximation (points) and numerically computed values (lines).
The behaviour follows that described in Figure~\ref{fig:ReflB1}.
However, the absorption of the $m<0$ increases quicker with $\sigma$ for larger $B$.
Also, the transition from superradiance to absorption occurs more abruptly.
} \label{fig:ReflB5}
\end{figure}

\noindent Both Eqs.~\eqref{waveeqnFx} and \eqref{waveeqnFy} are second order ordinary differential equations with a regular singular point at $x=1$ and $y=1$ respectively \cite{churilov2018scattering}.
A numerical solution requires initial conditions which are provided by the Frobenius expansion of $f$ about the regular singular point.
This is most easily obtained from Eq.~\eqref{waveeqnFy}, since the polynomials in front of $f$ and it's derivatives are of lower order.
We first write $f$ as,
\begin{equation}
f(y) = (y-1)^p\sum_{n=0}^\infty a_n(y-1)^n,
\end{equation}
and substitute into Eq.~\eqref{waveeqnFy}.
Demanding that the equation is satisfied for the lowest order in $(y-1)$, we find for the index,
\begin{equation}
p = \begin{cases}
0 \\
i(\sigma-mB)
\end{cases},
\end{equation}
The different values of $p$ correspond to the two linearly independent solutions.
The solution with $p=i(\sigma-mB)$ is an out-going mode which diverges on the horizon.
To see this, one can write the overall factor preceding the power series $(y-1)^p$ in the form \mbox{$e^{i(\sigma-mB)\log(y-1)}$} which has the form of a plane wave whose wave number diverges as $y\to1$.
Indeed, this solution corresponds to the mode we discarded in Eq.~\eqref{asymp_sols_b} when we imposed a purely in-going boundary condition on the horizon.
Hence, we discard it for the same reason here.
The next lowest order in $(y-1)$ gives an expression for $a_1$ in terms of $a_0$, but since Eq.~\eqref{waveeqnFy} is linear in $f$ we may set $a_0=1$.
Hence the first two terms in the expansion are,
\begin{equation} \label{frob}
f(y) = 1 - \frac{(\sigma-mB)^2-2imB-m^2}{4[1-i(\sigma-mB)]}(y-1) + \mathcal{O}\left((y-1)^2\right).
\end{equation}
From Eq.~\eqref{frob}, we convert back to the $x$ variable using \mbox{$f'(x=1)=2\partial_yf|_{y=1}$}, then compute initial conditions \mbox{$f(1+\epsilon)$} to \mbox{$\mathcal{O}(\epsilon^2)$} and \mbox{$f'(1+\epsilon)$} to $\mathcal{O}(\epsilon)$.
Better accuracy can be obtained by taking higher order terms in the expansion. 
However, we found this was not necessary since we were able to find consistent solutions for $\epsilon=10^{-4}$ and $\epsilon=10^{-6}$ (an error due to poor initial conditions would decrease with $\epsilon$).
These initial conditions are used to solve Eq.~\eqref{waveeqnFx} numerically over the range $x\in[1+\epsilon,x_n]$ with $x_n = 20(2\pi/\sigma)$, i.e. about $20$ (flat space) wavelengths away from the centre.
We used Matlab's inbuilt function \emph{ode45} (which is based on a forth order Runge-Kutta algorithm) to evolve from the starting point into the asymptotic region.

To extract the amplitudes in this region, we use the asymptotic solution to Eq.~\eqref{waveeqnFx}:
\begin{equation} \label{fSolInfty}
f(x\to\infty) = x^{i\sigma-1/2}\left(A_m^\mathrm{in}e^{-i\sigma x} + A_m^\mathrm{out}e^{i\sigma x} \right).
\end{equation}
Using Eq.~\eqref{fSolInfty} and it's derivative, we may solve for $A_m^\mathrm{in,out}$ in terms of our numerical solution at $x=x_n$, i.e. $(f_n,f'_n)$.
This gives,
\begin{equation}
\frac{A_m^\mathrm{out}}{A_m^\mathrm{in}} = -e^{-2i\sigma x_n}\frac{2x_nf'_n+f_n(1-2i\sigma(1-x_n))}{2x_nf'_n+f_n(1-2i\sigma(1+x_n))},
\end{equation}
and the reflection coefficient is then \mbox{$\mathcal{R}_m=|A_m^\mathrm{out}/A_m^\mathrm{in}|$}.
We display results for the $|m|\leq3$ modes for $B=1$ and $B=5$ in Figures~\ref{fig:ReflB1} and \ref{fig:ReflB5} respectively, finding good qualitative agreement with the WKB approximation.

\section{Detection in the laboratory} \label{sec:detectSR}

\begin{figure} 
\centering
\includegraphics[width=\linewidth]{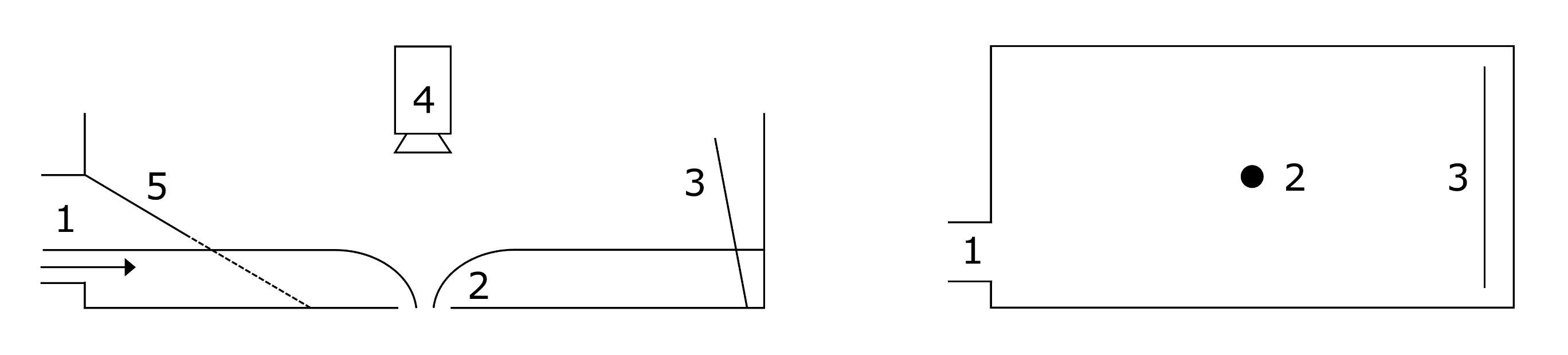}
\caption{A schematic of the experimental set up viewed from side-on (left side) and above (right side).
Diagram is not drawn to scale.
1: Water inlet. 
2: Drain hole. 
3: Wave generator.
4: Camera position.
5: Absorption beach (there is a mesh in front of the inlet to allow water to pass through).
The arrow indicates the flow of water and the shape of the water's surface is also indicated.
} \label{fig:apparatus}
\end{figure}

To verify the existence of superradiance in a DBT system, we conducted an experiment in a $2.9~\mathrm{m}$ long and $1.38~\mathrm{m}$ wide rectangular water tank. 
Water is pumped continuously in from one end corner, and is drained through a hole (radius $d=2\mathrm{cm}$) in the middle. 
The water flows in a closed circuit. 
A schematic representation of the system can be found in Figure~\ref{fig:apparatus}.

We first establish a stationary rotating draining flow by setting the flow rate of the pump to $Q = 37.5~\pm~0.5~\mathrm{\ell/min}$ and waiting until the depth (away from the vortex) is steady at $H_\infty = 6.25 ~\pm ~0.05 ~\mathrm{cm}$. 
These parameter's were required to make the fluid rotate fast enough for superradiance to occur.
The velocity field pertaining to this flow configuration is measured using a standard flow visualisation technique called Particle Imaging Velocimetry (PIV).

\begin{figure} 
\centering
\includegraphics[width=0.6\linewidth]{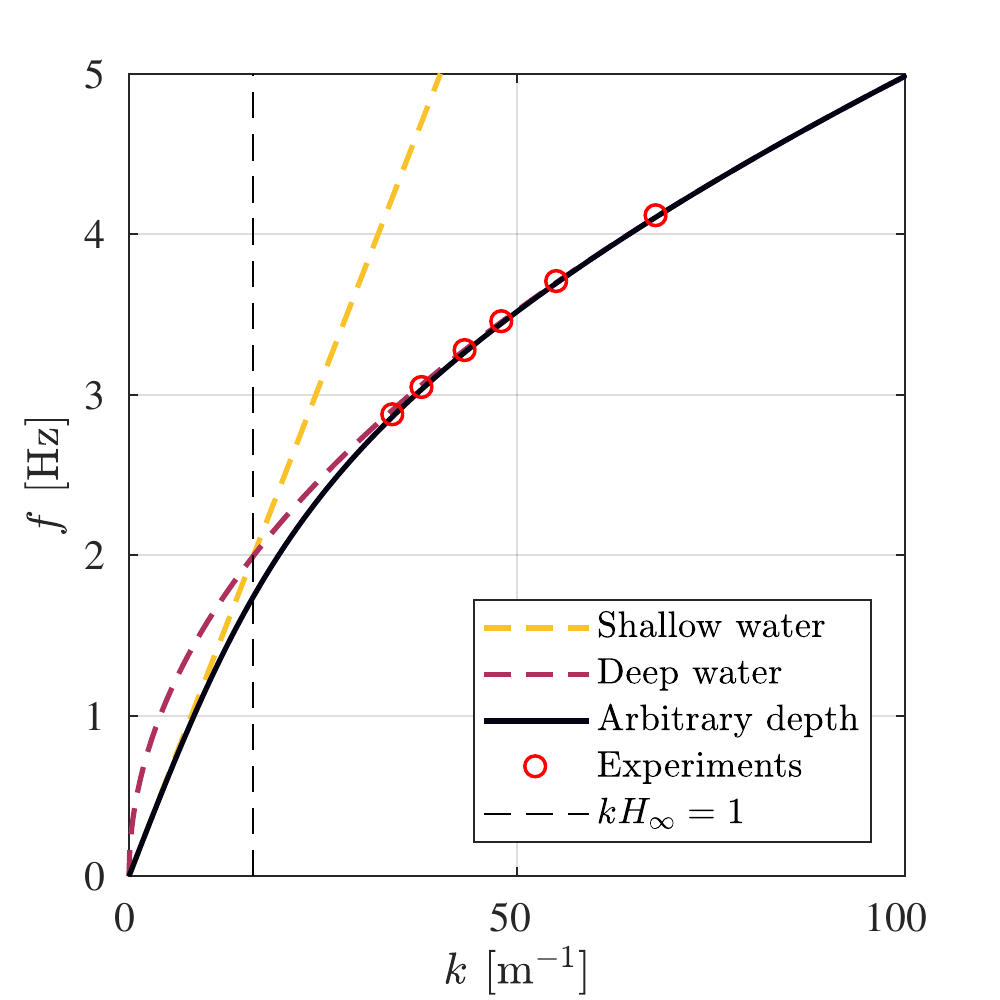}
\caption{The dispersion relation far away from the vortex where the velocity field is negligible.
We show where the frequencies used for experiment lie on the dispersion relation in Eq.~\eqref{dispersion2}, as well as the shallow water $kH_\infty\ll1$ and deep water $kH_\infty\gg1$ approximations.
This indicates that our experiments were performed closer to the deep water regime.} \label{fig:ExpDispersion}
\end{figure}

Once the background flow is determined, we generate plane waves using a mechanical piston from one side of the tank, with excitation frequencies $f_0$ ranging from $2.87$ to $4.11~\mathrm{Hz}$.
For our chosen water height, this frequency range does not lie on the linear part of the dispersion relation.
The dispersion relation in water of constant arbitrary depth is,
\begin{equation} \label{dispersion2}
\Omega^2 = (\omega - \mathbf{V}\cdot\mathbf{k})^2 = gk\tanh\left(H_\infty k\right), \qquad k = ||\mathbf{k}||,
\end{equation}
which reduces to Eq.~\eqref{dispersion} in the shallow water limit $kH_\infty\ll 1$. In the deep water limit $kH_\infty\gg 1$, this becomes,
\begin{equation}
\Omega^2 = (\omega - \mathbf{V}\cdot\mathbf{k})^2 = g|k|.
\end{equation}
On Figure~\ref{fig:ExpDispersion}, we show the dispersion relation far from the vortex where $\mathbf{V}\simeq 0$.
The frequencies used for experiments are in fact closer to the deep water regime than shallow water.

On the side of the tank opposite the wave generator, we have placed an absorption beach (we have verified that the amount of reflection from the beach is below $5\%$ in all experiments). 
We record the free surface with a high speed 3D air-fluid interface sensor. The sensor is a joint-invention~\cite{enshape2016verfahren} (patent No.~DE 10 2015 001 365 A1) between The University of Nottingham and EnShape GmbH (Jena, Germany). 
Finally, the images from the sensor are analysed to extract the scattering coefficients. A reflection coefficient greater than one indicates that the system superradiates.

\subsection{The velocity field} \label{sec:BgConfig}

\begin{figure*}[t!]
    \centering
    \begin{subfigure}[t]{0.5\textwidth}
        \centering
        \includegraphics[height=2in]{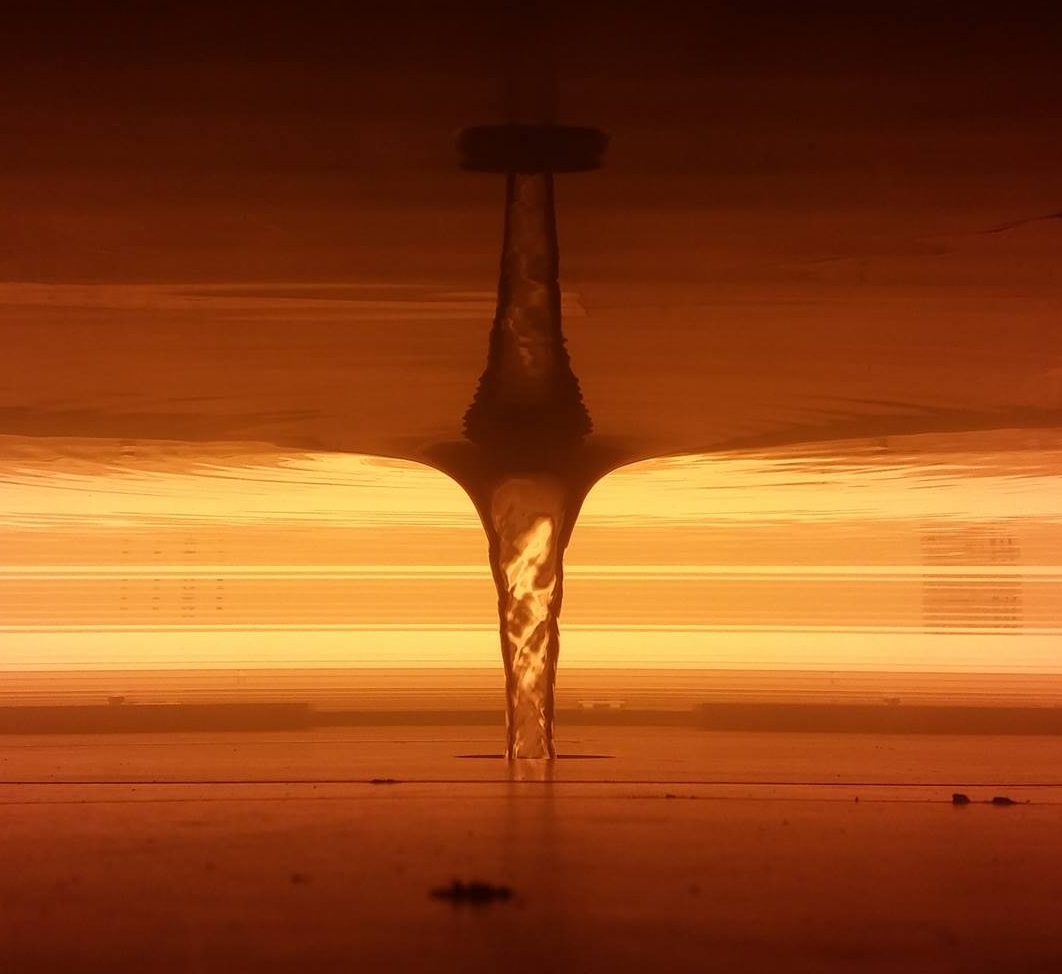}
    \end{subfigure}%
    ~ 
    \begin{subfigure}[t]{0.5\textwidth}
        \centering
        \includegraphics[height=2in]{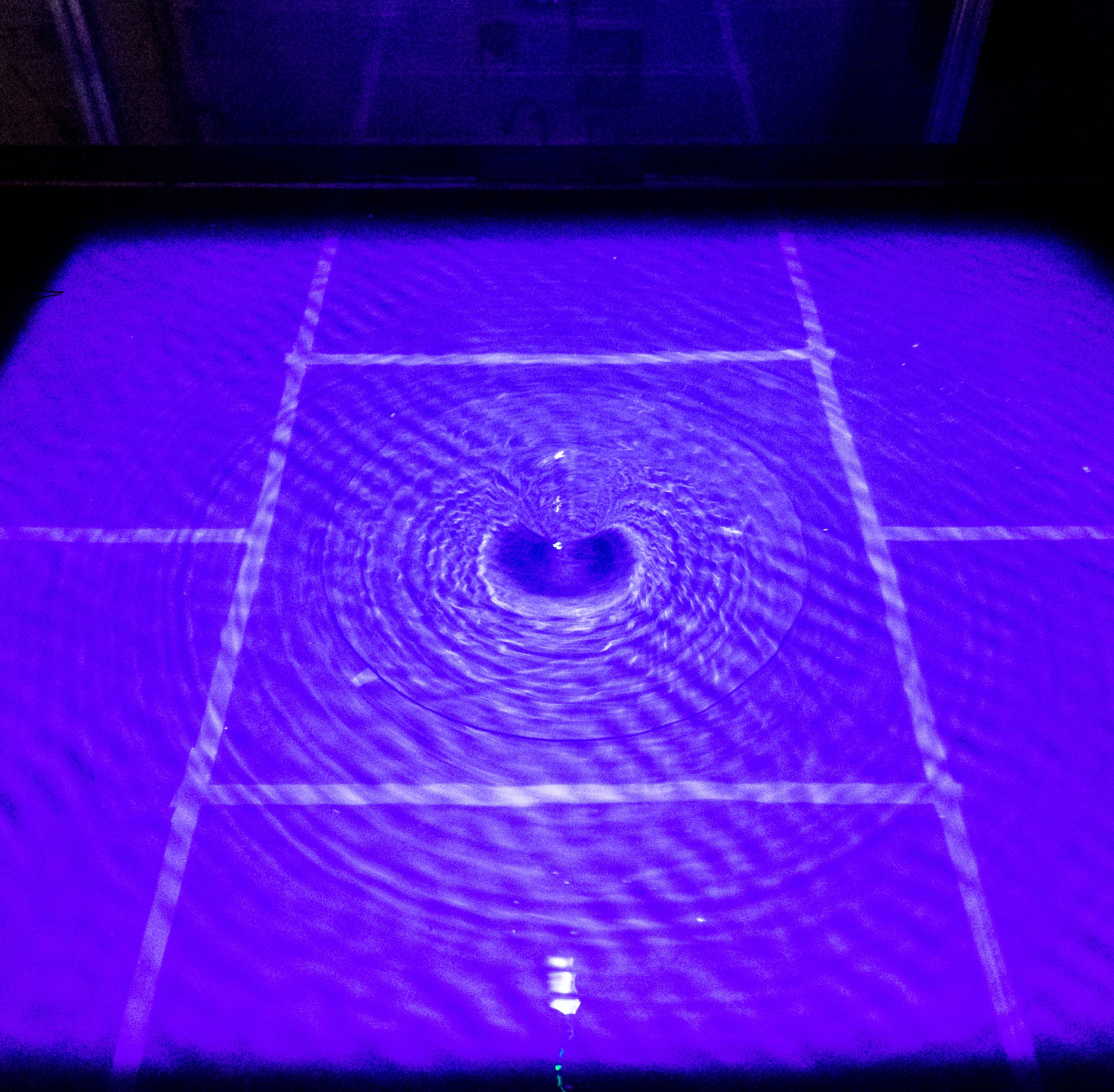}
    \end{subfigure}
    \caption{Typical images of the fluid flow in our experiment from side on (left side) and from above (right side). Far away the free surface is flat, but bends significantly over the drain hole. The hollow air core is known as the `throat' of the vortex.} \label{fig:ExpPhoto}
\end{figure*}

Images of a typical DBT-type vortex obtained using this set-up are shown in Figure~\ref{fig:ExpPhoto}. 
Before performing any measurements, we can already see how a real experiment compares to the simple 2 dimensional theory outlined in Section~\ref{sec:dbt}. 
The vertical component of the velocity field will be non-negligible close to the drain hole (due to the downward draining that takes place there), which results in large gradients in the free surface. 
Using Eq.~\eqref{height_profile} as an approximate form for the free surface profile $H(r)$, we expect our theoretical understanding of the system to be valid far from the core $r>r_\mathrm{min}\gg r_a$ where $H'\ll 1$.

Moreover, due to the Cartesian geometry of the apparatus in Figure~\ref{fig:apparatus}, the velocity field will be highly asymmetric close to the boundaries. 
However, we expect the flow to be cylindrically symmetric to a good approximation far from the boundaries $r<r_\mathrm{max}\ll L_\mathrm{sys}$, where $L_\mathrm{sys}$ is representative scale of the system size. 
This restricts our analysis to a band $r\in[r_\mathrm{min},r_\mathrm{max}]$.

Inside this region, the axisymmetric velocity field in polar coordinates can be expressed as,
\begin{equation} \label{profile}
\textbf{V}(r,z) = V_r(r,z) \bm{\mathrm{e}}_r + V_{\theta}(r,z) \bm{\mathrm{e}}_\theta + V_z(r,z) \bm{\mathrm{e}}_z.
\end{equation} 
We are specifically interested in the velocity field at the free surface $z = H(r)$.
Taking Eq.~\eqref{height_profile} for the free surface, $V_z$ can be approximated from $V_r$ via,
\begin{equation}
\begin{split}
V_z(r,H) = \ & V_r(r,H)\partial_rH \\
= \ & V_r(r,H)\frac{2H_\infty r_a^2}{r^3}.
\end{split}
\end{equation}
Hence, our task is reduced to the measurement of $V_r(r,H)$ and $V_\theta(r,H)$. 
We also need to measure the water height at the edge of the tank $H_\infty$ and the radius $r_a$ at which the free surface inserts the floor of the tank.

\subsection{Particle Imaging Velocimetry} \label{sec:PIV}

The velocity field is measured using a technique routinely applied to fluid systems called Particle Imaging Velocimetry (PIV) (see e.g. \cite{raffel1998PIV} for a review). 
The technique is implemented using the Matlab extension \textit{PIVlab} developed in \cite{thielicke2014flapping,thielicke2014pivlab,thielicke2014pivlab2} and can be summarised as follows.

The flow is seeded with flat paper particles of mean diameter $d_p = 2~\mathrm{mm}$. 
The particles are buoyant which allows us to evaluate the velocity field exclusively at the free surface. 
The amount by which a particle deviates from the streamlines of the flow is given by the velocity lag $U_s = d_p^2(\rho_p-\rho)a/18\mu$ \cite{raffel1998PIV}, where $\rho_p$ is the density of a particle, $\rho$ is the density of water, $\mu$ is the dynamic viscosity of water and $a$ is the acceleration of a particle. 
For fluid accelerations in our system $U_s$ is at most of the order $10^{-4}~ \mathrm{m/s}$, an order of magnitude below the smallest velocity in the flow. 
Thus we can safely neglect the effects of the velocity lag when considering the motions of the particles in the flow.

The surface is illuminated using two light panels positioned at opposite sides of the tank. 
The flow is imaged from above using a Phantom Miro Lab 340 high speed camera with a Sigma 24-70mm F2.8 lens.
The frame rate was set to $800~ \mathrm{fps}$ (frames per second) and the exposure time was $1200 ~\mathrm{\mu s}$.
The camera was focussed to the plane lying at $z=H_\infty$, where a calibration image was place in order to determine the spatial resolution $(\mathrm{meter}/\mathrm{pix})$.
This was about $\sim0.1~\mathrm{mm/pix}$ for our experiments.
The raw images are analysed using \textit{PIVlab} which executes the following routine:
\begin{enumerate} [noitemsep] 
\item A single image is divided into smaller regions called windows. These windows may overlap. The union of all windows covers the full image.
\item Each window is scanned over the next image in the sequence and the correlation at each location is computed, yielding a correlation table.
\item The maximum in the correlation table gives the position this window has moved to in the next image. The distance moved by the window in pixels is given by $\Delta\mathrm{pix}_x$ and $\Delta\mathrm{pix}_y$.
\item This process is repeated for all windows in an image and all images.
\end{enumerate}
To improve results, the analysis was performed over three iterations using window sizes $256\times256$, $128\times128$ and $64\times64$ pixels respectively, with $50\%$ overlapping. 
We also used \textit{PIVlab}'s spline deformation tool which deforms the window shape to find the best match \cite{raffel1998PIV}.
The Cartesian velocity components are obtained from,
\begin{equation}
V_{x,y}(t,x,y) = \left(\frac{\mathrm{meter}}{\mathrm{pix}}\right)\times\Delta\mathrm{pix}_{x,y}\times\mathrm{fps},
\end{equation}
which are then converted into polar components. 
The origin is defined as the centre of the drain hole.
Since the air-core has the effect of strongly pinning the vortex, we assume that the origin coincides with the centre of the velocity field\footnote{We checked the effect of choosing a different centre and found that the result was unchanged within error estimates.}.
Finally, we obtain the radial profiles $V_r(r)$ and $V_\theta(r)$ by averaging in $t$ and $\theta$.
The error is computed from the standard deviation at each stage of averaging.

To test the accuracy of this method, we simulated the motion of particles with 20 pixel diameter and a density of $0.1~\mathrm{particles/pixel}$ (these values were chosen to keep data as close as possible to that obtained in our experiments) through the velocity field of Eq.~\eqref{DBT} using values of $C$ and $D$ similar to those obtained in Section~\ref{sec:flow_values} below.
The fluid parameters obtained were correct to $\sim 5\%$ with the PIV method consistently over-estimating the true value.
The precise value of this systematic shift depends on the relative sizes of $V_\theta$ and $V_r$, see \cite{torres2019analogue} for specific examples.

\subsection{Velocity field modelling} \label{sec:Vmodelling}

Exact solutions to the Navier-Stokes equations are difficult to obtain due to their non-linearity and the requirement of specific boundary conditions.
Here, we briefly review a few known solutions to the Navier-Stokes equations, as well as some heuristic models that capture their essential features.
Although the precise form of the velocity field depends on the geometry of the set-up, we indicate which models provide a good description of our system.

The solutions in Eq.~\eqref{DBT} break down under experimental conditions when viscosity (which affects the angular component) and free surface gradients (which influence the radial component) cannot be neglected.
A well known heuristic model accounting for viscosity in the vortex core is the Rankine vortex \cite{lautrup2005exotic},
\begin{equation} \label{Rankine}
V^\mathrm{Rankine}_\theta = 
\begin{cases}
\Omega_0 r, \qquad r\leq r_0 \\
C/r, \ \ \quad r>r_0
\end{cases},
\end{equation}
describing a forced vortex inside the core and a free vortex outside the core, where $r_0$ is the core radius.
The core rotates as a solid body with frequency $\Omega_0=C/r_0^2$.
Although this model accounts for the shape of a realistic vortex flow, there is a discontinuity in the vorticity at $r=r_0$ which is non-physical. 
A more realistic approximation is provided by the Lamb vortex, which smoothly interpolates between the two extremes of Eq.~\eqref{Rankine},
\begin{equation} \label{Lamb}
V^\mathrm{Lamb}_\theta = \frac{C}{r}\left(1-e^{-r^2/r_0^2}\right).
\end{equation}
If we replace $r_0^2\to r_0^2+4\nu t$ in this expression, this is an exact solution to the Navier-Stokes equations called the Lamb-Oseen vortex.
It describes a free vortex which decays due to viscosity $\nu$ \cite{lautrup2005exotic}.

Although more realistic, the exponential in Eq.~\eqref{Lamb} makes it difficult to manipulate analytically.
A phenomenological model which captures the same features is the $n$-vortex \cite{mih1994intakes,mih1990comment,rosenhead1930spread,vatistas1989combined},
\begin{equation} \label{Nvortex}
V^{(n)}_\theta = \frac{Cr}{(r_0^{2n}+r^{2n})^{1/n}},
\end{equation}
which reduces to the Rankine vortex in the limit $n\to\infty$.
This solution provides a good fit to many vortices observed in experiment, e.g. \cite{vatistas1991model}.
We will make extensive use of the $n=1$ vortex throughout this thesis due to it's analytic simplicity.
Note that any model that interpolates between a solid body rotation in the core and a potential flow at infinity can be referred to as a Rankine-type vortex in the literature.

A more realistic model for $V_r$ in the presence of a drain can be obtained by estimating the vertical velocity there.
Applying Eq.~\eqref{vz_shallow} at the free surface gives,
\begin{equation}
\frac{1}{r}\partial_r(rHV_r) = V_z(z=0).
\end{equation}
Let the drain in this solution be centred on the origin with radius $a$. Assuming \mbox{$V_z(z=0)=-U$} for $r\leq a$ (where $U>0$ is a constant) and zero for $r>a$ gives,
\begin{equation} \label{sharpplug}
V^\mathrm{sharp}_r = 
\begin{cases}
-Ur/2H, \qquad r\leq a \\
-Ua^2/2Hr, \quad r>a
\end{cases}.
\end{equation}
The sharp transition at $r=a$ makes this unrealistic. A soft plug flow \cite{lautrup2005exotic} assumes a Gaussian form to $V_z(z=0)$ and yields for the radial flow,
\begin{equation} \label{softplug}
V^\mathrm{soft}_r = \frac{-Ua^2}{2Hr}\left(1-e^{-r^2/a^2}\right).
\end{equation}
Eqs.~\eqref{sharpplug} and~\eqref{softplug} have the same form as the Rankine and Lamb vortices if we assume a flat free surface $H\simeq H_\infty$, and reduce at large $r$ to the simple $1/r$ profile of Eq.~\eqref{DBT} with $D=Ua^2/2H_\infty$.

\subsection{Flow measurements} \label{sec:flow_values}

Using the PIV method, we performed 10 measurements of the velocity field in our system to check that the flow configuration was repeatable.
The radius of the aircore in our experiments was \mbox{$r_a=9\pm1~\mathrm{mm}$}.

\begin{figure} 
\centering
\includegraphics[width=\linewidth]{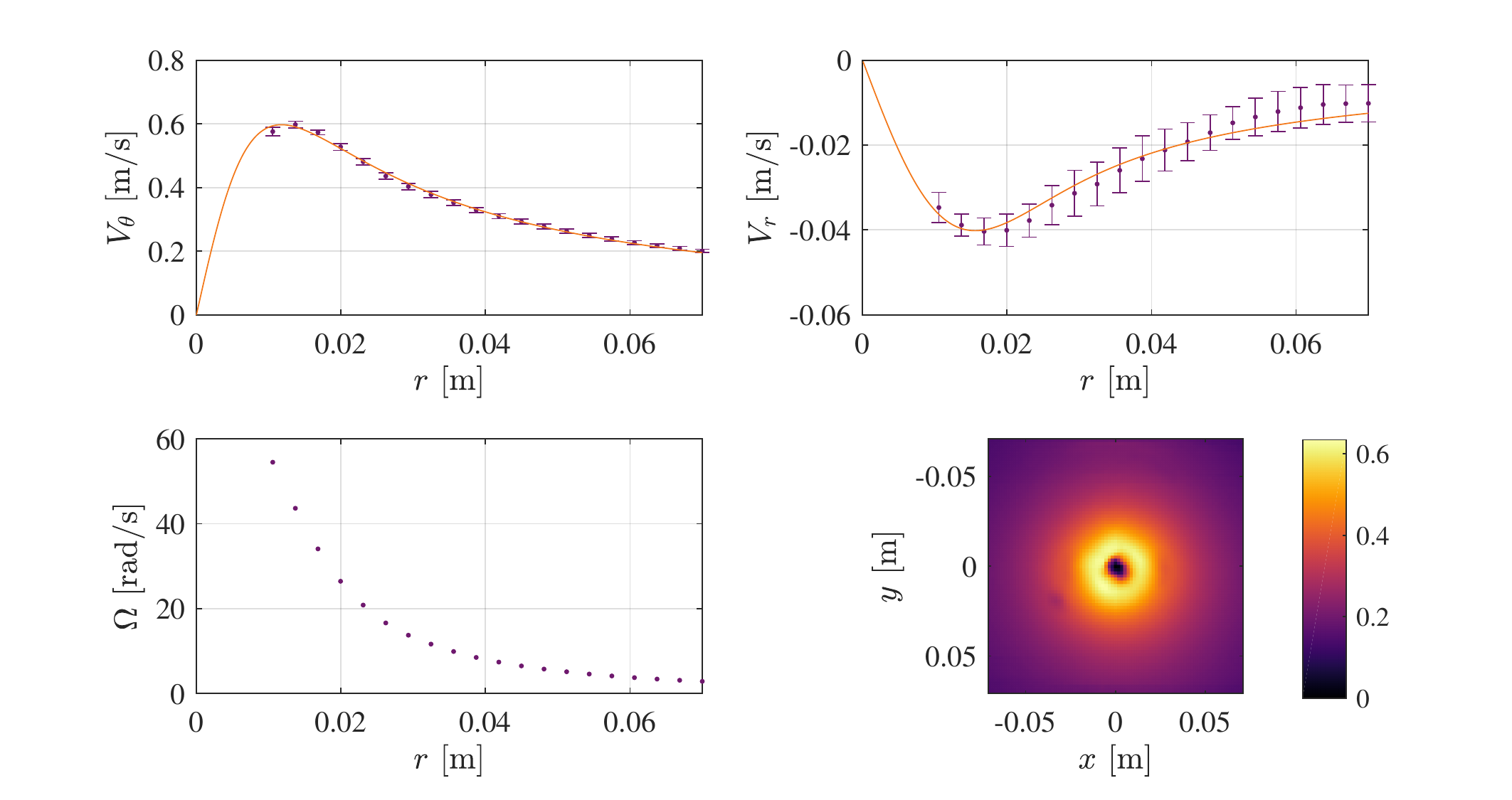}
\caption{PIV measurements of the velocity field averaged over $10$ experiments.
The errorbars represent the spread in the measured velocities. 
Data is absent in the region $r<r_a$ where there is no water.
Top left: $V_\theta(r)$ is fitted with the $n=1$ vortex in Eq.~\eqref{Nvortex}. Top right: $V_r(r)$ is fitted with a soft plug flow in Eq.~\eqref{softplug}. Bottom left: The angular frequency of the fluid. Bottom right: The norm of the velocity field. The colour axis is $||\mathbf{V}||$.} \label{fig:VelocityField}
\end{figure}

In Figure~\ref{fig:VelocityField}, we display the results of our measurements. 
We also show the angular frequency of the flow $\Omega = V_\theta/r$ (which is an important quantity in the discussion of superradiance) and the norm of the velocity field $||\mathbf{V}||$, which is notably axisymmetric to a good approximation.
The lack of data for $r<r_a$ is due to the absence of water in that region.

Using a least squares regression, we fit $V_\theta$ with the $n$-vortex model in Eq.~\eqref{Nvortex}, taking $n=1$.
The best fit parameters are \mbox{$C = 1.41\pm0.01\times10^{-2}~\mathrm{m^2/s}$} for the circulation and \mbox{$r_0 = 1.18\pm0.01~\mathrm{cm}$} for the core radius, where the errors are obtained from the residuals of the fit. 
The $R^2$ coefficient is a measure of how much variation in the data about it's mean is explained by the model.
The fit gives $R^2 = 0.9976$, which means it accounts for $99.76\%$ of the variation.
Recalling that the drain is of true radius \mbox{$2~\mathrm{cm}$}, we see that the viscous core is effectively confined over the drain.
Recalling that the only axisymmetric irrotational flow profile in 2D is that in Eq.~\eqref{DBT}, this means that vorticity is non negligible inside a radius of $r\sim r_0$ in our setup.
Consequences of this are explored further in Chapter~\ref{chap:vort}.

We fit $V_r$ with the soft plug profile in Eq.~\eqref{softplug}.
The best fit parameters are $D = -8.8\pm0.3\times10^{-4}~\mathrm{m^2/s}$ and \mbox{$a = 1.4\pm0.1~\mathrm{cm}$}, which give \mbox{$R^2=0.9639$}.
The quoted errors on the radial parameters are larger than those for angular component since $V_r$ is much slower than $V_\theta$.
Indeed, far away we have \mbox{$V_\theta/V_r \sim C/D \sim 16$}.
This makes $V_r$ harder to measure since a cluster of particles will be smeared out by the larger angular flow, making erroneous vector identifications in the radial direction more likely.

\subsection{Wave measurements} \label{sec:getWavesSR}

\quad \emph{Wave generation}. Waves are generated by the paddle board (labelled 3) in Figure~\ref{fig:apparatus}, which is connected to a motor by a simple piston-wheel mechanism.
Plane waves generated in this manner are not perfectly sinusoidal, but this is inconsequential since we perform frequency filter (see below) at the motor frequency $f_0$.
Details of wave generation can be found in \cite{hughes1993coastal}.

\emph{Data acquisition}. We start taking data once the first wavefront reaches the absorption beach. 
We record the free surface of the water in a region of $1.33~ \mathrm{m} \times 0.98~ \mathrm{m}$ over the vortex for $13.2~ \mathrm{s}$. 
From the sensor we obtain $248$ reconstructions of the free surface. 
These reconstructions are triplets $X_{ij}$, $Y_{ij}$ and $Z_{ij}$ giving the coordinates of $640 \times 480$ points on the free surface. 
Due to free surface gradients in the core and inherent noise in the system, parts of the free surface cannot be seen by our sensor, resulting in black spots on the images. 
Isolated black spots are corrected by interpolating neighbouring points. 
This procedure is not possible in the core of the vortex where there is a large shadow. We set these values to zero.

\emph{Frequency filter}. To filter the signal in frequency, we first crop the signal in time to keep an integer number of oscillations, reducing spectral leakage. 
We then select a single frequency corresponding to the excitation frequency $f_0>0$.
This allows us to filter out the background located at $f=0$, high frequency noise and non-linearities excited at the wave generator. 
These had a maximum of $14\%$ of the signal amplitude. 
There is an identical peak at $-f_0$ containing half the amplitude of the fluctuations since the data is real. We multiply the data by a factor of $2$ to account for this\footnote{Extracting the $f_0>0$ component of $\cos(2\pi f_0 t)$ gives $\frac{1}{2}\exp(2\pi if_0 t)$. Hence, we need to multiply by $2$ so that the real part of this corresponds to the original data, i.e. \mbox{$\mathrm{Re}[\exp(2\pi if_0 t)]=\cos(2\pi f_0 t)$}.}.

After this filter, we are left with a 2-dimensional array of complex values, encoding the fluctuations of the water height $h(X_{ij},Y_{ij})$ at the frequency $f_0$. $h(X_{ij},Y_{ij})$ is defined on the grids $X_{ij}$, and $Y_{ij}$, whose points are not perfectly equidistant.
This is due to the fact that the discretization is done by the sensor software in a coordinate system that is not perfectly parallel to the free surface.

\begin{figure} 
\centering
\includegraphics[width=\linewidth]{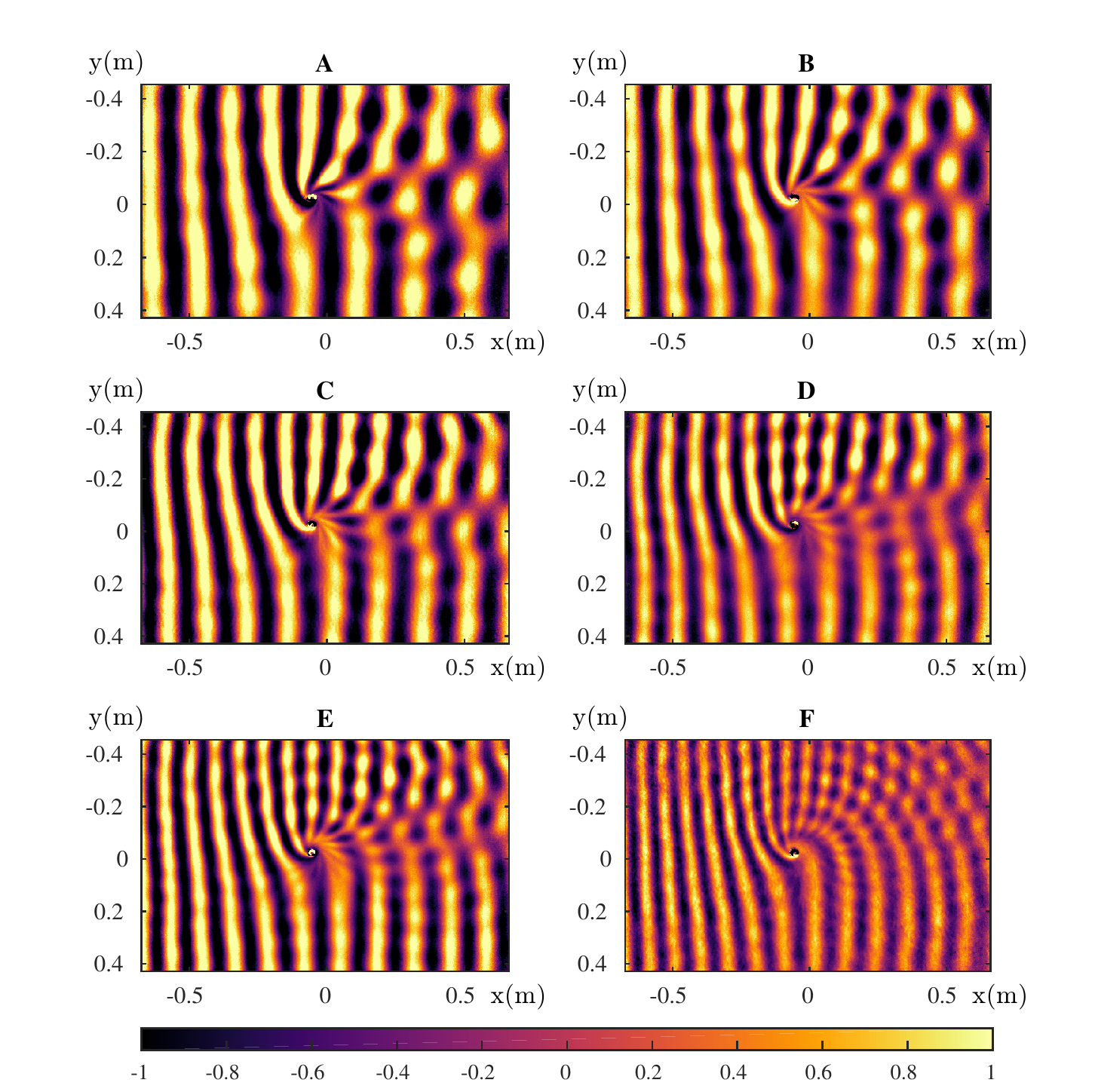}
\caption{Wave characteristics of the surface perturbation $h$, filtered at a single frequency $f_0$, for six different frequencies. The frequencies are $2.87 ~ \mathrm{Hz}$ \textbf{(A)}, $3.04 ~ \mathrm{Hz}$ \textbf{(B)}, $3.27 ~\mathrm{Hz}$ \textbf{(C)}, $3.45 ~\mathrm{Hz}$ \textbf{(D)}, $3.70 ~ \mathrm{Hz}$ \textbf{(E)}, and $4.11 ~\mathrm{Hz}$ \textbf{(F)}. The horizontal and vertical axis are in metres $[\mathrm{m}]$, while the color scale is in millimetres $[\mathrm{mm}]$. 
The patterns show the interfering sum of the incident wave with the scattered one. 
The waves are generated on the left side and propagate to the right across the vortex centred at the origin.} \label{fig:Characteristics}
\end{figure}

Some examples of the frequency filtered data are shown in Figure~\ref{fig:Characteristics}.
The obtained pattern shows a stationary wave of frequency $f_0$ scattering with the vortex, and has strong similarities to theoretical predictions in~\cite{dolan2011AB,dolan2013scattering} based on the simple flow model in Eq.~\eqref{DBT}. 
We also observe that incident waves have more wave fronts on the upper half of the vortex in comparison with the lower half.
For example, in panel D of Figure~\ref{fig:Characteristics}, there are 13 wavefronts on the top of the figure and 10 on the bottom.
This angular phase shift is analogous to the Aharonov-Bohm effect, and has been observed in previous water wave experiments~\cite{berry1980AB,vivanco1999surface}. 
Our detection method allows for a very clear visualization of this effect.

\emph{Azimuthal decomposition}. To extract the azimuthal components of $h$, we convert the signal from Cartesian to polar coordinates. 
For this we need to find the centre of symmetry of the background flow. 
We define our centre to be the centre of the vortex shadow averaged over time (the fluctuations in time are smaller than a pixel). 
To verify that this choice does not affect the end result, we performed a statistical analysis on different centre choices around this value, and added the standard deviation to the error bars. 

Once the centre is chosen, we perform a discrete Fourier transform on the irregular grid $(X_{i j}, Y_{i j})$. 
To do this, we create an irregular polar grid $(r_{i j}, \theta_{i j})$. 
We then collect the data into $N$ radial bins of width $\Delta r$ and find all the values $i_n,j_n$ in the $n$'th bin (where $n$ runs from $1$ to $N$).
The radius of $n$'th bin is defined as the mean of all the values in that bin, i.e. $r_n = \frac{1}{M_n}\sum_{i_n,j_n}r_{i_n,j_n}$ where $M_n$ is the number of points in the bin.
The $m$-components are then computed from,
\begin{equation}
h_{m}(r_n)= \frac{\sqrt{r_n}}{2\pi} \sum_{i_nj_n} h(r_{i_nj_n},\theta_{i_nj_n}) e^{-im\theta_{i_nj_n}} \Delta\theta_{i_nj_n},
\end{equation}
where $\Delta\theta_{i_nj_n} =\Delta X_{i_nj_n} \Delta Y_{i_nj_n}/r_n \Delta r$ is the irregular spacing between $\theta$ points at radius $r_n$. \\

\emph{Splitting in and out}. To extract the in-going ($B_m^-$) and out-going ($B_m^+$) amplitudes, we compute the radial Fourier transform $\hat{h}_m(k_r) = \frac{1}{2\pi}\int h_m(r)\exp(-i k_r r)dr$ over the window $[r_1,r_N]$.
This will contain two peaks: the peak at $k^-_r<0$ corresponds to the in-going mode and $k^+_r>0$ to the out-going mode.
The maximum radius is determined by the region filmed by the sensor.
The minimum radius must be far enough out in the asymptotic region for the signal to be approximately a sum of plane waves, but the window must contain enough oscillations to perform a radial Fourier transform.
We set \mbox{$r_1=15~\mathrm{cm}$} and \mbox{$r_N=40~\mathrm{cm}$}.
The number of oscillations contained within this window was typically between 1 and 3.

\begin{figure} 
\centering
\includegraphics[width=\linewidth]{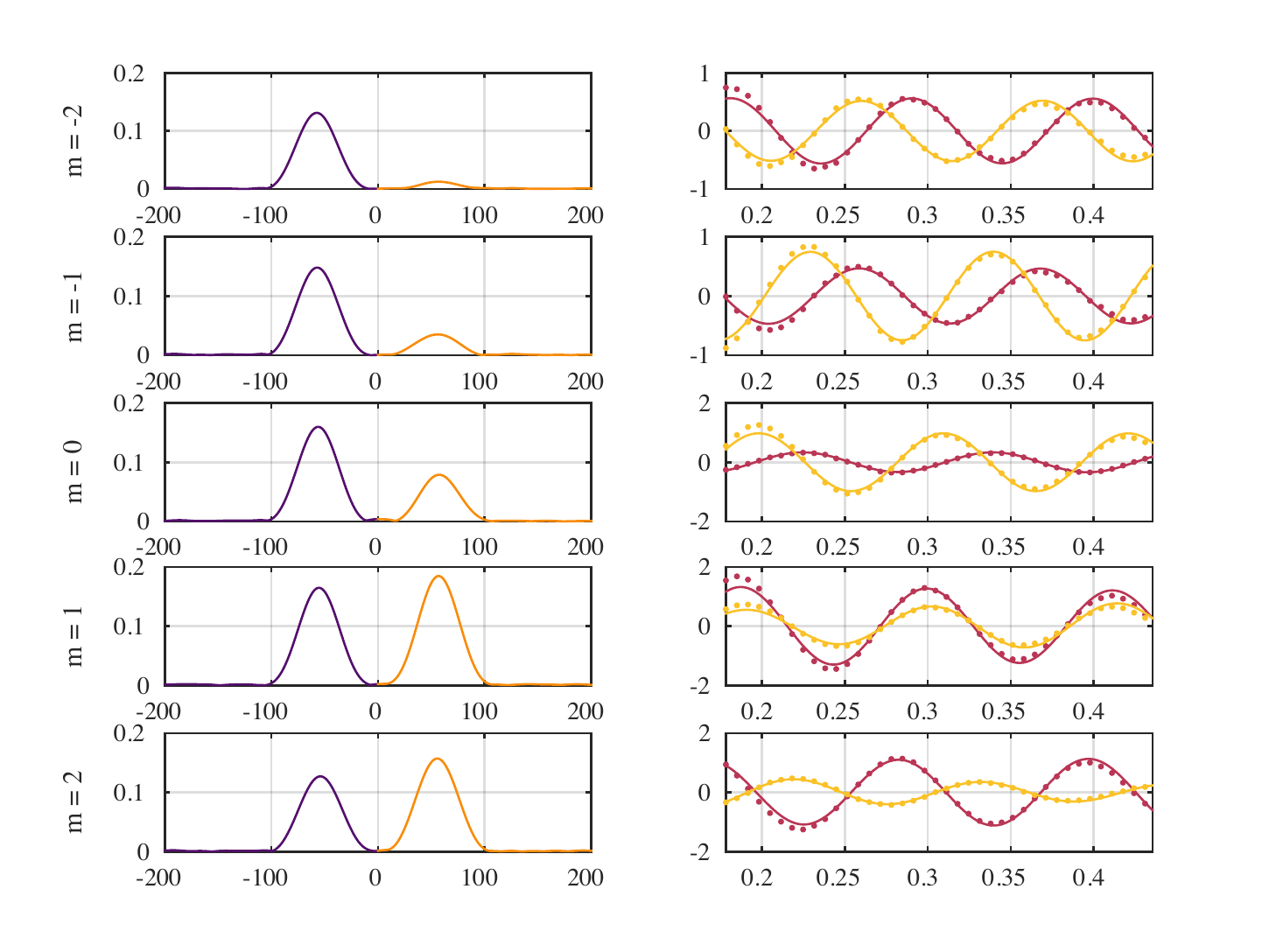}
\caption{Left side: modulus of the Fourier profiles $|\hat{h}_m(k_r)|^2$ for various $m$ modes. The horizontal axis is in $[m^{-1}]$ and the vertical is arbitrary.
Right side: radial profiles $h_m(r)$ for various $m$-modes (maroon: real part, yellow: imaginary part), with the horizontal axes in $[m]$ and again, the vertical is arbitrary.
The dots are the experimental data and we have represented only 1 in 3 points for clarity. 
The solid lines are \mbox{$B_m^+e^{ik_r^+r}+B_m^-e^{ik_r^-r}$}, which demonstrates that the WKB formula of Eq.~\eqref{WKBsol1} with constant wavelength is a good approximation in this region} \label{fig:RadProfiles}
\end{figure}

Due to the finite window, $\hat{h}_m$ contains broad peaks around the values $k_r^\pm(m)$. 
We assume that these peaks contain only one wavelength (no superposition of nearby wavelengths), which is corroborated by the fact that we have filtered in time, and the dispersion relation imposes a single wavelength at a given frequency.
To get the actual mode amplitudes, we note that the Fourier transform of a perfect sinusoid on a finite window is the cardinal sine function with amplitude $A_m(r_N-r_1)/2\pi$.
Therefore, we need to multiply the peak amplitude by a factor $2\pi/(r_N-r_1)$.
Furthermore, the secondary lobe of the cardinal sine function can have the effect of obscuring the smaller of the two peaks.
To account for this, we use a Hamming window function on $[r_1, r_N]$, 
\begin{equation}
W(n)=0.54 - 0.46\cos \left(\frac{2\pi n}{N}\right), 
\end{equation}
which is optimized to reduce the secondary lobe, hence allowing us to resolve the smaller of the two peaks~\cite{prabhu2013window}.
The cost of the Hamming window is to reduce the overall amplitude of the peak. 
We account for this by scaling the amplitudes using the value of the peak that we can resolve before $W(n)$ is applied.
In Figure~\ref{fig:RadProfiles}, we show some example radial profiles and their Fourier transforms. \\

\emph{Computing the reflection coefficient}. First of all, the reflection coefficient $\mathcal{R}$ defined in Eq.~\eqref{RTcoeffs} assumes that we have access to the amplitude of the modes at spatial infinity.
In an experiment of finite size, we measure the amplitudes at a finite radius $L$ which we take here to be $(r_1+r_N)/2$.
If the velocity field is small in this region, then the solutions are well approximated by the WKB modes in Eq.~\eqref{WKBsol1}.
The amplitude of these modes at infinity is $A_m^\mathrm{out,in}=\alpha_m^{\pm}/|c\omega|^{1/2}$, and at a finite radius $r=L$ we have $A_m^\pm(L)=\alpha_m^{\pm}/|v_{g,r}^\pm\Omega^\pm|^{1/2}$.
Hence the reflection coefficient can be expressed as,
\begin{equation}
\mathcal{R} = \sqrt{\frac{|v_{g,r}^+\Omega^+|}{|v_{g,r}^-\Omega^-|}}\frac{|A_m^+(L)|}{|A_m^-(L)|}.
\end{equation}
Finally, in the experiment we measure the amplitudes of the height perturbation. 
In the WKB approximation, these are related to the amplitudes of the velocity potential by $B_m^\pm=-\frac{1}{g}\Omega^\pm A_m^\pm(L)$, which gives,
\begin{equation} \label{ReflEnergy}
\mathcal{R} = \sqrt{\frac{|v_{g,r}^+\Omega^-|}{|v_{g,r}^-\Omega^+|}}\frac{|B_m^+|}{|B_m^-|}.
\end{equation}
The quantity under the square root is computed inserting $k_r^\pm$ in the dispersion relation of Eq.~\eqref{dispersion2} and using \mbox{$v_{g,r}^\pm = \partial_{k_r}\omega|_{k_r=k_r^\pm}$}.
This quantity is the ratio of the energies of the in-going and out-going waves, which can be seen by inserting WKB modes in Eq.~\eqref{WKBsol1} into the energy current of Eq.~\eqref{e_current} and showing it is conserved radially. 
This is explained further in \cite{coutant2016imprint}.

\subsection{Results} \label{sec:SRresults}

\begin{figure*}[t!]
    \centering
    \begin{subfigure}[t]{0.5\textwidth}
        \centering
        \includegraphics[height=3in]{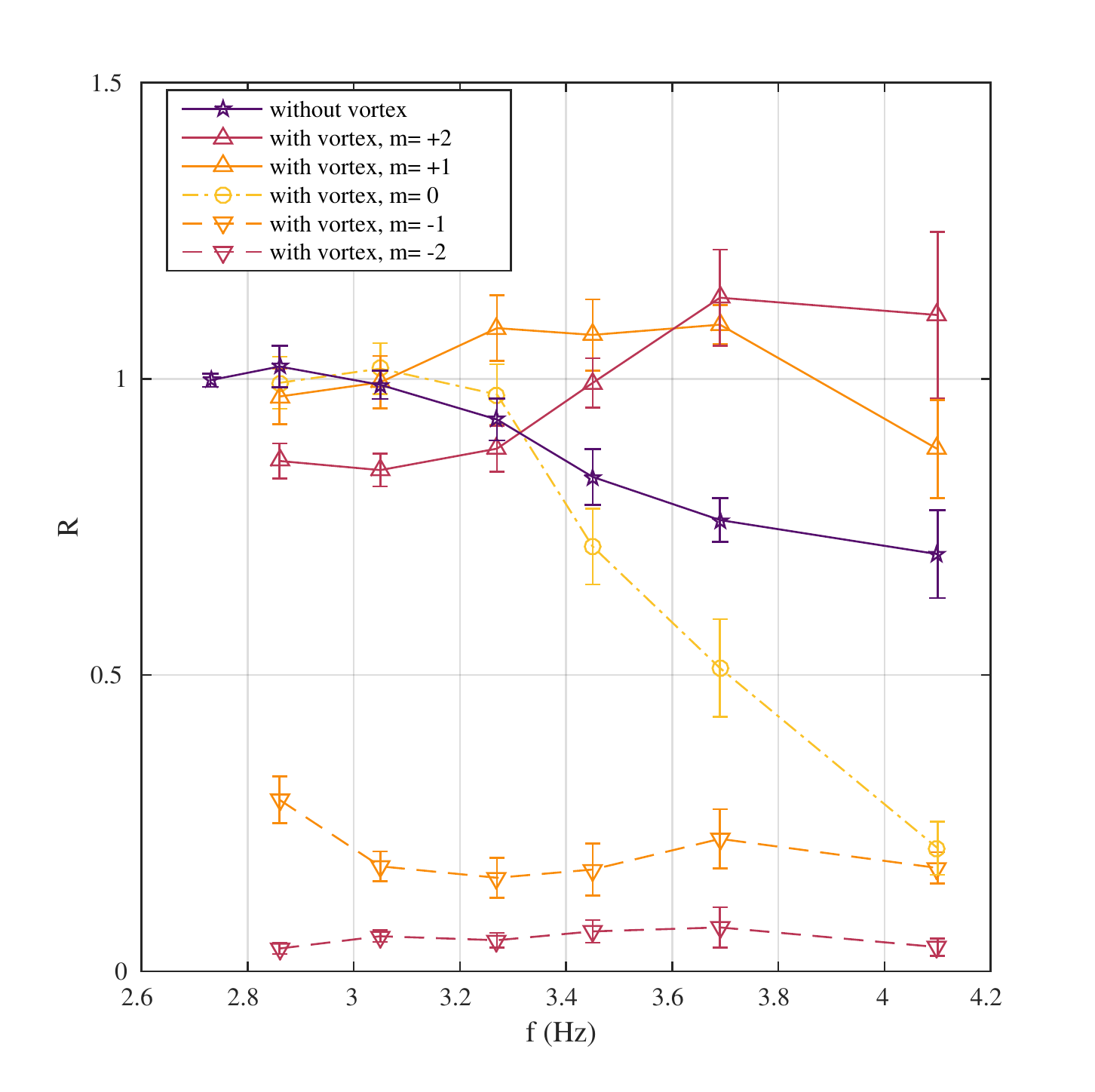}
    \end{subfigure}
    ~ 
    \begin{subfigure}[t]{0.5\textwidth}
        \centering
        \includegraphics[height=3in]{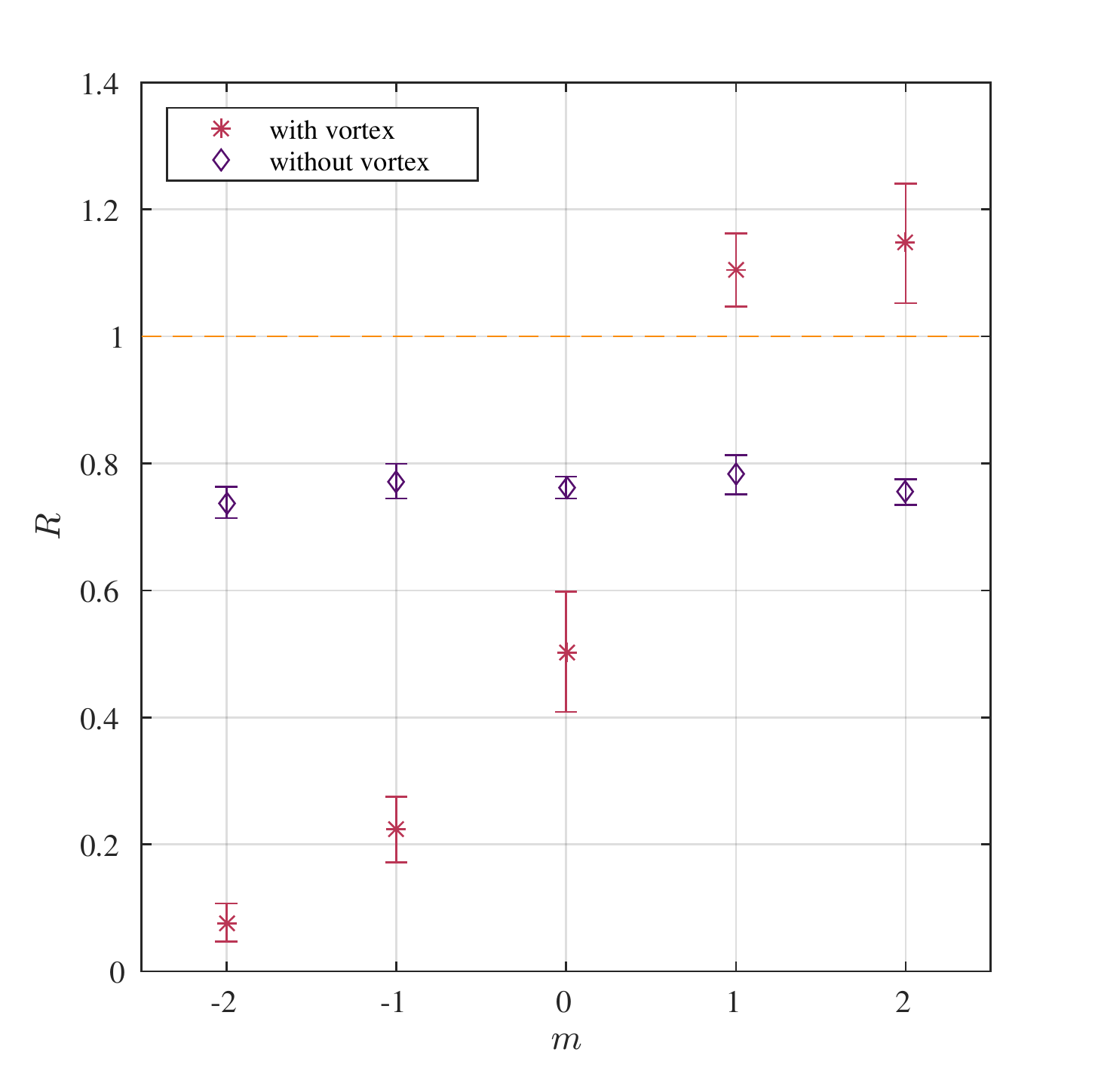}
    \end{subfigure}
    \caption{Left side: Reflection coefficients for $m\in[-2,2]$ as function $f$. Negative $m$'s are mainly absorbed by the vortex as \mbox{$\mathcal{R}<1$}. Positive $m$'s are mainly reflected and this reflection can be larger than the incident mode amplitude, indicated by \mbox{$\mathcal{R}>1$}. We also display the $m$ averaged reflection coefficient on standing water. Right side: Reflection coefficient for different $m$'s for the frequency $f = 3.70 ~ \mathrm{Hz}$. The amount of reflection when no vortex is present is independent of $m$} \label{fig:ReflExp}
\end{figure*}

On the left panel of Figure~\ref{fig:ReflExp} we represent, for several azimuthal numbers, the absolute value of the reflection coefficient $\mathcal{R}$ as a function of the frequency $f$.
This was the range in $m$ we could resolve with confidence, since $r_1$ increases with $m$ and thus, for larger $|m|$ there aren't enough wavelengths in the observation window to distinguish in and out-going modes reliably.
For the vortex experiments the statistical average is taken over 6 repetitions, except for $f=3.70 ~\mathrm{Hz}$ where we have 15 repetitions. 
The purple line (star points) shows the reflection coefficients of a plane wave in standing water ($\mathbf{V}=0$) of the same height. 
We observe a significant damping for the frequencies above 3~Hz, which could be reduced in future by working with purer water~\cite{przadka2012fourier}. 
Experiments for no vortex are averaged over 5 repetitions and over $m=[-2,2$] (the reflection coefficient of a plane wave on standing water is in theory independent of $m$). 
The errors bars indicate the standard deviation over these experiments, the energy uncertainty due to the prefactor in Eq.~\eqref{ReflEnergy} and the standard deviation over several choices of the coordinate origin. 
The main contribution comes from the variability of the value of the reflection coefficient for different repetitions of the experiment. 
We have also extracted the signal-to-noise ratio for each experiment, and its contribution to the error bars is negligible.

We observe two distinct behaviours, depending on the sign of $m$.
Negative $m$'s (waves counter-rotating with the vortex) have a low reflection coefficient, which means that they are essentially absorbed in the vortex hole. 
On the other hand, positive $m$'s have a reflection coefficient close to 1. 
In some cases this reflection is above one, meaning that the corresponding mode has been amplified by scattering with the vortex. 
To confirm this amplification we have repeated the same experiment 15 times at the frequency $f = 3.8~\mathrm{Hz}$, for which the amplification was the highest.
We present the result on the right panel of Figure~\ref{fig:ReflExp}.
On this figure we clearly observe that the modes $m=1$ and $m=2$ are amplified, with the precise values being \mbox{$\mathcal{R}_{m=1} = 1.09 \pm 0.03$}, and \mbox{$\mathcal{R}_{m=2} = 1.14 \pm 0.08$} respectively. 
We also display the reflection coefficients obtained for a plane wave propagating on standing water of the same depth. Unlike what happens in the presence of a vortex, the reflection coefficients are all below 1 (within error bars). For low frequencies it is close to 1, meaning that the wave is propagating without losses, while for higher frequencies it decreases due to a loss of energy during the propagation, i.e.~damping. 

The main check for consistency with superradiance is that $\tilde{\omega}_h<0$, for which we need to know the location of the system's horizon $r_h$, and the angular frequency of the flow at that location $\Omega_h$.
The angular frequency is displayed in the bottom left panel of Figure~\ref{fig:VelocityField}.
To know the location of the horizon, one would need an effective field theory for the waves including dispersive effects, non-constant height and also vorticity (since $\bm\nabla\times\mathbf{V}\neq0$ for $r<r_0$).
In other words, we would need to rederive Eq.~\eqref{waveeqn} without the shallow water and constant height approximations and also account for rotational flow. 
Since this is beyond the scope of this thesis, we provide an estimate for the horizon based on the following argument.

\begin{figure} 
\centering
\includegraphics[width=\linewidth]{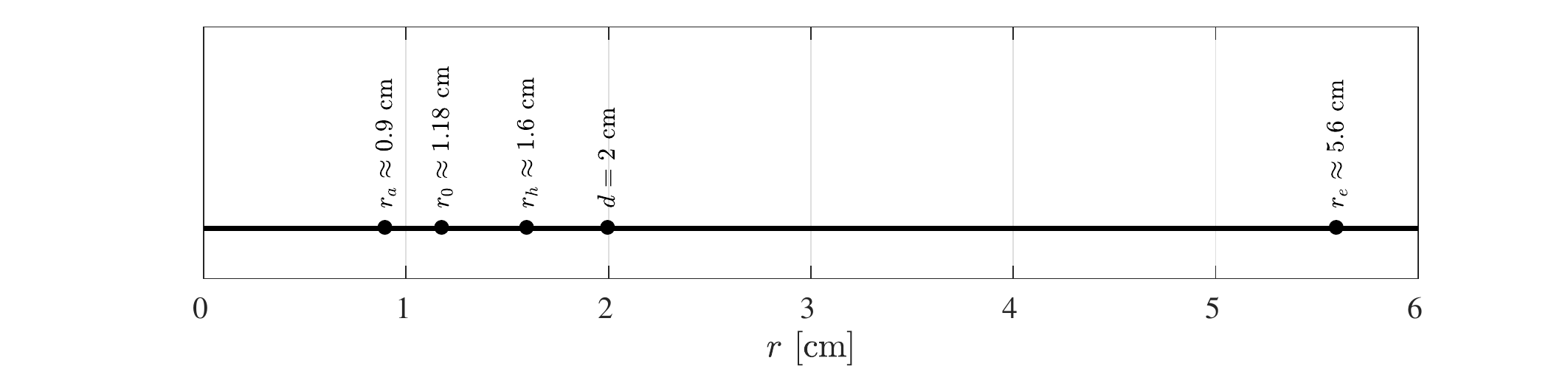}
\caption{Our estimations of important locations in the flow: the air-core $r_a$, the viscous core $r_0$, the horizon $r_h$, the drain hole radius $d$ and the ergosphere $r_e$, which is approximated by using the average wave speed far away $\tilde{c}\approx0.25~\mathrm{m/s}$ and computing $r_e=C/\tilde{c}$.} \label{fig:radlocat}
\end{figure}

As a first approximation, we assume that the wave equation is not modified by the presence of vorticity in the background (this assumption is readdressed in Chapter~\ref{chap:vort}).
Then using the deep water dispersion relation of Eq.~\eqref{dispersion2}, the location where the out-going $m=0$ mode has $|\mathbf{v}_g^\mathrm{lab}|=0$ is given by,
\begin{equation} \label{DeepHorizon}
|V_r(r_h)| = \frac{g}{4\omega}.
\end{equation}
Now, it was shown in \cite{richartz2015rotating} that for small height deformations in the shallow water regime, the perturbations obey the wave equation of Eq.~\eqref{waveeqn} with $H=H(r)$ and $g$ replaced by the effective gravitational acceleration,
\begin{equation}
\tilde{g} = g + V_rV_r'H' + V_r^2H'',
\end{equation}
where prime denotes derivative with respect to $r$.
For $H$ we take the simple height profile in Eq.~\eqref{height_profile} and take \mbox{$V_r\sim1/r$}, which according to Eq.~\eqref{sharpplug} is valid for $r>a$.
Replacing $g\to\tilde{g}$ in Eq.~\eqref{DeepHorizon}, we find \mbox{$r_h\approx1.6~\mathrm{cm}$} for all values of $f_0$ in our experiments, which is just outside our estimate of the drain radius $a$ we found in Section~\ref{sec:flow_values}.
Hence, the horizon is located inside the vortex throat\footnote{This is further corroborated by the fact that one can create \textit{sound cones} (analogous to light cones) by touching the free surface with a pin. The small ripples that are created sketch out a conic shaped area which is the region they are able to influence. The sound cones point toward the drain inside the vortex throat, indicating the horizon is somewhere in this region.}.

Using Figure~\ref{fig:VelocityField}, we estimate $\Omega_h\approx 40~\mathrm{rad/s}$ in this region, which is consistent with the observed amplification of the $m=1$ and $m=2$ modes.
As a final note, it was shown in \cite{richartz2009generalized} that the minimum conditions for superradiance are the presence of an ergosphere and an absorbing boundary condition.
The wave speed in our experiments was roughly \mbox{$[0.2,0.3]~\mathrm{m/s}$}. 
These speeds are clearly exceeded by $V_\theta$ in Figure~\ref{fig:VelocityField}, hence the condition for an ergosphere is met.
The absorbing boundary condition could either be provided by a horizon, which we have estimated, or simply by the fact that there is a hole in our set up where negative wave energy can be deposited.
Another possibility is that the negative wave energies are dissipated away in the vortex throat, similar to superradiant scattering off a rotating cylinder \cite{zeldovich1972amplification,cardoso2016detecting}.

Lastly, we performed several additional checks to verify that the amplification was consistent with superradiant scattering:

\begin{enumerate} [noitemsep]
\item The second harmonic at $2f_0$ has a relative amplitude to the fundamental of below $14\%$ in all experiments, which is of the order of the noise in the sensor data.
If the amplification were a result of a non-linear wave interaction, then an increase in energy at $f_0$ would be accompanied by a decrease in energy at $2f_0$.
We checked to make sure that this was not the case.
\item We extracted the signal-to-noise ratio by comparing the standard deviation of the noise to the value of our signal. 
It is sufficiently high to exclude the possibility that the amplification we observed is due to a noise fluctuation, and its contribution is negligible compared to other sources of error.
\item We estimate the coupling of waves with $m \neq m'$ through asymmetry. The change of the reflection coefficient due to this coupling is of the order of $|\tilde v^l/v_g|$, where $\tilde v^l$ is the angular Fourier component of azimuthal number $l=m-m'$ of the velocity field. This ratio is smaller than $3\%$ in all experiments.
\end{enumerate}

\section{Summary}

In this chapter, we have demonstrated that the simple bathtub vortex exhibits rotational superradiance, a phenomenon perhaps most commonly associated with astrophysical black holes.
We predicted the form of the superradiance spectrum (see Figure~\ref{fig:ReflB5} in particular) using the shallow water theory outlined in the previous chapter.
We then conducted an experiment to verify that this behaviour was present in a real system where the ideal conditions used for the theory are not achievable.
Particularly, to drive a fast enough fluid flow we required a certain water height, which for the frequency range allowed by our apparatus meant we were in the deep water regime.
We also found from our measurement of the velocity field that vorticity is non-negligible close to the drain, and the surface of the water is not flat in the region where we expect the horizon to be.

Despite these experimental realities, we were still able to observe superradiant amplification in the laboratory.
Particularly, we showed that a wave scattering on a rotating vortex flow can extract energy from the system. 
More generally, our results show that the phenomenon of superradiance is very robust and requires few ingredients to occur: namely high angular velocities, allowing for negative energy waves, and a mechanism to absorb these negative energies. 
Furthermore, the experiment revealed that superradiance was still observable despite significant damping of the waves at large frequencies.
It seems therefore that superradiance is not only observable, but can withstand the less-than-ideal experimental conditions which we haven't accounted for in our theoretical analysis. 

However, the experiment did reveal some key differences to the spectrum for $\mathcal{R}$ in Figure~\ref{fig:ReflB5}, which (like our experiment) has parameters satisfying $C>D$.
One particularly interesting difference is that, in experiment, the $m=2$ mode had a higher reflection coefficient than that of $m=1$.
This is in stark contrast to the irrotational, shallow water prediction where \mbox{$\mathrm{max}(\mathcal{R})$} decreases with $m$.
To pinpoint the reason for this difference, one would require an effective field theory which incorporates more realistic conditions; specifically non-constant water height, dispersion and vorticity.
The effect of small height deformations has already been investigated in \cite{richartz2015rotating} where it was found that maximum amplification is still achieved for $m=1$.
However, the effect of larger deformations remains unclear.

Weak dispersion has been studied in a range of 1D analogue black hole models \cite{corley1996spectrum,coutant2012smatrix,coutant2016imprint,coutant2018kdv}, 
usually in the context of Hawking radiation.
Generalised superradiance was discussed in \cite{richartz2013dispersive} and results were derived in the weak dispersive case.
However the nature of superradiance when dispersion becomes significant remains unclear, and this represents a possible avenue for future research.
The problem in the deep water regime (strong dispersion) is more difficult to address since the usual trick is to expand the dispersion relation (see e.g. \cite{coutant2016imprint}) to the next order, but the expansion of Eq.~\eqref{dispersion2} does not converge in this regime. 
A possible way to explore superradiance in the deep water regime could be to use the WKB analysis in Section~\ref{scatterO1}.

Finally, the effect of vorticity is to couple the usual scalar perturbations $\phi$ to additional degrees of freedom, thereby greatly increasing the difficulty of the problem.
In Chapter \ref{chap:vort}, we will demonstrate the effect of vorticity in a purely rotating system on it's characteristic frequency spectrum.
To learn about the scattering coefficients, one possibility would be to perform the numerical simulations of Section~\ref{sec:Rnums} for Eqs.~\eqref{EqMot2} and \eqref{EqMot3} of Chapter \ref{chap:vort}.
However, computation of the Frobenius expansion to obtain the initial conditions is more delicate in this case since there are extra rotational modes to solve for.

\chapter{Relaxation} \label{chap:qnm}
The relaxation of a perturbed system back to equilibrium is well approximated at late times by a set of damped resonances called quasinormal modes (QNMs).
Mathematically, these are solutions to the governing wave equation obeying purely out-going boundary conditions on the system's outer boundary (or in-going on an inner boundary, such as a horizon).
Their application spans a wide variety of disciplines, from molecular physics and seismology, to fluid dynamics and black hole physics \cite{berti2009review}.
Sometimes called eigenmodes of dissipative systems\footnote{This name is slightly deceptive since they do not form a complete basis from which an arbitrary perturbation can be constructed. It does however convey the idea that they are the characteristic modes of the system.}, they are used to approximate how a system responds to a disturbance at late times when the linear approximation of the perturbation becomes valid.
Quasinormal (QN) ringing of a field $\Psi$ is of the form,
\begin{equation}
\Psi \sim e^{-i\omega_\mathrm{QN}t}, \qquad \omega_\mathrm{QN} = \omega + i\Gamma,
\end{equation}
where the real part of $\omega_\mathrm{QN}$ gives the oscillation frequency and the imaginary part determines the decay rate of the oscillation.

Perhaps the most common example of QNMs in everyday experience is that of a ringing bell, and it is this analogy that gives rise to the name \textit{ringdown} when talking about the response of a black hole to perturbations.

QN frequencies are determined completely by the parameters of a system, a fact which has led researchers to ponder whether one can ``hear the shape of a system'' just by listening to it's characteristic frequencies \cite{moss2002bell}.
The ``shape'' in the case of a black hole (in a vacuum) refers to it's mass, angular momentum and charge, and the fact that there are only a handful of parameters to deal with makes the computation of black hole QNMs a relatively simple task \cite{konoplya2011qnms}, compared to other system's which may have a much more complicated ``shape''.
In reality, black holes are not found in isolation and are surrounded by distributions of matter, leading to a more complicated response of the system to disturbances at late times. 
It turns out, however, that the vacuum QNMs still dominate the response at intermediate times \cite{barausse2014environment} and hence their study is a worthwhile endeavour.

It was originally proposed in the 1970s to use the QNM spectrum to measure the mass of a black hole, \cite{press1972gravitational,schutz2009physics,echeverria1989gravitational}. 
With recent advancements in gravitational wave astronomy, this long held goal has started to become a reality \cite{abbott2016merger,abbott2016properties}.
Specifically, binary black hole mergers produce large enough gravitational waveforms for measurement here on earth, thereby allowing us to test GR in the strong gravity regime \cite{yunes2016implications}. 
The waveform consists of three phases: inspiral, merger and ringdown, where the ringdown phase is comprised of the QNMs.
It is not an understatement, therefore, to say that QNMs have played a key role in providing us with a window into the distant universe, through which we may glimpse some of the most elusive objects that nature has to offer.

Having seen in the previous chapter that superradiance (which is usually associated with black holes) can also be observed in surface gravity wave experiments in the laboratory, a natural next step is to ask whether the successes of QNMs can also be carried over to fluids.
In this chapter, we will use techniques developed in black hole physics \cite{goebel972vibrations,cardoso2009geodesic} to measure the parameters of the DBT experiment outlined in Section~\ref{sec:detectSR} through the observation of its characteristic frequency spectrum.
The QNM spectrum of the simple DBT model of Eq.~\eqref{DBT} in the shallow water regime has already been thoroughly explored in the literature \cite{cardoso2004qnm,berti2004qnm,dolan2012resonances}. 
We begin by reviewing some of the techniques used compute QNMs outlined therein, paying particular attention to the WKB method which provides an intuitive picture of QN ringing.
We will then adapt these techniques so that they can be applied to our experiment, which as we saw in the previous chapter creates deep water gravity waves.

Lastly, we note that in this chapter, we mainly focus on methods of approximating QNMs. There exists a range of techniques to compute their frequencies more precisely \cite{cardoso2004qnm,dolan2012resonances}, including the continued fraction method and direct simulation of the wave equation.
We will make use of the latter toward the end of this chapter.

\section{QNM theory} \label{sec:qnm_est}



\begin{figure} 
\centering
\includegraphics[width=0.5\linewidth]{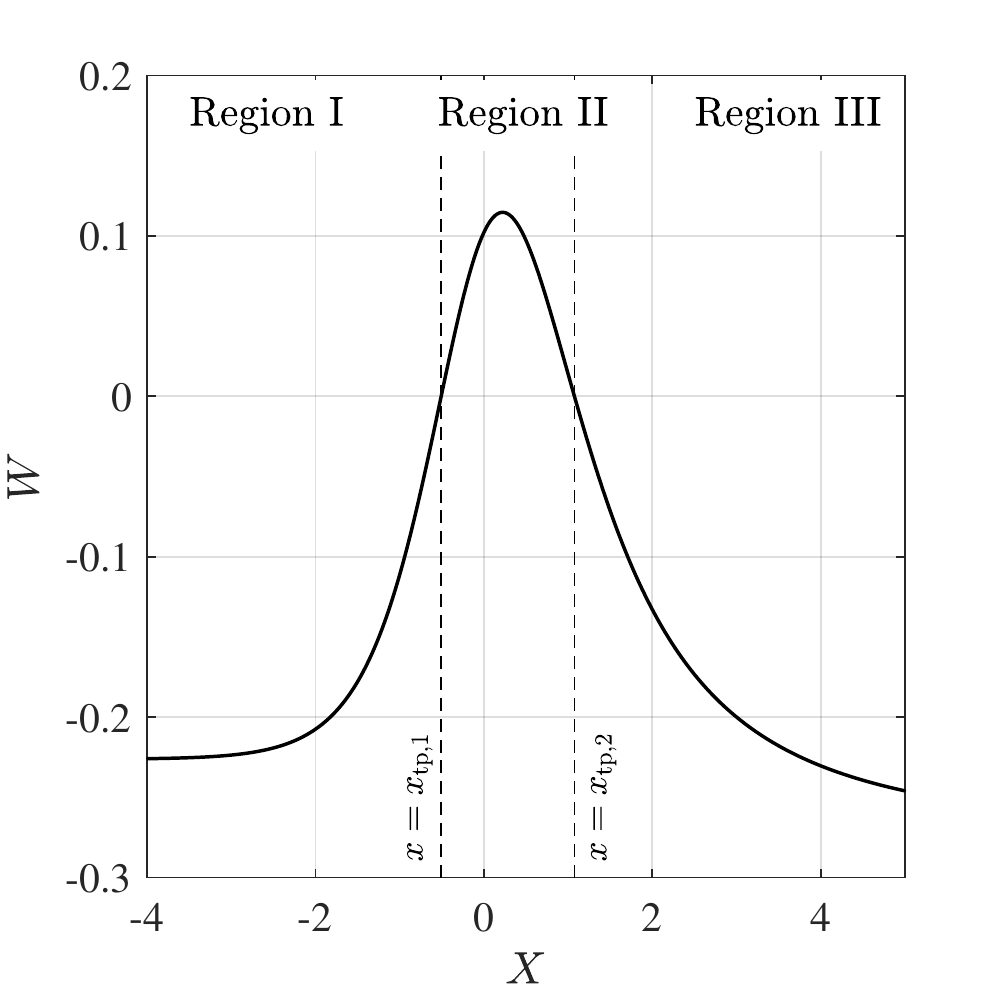}
\caption{An example of the effective potential in Eq.~\eqref{potential2} in terms of the dimensionless tortoise coordinate $X=x+\frac{1}{2}\log\frac{x-1}{x+1}$.
$W$ has a maximum of two turning points, which divides the $X$ range into three regions.
For QNMs, the two turning points are close together, and region II is a narrow strip where Eq.~\eqref{modeeqn2} can be solved exactly by approximating $W$ as a parabola.} \label{fig:qnm_rstar}
\end{figure}

The WKB approximation provides a simple way to compute the QNM frequencies, which is where we begin our discussion.
We will focus on the DBT model of Eq.~\eqref{DBT}, which we recall is independent of $t$ and $\theta$.
Therefore, each $m$ and $\omega$ component will evolve independently according to the mode equation, which we write here in dimensionless form using the variables of Eq.~\eqref{adim},
\begin{equation} \label{modeeqn2}
-\psi_m'' + W\psi_m = 0,
\end{equation}
where prime denotes derivative with respect to the rescaled tortoise coordinate $X=r_*c/D$, and the effective potential in dimensionless form is,
\begin{equation} \label{potential2}
W(x) = -\left(\sigma-\frac{mB}{x^2}\right)^2 + \left(1-\frac{1}{x^2}\right)\left(\frac{m^2-1/4}{x^2}+\frac{5}{4x^4}\right).
\end{equation}
As we saw in the previous chapter, the effective potential for the DBT flow contains a maximum of two turning points $x_\mathrm{tp}$, and WKB modes are valid everywhere except in the vicinity of these.
In Figure~\ref{fig:qnm_rstar}, we give an example of $W$ and indicate the three different regions: regions I ($1<x<x_{\mathrm{tp},1}$) and III ($x>x_{\mathrm{tp},2}$) are classically allowed and hence the modes are oscillatory there, whereas region II ($x_{\mathrm{tp},1}<x<x_{\mathrm{tp},2}$) is classically forbidden and results in growing/decaying modes.
In the previous chapter, our approach was to use WKB modes in all three regions and then solve the limiting form of the mode equation at the turning points to connect the solutions.
However, it is known from exact numerical solutions that for $\sigma_\mathrm{QN}$, the two turning points $x_{\mathrm{tp},1}$ and $x_{\mathrm{tp},2}$ are close together \cite{iyer1987black1,iyer1987black2}, and thus it is not appropriate to use WKB modes in region II. 
Instead, we may solve the mode equation exactly in region II by expanding the potential in the vicinity of the stationary point $x=x_\star$ (i.e. the peak), which is determined by the condition $\partial_xW(x_\star)=0$\footnote{Since there is a one-to-one correspondence between $x$ and $X$ for $x>1$, this is equivalent to $W'(X_\star)=0$.}.
To quadratic order, $W$ is given by,
\begin{equation} \label{W_expansion}
W(X) = W_\star + \frac{1}{2}W_\star''(X-X_\star)^2 + \mathcal{O}\left((X-X_\star)^3\right),
\end{equation}
where subscript $\star$ indicates that a quantity is evaluated at $X=X_\star$.
Defining the rescaled coordinate $Z = (-2W_\star'')^{1/4}e^{i\pi/4}(X-X_\star)$, Eq.~\eqref{modeeqn2} takes on the form of the parabolic cylinder equation,
\begin{equation} \label{cylinderEqn}
\partial_Z^2\psi_m + \left(\nu+\frac{1}{2}-\frac{1}{4}Z^2\right)\psi_m = 0, \qquad \nu = \frac{iW_\star}{\sqrt{-2W_\star''}}-\frac{1}{2},
\end{equation}
whose solutions are the parabolic cylinder functions $D_\nu(Z)$ \cite{abramowitz1965handbook}. 
Hence, the general solution in region II can be written as a combination of the two linearly independent solutions,
\begin{equation}
\psi_m = A_1D_\nu(Z)+A_2D_{-\nu-1}(iZ).
\end{equation}
This is then matched to the WKB solutions in regions I and III.
On the horizon (region I) the mode is purely in-going, and at infinity (region III) we demand a purely out-going solution according to the definition of a QNM.
Using the asymptotics of $D_\nu$, a requirement for the boundary conditions to be met is $\Gamma(-\nu)^{-1}=0$ \cite{berti2009review} (where $\Gamma(\nu)$ is the gamma function, not to be confused with the decay rate introduced at the start of this chapter\footnote{This is the only time we refer to the gamma function and reserve $\Gamma$ for the decay time from here on.}).
This is satisfied for $\nu=n$ where $n\in\mathbb{Z}$ is called the overtone number (which we will see essentially classifies QNMs according to their lifetime).
Hence the QN frequencies are those satisfying the condition,
\begin{equation} \label{QNMcond}
n+\frac{1}{2}=\frac{iW_\star}{\sqrt{-2W_\star''}}\Bigg|_{\sigma_\mathrm{QN}}.
\end{equation}
Eq.~\eqref{QNMcond} is solved by complex frequencies $\sigma_\mathrm{QN}\in\mathbb{C}$ which can be found numerically.
However, in light of the fact that this is already an approximation, it is more illuminating to apply further approximations to obtain expressions for the real and imaginary parts of $\sigma_\mathrm{QN}$ respectively.

Given that we know $x_{\mathrm{tp},1}$ and $x_{\mathrm{tp},2}$ are close together, it is not a drastic leap to assume that they are sufficiently close that they may be approximated by a single turning point.
For this to occur, the turning points of $W$ must coincide with the stationary point at $x=x_\star$.
This means we require $W_\star=0$, which only occurs for a specific frequency $\sigma=\sigma_\star$.
Hence the oscillatory part of the QN frequency can be approximated as \mbox{$\mathrm{Re}[\sigma_\mathrm{QN}]\simeq\sigma_\star$}.

Eq.~\eqref{QNMcond} can then be solved for imaginary component, which we assume is contained in a small perturbation $\delta\sigma$ about $\sigma_\star$.
Since $W_\star=0$ in this approximation, we replace $W_\star$ in Eq.~\eqref{W_expansion} with \mbox{$(\partial_\sigma W_\star)\delta\sigma$}, which can be solved to find,
\begin{equation} \label{QNMimag}
\mathrm{Im}[\sigma_\mathrm{QN}] \simeq -\left(n+\frac{1}{2}\right)\sqrt{\frac{-2W_\star''}{(\partial_\sigma W_\star)^2}},
\end{equation}
This supports the interpretation of the overtone number $n$ as indexing the size of the $\mathrm{Im}[\sigma_\mathrm{QN}]$. 
Usually only the $n=0$ mode is studied since modes with $n>0$ decay more quickly.

This single turning point approximation for the QN frequencies means we are approximating the QNMs as modes which orbit the (analogue) black hole on \textit{light-rings} \cite{cardoso2009geodesic}.
The light-rings are null geodesics of the underlying effective metric which form closed (circular) orbits \cite{goebel972vibrations}.
They correspond to stationary points $x=x_\star$ in the geodesic potential in Eq.~\eqref{GeoPotential} which, as we showed in Section~\ref{sec:Eik}, is identical to the effective potential in Eq.~\eqref{potential2} in the limit of large $m$.
The light-ring conditions are,
\begin{equation} \label{potential_LRconds}
W_\star=0, \qquad W_\star'=0,
\end{equation}
where $W_\star''>0$ ($W_\star''<0$) indicates a stable (unstable) orbit, which can be solved to find the light-ring frequency $\sigma=\sigma_\star$\footnote{Strictly, when applying these conditions, the terms $-1/4x^2$ and $5/4x^4$ should be discarded from Eq.~\eqref{potential2} to cast it in the form of the geodesic potential, discussed further in Appendix~\ref{app:geos}. However, the first term is always small for $m$ not too close to to zero (the results are barely distinguishable for $m>2$) and the second is negligible everywhere except near the horizon.}.
When applying this approximation, it is important to bare in mind that null geodesics are the paths taken by massless particles and since waves have finite extent, a particle description does not capture all of the physics.
This means that the approximation of QNMs as light-ring modes is also not exact.
However, the particle description improves for shorter wavelengths as the wave becomes more localised, and in the limit of vanishingly small wavelength (i.e. the eikonal approximation) the waves would follow geodesics exactly.
As stated above, this is achieved formally by setting $|m|\to\infty$ in the potential.
Hence, the approximation $\mathrm{Re}[\sigma_\mathrm{QN}]\simeq\sigma_\star$ works best in the large $m$ limit (this same reasoning applies to photons orbiting a black hole \cite{mendoza2009connection}).

In the geodesic description, the spread of neighbouring trajectories toward or away from one another is encoded in a quantity $\Lambda$ called the Lyapunov exponent, where $\Lambda<0$ ($\Lambda>0$) indicates convergent (divergent) trajectories, see \cite{cardoso2009geodesic} for a discussion.
Computation of the Lyapunov exponent leads to the expression in Eq.~\eqref{QNMimag} upto the factor $-(n+1/2)$, with the geodesic potential replacing the effective potential \cite{cardoso2009geodesic}.
This provides an intuitive interpretation QNM decay.
A mode with the correct amount of energy (determined by the frequency) will orbit the black hole on the light-ring.
However, due to the spreading of geodesics there, the mode will gradually leak away to infinity, which can be envisaged as rolling down the peak of the effective potential barrier.

Despite the close connection between QNMs and the light ring modes, there is a key difference between them in principle:
the QNMs are modes which satisfy specific boundary conditions at infinity, whereas the light ring modes are determined by local properties of the underlying geometry.
Therefore, whilst the light-ring modes are a useful tool in approximating the QNMs, a wave only really ``knows'' whether it is exactly a QNM or not once it reaches infinity.
This is linked to the the reason why the vacuum QNMs provide a good description of black hole ringdown at intermediate times even in the presence of matter distributions: 
the light ring modes can still be present in the signal even if they are not precisely QNMs of the whole system.

In general, one must be careful when using light-rings to approximate QNMs. Specifically, in potentials with more complicated structure there can be a significant differences between the two \cite{khanna2017ringing}.

\section{Generalising the notion of trajectories}

We saw in the previous chapter that dispersion plays a role in gravity wave experiments when the water height is not shallow.
The wave equation in water of arbitrary depth is given by Eq.~\eqref{waveeqndisp}, which cannot be written in the form of the KG equation, meaning that we no longer have the notion of an analogue geometry.
Recalling that our ultimate goal is to observe QN ringing in our analogue black hole experiment, we need to find a method of approximating the QNMs that does not rely on the effective metric. 
Fortunately, the notion of wave-trajectories (and therefore light-rings) can be extended to the dispersive regime in such a way that there is no reference to the effective metric, and one only requires knowledge of the dispersion relation.
We demonstrate how to do this now.

Before proceeding, we write the dispersive wave equation of Eq.~\eqref{waveeqndisp} in the form,
\begin{equation} \label{waveeqnabstract}
D_t^2\phi + \mathcal{D}(-i\bm\nabla)\phi = 0,
\end{equation}
where $\mathcal{D}$ is the dispersion function, which allows us in principle to apply our formalism to arbitrary dispersion relations.
For example, for the dispersion relation of gravity waves in Eq.~\eqref{dispersion2} we have,
\begin{equation}
\mathcal{D}(\mathbf{k}) = gk\tanh(kH), \qquad k=||\mathbf{k}||,
\end{equation}
which in the shallow water regime $kH\ll1$ reduces to,
\begin{equation}
\mathcal{D}_\mathrm{shal}(\mathbf{k}) = gHk^2,
\end{equation}
and for deep water $kH\gg1$ is,
\begin{equation}
\mathcal{D}_\mathrm{deep}(\mathbf{k}) = gk.
\end{equation}
For simplicity, we assume that the physics we care about in this section takes place in the region where \mbox{$H\simeq H_\infty$}, which we will see turns out to be a good approximation.

We start by substituting the WKB ansatz $\phi\simeq A\exp(iS/\epsilon)$ into Eq.~\eqref{waveeqnabstract}, where $A$ is a slowly varying amplitude and $S$ is the phase of the wave.
We saw in Section~\ref{sec:Eik} for the shallow water case that, at leading order in the expansion parameter $\epsilon$, this yields the dispersion relation when we define $-\omega=\partial_tS_0$ and $\mathbf{k}=\bm\nabla S_0$.
In Appendix \ref{app:geos}, we show that the dispersion relation is related to an effective Hamiltonian for the geodesics, which for a general dispersion relation is given by,
\begin{equation} \label{RayHamiltonian}
\mathcal{H} = -\frac{1}{2}\left(\omega-\mathbf{V}\cdot\mathbf{k}\right)^2 + \frac{1}{2}\mathcal{D}(\mathbf{k}).
\end{equation}
This is discussed in further detail in \cite{torres2017rays}.
The dispersion relation is recovered from Eq.~\eqref{RayHamiltonian} by enforcing the Hamiltonian constraint $\mathcal{H}=0$.

The trajectories are obtained from Eq.~\eqref{RayHamiltonian} by applying Hamilton's equations, which in polar coordinates are,
\begin{equation}
\begin{split}
\dot{t} = -\partial_\omega\mathcal{H}, \qquad \dot{r}  =\partial_{k_r}\mathcal{H}, \qquad \dot{\theta} = \partial_m\mathcal{H}, \\
\dot{\omega} = \partial_t\mathcal{H}, \qquad \dot{k_r} = \partial_r\mathcal{H}, \qquad \dot{m} = \partial_\theta\mathcal{H},
\end{split}
\end{equation}
where overdot signifies derivative with respect to $\lambda$, which parametrises the path along the trajectories.
In the shallow water regime, these equations yield precisely the geodesics.
We also see directly from these equations that $\omega$ and $m$ are conserved for a system whose Hamiltonian does not depend on $t$ or $\theta$.

In the Hamiltonian formalism, the light-ring is the location where a mode does not propagate in the radial direction ($\dot{r}=0$) and does not move away from that location ($\dot{k_r}=0$), which leads to the conditions,
\begin{equation} \label{HJ_LRconds}
\partial_{k_r}\mathcal{H}=0, \qquad \partial_r\mathcal{H}=0.
\end{equation}
These are of course equivalent in shallow water to the conditions on the effective potential in Eq.~\eqref{potential_LRconds}.
The conditions in Eq.~\eqref{HJ_LRconds} are more general, however, since they can be applied to systems of arbitrary water height.
Below, we illustrate the application of both the potential and the Hamiltonian formulations in finding the light-ring modes of a DBT flow in shallow and deep water respectively.

Finally, it is also possible to extend the notion of the Lyapunov exponent beyond the effective metric, thereby allowing the computation of decay times in the dispersive regime.
However, since the decay of the characteristic modes was not observable in the experiment we present later on, we do not give a derivation of this here and refer the reader instead to \cite{torres2017rays}.

\section{Shallow water}

Simple analytic formula for the light-ring frequencies can be obtained in the shallow water regime, which we derive here using the potential formulation with the conditions in Eq.~\eqref{potential_LRconds}.
Since the approximation $\mathrm{Re}[\sigma_\mathrm{QN}]\simeq\sigma_\star$ works best in the eikonal limit, we apply the large $m$ approximation directly to Eq.~\eqref{potential2},
\begin{equation} \label{potential3}
W(x) = -\left(\sigma-\frac{mB}{x^2}\right)^2 + \left(1-\frac{1}{x^2}\right)\frac{m^2}{x^2}.
\end{equation}
The condition $W'(x_\star)=0$ can be solved to find the location of the light-ring,
\begin{equation}
\frac{1}{x_\star^2} = \frac{\lambda B+1/2}{B^2+1},
\end{equation}
where we have defined $\lambda = \sigma/m$ (this parameter is the ratio of wave energy to angular momentum discussed in Section~\ref{sec:currents}).
Inserting into $W(x_\star)=0$ gives the real part of the light ring frequency,
\begin{equation}
\lambda_\star^\pm = \frac{B}{2}\pm\frac{\sqrt{B^2+1}}{2},
\end{equation}
and since we are interested in positive frequencies, we take $+$ for the $m>0$ modes and $-$ for $m<0$.
In terms of dimensional quantities, the light-ring frequency and location are given by,
\begin{equation} \label{LRfreqsShallow}
\omega_\star(m) = \frac{mc^2}{2D^2}\left(C\pm\sqrt{C^2+D^2}\right), \qquad r_\star = \frac{D}{c}\left(\frac{2\sqrt{C^2+D^2}}{\sqrt{C^2+D^2}\pm C}\right)^{\frac{1}{2}}.
\end{equation}
Hence in shallow water, $\omega_\star$ is linear in $m$ and $r_\star$ is $m$-independent.
Taking the second derivative of $W$ and inserting into Eq.~\eqref{QNMimag}, we find for the imaginary part,
\begin{equation}
\Gamma = -\left(n+\frac{1}{2}\right)\frac{c^2}{D}\sqrt{\beta_\pm^2-\beta_\pm}, \qquad \beta_\pm = 2\left(B^2+1\pm B\sqrt{B^2+1}\right),
\end{equation}
which is also independent of $m$.

\section{Deep water}

It is also possible to obtain analytic expressions for the light-ring modes in the deep water regime, where the Hamiltonian is given by,
\begin{equation} \label{HamiltonianDeep}
\mathcal{H} = -\frac{1}{2}\left(\omega - \frac{mC}{r^2} - \frac{k_rD}{r}\right)^2 + \frac{1}{2}g\sqrt{k_r^2+\frac{m^2}{r^2}}.
\end{equation}
For fixed background properties $(C,D)$ there are four free parameters, $(\omega,m,k_r,r)$.
We may then use the conditions in Eq.~\eqref{HJ_LRconds} with the Hamiltonian constraint $\mathcal{H}=0$ to eliminate two free parameters, resulting in the following expressions for the light-ring frequency and radius:
\begin{equation} \label{LRfreqsDeep}
\omega_\star(m) = \frac{3}{8}\left(\frac{4g^2/D}{\sqrt{\beta_\mp-1}}\right)^{\frac{1}{3}}m^{\frac{1}{3}}, \qquad r_\star(m) = \beta_\mp\left(\frac{4D^2/g}{\sqrt{\beta_\mp-1}}\right)^{\frac{1}{3}}m^{\frac{1}{3}}.
\end{equation}
Crucially, the dependence of $\omega_\star$ on $m$ has changed from its linear form in shallow water.
Furthermore, the location of the light-ring is no longer independent of radius as it was in Eq.~\eqref{LRfreqsShallow}.
Both of these features are a result of the dispersive nature of gravity waves in the deep water regime \cite{torres2017rays}.


\section{The general case} \label{sec:LRmethod}

Analytic expressions for the light-ring frequencies cannot be obtained in the general case, meaning we have to search for solutions numerically for a particular background state given by $(C,D,H)$. 
The simplest way to implement the light ring conditions is by using the Hamiltonian constraint to obtain the dispersion relation,
\begin{equation}
\omega = \omega_d(\mathbf{k}) = \mathbf{v}\cdot\mathbf{k} \pm \sqrt{\mathcal{D}(k)},
\end{equation}
from which it follows that,
\begin{equation}
\partial_{k_r}\mathcal{H}=\pm\sqrt{\mathcal{D}}\partial_{k_r}\omega_d, \qquad \partial_r\mathcal{H}=\pm\sqrt{\mathcal{D}}\partial_r\omega_d.
\end{equation}
Therefore, the conditions on the Hamiltonian of the system maybe re-expressed as conditions on the dispersion relation.
A numerical solution is obtained from the following three step procedure:
\begin{enumerate}
\item Solve $\mathcal{H}=0$ to obtain the dispersion relation $\omega = \omega_d(m,k_r,r)$. 
\item Solving \mbox{$\partial_{k_r}\omega_d = 0$} yields $k_r^\pm$ in terms of the other variables (the $\pm$ corresponds the two branches of the dispersion relation).
Substitution back into $\omega_d$ results in two curves \mbox{$\omega = \omega_\pm(m,r)$}.
\item Finally, $\partial_r\omega_\pm = 0$ gives the location of the extremum of $\omega_\pm$.
This means that the value of $\omega$ at the extremum gives the light ring frequency $\omega = \omega_\star(m)$.
\end{enumerate}
This rewriting allows us to give a different interpretation to the light-ring conditions.
The condition \mbox{$\partial_{k_r}\omega_d = 0$} means that a mode has zero group velocity in the radial direction.
The curves resulting from this condition, $\omega = \omega_\pm(r)$, determine the minimum frequency required to propagate at a particular radius. 
A frequency just below this curve would not be able to propagate, and the mode would be evanescent.
In the shallow water regime, these curves correspond to $\omega_\pm$ displayed in Figure~\ref{fig:PotentialPlots} of Chapter~\ref{chap:super}.
The frequency at the extremum of $\omega_\pm$ is therefore the \textit{minimum frequency required to propagate at all radii.}
This makes the light-ring frequencies the most efficient frequency band to transfer energy across the entire system, giving a natural interpretation as to why they are in close correspondence with the QNMs, i.e. the modes of energy dissipation.


\section{Detection in the laboratory}

This section is based on our work in \cite{torres2018application}. 
To investigate the application of light-ring modes to fluids, we performed an experiment in a draining bathtub type system.
The aim of this experiment is two-fold.
\begin{enumerate}
\item Firstly, we demonstrate the presence of the light-ring modes, finding excellent agreement with our theoretical predictions above when using flow parameters obtained from PIV measurements.
This constitutes the first detection of light-ring modes in an analogue black hole set-up.
\item Secondly, inspired by black hole spectroscopy in GR, we use the light-ring modes to determine properties of our fluid flow (in particular, the circulation parameter) which agrees with the PIV value within $8\%$. 
Hence, our method presents a viable non-invasive alternative to standard flow visualisation techniques, such as PIV, where the flow is disturbed by the presence of tracer particles.
We call this new method \textit{Analogue Black Hole Spectroscopy} (ABHS).
\end{enumerate}

\subsection{Experimental overview}

\begin{figure*}[t!]
    \centering
    \begin{subfigure}[t]{0.5\textwidth}
        \centering
        \includegraphics[height=3in]{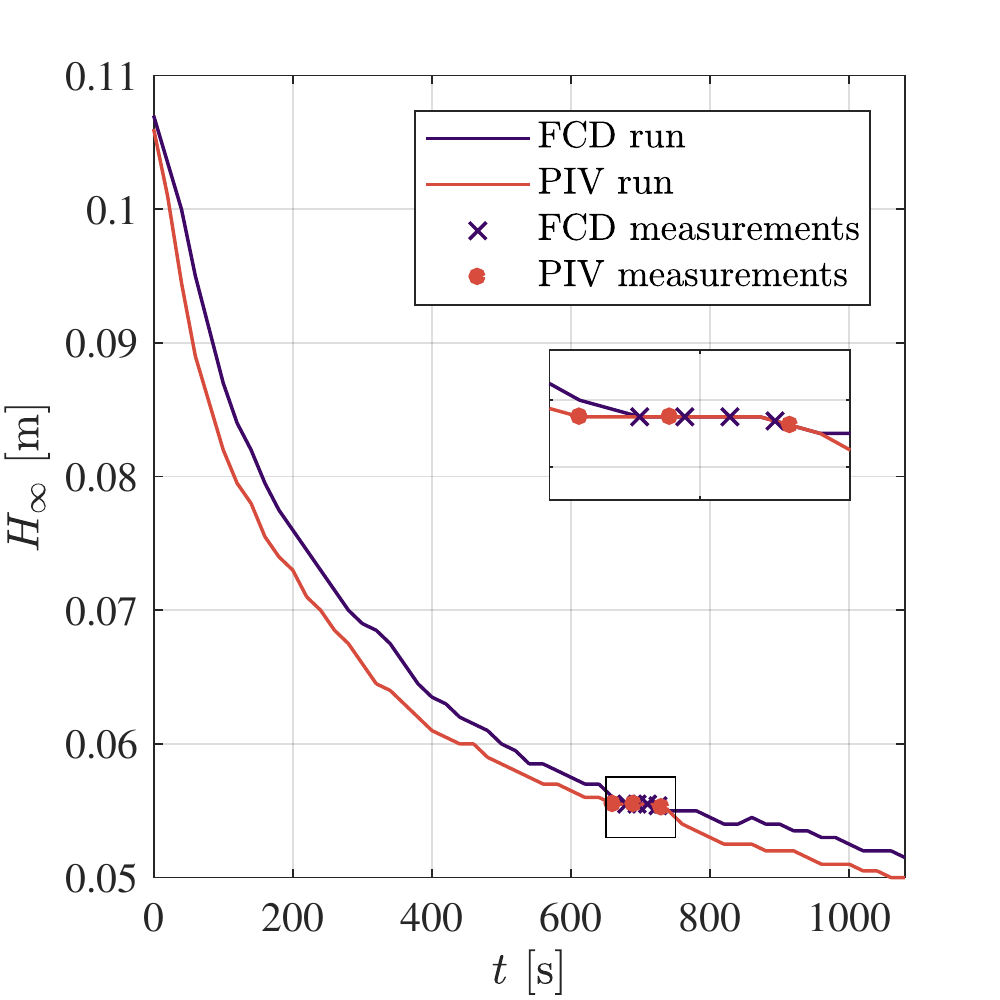}
    \end{subfigure}
    ~ 
    \begin{subfigure}[t]{0.5\textwidth}
        \centering
        \includegraphics[height=3in]{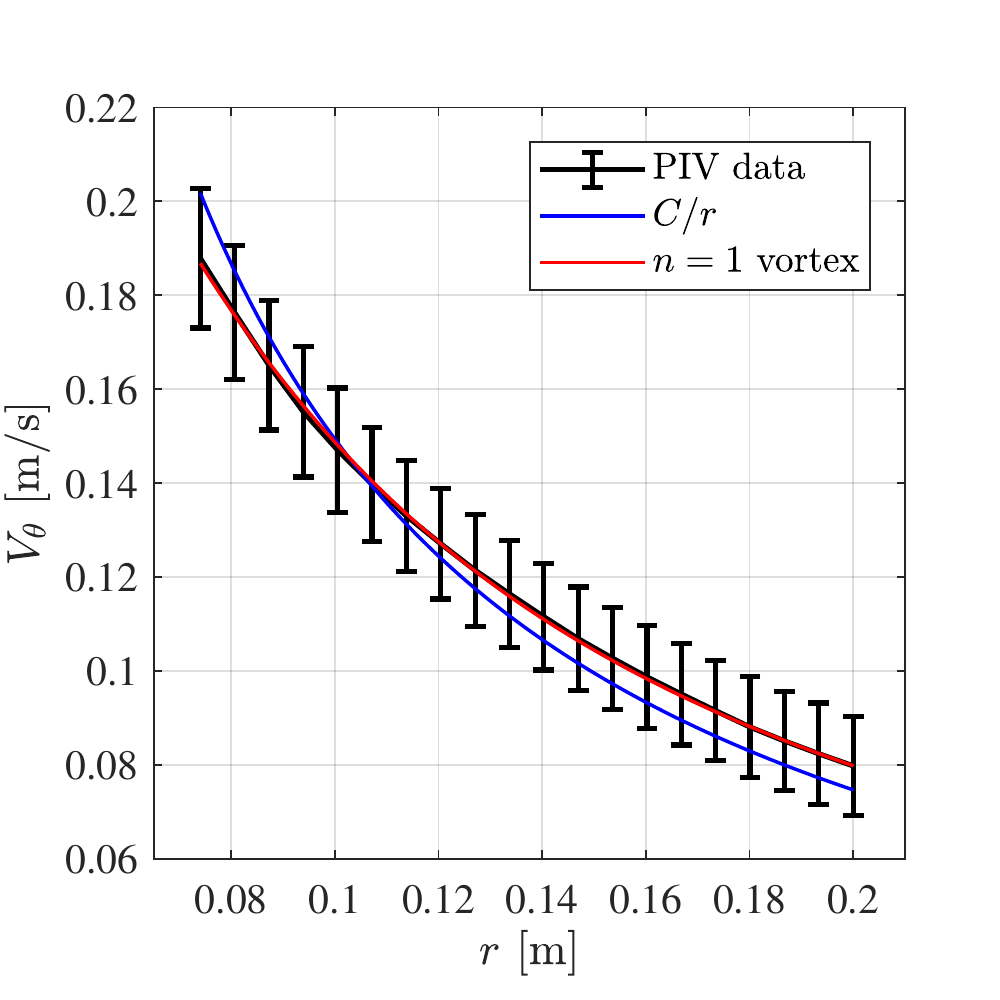}
    \end{subfigure}
    \caption{Left side: an example of how the height changed over the course of the experiment. 
The apparatus was set-up to take either PIV or FCD measurements and data was taken once the free surface reached a height of \mbox{$5.55~\mathrm{cm}$}.
For this particular example, we were able to take 4 measurements on the FCD run and 3 measurements on PIV run.
For clarity, we have enlarged the window where measurements were taken and plotted this region in the inset.
Right side: angular velocity profile determined via PIV. The errorbars indicate the spread of the measured velocities at a particular radius. 
These are larger compared to those in Figure~\ref{fig:VelocityField} due to larger spread in velocity vectors when using smaller particles with a laser sheet. We fit the data with the irrotational profile of Eq.~\eqref{DBT} (blue) and the $n=1$ vortex described in Eq.~\eqref{Nvortex} (red). Although the $n=1$ model provides a better fit, our theory in the previous section assumes an irrotational fluid.} \label{fig:HeightVariation}
\end{figure*}

The experiment was performed using the set-up described in Chapter \ref{chap:super} (see Section~\ref{sec:detectSR} for a description of the apparatus and specifically Section~\ref{sec:BgConfig} for a discussion of the type of fluid flow produced).
The principle difference between this experiment and that of the previous chapter is that we did not excite waves using a mechanical wave generator.
This is because, in a finite sized system, the excited wave will reflect off the boundary, thereby obscuring the natural resonances of the system we are trying to measure.
Instead, our approach was to set up a system out of equilibrium and measure it's oscillations during the relaxation process.

To do this, we pump water continuously from the inlet at a flow rate of $15\pm 1~\ell/\mathrm{min}$, and keep the drain hole covered until the water raises to a height of \mbox{$10\pm 1~\rm{cm}$}. 
The drain is then unplugged causing the water height to decrease.
The system is quasi-stationary at a height of \mbox{$H_\infty = 5.55 \pm 0.05~\mathrm{cm}$}, where the velocity field and perturbations on the free surface are recorded.
On the left panel of Figure~\ref{fig:HeightVariation}, we show the typical variation in height during this process.
Once the water height drops below this value, we allow the tank to drain completely before repeating the full procedure.
This was necessary to minimise the variation in the background flow between measurements.

\subsection{Flow measurements} \label{sec:flow_values2}

The velocity field was measured using the PIV technique detailed in Section~\ref{sec:PIV}.
There were however some important differences in our method for this experiment, which we expand upon now.

Rather than seed the flow with buoyant paper particles like in the previous chapter, we used much smaller plyolite particles with an average diameter of \mbox{$1~\mathrm{mm}$} (this type of particle is much more common in standard PIV studies, e.g. \cite{cristofano2016bathtub}).
To visualise the particles, a 2 dimensional cross section of the tank was illuminated with a laser sheet, which was oriented to be in the same plane as the waters surface, i.e. the $(x,y)$ plane.
We used a Yb-doped laser of characteristic wavelength $523~{\rm nm}$. 
The laser was aligned in all cases to be pointing directly at the vortex and the height of the sheet was adjusted to $5.4~{\rm cm}$, i.e.~just below the free surface. 
Due to the presence of a shadow in the camera's field of view (where the laser sheet intersects the vortex core), the laser source was placed at three different positions: half way along the length of the tank and offset by $1.2~{\rm m}$ to either side.
Two measurements were taken for each laser position, giving a total of 6 experimental runs. 
Since PIV requires the presence of tracer particles, it was not possible to measure the flow and the waves in the same run, requiring a full restart of the procedure outlined above.

The benefit of this method over that in the previous chapter is that the smaller particles size means they experience less of a velocity lag, hence the particle motion should be closer to the true motion of the fluid. 
The disadvantage is that the particles don't really 
move on 2-dimensional cross sections due to the 3D nature of the flow near the core.
This can cause particles to move out of the laser sheet, resulting in more erroneous vector identifications than when using the buoyant paper particles.
The blind spot due to the vortex shadow in the laser sheet is an additional problem we didn't have in the previous experiment.

In each experimental run the particles were recorded for $5.445~{\rm s}$ using a Phantom Miro Lab 340 camera, with a Sigma 24-70mm F2.8 lens attached, at a resolution of 1600x1600 pixels and a frame rate of 200 frames per second (corresponding to 1090 images). 
The camera was positioned such that the pixel size was $0.32~{\rm mm}$. 
We obtain the velocity field components $V_r(r)$ and $V_\theta(r)$, using the same method described in Section~\ref{sec:PIV}. 
The window of observation was $r \in [7.4,20]~\mathrm{cm}$, which was further out than the previous experiment since the light rings are located further from the origin than the horizon.
In this window $V_r\ll V_\theta$, and the maximum value of the radial velocity is approximately a tenth of the angular one\footnote{In fact, it is expected that most of the draining in free surface vortices occurs through the bulk and through the boundary layer at the bottom of the tank, with a negligible radial flow at the surface, especially far from the vortex core~\cite{andersen2003anatomy,andersen2006container,cristofano2016bathtub}. 
In our simplified theoretical treatment, we do not account for $z$ dependence of $V_r$ and $V_\theta$, and this is one way theoretical studies could be improved in future.}. 
More importantly, the bias in our PIV method (estimated by applying the method to simulated data) has the same order of magnitude as the measured radial velocities, implying that we cannot perform a reliable measurement of the radial velocity in the region of interest. 

We can estimate the parameter $D$ in Eq.~\eqref{DBT}, using the fact that the flow rate is approximately the volume flux through a surface surrounding the drain when the system is quasistationary.
If we take this to be a radial surface and assume uniform draining through the bulk then,
\begin{equation}
Q \simeq -\int^{2\pi}_0\int^{H_\infty}_0\mathbf{V} \cdot\mathbf{r}dz d\theta = 2\pi D H_\infty,
\end{equation}
which gives \mbox{$D\approx 7\times10^{-4}~\mathrm{m^2/s}$}.
Furthermore, we can determine an upper-bound for $D$ by computing \mbox{$\mathrm{\max}_{ij} [r_{ij}V_r(r_{ij},\theta_{ij})]$} using our PIV data, where $i,j$ are the indices over the Cartesian grid in the raw data. 
This gives \mbox{$D_\mathrm{max}=3.9\times10^{-3} \mathrm{m^2/s}$}.

The angular velocity profile is displayed on the right panel of Figure~\ref{fig:HeightVariation}.
The errorbars are computed from standard deviation when averaging over $t$, $\theta$ and the 6 experiments.
The errorbars are larger than in Figure~\ref{fig:VelocityField} since the method used here results in a wider spread of velocity vectors.
We fit the data with two models: the first is the irrotational profile of Eq.~\eqref{DBT}, which yields \mbox{$C=1.51\pm0.3~\mathrm{m^2/s}$} for the circulation parameter, with a quality of fit given by \mbox{$R^2=0.9674$}.
The second is the $n=1$ vortex described in Eq.~\eqref{Nvortex}, which gives \mbox{$C = 1.64\pm0.01\times10^{-2}~\mathrm{m^2/s}$} and a viscous core of radius of \mbox{$r_0 = 3.2\pm0.1~\mathrm{cm}$}. 
The quality of fit is \mbox{$R^2=0.9997$}.
Note that although the quality of fit is better for the latter, our theory assumes an irrotational fluid and thus we must take first value of $C$ to model the light-ring spectrum. 
We estimate that this should introduce a relative error of $\sim 10\%$ in the region we care about.

To check the axisymmetry assumption, we compared the energy (proportional to the square of the velocity) of $m\neq0$ with the $m=0$ component (i.e. the angular average).
The maximum of this ratio for $V_\theta$ was $1.3\%$, indicating our flow is highly symmetric.
Extracting the $m\neq0$ components is complicated by the fact that we have no data for part of the $\theta$ range due to the vortex shadow.
To overcome this, we take half the image which does not contain the shadow and compute only the $m=\mathrm{even}$ components. We note that this leads to contamination of $m=\mathrm{even}$ by the $m=\mathrm{odd}$ parts which do not integrate to zero. However, we do not believe this to be of consequence since we are only interested in estimating the upper bound on asymmetry, which we clearly see is small.

To check the stationarity of the flow, we compare the angular velocity vector fields in the first and final seconds of our experimental runs, finding that the maximum difference at any point is smaller than the uncertainty in vector identification inherent in the PIV method. Hence, we deem the velocity field to be stationary within the error of our method.

\subsection{Wave measurements}

Our method to extract the light-ring frequencies closely parallels the technique outlined in Section~\ref{sec:getWavesSR}.
The main difference is that here we are looking for the natural frequencies of the system, i.e. a range of frequencies, rather than one specific frequency band which we excite. \\

\textit{Data acquisition.}
The free surface of the water is obtained using the Fast-Chequerboard Demodulation (FCD) method~\cite{sanders2018schlieren}, which has a greater spatial and temporal resolution than the sensor used in Chapter~\ref{chap:super} to detect superradiance.
To implement this method, a periodic pattern is placed at the bottom of the tank over a region of \mbox{$59~ \mathrm{cm} \times 84~ \mathrm{cm}$}. 
The pattern is composed of two orthogonal sinusoidal waves with wavelengths of $6.5~\mathrm{mm}$ each.
Deformations of this pattern due to free surface fluctuations are recorded in a region of \mbox{$58~ \mathrm{cm} \times 58~ \mathrm{cm}$} over the vortex using a Phantom Camera Miro Lab 340 high speed camera at a frame rate of 40 $\mathrm{fps}$ over $16.3~ \mathrm{s}$ with an exposure time of 24000 $\mathrm{\mu s}$.
For each of the 651 pictures of the deformed pattern, we reconstruct the free surface in the form of an array $h(t_k,x_i,y_j)$ giving the height of the water at the $1600 \times 1600$ points on the free surface $(x_i,y_j)$ at every time step $t_k$\footnote{The MATLAB code used for this is available at: https://github.com/swildeman/fcd}. 
Note that the Cartesian grid of the camera has equidistant points in contrast to the grid $(X_{ij},Y_{ij})$ when using the sensor.
This is an added benefit of the method employed here.
The typical amplitude of the free surface deformations corresponds, at most, to 2\% of the unperturbed water height. 
This justifies the linear treatment of the perturbations. \\

\textit{Azimuthal decomposition.} To select specific azimuthal numbers, we choose the centre of our coordinate system to be the centre of the hole and convert the signal from Cartesian to polar coordinates. 
In addition to this change of coordinates, we discard all data points within a minimal radius $r_\mathrm{min}\approx 7.4~\mathrm{cm}$. 
This cropping is necessary since the curvature of the free surface near the drain deforms the pattern beyond what can be reconstructed by the FCD method.
We also discard points above a radius $r_\mathrm{max} \approx 25~\mathrm{cm}$, to avoid errors coming from the edge of the images.
After this step, our data is in the form $h(t_k,r_i,\theta_j)$ with $r_i\in[r_\mathrm{min},r_\mathrm{max}]$.

Before selecting azimuthal components, we use the fact that for real valued data, the information contained in the negative frequency range of the time Fourier transform is redundant and can be discarded, resulting in a new array \mbox{$h_\mathbb{C}$}.
This step allows us to identify $m>0$ as those co-rotating with the flow and $m<0$ as counter-rotating.
The real part of the filtered data corresponds to the original array after multiplying by a factor of $2$, i.e. \mbox{$2\mathrm{Re}[h_\mathbb{C}]=h$}, for the reason explained in Section~\ref{sec:getWavesSR}. 
In the following we will discard the index $\mathbb{C}$ and keep in mind that we are now dealing with a complex array.

We then perform a discrete Fourier transform in the angular direction, to separate the various azimuthal components, according to,
\begin{equation}
h_m(t_k,r_i) = \sum_j h(t_k,r_i,\theta_j) e^{-im\theta_j} \Delta \theta.
\end{equation} 

\begin{figure} 
\centering
\includegraphics[width=0.85\linewidth]{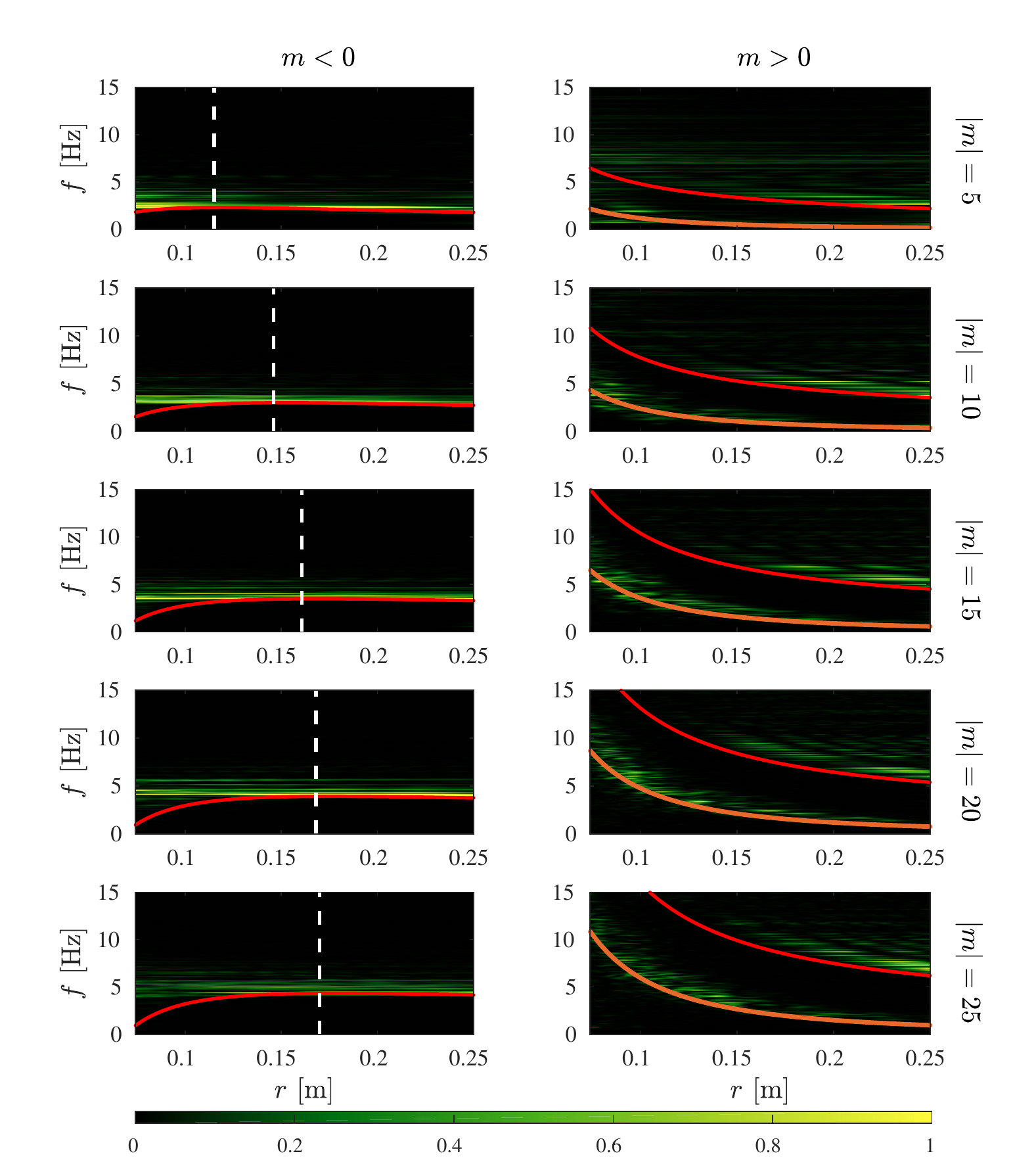}
\caption{The power spectral density is compared with the minimum energy curve $f_\mathrm{min}^+(m,r)$, plotted in red, for various $m$. The maxima of $f_\mathrm{min}^{+}(m,r)$ indicate the location of the light-rings, $r_\star(m)$, which are shown in dashed white lines.
For $m<0$, the spectral density peaks and the minimum energy curve are clearly distinguishable for small radii. 
This means that the frequency peak is not noise generated locally with the minimum energy, but rather a signal which propagates across the whole flow.
We show in Figure~\ref{fig:LRspectrum} that this signal coincides with the light-ring frequencies.
For $m>0$, we observe two signals whose peaks are radius-dependent. 
The upper one follows the minimum energy curve and corresponds, most probably, to random noise generated locally. 
The lower one follows the angular velocity of the fluid flow according to $f_\alpha(m,r)=mV^\theta(r)/2\pi r$ (orange curve) and is likely sourced by potential vorticity perturbations, which are convected by the fluid in an irrotational flow \cite{churilov2018scattering}.} \label{fig:PSD}
\end{figure}

\begin{figure*}[t!]
    \centering
    \begin{subfigure}[t]{0.5\textwidth}
        \centering
        \includegraphics[height=3in]{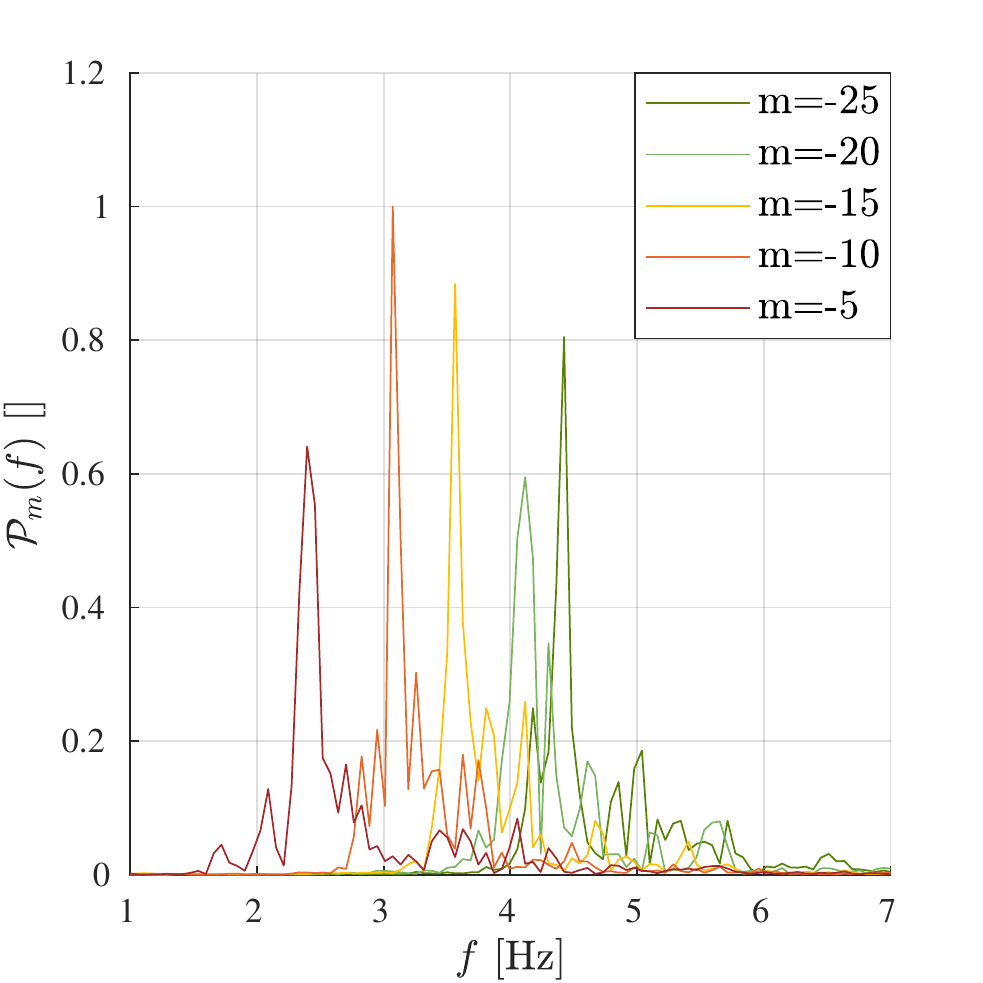}
    \end{subfigure}
    ~ 
    \begin{subfigure}[t]{0.5\textwidth}
        \centering
        \includegraphics[height=3in]{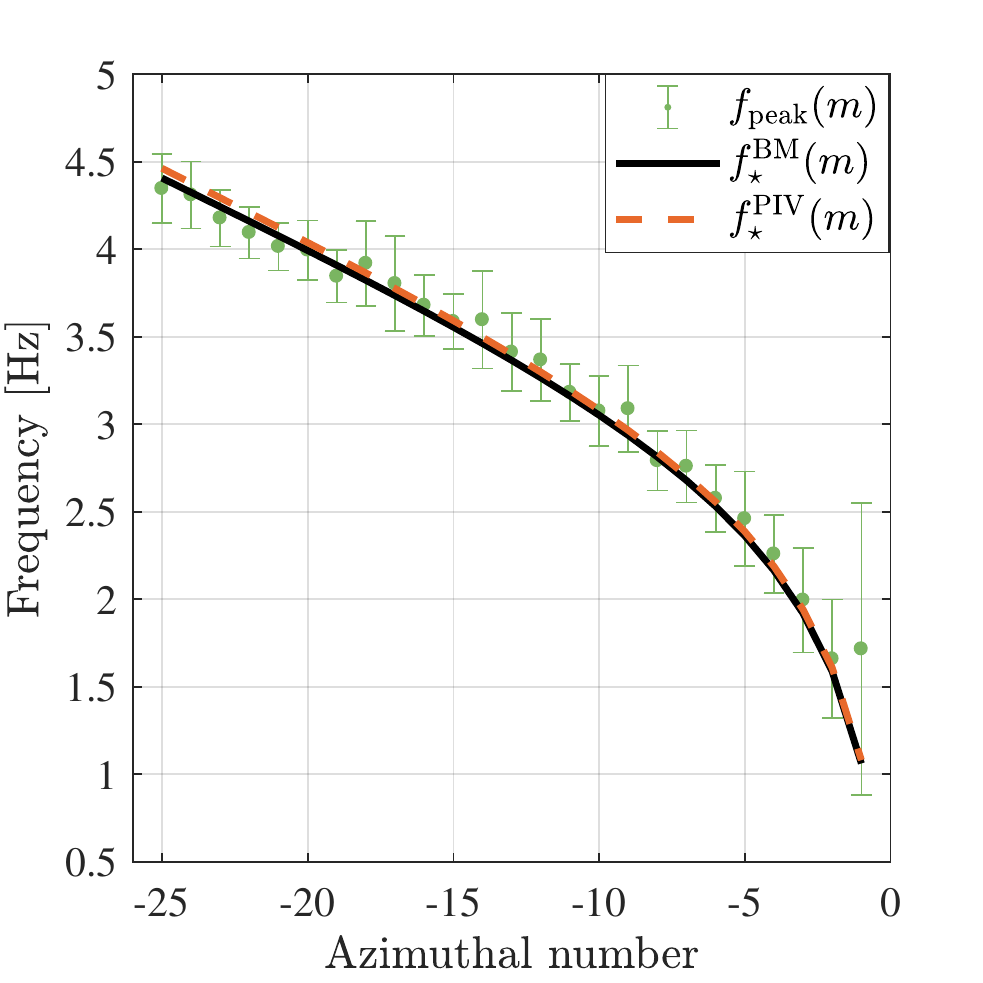} 
    \end{subfigure}
    \caption{Left side: typical power spectra $\mathcal{P}_m(f)$ obtained from averaging the PSDs over radius. The scale on the vertical axis is arbitrary since we are only interested in the peak location $f_\mathrm{peak}(m)$. This corresponds to the oscillation frequencies of the light-ring modes. 
Right side: the frequency spectrum $f_{\mathrm{peak}}(m)$, extracted from the experimental data and represented by green dots, is compared with the theoretical prediction for the light-ring frequencies, $f_{\star}(m)$. 
The error bars indicate the standard deviation over 25 experiments, and thus represent the spread of the frequencies obtained. 
The dashed orange curve is the predicted spectrum computed from the flow parameters extracted from PIV measurements, \mbox{$C = 1.51\times 10^{-2}~\mathrm{m^2/s}$} and \mbox{$D = 0~\mathrm{m^2/s}$}.
The two spectra agree, confirming the detection of light-ring mode oscillations. 
The solid black curve is the non-linear regression of the experimental data to DBT model in Eq.~\eqref{DBT}, and provides the values for $C$ and $D$ presented in Figure~\ref{fig:Avocado}
} \label{fig:LRspectrum}
\end{figure*}

\textit{Frequency content.} Using the $h_m$'s, it is possible to estimate the Power Spectral Density $\mathrm{PSD}(f,r_i,m)$ of the waves emitted by the vortex for every azimuthal number $m$ at each radius $r_i$,
\begin{equation}
\mathrm{PSD}(f,r_i,m) \propto \big| \hat{h}_m(f,r_i) \big|^2,
\end{equation}
where $\hat{h}_m$ is the time Fourier transform of $h_m$ at fixed radius and $f=\omega/2\pi$ is the frequency in Hz. 
In Figure~\ref{fig:PSD}, we display the PSDs for a range of $m$ values. 
Finally, the power spectrum $\mathcal{P}_m(f)$ is obtained by averaging the PSDs over radius.
From $\mathcal{P}_m(f)$, we can look at the $r$-independent frequency content, which based on the argument at the end of Section~\ref{sec:LRmethod} should coincide with \mbox{$\omega_\star$}.
We display the power spectrum of various $m$ modes on the left panel of Figure~\ref{fig:LRspectrum}.
The location of the peak $f_\mathrm{peak}(m)$ is obtained by parabolic interpolation of the maximum of $\mathcal{P}_m(f)$ and its nearest neighbouring points. 
The value of $f_\mathrm{peak}(m)$ for $m\in[-25,-1]$ is plotted in the right panel of Figure~\ref{fig:LRspectrum}.
By repeating this procedure using a different choice for the center, located 10 pixels ($\approx 4\mathrm{mm}$) away from our original choice, we observed that the maximum deviation in the location of a frequency peak is approximately $2\%$.
This indicates our analysis is robust against an inaccurate choice of the centre of symmetry in our system.

\subsection{Results}

\begin{figure} 
\centering
\includegraphics[width=0.85\linewidth]{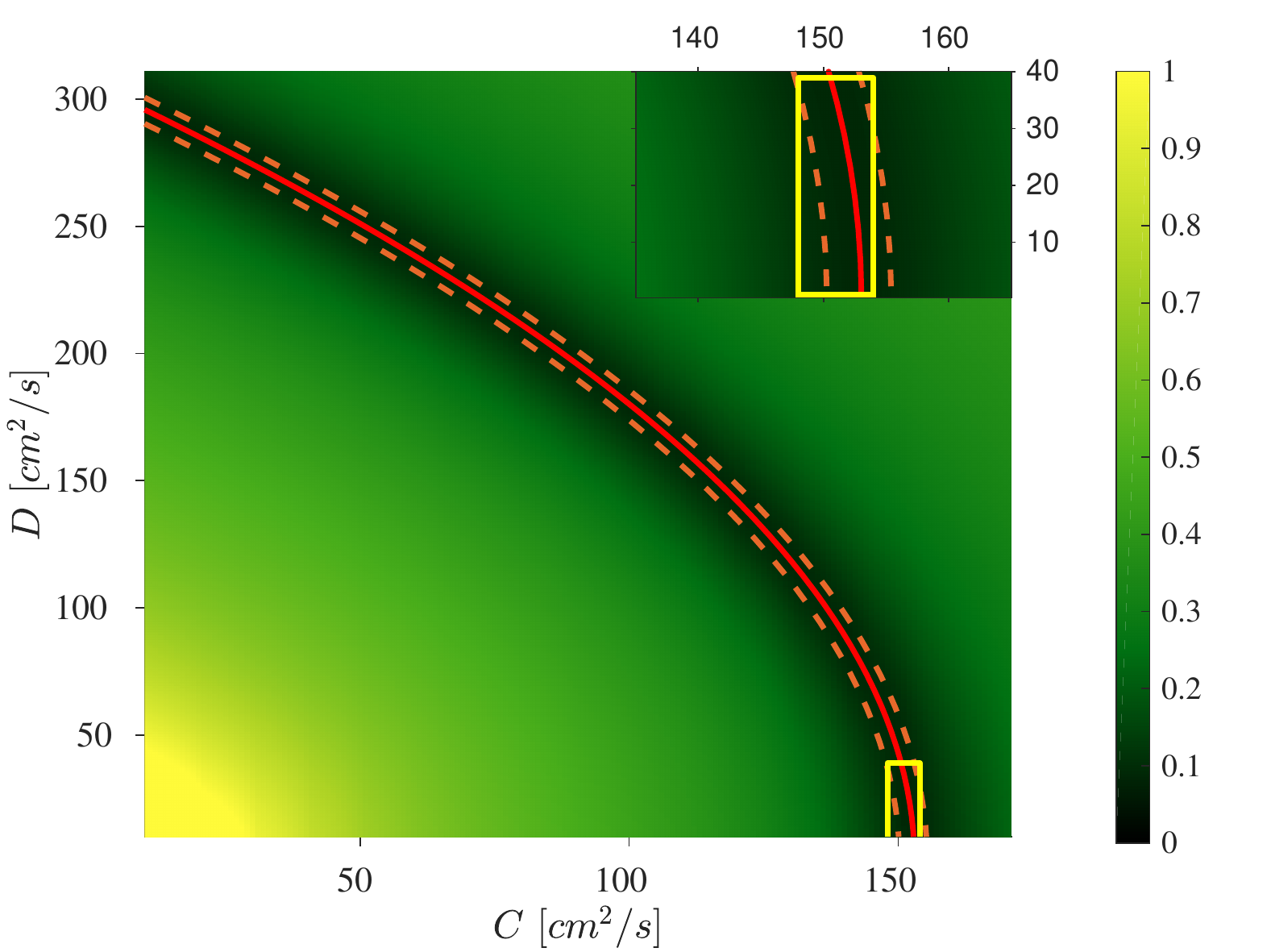}
\caption{The intensity of the background image represents the normalised sum of squared residuals between the experimental spectrum, $f_{\mathrm{peak}}(m)$, and the theoretical prediction for the light-ring frequencies, $f_{\star}(m)$, as a function of the flow parameters. 
The red curve represents the family of possible values for $C$ and $D$ that best match the experimental data (using the method of weighted least squares). 
The area delimited by the dashed orange curves represents the 95\% confidence interval. 
It overlaps with the yellow rectangle on the bottom right corner, which corresponds to the flow parameters obtained using Particle Imaging Velocimetry. 
The spread along the $C$-axis represents the 95\% confidence interval estimated using the covariance matrix. 
The spread along the $D$-axis represents the extracted upper bound for $D$. 
In the top right corner we present a detailed view of the parameter space where the two flow measurements overlap.} \label{fig:Avocado}
\end{figure}

We can identify two different behaviours depending on the sign of $m$.
For negative $m$'s, the PSDs are approximately constant over the window of observation. 
The spectral density is peaked around a single frequency, which allows us to define the position-independent spectrum $f_\mathrm{peak}(m)$ shown in Figure~\ref{fig:LRspectrum}.
This spectrum can be used in two different ways.

Firstly, using the values for $C$ and $D$ obtained from the PIV method, we can predict the characteristic mode spectrum $f_{\star}^\mathrm{PIV}(m)$ following the method outlined in Section~\ref{sec:LRmethod}.
We take as our dispersion function \cite{whitham2011linear},
\begin{equation} \label{dispersionFullCap}
\mathcal{D}(\mathbf{k})=\left(gk+\frac{\gamma}{\rho}k^3\right)\tanh(Hk),
\end{equation}
to account for the effects of surface tension, where \mbox{$\gamma = 0.0728~\mathrm{kg/s^2}$} and \mbox{$\rho=997~\mathrm{kg/m^3}$} are the surface tension and density of water respectively.
The PIV prediction is shown by the dashed orange curve on the right panel of Figure~\ref{fig:LRspectrum}.
We observe that the model describing the characteristic oscillations of our system as light-ring modes is consistent with the data.
This is the first experimental observation of the oscillatory part of the light-ring spectrum in an analogue black hole set-up. 

Secondly, having validated our approach, we propose to use the light-ring spectrum to determine the fluid flow parameters as an alternative to PIV. 
To do this, we perform a non-linear regression analysis to look for the best match between the experimental spectrum $f_\mathrm{peak}(m)$ of counter-rotating modes and the corresponding theoretical predictions for the light-ring spectrum $f_\star(m)$, computed from the method in Section~\ref{sec:LRmethod}.
To find the best match, we compute the mean square error,
\begin{equation} \label{MSE}
\mathrm{MSE}(C,D) = \frac{1}{M}\sum_m\left(f_\mathrm{peak}(m)-f_\star(m,C,D)\right)^2,
\end{equation}
where $M$ is the number of $m$ modes in the data, and look for the minimum of $\mathrm{MSE}$\footnote{Note, the $\mathrm{MSE}$ will be a function of the parameters that enter into $f_\star$.
Here, $f_\star$ depends only on $C$ and $D$ since we are working with the irrotational field theory for the perturbations. However, one could in theory use this method to test alternative field theories that predict different parameter dependences of $f_\star$.
This is explored further in the next chapter.}.
This reduces the DBT parameter space from two dimensions to one, constraining the flow parameters $C$ and $D$ to lie on the red curve shown in Figure~\ref{fig:Avocado}. 
Any pair of points along this curve will give the same spectrum, $f_{\star}^\mathrm{BM}(m)$, represented by the solid black curve on the right panel of Figure~\ref{fig:LRspectrum}. 
The region between the dashed orange curves shown in Figure~\ref{fig:Avocado} represents the 95\% confidence intervals for the values of $C$ and $D$. 
This region overlaps the yellow rectangle which represents the possible flow parameters found using PIV. Note that, in this case, the black hole spectroscopy method imposes a slightly stronger constraint on the circulation parameter than our PIV measurements.

We highlight that in order to uniquely determine $C$ and $D$, the positive $m$ part of the light-ring spectrum is also needed.
However, when the flow is characterised by only one parameter (e.g.~purely rotating superfluids), the counter-rotating LR modes contain all the information about the fluid velocity. 
We note that this is effectively the case in our experiment. 
Since $D \ll C$ in our window of observation, the vortex flow can be considered to be purely rotating and our observations are sufficient to fully characterise the flow in this region.

Although the LR modes are absent in the PSDs of the co-rotating modes shown in Figure~\ref{fig:PSD}, two distinct, radius-dependent, signals are present.
We can understand their origin using the flow parameters previously obtained. 
By computing the minimum energy line, $f_\mathrm{min}^{+}(m,r)$ (shown as the red curve), we observe that one of the signals corresponds to random noise generated locally with the minimum energy. 
The other signal is related to the angular velocity of the fluid flow and can be matched with the curve $f_\alpha(m,r)=mV^\theta(r)/2\pi r$, shown in orange. 
This peak lies below the minimum energy and, hence, corresponds to evanescent modes.
A possible explanation for their appearence is that \textit{potential vorticity} (PV) perturbations act as a source for them \cite{churilov2018scattering}.
In irrotational flows (which is approximately the regime in which our observations are made), PV is carried by the flow as a passive tracer. 
Various $m$ components of PV will therefore source free-surface deformations which are transported at frequencies $f_\alpha(m,r)$. 
These observations strengthen our confidence in the flow parameters obtained.

\section{Analogue black hole spectroscopy} \label{sec:ABHSmethod}

This section is based on our work in \cite{torres2019analogue}. 
To further demonstrate the applicability of the ABHS method in determining the properties of a fluid flow, we apply the technique to numerically simulated data.
In particular, we saw in the experiment that since the circulation parameter was much larger than that of the drain, it was not possible to extract both parameters uniquely from the light-ring spectrum.
We now turn to a situation where the circulation is slower than the drain rate, demonstrating how both sides of the light-ring spectrum can be used to characterise the flow in this case.

The shallow water wave equation of Eq.~\eqref{waveeqn} is simulated using the DBT profiles in Eq.~\eqref{DBT}, with parameters \mbox{$c=1~\mathrm{m/s}$}, \mbox{$C=2.2\times10^{-3}~\mathrm{m^2/s}$} and \mbox{$D=2.2\times10^{-2}~\mathrm{m^2/s}$}.
The wave equation is simulated using the Method of Lines for each $m$-component individually, using a hard wall for the outer boundary at $r=r_N$ and the inner boundary is left free inside the horizon at $r=0.1~r_h$. 
The outer boundary is sufficiently far away that no reflections occur over the simulation time.
The initial condition is a Gaussian pulse centred on $\bar{r}=r_N-5\varpi$ where \mbox{$\varpi=0.9r_h$} is the width,
\begin{equation}
\phi_m(r,t=0) =  \frac{1}{\sqrt{2\pi\varpi^2}}\exp\left(-\frac{(r-\bar{r})^2}{2\varpi^2}\right),
\end{equation}
and \mbox{$\partial_t\phi_m|_{t=0}$} is chosen such that the pulse propagates toward the vortex.
More details can be found in Appendix~\ref{app:RK4}.

To extract the frequency response of the vortex, we first compute the time Fourier transform \mbox{$\hat{\phi}_m(\omega)=\mathrm{FT}[\phi_m(t,r=5r_h)]$}. 
The location of the peaks in $\hat{\phi}_m$ determines the real part of the characteristic frequency spectrum.
Since the wave equation is symmetric under the inversion \mbox{($\omega\to-\omega,m\to-m$)}, we only need to simulate for $m>0$. Then, the peak at $\omega>0$ gives $\omega_\mathrm{num}(m)$ and the peak at $\omega<0$ gives \mbox{$\omega_\mathrm{num}(-m)$}.

\begin{figure} 
\centering
\includegraphics[width=0.5\linewidth]{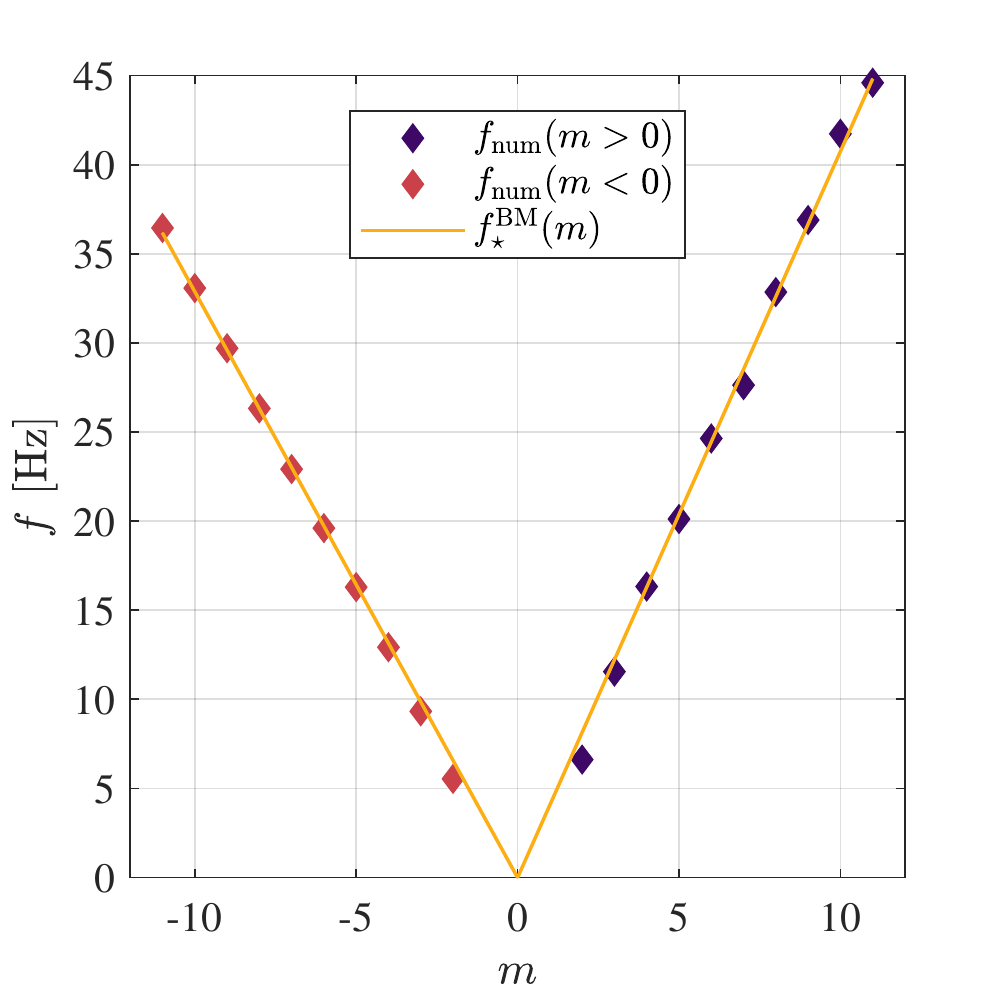}
\caption{Numerically obtained characteristic frequencies of the DBT (diamonds) are matched with predictions for the light-ring frequencies in Eq.~\ref{LRfreqsShallow}.
The best match is found by minimising the MSE in Figure~\ref{fig:MSEshallow} for both sides of the spectrum, resulting in \mbox{$C^\mathrm{BM} = 2.3\pm1\times 10^{-3}~\mathrm{m^2/s}$} and \mbox{$D^\mathrm{BM} = 2.2\pm0.1\times 10^{-2}~\mathrm{m^2/s}$}.} \label{fig:LRshallow}
\end{figure}

\begin{figure} 
\centering
\includegraphics[width=\linewidth]{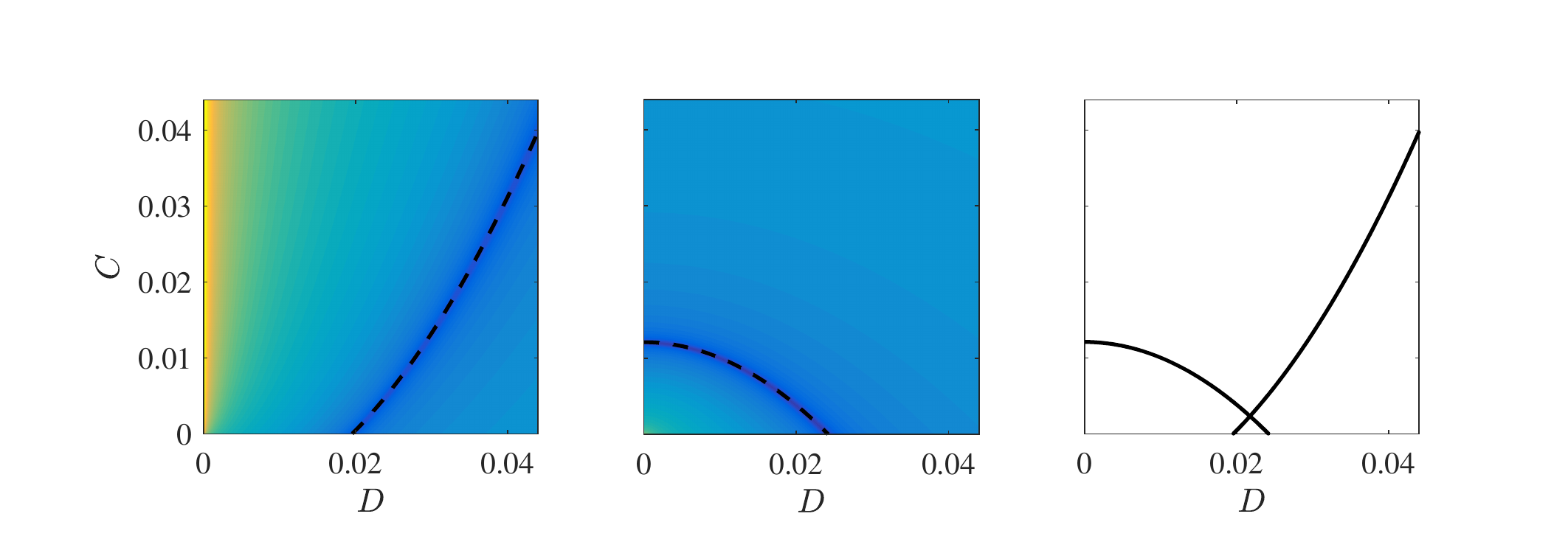}
\caption{MSE for $m>0$ (left side) and $m<0$ (center). 
The colour axis is log(MSE) where a darker region indicates a smaller value, and the black dashed curves indicate where the MSE is minimised.
The axes are in $[\mathrm{m^2/s}]$.
There is an equivalence class of $C$ and $D$ values that will produce the best match in Figure~\ref{fig:LRshallow} when considering $m<0$ and $m>0$ independently.
However, $C$ and $D$ are uniquely defined by considering both sides simultaneously, and their values are given by the intersection of the two curves (right side).} \label{fig:MSEshallow}
\end{figure}

In Figure~\ref{fig:LRshallow} we display \mbox{$f_\mathrm{num}=\omega_\mathrm{num}/2\pi$}.
At large $m$, the co-rotating modes are harder to excite due to their large frequency, hence the peak at $\omega>0$ becomes difficult to resolve beneath the tail of the $\omega<0$ peak. 
This accounts for the poor behaviour of the purple diamonds for large $m>0$, which should be perfectly linear.
To apply the ABHS method, we compute $f_\star$ using Eq.~\eqref{LRfreqsShallow} for a range of $C$ and $D$ values.
We then compute the MSE in Eq.~\eqref{MSE} using $f_\star$ and $f_\mathrm{num}$ for $m<0$ and $m>0$ separately.
We display the logarithm of the MSE in Figure~\ref{fig:MSEshallow} for the $m<0$ and $m>0$ independently.
For each side of the spectrum, there is a family of $C$ and $D$ values that minimise the MSE indicated by the black dashed curves.
The intersection of these two curves provides the values that simultaneously matches both sides of the spectrum with the prediction.
The values provided by this intersection are \mbox{$C^\mathrm{BM} = 2.3\pm 1\times 10^{-3}~\mathrm{m^2/s}$} and \mbox{$D^\mathrm{BM} = 2.2\pm0.1\times 10^{-2}~\mathrm{m^2/s}$}, which are within $6\%$ and $1\%$ of the true value respectively.
The errors on the parameters represent the $95\%$ confidence interval obtained via the covariance matrix.

We now compare these values to ones obtained from PIV.
To create test data for the PIV method, we seed an initial image with particles of 4 pixel diameter such that the density is $0.013$ particles/pixel (these values were chosen to keep data as close as possible to that obtained in experiment). 
We then evolve the position of the particles using the velocity field in Eq.~\eqref{DBT} to create a second image. 
These images are analysed in \textit{PIVlab} using the method outlined in Section~\ref{sec:PIV}.
The best fit to the velocity profiles obtained is provided by \mbox{$C_\mathrm{PIV} = 2.20\pm0.05\times 10^{-3}~\mathrm{m^2/s}$} and \mbox{$D_\mathrm{PIV} = 2.31\pm0.01\times 10^{-2}\times 10^{-3}~\mathrm{m^2/s}$}. 
These represent a relative error of $2\%$ and $5\%$ on $D$ and $C$ respectively. 
We note that while the confidence interval are significantly smaller for the PIV method, a systematic error is present in the method. 
We further remark than the relative error on the parameters obtained using both methods are of the same order, proving the validity of ABHS as a flow measurement technique.

\section{Summary}

In this chapter, we have demonstrated that the response of a perturbed vortex flow is encapsulated by characteristic modes called light-ring modes, which are frequently used in black hole physics to approximate the QNMs.
Our experiment exhibits a new facet of analogue gravity.
Besides providing the first observation of light-ring mode oscillations, this successful demonstration of the principle behind black hole spectroscopy paves the way for real-life applications of the fluid-gravity analogy.
This method, which we call analogue black hole spectroscopy, can be used as an alternative to standard fluid flow visualisation techniques, such as particle imaging velocimetry, that require tracer particles. In particular, this is a promising non-invasive method to characterise fluid flows when suitable tracer particles are difficult to find, e.g. in superfluids~\cite{chopra1957suspension}.

The ABHS method we have described relies on knowledge of the effective field theory governing the perturbations and fortunately for many systems, the equations of motion for the waves are known exactly. 
In our experiment, we have demonstrated good agreement between PIV flow parameters and those obtained assuming an effective field theory for irrotational perturbations on a flat background with the dispersion relation in Eq.~\eqref{dispersionFullCap}.
This is because the light-ring in deep water is located far away from the vortex core, where the vorticity of the background is negligible and the height is very close to flat.
However, in the vortex core, other effects are expected to influence the equations of motion governing the perturbations, e.g. in our experimental set-up, we expect free surface deformations and vorticity to play a particularly important role.
Therefore we propose that the light-ring spectrum can be used in conjunction with known fluid parameters (measured using other techniques, e.g. PIV) to learn about the effective field theory governing the perturbations in set-ups where the equations of motion are unknown.

In the next chapter, we discuss possible modifications to the light-ring spectrum resulting from vorticity. 
Rotational vortex flows typically have an extra parameter characterising the background (see e.g. Eq.~\eqref{Nvortex}) which can lead to additional structure in the spectrum.
If one were able to pin down the effective field theory in such a scenario by performing a similar experiment to that explained here, the ABHS scheme could in principle be used to measure these extra parameters too.

\chapter{Vorticity} \label{chap:vort}

It is widely accepted that bathtub vortices which form under experimental conditions contain a vortex core where vorticity is non-negligible, resulting from the fluid's viscosity \cite{batchelor1967introduction,lautrup2005exotic}.
Many authors in the literature have studied this type of fluid flow; for example see
\cite{andersen2003anatomy,andersen2006container,kumar2018vortex} for experimental studies,
\cite{forbes1995bath,bohling2010structure,stepanyants2008stationary} 
for numerical studies and
\cite{odgaard1986free,forbes1995bath,tyvand2005impulsive} 
for analytical studies.
Since most of our theoretical discussion has been centred on rotating fluids in shallow water, we will be interested in the component of vorticity which is perpendicular to the free surface.
We briefly discuss how such vorticity comes into existence in an experiment.

Since vorticity ultimately arises in real fluids due to viscosity, it is useful to introduce a dimensionless quantity called the \textit{Reynolds number}.
The Reynolds number is the ratio of inertial forces to viscous forces in a fluid and is defined,
\begin{equation}
\mathrm{Re} = \frac{UL}{\nu},
\end{equation}
where $U$ is a characteristic velocity, $L$ is a characteristic length and $\nu$ is the fluid's viscosity.
In the irrotational DBT model, the angular Reynold's number is \mbox{$\mathrm{Re}_\theta=C/\nu$}, which for the values of $C$ obtained in Sections~\ref{sec:flow_values} and \ref{sec:flow_values2} is $\sim10^4$.
The viscous term in the Navier-Stokes equation is of order $\mathrm{Re}^{-1}$ and is thus usually neglected in this regime, leading to the Euler equation in Eq.~\eqref{euler} where we began our original discussion.

Whilst this is reasonable approximation in the bulk of the fluid, it necessarily breaks down at solid boundaries where the fluid must satisfy the no slip condition \cite{batchelor1967introduction}.
The fact the viscosity becomes important at the boundary is the reason that many fluid flows are not irrotational in nature.
To see this in the case of the bathtub vortex, consider the $\theta$-component of the steady Navier-Stokes equations for an axisymmetric fluid,
\begin{equation}
V_r\zeta + \nu\partial_r\zeta = -V_z\partial_zV_\theta - \nu\partial_z^2V_\theta,
\end{equation}
where $\zeta$ is the $z$-component of the vorticity vector (defined properly below in Eq.~\eqref{zeta}) and we have retained $z$-derivatives since $V_\theta$ must go to zero at the floor of the container.
This states that vorticity is present when velocity gradients in $z$ are non-negligible, which occurs in particular close to the drain where $V_\theta$ reaches high values.
In an axisymmetric flow, vorticity manifests as a deviation from $V_\theta\propto1/r$, which we have already seen in our flow measurements in Chapters~\ref{chap:super} and~\ref{chap:qnm}.

The standard analogue gravity formalism assumes an irrotational flow at the outset, allowing perturbations to be written in terms of a single scalar function which describes the surface gravity waves we have been studying.
However, the presence of vorticity allows for the excitation of additional degrees of freedom (i.e. other types of waves) thereby greatly increasing the complexity of the problem.
This means that in general it is not possible to define an effective geometry for disturbances to a fluid with vorticity.
Possibilities for other types of perturbations include inertial waves (or inertia-gravity waves), a broad class of oscillations whose restoring force is the Coriolis force \cite{mougel2014waves,buhler2014waves,batchelor1967introduction},
and Rossby waves \cite{reznik2001nonlinear}, resulting from gradients in the potential vorticity (PV)\footnote{In shallow water \mbox{$\mathrm{PV}=\mathbf{\Omega}/H$}, which coincides with the usual vorticity when the free surface is flat \cite{buhler2014waves,churilov2018scattering}.}.
Additionally, non-zero PV can lead to more complicated scattering cross-sections of usual gravity waves \cite{ward2010scattering,churilov2018scattering} since the potential and rotational disturbances couple to one-another \cite{favraud2013acoustic}.

In spite of such difficulties, we argue in this chapter that, in certain regimes, the effects of vorticity can be encoded in a quantity that preserves the causal structure of the geometry; namely, an effective local mass.
We then justify our model by exploiting an exact wave equation derived from the fluid equations in \cite{visser2004vorticity}, which governs the propagation of gravity waves and their interaction with the rotational degrees of freedom.
Specifically, using a Rankine type vortex as the fluid flow, we demonstrate good agreement of the effective potentials and the characteristic modes (i.e. the resonance spectra) of the two different wave equations.
One particularly interesting finding is that the spectrum exhibits additional modes to those found in an irrotational vortex, which are similar in origin to the bound states of a hydrogen atom.
Other studies of waves on a Rankine vortex include 
\cite{kopiev2010rankine,pritchard1970solitary,ford1994instability}.
The results of this chapter are based our work in \cite{patrick2018QBS}; the analysis there is the same, although the system under consideration differs importantly as we shall explain later.

\section{Wave equation in shallow water}

To find the equation of motion of the perturbations, it is helpful to define the splitting,
\begin{equation} \label{RotPerturb}
\mathbf{v} = \bm\nabla\phi + \bm\xi,
\end{equation}
where $\phi$ is the usual irrotational degree of freedom and $\bm\xi$ encodes the rotational part of the wave motion.
This decomposition in it's current form is not unique since we have the freedom to redefine the fields \mbox{$\phi\to\phi+\phi'$}, \mbox{$\bm\xi\to\bm\xi-\bm\nabla\phi'$}.
Two cases which are of particular interest to us are the  Helmholtz decomposition and the Clebsch representation, which we discuss separately in a moment.
In general, the two decompositons are not equivalent to one another; in particular the second term of the Clebsch representation is not solenoidal (see \cite{wu2007vorticity} around page 43 for a discussion of this).

In the following, we consider perturbations to an effectively two dimensional shallow fluid flow which is only a function of $\mathbf{x}=(x,y)$.
The vorticity of a 2D velocity field $\mathbf{V}$ is a scalar quantity,
\begin{equation} \label{zeta}
\zeta = \bm{\mathrm{e}}_z\cdot(\bm\nabla\times\mathbf{V}) = \bm{\tilde{\nabla}}\cdot\mathbf{V},
\end{equation}
where we have defined the co-gradient (co-grad) operator,
\begin{equation}
\bm{\tilde{\nabla}} = \bm{\mathrm{e}}_z\times\bm\nabla = \begin{pmatrix}
-\partial_y \\ \partial_x
\end{pmatrix},
\end{equation}
which is an anti-clockwise rotation of the 2D gradient operator by $90\degree$.

Since we are no longer dealing with an irrotational field, the Euler equation in Eq.~\eqref{euler} cannot be reduced to the Bernoulli equation in Eq.~\eqref{bernoulli}.
Thus the equations of motion for the fluctuations in shallow water are the perturbed Euler and continuity equations, which for a fluid with a flat free surface are,
\begin{subequations} 
\begin{align}
D_t\mathbf{v} + (\mathbf{v}\cdot\bm\nabla)\mathbf{V} + g\bm\nabla h = \ & 0, \label{PerturbEuler_a} \\
D_t h + H\bm\nabla\cdot\mathbf{v} = \ & 0. \label{PerturbEuler_b}
\end{align}
\end{subequations}

\subsection{Helmholtz decomposition}

The first way that we propose to decompose the velocity vector field is by writing it in terms of it's curl-free (potential) and divergence free (solenoidal) parts.
This representation is the well known Helmholtz-Hodge decomposition \cite{wu2007vorticity,lin2009study}.
For a perturbation in two dimensions, we may write it using the co-grad operator as,
\begin{equation} \label{Helmholtz}
\mathbf{v} = \bm\nabla\phi + \bm{\tilde{\nabla}}\psi.
\end{equation}
With this notation, we may identify $\psi$ with the \textit{stream function}, which is frequently used for classifying incompressible fluids in two dimensions.

This representation is particularly well-suited to writing a wave equation for the perturbations for the case of an axisymmetric solid body rotation, which is of course what we have in the core of the Rankine vortex.
Inserting Eq.~\eqref{Helmholtz} into the perturbed shallow water equations, Eq.~\eqref{PerturbEuler_a} becomes,
\begin{equation} \label{EqMot4}
\begin{split}
\bm\nabla\left(D_t\phi- \zeta\psi+gh\right) + & \ \bm{\tilde{\nabla}}\left(D_t\psi+\zeta\phi\right) \\ 
+ \psi\bm\nabla\zeta - \phi\bm{\tilde{\nabla}}\zeta + & \ (\partial_iV_j)\tilde{\partial}_j\psi - (\tilde{\partial}_iV_j)\partial_j\psi = 0.
\end{split}
\end{equation}
In an axisymmetric system, the last two terms vanish either if $\psi=0$ (purely potential perturbations) or \mbox{$\partial_rV_{r,\theta} = V_{r,\theta}/r$} is satisfied.
The condition on $V_\theta$ restricts us to a solid body rotation, i.e. $\zeta=\mathrm{const}$, in which case the terms proportional to the gradients of the vorticity also vanish.
The condition on the radial component restricts us to $V_r\propto r$.
Although in this chapter our system will not have a radial flow, we note that this condition is approximately satisfied by the function in Eq.~\eqref{softplug} which we used to fit the radial flow in the experiment of Chapter~\ref{chap:super} (see discussion in Section~\ref{sec:flow_values} for further details).

We are left then with the first two terms in Eq.~\eqref{EqMot4}, which is an equation of the form \mbox{$\bm\nabla P_1 + \bm{\tilde{\nabla}}P_2 = 0$}.
Taking the divergence of this equation gives $\nabla^2P_1=0$, and taking the curl (i.e. the co-divergence) yields $\nabla^2P_2=0$.
Now, since we will later assume an axisymmetric background, we may write \mbox{$P_k=P_{k,m}\exp(im\theta)$} where $k=1,2$.
In two dimensions the Laplace equation assumes the form of the Cauchy-Euler equation, which is solved by,
\begin{equation}
\partial_r^2P_{k,m} + \frac{1}{r}\partial_rP_{k,m} - \frac{m^2}{r^2}P_{k,m} = 0 \quad \to \quad P_{k,m} = \alpha_{k,m}r^m + \beta_{k,m}r^{-m}.
\end{equation}
Clearly, this diverges at the origin and at infinity and since we are searching for wave solutions, we only obtain a non-divergent solution when  $\alpha_{k,m}=\beta_{k,m}=0$.
Therefore \mbox{$P_{1,2}$} must both vanish respectively.

We now arrive at the three equations for our three fields \mbox{$\phi,\psi,h$}. These are,
\begin{equation}
\begin{split}
D_t\phi-\zeta\psi+gh = & \ 0, \\
D_t\psi+\zeta\phi = & \ 0, \\
D_th+H\nabla^2\phi = & \ 0,
\end{split}
\end{equation}
which may be combined into a wave equation for a single field,
\begin{equation} \label{waveeqnMass}
D_t^2\phi - c^2\nabla^2\phi + \zeta^2\phi = 0.
\end{equation}
This is the KG equation of motion for a free field with an effective mass which is proportional to the background vorticity.
We emphasize again that this equation only respects it's fluid dynamical origins either when either \mbox{$V_{r,\theta}\propto r$} or when $\zeta=0$.
Specifically, for a fluid with no radial flow, the equation is exact for either solid body rotation or irrotational flow.

Since these two cases are the extreme limits of a Rankine type vortex, we propose to use Eq.~\eqref{waveeqnMass} over the whole $r$ range as a toy model to predict the effects of vorticity.
This toy model neglects extra terms that contribute to wave equation around $r\sim r_0$, i.e. in the region of interpolation between solid body and irrotational flow.
Although this proposal seems fairly ad hoc, one of the aims of this chapter will be to show that, at least in some regimes, Eq.~\eqref{waveeqnMass} captures the essence of how the perturbations see the background flow.

Performing the usual coordinate transformations (see Eqs.~\eqref{fields} and \eqref{tortoise}) we may cast Eq.~\eqref{waveeqnMass} into the form \mbox{$-\partial_{r_*}^2\psi_m+V_\mathrm{mass}\psi_m=0$}, where the potential is,
\begin{equation} \label{potentialMass}
V_\mathrm{mass}(r) = - \left(\omega-\frac{mV_\theta}{r}\right)^2 +
\big(c^2- V_r^2 \big) \left(\frac{m^2-1/4}{r^2} + \frac{V_r^2}{4c^2r^2} - \frac{V_r\partial_rV_r}{c^2r} + \frac{\zeta^2}{c^2}\right).
\end{equation}
To show that this toy model potential is a good approximation, we now derive another wave equation for the perturbations based on a different vector decomposition and compare the two.

\subsection{Clebsch representation}

The second vector field representation we will discuss is the Clebsch decomposition, which we will use to derive a wave equation that is both fully consistent with the fluid equations and justifies the toy model derived above.

Any vector field (in up to three dimensions) can be decomposed into the Clebsch representation \cite{Clebsch1859,whitham1968variational,wu2007vorticity},
\begin{equation} \label{clebsch}
\mathbf{V} = \bm\nabla X_1 + X_2\bm\nabla X_3,
\end{equation}
where $X_1,X_2,X_3$ are scalar functions of $x,y,z$, provided $\mathbf{V}$ is continuous, smooth and decays sufficiently quickly toward infinity.
Let $\chi_1,\chi_2,\chi_3$ be linear perturbations to $X_1,X_2,X_3$.
A linear perturbation of the velocity field may then be written in the form,
\begin{equation} \label{clebschPert}
\mathbf{v} = \bm\nabla\chi_1 + \chi_2\bm\nabla X_3 + X_2\bm\nabla\chi_3.
\end{equation}
A particular combination of the $X$ and the $\chi$ potentials yields simple equations of motion for the perturbations, which can be identified most readily by direct insertion of Eq.~\eqref{RotPerturb} into the shallow water equations in Eqs.~\eqref{PerturbEuler_a} and~\eqref{PerturbEuler_b}.
This yields,
\begin{subequations}
\begin{align}
\bm\nabla(D_t\phi+gh) + \left[D_t\bm{\xi} + (\bm\xi\cdot\bm\nabla)\mathbf{V} + \zeta\bm{\tilde{\nabla}}\phi\right] = 0, \label{EqMot1_a} \\
D_th + H\nabla^2\phi + H\bm\nabla\cdot\bm\xi = 0, \label{EqMot1_b}
\end{align}
\end{subequations}
where this $\phi$ is not the same as that used for the Helmholtz decomposition of the previous subsection.
Although it is not immediately obvious, both the term under the $\bm\nabla$ and the term under square brackets in Eq.~\eqref{EqMot1_a} are respectively equal to zero when $\phi$ and $\bm\xi$ chosen such that,
\begin{equation} \label{PhiClebsch}
\phi = \chi_1 + X_2\chi_3, \qquad \bm\xi = \chi_2\bm\nabla X_3 - \chi_3\bm\nabla X_2,
\end{equation}
which is demonstrated in \cite{visser2004vorticity} for acoustic perturbations. 
In Appendix~\ref{app:clebsch}, we show that their analysis also holds for surface waves.
Combining the first term of Eq.~\eqref{EqMot1_a} with Eq.~\eqref{EqMot1_b}, we obtain the wave equation for $\phi$,
\begin{equation} \label{EqMot2}
D_t^2\phi - c^2\nabla^2\phi = c^2\bm\nabla\cdot\bm\xi,
\end{equation}
which is sourced by the divergence of the rotational part of the wave.
The equation of motion for $\bm\xi$ is,
\begin{equation} \label{EqMot3}
D_t\bm{\xi} + (\bm\xi\cdot\bm\nabla)\mathbf{V} + \zeta\bm{\tilde{\nabla}}\phi = 0.
\end{equation}
The benefit of defining the splitting this way is that both $\phi$ and $\xi$ are related to physical variables and are hence gauge invariant quantities:
\begin{enumerate} [noitemsep]
\item $\phi$ is related to height fluctuations in the usual manner, $gh=-D_t\phi$.
\item $\bm\xi$ is the product of the vorticity with a rotation of the particle displacement $\mathbf{x}_1$ due to the wave \cite{visser2004vorticity}, i.e. \mbox{$\bm\xi=\zeta\mathbf{x}_1\times\mathrm{\mathbf{e}}_z$}.
\end{enumerate}
Since \mbox{$\bm\nabla\phi\sim\omega\mathbf{x}_1$}, $\bm\xi$ forms only a small correction to the perturbed velocity and can be neglected in scenarios where the vorticity is much smaller than the wave frequency \cite{liberati2019vorticity}, i.e. $\zeta\ll\omega$.
In this case, Eq.~\eqref{EqMot2} reduces to the usual wave equation in Eq.~\eqref{waveeqn}.
For $\omega\sim\zeta$ however, the source term must be taken into account and additional degrees of freedom become important in describing the linear dynamics.


\subsection{Purely rotating fluid}

Eq.~\eqref{EqMot3} is exactly solvable in the case of an axisymmetric, stationary fluid which only has an angular component, i.e. \mbox{$\mathbf{V}=V_\theta(r)\mathrm{\mathbf{e}}_\theta$}.
The fields may be decomposed in the usual manner \mbox{$(\phi,\bm\xi) = (\phi_m,\bm\xi_m)\exp(im\theta-i\omega t)$}.
Then, since $D_t\to-i\tilde{\omega}$ and there are no $r$ derivatives acting on $\bm\xi$, Eq.~\eqref{EqMot3} becomes a set of algebraic equations for the components of $\bm\xi_m$ which (following \cite{richartz2009generalized}) can be solved for in terms of the potential field,
\begin{equation} \label{XiRotating}
\bm\xi_m(r) = \frac{\zeta}{K\tilde{\omega}}\left[\frac{\mathrm{\mathbf{e}}_r}{r}\left(-m+\frac{2V_\theta}{\tilde{\omega}}\partial_r\right)+i\mathrm{\mathbf{e}}_\theta\left(\frac{m\zeta}{r\tilde{\omega}}-\partial_r\right)\right]\phi_m(r),
\end{equation}
where $K$ is defined,
\begin{equation}
K = 1 - \frac{2\zeta V_\theta}{\tilde{\omega}^2r}.
\end{equation}
Using Eq.~\eqref{XiRotating}, we may write Eq.~\eqref{EqMot2} as a second order ODE for a new field $R_m = (r/K)^{1/2}\phi_m$,
\begin{equation} \label{EqMot5}
-\partial_r^2R_m + \mathcal{V}(r)R_m = 0,
\end{equation}
where the potential is given by,
\begin{equation} \label{potentialExact}
\mathcal{V}(r) = -\tilde{\omega}^2K\left[\frac{1}{c^2}- \frac{m}{r\tilde{\omega}}\partial_r\left(\frac{\zeta}{\tilde{\omega}^2K}\right)\right] - \frac{m^2}{r^2}
+ \frac{1}{2}\partial_r^2\log\left(\frac{r}{K}\right) - \frac{1}{4} \left(\partial_r\log\left(\frac{r}{K}\right)\right)^2.
\end{equation}
Note that when $\zeta=0$, we recover the usual potential for irrotational perturbations in Eq.~\eqref{potential}, and when $\zeta=\mathrm{const}$, $\mathcal{V}$ reduces to the toy model potential in Eq.~\eqref{potentialMass}.
However, for $\zeta=\zeta(r)$, $\mathcal{V}$ can in general have complicated behaviour.
In particular, if $\tilde{\omega}$ or $K$ are equal to zero at some radius then $\mathcal{V}$ will diverge.

Since Eq.~\eqref{EqMot5} is exact, understanding the properties of $\mathcal{V}$ may provide useful insights into the effect of vorticity on wave propagation. 
For our purposes, it serves to illustrate that the toy model potential of Eq.~\eqref{potentialMass} is a good approximation at least in a certain regimes.
Specifically, if $\tilde{\omega}>0$ and $K>0$ everywhere (the first of which is guaranteed for $m<0$), $\mathcal{V}$ does not diverge anywhere except for the origin due to the usual angular momentum barrier.

\begin{figure} 
\centering
\includegraphics[width=0.5\linewidth]{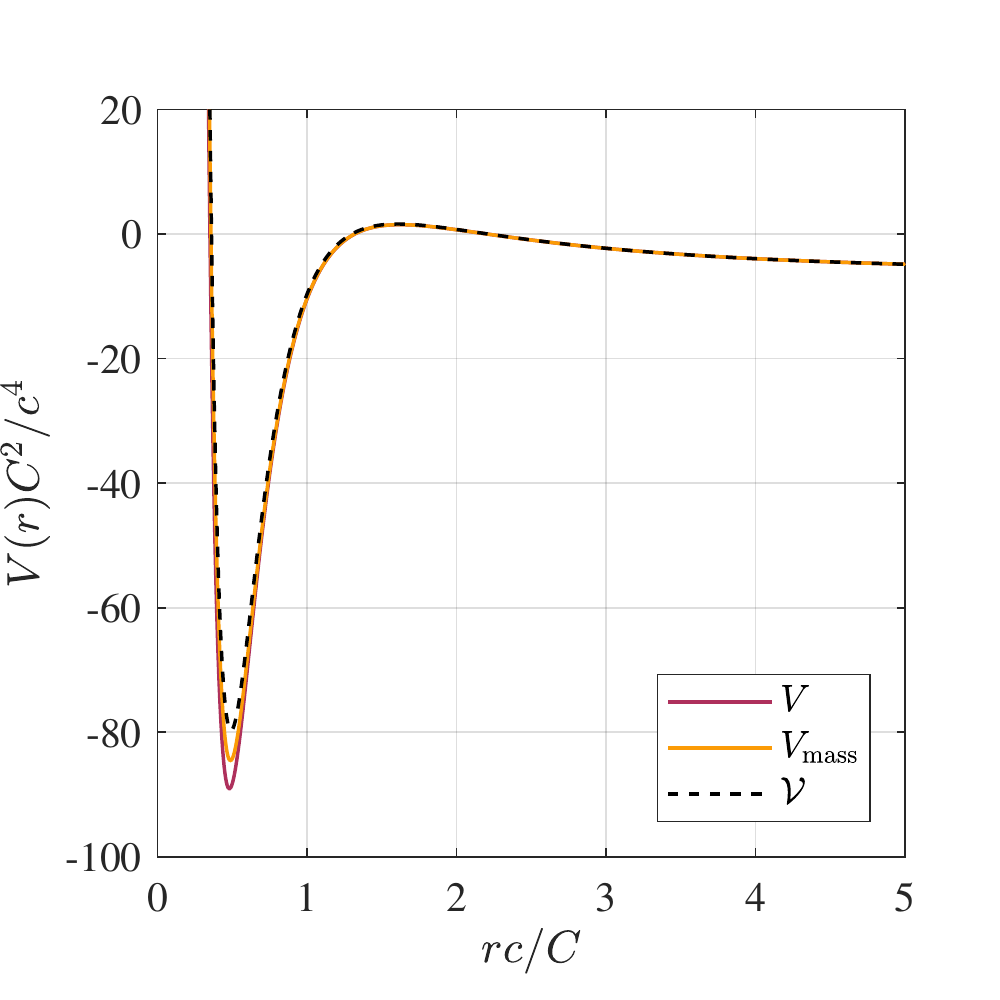}
\caption{Comparison of the different effective potentials for a rotating fluid with vorticity.
$\mathcal{V}$ is the exact potential in Eq.~\eqref{potentialExact}, $V$ is the usual potential in Eq.~\eqref{potential} for $V_r=0$ and $V_\mathrm{mass}$ is the toy model potential in Eq.~\eqref{potentialMass}.
In all three cases, we use an $n=1$ vortex with \mbox{$r_0c/C=1/1.95$} at a frequency \mbox{$\omega C/c^2 = 2.85$} for the $m=-10$ mode.
The presence of a potential maximum leads to the existence of the light ring modes we studied in the previous chapter.
The potential well centred on $\sim r_0$ suggests the system can also exhibit trapped resonances called \emph{bound states}.} \label{fig:compare}
\end{figure}

Using the $n=1$ vortex from Eq.~\eqref{Nvortex}, Figure~\ref{fig:compare} demonstrates that in this regime, $\mathcal{V}$ and $V_\mathrm{mass}$ for $V_r=0$ are qualitatively similar.
In particular, they both exhibit a local minimum around $r\sim r_0$.
We also display the usual potential without the effective mass term but using rotational velocity profile for $V_\theta$.
This also captures the qualitative features of $\mathcal{V}$, although the local minimum of $V_\mathrm{mass}$ is closer to that of $\mathcal{V}$.
As a consequence, the behaviour of the perturbations can be modelled using Eq.~\eqref{waveeqnMass}, which is much simpler to solve than the exact equations in Eqs.~\eqref{EqMot2} and~\eqref{EqMot3}.
In particular, we expect the characteristic frequency (or resonance) spectrum, which is determined by the features of the potential as we saw in the previous chapter, for both $V_\mathrm{mass}$ and $\mathcal{V}$ to be qualitatively similar.

\section{Bound states}

Bound states (BSs) are characteristic modes of a system that can occur when it's effective potential contains a well, and like quasinormal modes, they are determined by purely out-going conditions in the asymptotic region.
They appear as poles in the Green's function associated to the wave equation \cite{dolan2012resonances}, and can therefore be seen as the resonant frequencies of the system.
Perhaps the most renowned example of bound states are those in the Coulomb potential of the hydrogen atom \cite{loudon1959one,haines1969one,hainz1999centrifugal}.

If the energy of the bound state is higher than the potential energy at infinity, then it is possible for the mode to tunnel out of the confining well.
Modes that undergo this process are sometimes known as quasibound states, although we will stick to the generic name bound states in this chapter. 
These are known to occur, for example, in quantum systems (e.g. \cite{schomerus2004quantum}) and for massive fields around black holes \cite{hodd2015kerr,rosa2012massive}.
A semiclassical WKB analysis shows that the bound state energy levels in the potential well are determined by a Bohr-Sommerfeld condition \cite{berry1972semiclassical,dolan2015dirac}.

Having seen in Chapters~\ref{chap:super} and \ref{chap:qnm} that the angular velocity is the dominant component in our experimental set-up, we propose to study the resonant frequencies of a purely rotating fluid with vorticity.
As we shall see, it is possible for perturbations to become trapped in a region close to the core radius, due to the local minimum in the effective potential in Figure~\ref{fig:compare}.
We proceed in this section as follows:
\begin{enumerate}
\item First we illustrate, with an example in flat space ($\mathbf{V}=0$), how to deal with a divergence in the potential due to the angular momentum barrier at the origin using the WKB approximation.
\item To predict the characteristic modes, we perform a WKB analysis using the toy model potential in Eq.~\eqref{potentialMass}, finding that in addition to the usual light-ring modes, $V_\mathrm{mass}$ also exhibits bound states.
We choose to work with $V_\mathrm{mass}$ over $\mathcal{V}$ due to it's analytic simplicity.
\item We simulate the toy model in Eq.~\eqref{waveeqnMass} to check how the full numerical solution compares to the WKB predictions.
We then simulate the exact equations in Eqs.~\eqref{EqMot2} and~\eqref{EqMot3} to check that the interesting features of the toy model are also present in the true system.
\end{enumerate}

\subsection{Scattering in flat space} \label{sec:BesselR0}

In Appendix~\ref{app:polar}, we discuss the $m$-components of a wave propagating in flat space, i.e. $\mathbf{V}=0$, observing that there is a single turning point in the effective potential located at $r_\mathrm{tp}=|m|c/\omega$.
There, we show that the WKB approximation is in good agreement with the exact solution (which is a Bessel function) in the region $r>r_\mathrm{tp}$.
The WKB solution is also valid for $r<r_\mathrm{tp}$ and the two regions can be connected using the transfer matrix in Eq.~\eqref{1tp_cf}.

To do this, we must first evaluate the WKB solution at the origin and throw away the divergent mode.
Recall that the WKB solution is of the form, 
\begin{equation}
\phi^\pm_m\sim e^{\pm i\int k_r dr}/|k_rr|^{1/2},
\end{equation}
where the wave vector is given by,
\begin{equation}
k_r=\sqrt{\omega^2/c^2-m^2/r^2}.
\end{equation}
Clearly, in the limit that $r\to 0$, the amplitude factor becomes $|k_rr|^{-1/2}\to|m|^{-1/2}$.
The phase integral is,
\begin{equation}
\begin{split}
\int k_r dr & \ = k_rr + m\tan^{-1}(m/k_rr) \\
& \ = i|m|\sqrt{1-r^2/r_\mathrm{tp}^2} + m\tan^{-1}\left(-i~\mathrm{sgn}(m)\left(1-r^2/r_\mathrm{tp}^2\right)^{-\frac{1}{2}}\right) \\
& \ \mathrel{\overset{r\to0}{\sim}} i|m| + m\tan^{-1}\left(-i~\mathrm{sgn}(m)\left(1+\frac{r^2}{2r_\mathrm{tp}^2}\right)\right).
\end{split}
\end{equation}
To reduce this further, we use the identity,
\begin{equation}
\tan^{-1}(z) = \frac{i}{2}\log\left(\frac{1-iz}{1+iz}\right),
\end{equation}
for $z\in\mathbb{C}$, which gives,
\begin{equation}
\int k_r dr \mathrel{\overset{r\to0}{\sim}} i|m| + \frac{i|m|}{2}\log\left(\frac{r^2}{4r_\mathrm{tp}^2}\right).
\end{equation}
Hence the WKB solution at the origin is of the form,
\begin{equation} \label{BesselLim0}
\phi_m^\pm(r\to0) \sim \frac{1}{|m|^{1/2}}\left(\frac{er}{2r_\mathrm{tp}}\right)^{\mp |m|}.
\end{equation}
Since the $+$ mode diverges at the origin, it is unphysical and we discard it.
The $-$ mode however scales with $r^{|m|}$ and remains finite.
Using the transfer matrix in Eq.~\eqref{1tp_cf}, we may write,
\begin{equation}
\begin{pmatrix}
A_m^0\\ 0 
\end{pmatrix} = e^{i\frac{\pi}{4}}\begin{pmatrix}
1/2 & -i/2\\
-i & 1
\end{pmatrix} \begin{pmatrix}
A_m^\mathrm{out}\\ A_m^\mathrm{in}
\end{pmatrix},
\end{equation}
which tells us that the in and out-going components are related by a phase shift $A_m^\mathrm{out} = e^{-i\pi/2}A_m^\mathrm{in}$, i.e. the magnitudes are the same as expected.

Although nothing interesting happens in this simple example, the form of $V_\mathrm{mass}$ for a fluid which rotates as a solid body near the origin is the same as the effective potential used here, with the replacement $\omega^2\to(\omega-m\zeta/2)^2-\zeta^2$, which is not a function of $r$.
Hence the solution is the same in the limit $r\to0$ and will be of the form of Eq.~\eqref{BesselLim0}.

\subsection{Resonances of a Rankine vortex}

Now that we know how to deal with the divergence in the potential at the origin, we are in a position to consider the scattering of a wave with a potential resulting from a Rankine-type velocity field, particularly that of the $n=1$ vortex in Eq.~\eqref{Nvortex}. 
We concern ourselves with two possibilities. 
The potential either has one turning point, in which case the leading order contribution in the WKB method is a pure refection as explained above, or there are three turning points\footnote{Again, one could better estimate the scattering coefficients in the limit that the second and third turning points become close by expanding about the stationary point. We do not do this here, however this approach was followed in Section~\ref{sec:qnm_est} to estimate the QNMs (light-ring modes) which we will exploit later on.}.
We label these $1$ to $3$ moving outward from the origin.

Following in the spirit of Section~\ref{scatterO1}, we can relate the in-going and out-going amplitudes using transfer matrices at the turning points,
\begin{equation} \label{RankineScatter}
\begin{pmatrix}
A_m^0 \\ 0
\end{pmatrix} = \left|\frac{p_\infty}{p_0}\right|^{\frac{1}{2}}(T^*)^{-1} J_{1,2} T J_{2,3} (T^*)^{-1} \begin{pmatrix}
A_m^\mathrm{out} \\ A_m^\mathrm{in}
\end{pmatrix},
\end{equation}
where,
\begin{equation}
J_{1,2} = \begin{pmatrix}
e^{-iS_{1,2}} & 0 \\
0 & e^{iS_{1,2}}
\end{pmatrix}, \qquad J_{2,3} = \begin{pmatrix}
0 & e^{S_{2,3}} \\
e^{-S_{2,3}} & 0
\end{pmatrix}, \qquad S_{a,b} = \int^{r_{\mathrm{tp},b}}_{r_{\mathrm{tp},a}}|p_0(r)|\frac{dr}{c},
\end{equation}
where again, $p_0=(-V)^{1/2}$ from the leading order WKB solution. 
For $V$, we will use the toy model potential $V_\mathrm{mass}$ since it is easier to work with than the exact potential $\mathcal{V}$.
Furthermore, we will only consider $m<0$ modes for $\omega>0$ so that we don't have have the problem of superradiant instabilities.

Using the non-superradiant reflection and transmission coefficients in Eqs.~\eqref{ReflWKB} and \eqref{TransWKB}, we may write,
\begin{equation}
T J_{23} (T^*)^{-1} = \frac{1}{\mathcal{T}_{2,3}} \begin{pmatrix}
1 & -\mathcal{R}_{2,3} \\
\mathcal{R}_{2,3} & 1
\end{pmatrix},
\end{equation}
where subscript $2,3$ denote that the coefficient is associated to the peak between the second and third turning points.
For conciseness, we drop this subscript in the following since we will not be considering the overall scattering coefficients.

The resonant frequencies of the system can be found from Eq.~\eqref{RankineScatter} by discarding the in-going mode (\mbox{$A_m^\mathrm{in}=0$}) and normalising with respect to the out-going mode (which amounts to setting \mbox{$A_m^\mathrm{out}=1$}).
The bottom line of the vector equation in Eq.~\eqref{RankineScatter} can be solved to show that \mbox{$S_{1,2}(\omega_\mathrm{BS})$} satisfies
\begin{equation} \label{BScond0}
e^{-2iS_{1,2}}=-|\mathcal{R}|,
\end{equation}
where $\omega_\mathrm{BS}\in\mathbb{C}$ are the complex frequencies of the bound states.

Like we did with QNMs in Section~\ref{sec:qnm_est}, we write $\omega_\mathrm{BS}=\omega_\star+i\Gamma$ and assume the characteristic damping time is much longer than the oscillation period of the mode, i.e. $\Gamma\ll\omega_\star$.
This allows us to expand,
\begin{equation}
S_{1,2}(\omega_\mathrm{BS}) = S_{1,2}(\omega_\star) + i\Gamma\partial_\omega S_{1,2}\big|_{\omega=\omega_\star} + \mathcal{O}\left((\Gamma/\omega_\star)^2\right),
\end{equation}
Inserting this into Eq.~\eqref{BScond0} gives us two conditions.
The first is the Bohr-Sommerfeld condition \cite{berry1972semiclassical} that determines the different energy levels (indexed by $p$) of the bound states,
\begin{equation} \label{BScond1}
S_{1,2}(\omega_p) = \pi\left(p+\frac{1}{2}\right),
\end{equation}
where $\omega_\star=\omega_p$ is the oscillation frequency of the energy level.
The second condition gives the lifetime of the mode,
\begin{equation} \label{BScond2}
\Gamma_p = \frac{\log|\mathcal{R}|}{2\partial_\omega S_{1,2}}\bigg|_{\omega=\omega_p},
\end{equation}
where $\partial_\omega S_{1,2}$ is the classical time taken to propagate between turning points $r_{\mathrm{tp},1}$ and $r_{\mathrm{tp},2}$.
We saw in Section~\ref{scatterO1} that, provided turning points $r_{\mathrm{tp},2}$ and $r_{\mathrm{tp},3}$ were not too close to a stationary point, \mbox{$\mathcal{R}\sim 1$} and \mbox{$\mathcal{T}\sim e^{-S_{2,3}}$}.
By writing the reflection coefficient in terms of the transmission coefficient and expanding the $\log$ in Eq.~\eqref{BScond2}, we find,
\begin{equation}
\Gamma_p\sim e^{-S_{2,3}(\omega_p)}.
\end{equation}
In other words, the lifetime of the bound states is exponentially long.
This is in stark contrast to the QNM's whose lifetime is determined by the curvature at the stationary point of the potential, which is not exponential.

In Figure~\ref{fig:BSspec}, we display the WKB approximation for resonances of the $n=1$ vortex with $r_0c/C = 1/1.95$, using the toy model potential $V_\mathrm{mass}$\footnote{In principle, one could also use the exact potential $\mathcal{V}$ provided it doesn't blow up.
We chose to use $V_\mathrm{mass}$ since we locate the turning points analytically (to save computational time) before evaluating the WKB integrals numerically.}.
The QN branches (dark purple diamonds) are determined by Eq.~\eqref{potential_LRconds} for the real part of the spectrum and Eq.~\eqref{QNMimag} for the imaginary part.
The bound states (orange diamonds) are computed using Eq.~\eqref{BScond1} for the real part and Eq.~\eqref{BScond2} for the imaginary part, and the structure of the resonance spectrum is such that the different $\omega_{\mathrm{BS}}$ branches emanate from the $\omega_\mathrm{QN}$ branch.
When multiple orange diamonds are present for a given $m$, the potential can support multiple BSs.

The spectrum is characterised by a single parameter $r_0 c/C$, which controls gradient in $m$ of the BS branches (the branches are flat for the $n=1$ vortex for around $r_0 c/C\sim1/2$).
The further down a point is on the BS branch with respect to the QN branch, the deeper it is in the potential well.
These low lying modes are hard to generate since an external perturbation must tunnel through a wide barrier to excite them, and even once they are excited their long decay time means that only a small fraction is able to escape.
Hence, in an experiment or numerical simulation, the most promising candidates for observation are the BSs at the top of the branches.

\subsection{Numerical verification} \label{sec:BSsims}

\begin{figure} 
\centering
\includegraphics[width=\linewidth]{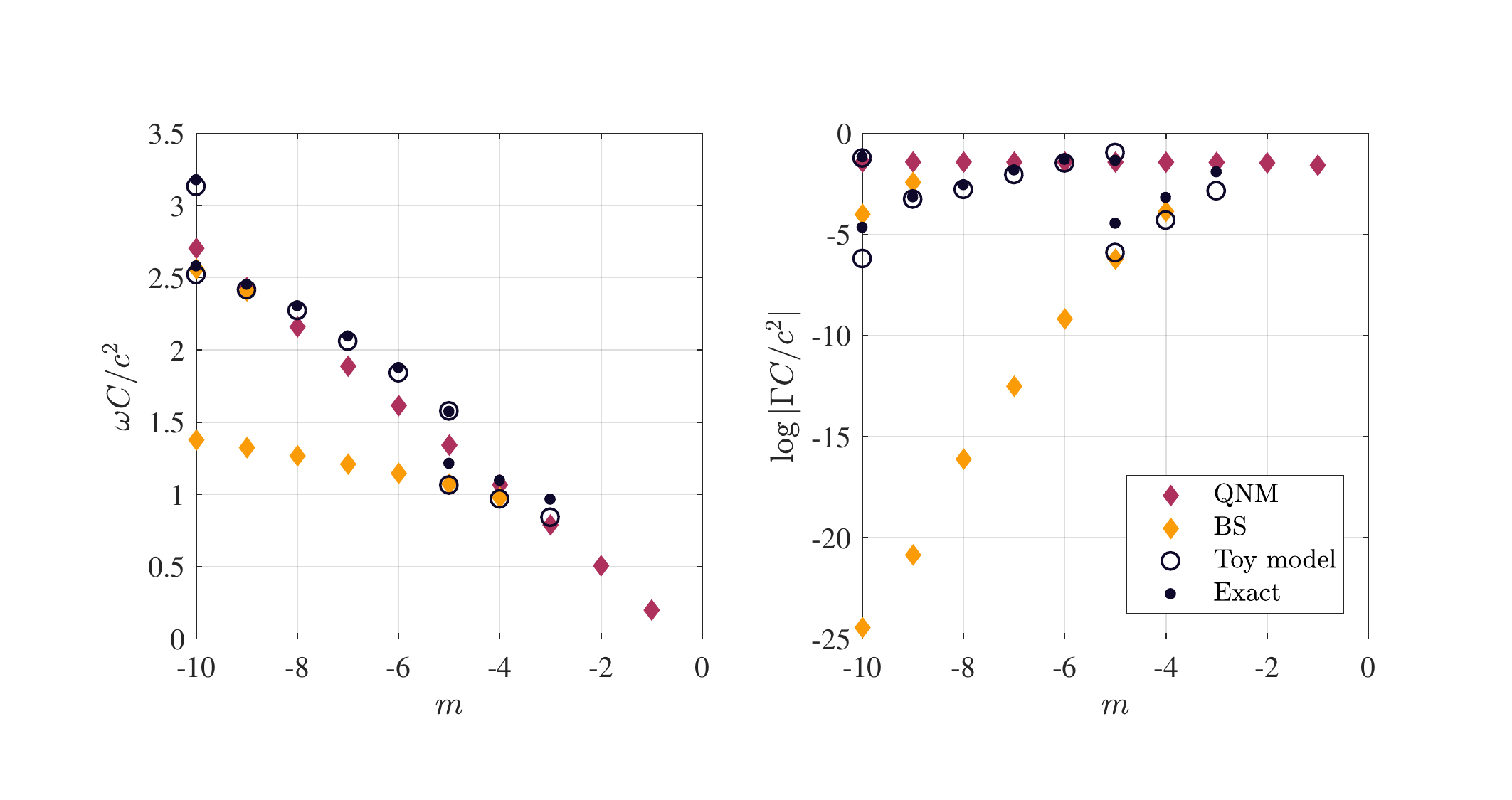}
\caption{The real part (left side) and imaginary part (right side) of the resonant frequency spectrum for the $n=1$ vortex in Eq.~\eqref{Nvortex} with $r_0c/C=1/1.95$.
WKB estimates are shown as diamonds, and predict that the spectrum should follow a branching structure, in contrast to an irrotational vortex whose spectrum is linear in $m$.
Frequencies extracted from numerical simulations confirm this structure.
In particular, the imaginary part of the spectrum demonstrates that resonant frequencies can linger around a vortex for much longer when the background has vorticity.
We plot the logarithm of $|\Gamma|$ to show clearly that the BS modes follow the yellow diamonds.} \label{fig:BSspec}
\end{figure}

To confirm the existence of BSs in a purely rotating fluid with vorticity, we perform a numerical simulation of both the toy model in Eq.~\eqref{waveeqnMass} and the exact system in Eqs.~\eqref{EqMot2} and~\eqref{EqMot3} including the rotational mode $\bm{\xi}$.

To do this, we initialise the simulation (in the same manner as Section~\ref{sec:ABHSmethod}) with a gaussian pulse for \mbox{$\phi_m(r,t=0)$} in the irrotational part of the flow, and \mbox{$\bm{\xi}_m(r,t=0)=0$}.
The inner boundary condition is \mbox{$\phi_m(r=\epsilon,t)=0$}, where epsilon is a small distance from the origin.
This is motivated by Eq.~\eqref{BesselLim0} which says that \mbox{$\phi_m\sim\mathcal{O}(\epsilon^{|m|})$} when epsilon is small enough.
According to Eq.~\eqref{XiRotating}, $\bm{\xi}_m$ is at most \mbox{$\mathcal{O}(\epsilon^{|m|-1})$} in this region, so we also set \mbox{$\bm\xi(r=\epsilon,t)=0$}.
The outer boundary is a hard wall placed far enough away that our perturbations do not reflect from it.

\begin{figure} 
\centering
\includegraphics[width=0.5\linewidth]{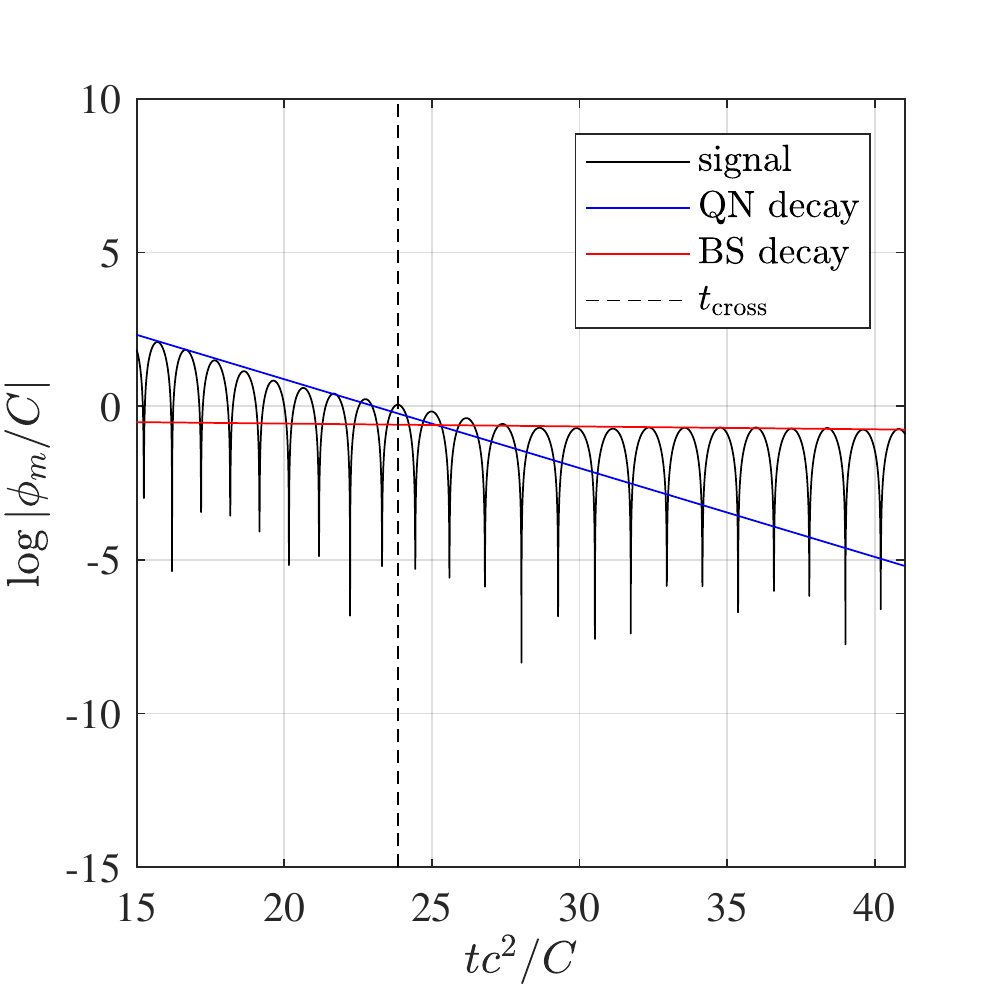}
\caption{Oscillations in the $m=-10$ mode of the $n=1$ vortex after being perturbed by a Gaussian pulse.
The background is characterised by the parameter $r_0c/C=1/1.95$.
These plots are similar to Figures.~2 and~3 of \cite{dolan2012resonances} which display the response of an irrotational vortex, except here we clearly see two modes at single $m$ which is due to vorticity of the flow.
The first mode is the usual QNM and decays quickly relative to the second mode, the BS, which has a much longer lifetime.
The time at which the behaviour switches from QN to BS is estimated by $t_\mathrm{cross}$ in Eq.~\eqref{tcross}.} \label{fig:ResonanceDecay}
\end{figure}

We then evolve the wave equation using the Method of Lines discussed in Appendix~\ref{app:RK4}.
The frequency response of the system is extracted in the same manner as Section~\ref{sec:ABHSmethod}, except this time we first perform the Fourier transform at early times (after the pulse has passed) to find the QNM, then at late times to find the BS (see Figure~\ref{fig:ResonanceDecay}).

In Figure~\ref{fig:BSspec}, we display the frequency response for $m\in[-3,-10]$ for an $n=1$ vortex with $r_0c/C=1/1.95$.
This value was chosen since at each value of $m$ the frequency response was dominated by a single frequency (except for $m=-5,-10$ where there are two) and the absence of interference between different modes allows us to extract the values of $\omega$ and $\Gamma$ more reliably.
In general, multiple frequencies can be present in the signal.
Results from the toy model are depicted as dark circles and the exact equation as dark points, indicating that the two are in good agreement.
The approximation gets better for larger $m$, except when interference occurs.

We find that the real part of the spectrum is no longer just linear (as was the case for the irrotational flow in Figure~\ref{fig:LRshallow}) but now follows the branching structure seen in the WKB estimates.
We see particularly for the $m=-5,-10$ modes that the WKB estimate has a tendency to underestimate the real part of $\omega_\mathrm{QN}$.
This means the red diamonds on the left of Figure~\ref{fig:BSspec} should be shifted up slightly\footnote{A potential reason for this is that the WKB approximation for the QNMs in Eq.~\eqref{QNMcond} assumes a purely in-going mode in the region close to the origin, whereas in the present case the modes reflect off the angular momentum barrier there. Hence, a possible improvement could be to modify Eq.~\eqref{QNMcond} with the correct boundary condition at the origin (obtained in Section~\ref{sec:BesselR0}) and solve for complex QN frequency, as opposed to using the light-ring approximation in Eqs.~\eqref{potential_LRconds} and~\eqref{QNMimag} as we have done.}, leaving room for extra modes on the BS branches.
This agrees with results in \cite{patrick2018QBS}, where it was seen using the toy model potential for the same flow with a $1/r$ radial component that the WKB approximation tended to underestimate the QN branch.
This was further corroborated by the continued fraction method, which allows for precise numerical computation of the resonant frequencies.

The character of the modes is perhaps more evident from the imaginary part of the spectrum on the right of Figure~\ref{fig:BSspec}, which makes it clear which modes are long-lived (those with small $\Gamma$).
All values of $\Gamma$ obtained from the numerics were negative.
However, this does not necessarily mean the system is stable, see e.g. \cite{mougel2017freesurface,oliveira2014ergoregion,mougel2014waves,ford1994instability} for instabilities in purely rotating vortex flows.
It is difficult to say for some modes (particularly $m=-6$ here) whether they are QN or BS in nature, since at the top of the potential well, there is a band in $\omega$ where the two regimes over lap.
The distinction becomes clearer when there are two modes in the spectrum (as in the case for $m=-5,-10$) which we discuss in more detail now.

Since we are exciting the system by sending a pulse from far away, the BS modes have to tunnel twice through the barrier between $r_{\mathrm{tp},2}$ and $r_{\mathrm{tp},3}$ before they appear in the response signal, meaning they will be inherently lower amplitude than the QNMs.
Hence, the QNMs will initially dominate the response and only after they have decayed sufficiently will there be a cross-over to the BS regime.
We estimate the cross-over time $t_\mathrm{cross}$ with the following argument.

Let $A_\mathrm{QN}$ and $A_\mathrm{BS}$ be the amplitudes of the $\omega=\omega_\mathrm{QN,BS}$ components in the initial signal respectively, and $t=t_0$ the instant at which the modes start decaying.
Since the QNM does not have to tunnel, it's amplitude at $t_0$ will be $\sim A_\mathrm{QN}$.
The BS has to transmit through the outer barrier twice, hence it's amplitude is $\sim|\mathcal{T}|^2A_\mathrm{BS}$ at $t_0$.
The evolution of a single decaying mode on a logarithmic plot of the response signal is a straight line with negative gradient (see the blue line on Figure~\ref{fig:ResonanceDecay} for the QNM  and red for BS).
The time at which a cross-over occurs is approximately the time at which these two curves intersect, i.e.
\begin{equation}
\log(A_\mathrm{QN})+\Gamma_\mathrm{QN}(t_\mathrm{cross}-t_0) = \log(|\mathcal{T}|^2A_\mathrm{BS})+\Gamma_\mathrm{BS}(t_\mathrm{cross}-t_0).
\end{equation}
Using the approximation $\mathcal{T}\sim e^{-S_{2,3}}$, we find,
\begin{equation} \label{tcross}
t_\mathrm{cross} \simeq t_0 + \frac{2S_{2,3}+\log(A_\mathrm{QN}/A_\mathrm{BS})}{\Gamma_\mathrm{BS}-\Gamma_\mathrm{QN}}.
\end{equation}
This can be further simplified by assuming $A_\mathrm{QN}\sim A_\mathrm{BS}$, which is satisfied for a broad initial pulse, in which case the $\log$ term may be neglected.
Thus, the cross-over time depends on the decay time of the two modes as well as the size of the barrier seen by the BS\footnote{Note, that for $\omega$ close to $\omega_\star^\mathrm{irrot}$ (where \mbox{$\omega_\star^\mathrm{irrot}=|m|c^2/4C^2$} is the light-ring frequency for the irrotational flow) we find \mbox{$S_{2,3}~\sim\mathcal{O}\left(m(1-\omega/\omega_\star^\mathrm{irrot})\right)$}, which is obtained by expanding the potential about the peak to quadratic order.
Since the peak of $V_\mathrm{mass}$ is in the region where vorticity is small, $\omega_\star^\mathrm{irrot}$ basically coincides with the red diamonds in Figure~\ref{fig:BSspec} (left panel).
Hence the size of $S_{2,3}$ increases with $m$ and the distance of a mode from the QN branch.}.
Using the WKB values for $\Gamma_\mathrm{QN,BS}$ we may estimate $t_\mathrm{cross}$ using Eq.~\eqref{tcross}.
This is shown as the black dashed line in Figure~\ref{fig:ResonanceDecay}.
Since it is an approximation, it does not coincide exactly with the crossing of the red and blue lines.
However, having an estimate a priori for the crossing time is useful for automated data analysis, in both numerics and (potentially) experiment, since it allows us to separate the two regimes.
From $t_\mathrm{cross}$ we can also see that if we want to only excite the bound states, then the initial amplitude of the QNM must be exponentially smaller, i.e. \mbox{$A_\mathrm{QN} <A_\mathrm{BS}e^{-2S_{2,3}}$}.

\begin{figure} 
\centering
\includegraphics[width=0.5\linewidth]{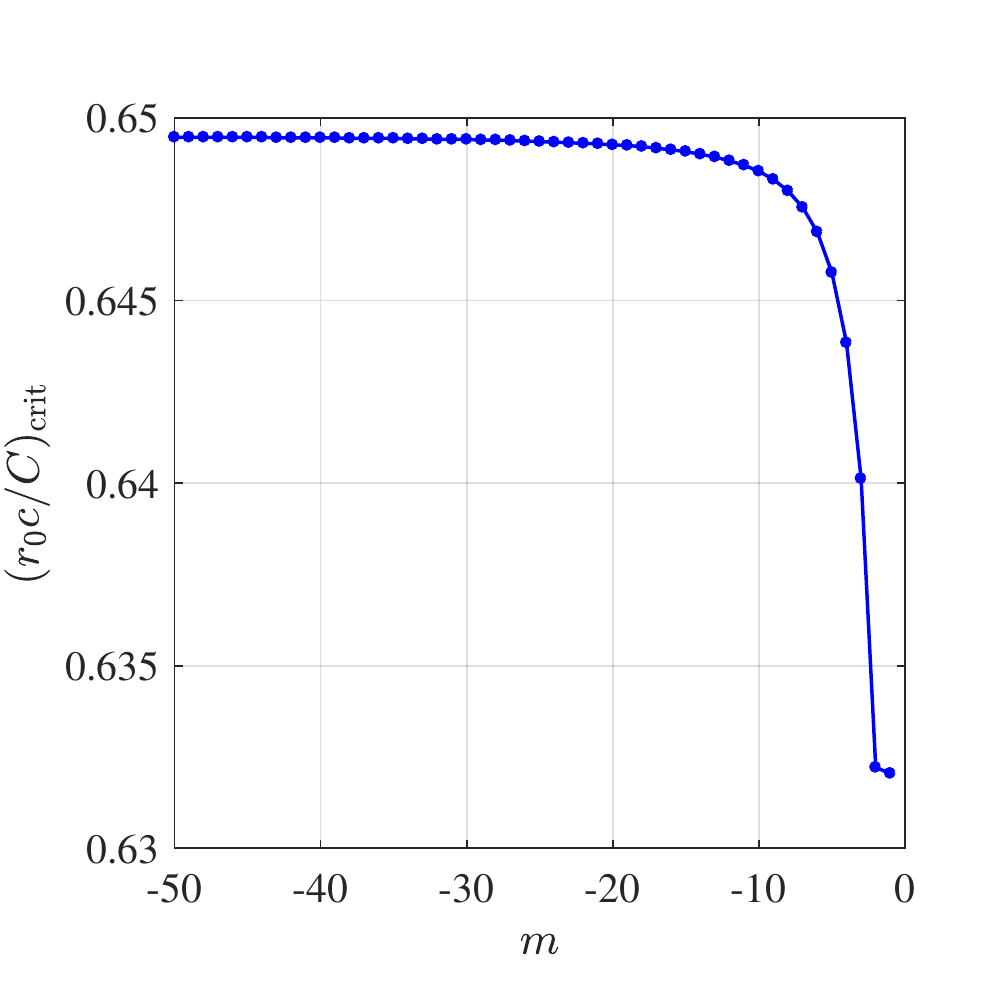}
\caption{The critical value of $r_0c/C$ as a function of $m$ for the $n=1$ vortex, above which the system can no longer support bound states.
This value was computed from a numerical search of $V_\mathrm{mass}$, and tends to $\sim 0.6495$ for large $m$.} \label{fig:CriticalParam}
\end{figure}

Finally, we estimate the regime in which one expects to find bound states using the toy model potential in the form \mbox{$V_\mathrm{mass} = -(\omega-\omega_+)(\omega-\omega_-)$}.
A minimum condition for the existence of bound states (for counter rotating modes with positive frequency) is that $\omega_+$ contains one maximum and one minimum.
There is a critical value $(r_0c/C)_\mathrm{crit}$ at which the stationary points merge to become an inflection point. Above this value, $\omega_+$ has no stationary points and no bound states can exist. 
We plot the value of $(r_0c/C)_\mathrm{crit}$ in Figure~\ref{fig:CriticalParam} for various $m$ values.

\section{Summary}


In this chapter, we have demonstrated that the core of a vortex can significantly alter its QNM spectrum, providing a fluid analogue of the problem of spectral stability in black hole physics \cite{barausse2014environment}.
Specifically, we have shown for a purely rotating fluid that the influence of viscosity on the velocity field in the core can lead to a local minimum in the scattering potential, which can support bound state resonances.
These modes are similar in origin to the bound states of the electron around a hydrogen atom, except the behaviour of the potential at infinity means that the modes we find have a finite lifetime.

Furthermore, we have seen that although the background vorticity can significantly complicate wave scattering, there are two regions in a purely rotating fluid where the equations of motion can be written in the form of the KG equation, thereby preserving the analogy.
The first is far out in the irrotational region, where we recover the usual analogy introduced in Chapter~\ref{chap:theory}.
The second is close to the origin where the fluid rotates as a solid body, in which case the waves obey a KG equation with an effective mass term.
By interpolating between these two regions, we argued that the main effect of vorticity on the propagation of waves over the whole region can be encoded in a quantity which preserves the causal structure of the geometry, namely a effective local mass (see Eq.~\eqref{waveeqnMass}).
This was demonstrated by good agreement of the resonance spectrum of this toy model with that of the exact equations of motion in Eqs.~\eqref{EqMot2} and \eqref{EqMot3}.
A similar study was performed in \cite{patrick2018QBS} using the same toy model with a small radial flow proportional to $1/r$, and similar conclusions to those presented here were also drawn.
However, by numerical simulation, we found that the toy model exhibits vastly different behaviour to the exact equations when a radial flow is included.
Hence, our use of the toy model in a purely rotating fluid is more consistent with its fluid dynamical origins, as we have demonstrated in Figure~\ref{fig:BSspec}.

Based on the experiment of Chapter~\ref{chap:qnm}, observation of the bound states in an experimental set-up may prove challenging.
Since the system there was out of equilibrium, the QNMs (or light-ring modes) are continually being excited, and this was the reason that we were able to able to measure their oscillation frequency but not their lifetime.
However, since the bound states have to tunnel through a barrier, they are inherently lower amplitude than the QNMs and only become observable in the response after the QNMs have decayed (see Eq.~\eqref{tcross}).
Thus, we clearly have a problem if the system does not settle into an equilibrium state since the QNMs will never decay.
One possible way to circumvent this issue could be to construct a larger set-up with a purely rotating flow driven by a rotating plate on the floor of the tank.
The absence of the radial flow negates the requirement for a pump, which was the main source of noise in our experiment, hence an equilibrium should be easier to establish.
Then the dominant bound states would be observable a time $\sim\Delta t=t_\mathrm{cross}-t_0$ after the initial excitation at $t_0$ and before the QNMs have reflected from the outer boundary.
Another possibility to observe the bound states is by looking in a region near $r_0$ where their amplitude is most intense, i.e. in the potential well.

We also note that our analysis focussed only on the counter-rotating modes, to avoid having to deal with complicated behaviour of the potential for those co-rotating. 
Hence, a natural extension to the theory would be to investigate the properties of these modes.
For example, it is known that instabilities (related to superradiance) can occur in purely rotating fluid flows \cite{mougel2017freesurface,oliveira2014ergoregion,mougel2014waves,ford1994instability}.
This would be important knowledge when performing experiments, since instabilities may dominate the response at late times if the damping timescale (inherent in any real fluid) is longer than the instability timescale.

As a final note, the ABHS method discussed at the end of the previous chapter could be used, in principle, with the resonance spectrum of a purely rotating fluid measured in experiment.
Since the light-ring is essentially in the irrotational region of the flow, the usual QNM part of the spectrum would be unaffected by the presence vorticity, and can be used to determine the circulation parameter.
By looking nearer the centre of the flow, one may be able to resolve the bound states in the frequency spectrum, which could be either used to determine the radius of the vortex core, or to test the effective field theory described here if the core radius is already known.

\chapter{Backreaction} \label{chap:back}


In many systems, the path of an object can be described accurately by assuming the environment it moves through is fixed. 
In this case, the trajectory of the object, which could be a particle or a wave for example, is completely determined by the nature of the environment, which can usually be described as a type of field.
Implicit in this assumption is that the object behaves as a `test-particle', whose presence does not alter the configuration of the background field that it sees.
In many scenarios however (e.g. with gravitational fields) the object has it's own associated field which induces corrections in the environment, and hence it's trajectory becomes modified.
Such modifications are known as the \textit{backreaction} of the object on it's environment.
When these corrections are negligible, the test-particle approximation is valid.
However, under certain circumstances, corrections can become large and the backreaction must be taken into account.
In this chapter we consider such a scenario, where the object and it's environment will be surface waves and a DBT flow respectively.

The backreaction is a general effect that occurs when considering the interaction of an object and environment in any field theory.
In linear field theories, e.g. electromagnetism, the self-force of a particle moving under the influence of it's own field causes deviations from the test-particle trajectory.
In non-linear field theories, e.g. general relativity, perturbations interact with the background through the non-linear terms in the governing equations.
For the latter, an expansion of the equations can be used to perturbatively compute corrections, with the leading backreaction term entering at second order~\cite{balbinot2006hawking}.

As we explain at the end of this chapter, the backreaction is expected to play a key role in the evaporation of black holes.
Backreaction type effects are also studied in fluid mechanics under the name \textit{wave-mean interaction} theory \cite{buhler2014waves}.
The theory is developed in the context of atmospheric physics and oceanography 
\cite{hartmann1984observations,dunkerton1980lagrangian} and is concerned with estimating the dynamical implications of small waves on large-scale flows \cite{buhler2005wave}.
For example, wave-mean interaction is responsible for the quasi-biennial oscillation, equatorial winds which oscillate with a period of just over two years \cite{lindzen1968theory,mcintyre1980introduction}.
The same type of mechanism is responsible for mass transport of gravity waves, which results in sediment transport, e.g. in rivers and on beaches \cite{deigaard1992mechanics,van1993principles,ng2004mass}.
Further examples of backreaction effects come from acoustics \cite{esposito2019gravitational} and cosmology \cite{brandenberger2000back,brandenberger2002back,behrend2008cosmological},
with the intriguing possibility that the backreaction of cosmological perturbations could provide an explanation for dark energy \cite{mukhanov1997backreaction,rasanen2004dark,buchert2012backreaction}.



We will see shortly that the backreaction of surface waves in a water tank experiment manifests itself in a particularly simple manner when the system has an open boundary; namely through a change in the water height.
There does not appear to be any mention of this effect in the mainstream literature, we assume because the open boundary of the system means that the total mass is not conserved, and the starting point for many fluid studies is the global conservation of mass.
It is the fact that mass is not conserved globally that allows the water height to change, and it is in estimating this effect that our analysis differs from the standard procedure in the literature.
In particular, we will neglect changes in the velocity field to obtain simple estimates for the evolution of the water height, stressing that these estimates are only valid at early times before any large scale changes in the background have occurred.

This chapter is based on work we initiated in \cite{goodhew2019backreaction} and is structured as follows.
First, we outline the theory for how waves in a shallow water flow can backreact to alter the water height.
We then detail an experiment that was performed to confirm this effect, finding good agreement with the theory.
We further elaborate on additional observations from the experiments of Chapter~\ref{chap:super} in light of this new theory.
Finally, we conclude with a discussion of the backreaction problem for black holes, indicating how our results might be interesting for the gravitational and analogue gravity communities.

\section{Theory}
For an incompressible fluid (density $\rho=\mathrm{const}$) the total mass of the fluid $M$ in a given volume $V$ can only change (in the absence of sources or sinks) if there is a flow of mass $\mathcal{M}$ across the boundary of the system $\delta V$. 
This is expressed mathematically as,
\begin{equation} \label{mdot}
\dot{M} = -\mathcal{M}\big|_{\delta V},
\end{equation}
where overdot denotes derivative with respect to time and the rate of flow of mass through a surface (also called the momentum per unit length) $\mathcal{M}$ is defined as the integral of the mass flux $\mathbf{J}$ over the surface $S$:
\begin{equation} \label{massflow}
\mathcal{M} = \iint_S \mathbf{J}\cdot\hat{\mathbf{n}}dS,
\end{equation}
where $\hat{\mathbf{n}}$ is the vector normal to $S$. 
Consider now a water tank of fixed cross sectional area $\mathcal{A}$ in which water enters the tank via an inlet and exits through a drain in a continuous cycle. 
In complete generality, the fluid state vector is $(H,\mathbf{V})$, where $H(t,\mathbf{x})$ is the height of water height, $\mathbf{V}(t,\mathbf{x})$ is the 3 component velocity field and $(t,\mathbf{x})$ are time and spatial coordinates.
The mass flux for the fluid is in fact the same as it's momentum density,
\begin{equation} \label{MomDensity}
\mathbf{J}=\rho\mathbf{V}.
\end{equation} 
If $\dot{M}\neq0$, the system can only respond by changing the height of the water since the density is constant. 
Our idea that even though local variations in the height change may be complicated, we can estimate the mean height change over the whole system using,
\begin{equation} \label{mhdot}
\dot{M} = \int_\mathcal{A}\int^{\dot{H}}_0\rho~dzdS = \rho\mathcal{A}\langle\dot{H}\rangle,
\end{equation}
where we have defined the spatial average of a quantity (say $f$) over the area of the system,
\begin{equation} \label{DefMean}
\langle f \rangle = \frac{1}{\mathcal{A}}\int_\mathcal{A} f dS.
\end{equation}
Combining Eqs.~\eqref{mdot}, \eqref{massflow}, \eqref{MomDensity} and~\eqref{mhdot} gives,
\begin{equation}
\langle\dot{H}\rangle = -\frac{1}{\mathcal{A}}\iint_{\delta V}\mathbf{V}\cdot\hat{\mathbf{n}}dS.
\end{equation}
Hence, the mean height change throughout the system is determined only by effects at the boundary $\delta V$. 
We will consider a wave scattering experiment so for our purposes, it will be the waves that create a disequilibrium between $\mathcal{M}_\mathrm{in}$ and $\mathcal{M}_\mathrm{out}$. 
Put simply, the extent of the height change during a wave scattering experiment is determined by the amount of fluid pushed out of the system by the waves.


\subsection{The set-up}

\begin{figure}
\centering
\includegraphics[width=0.4\linewidth]{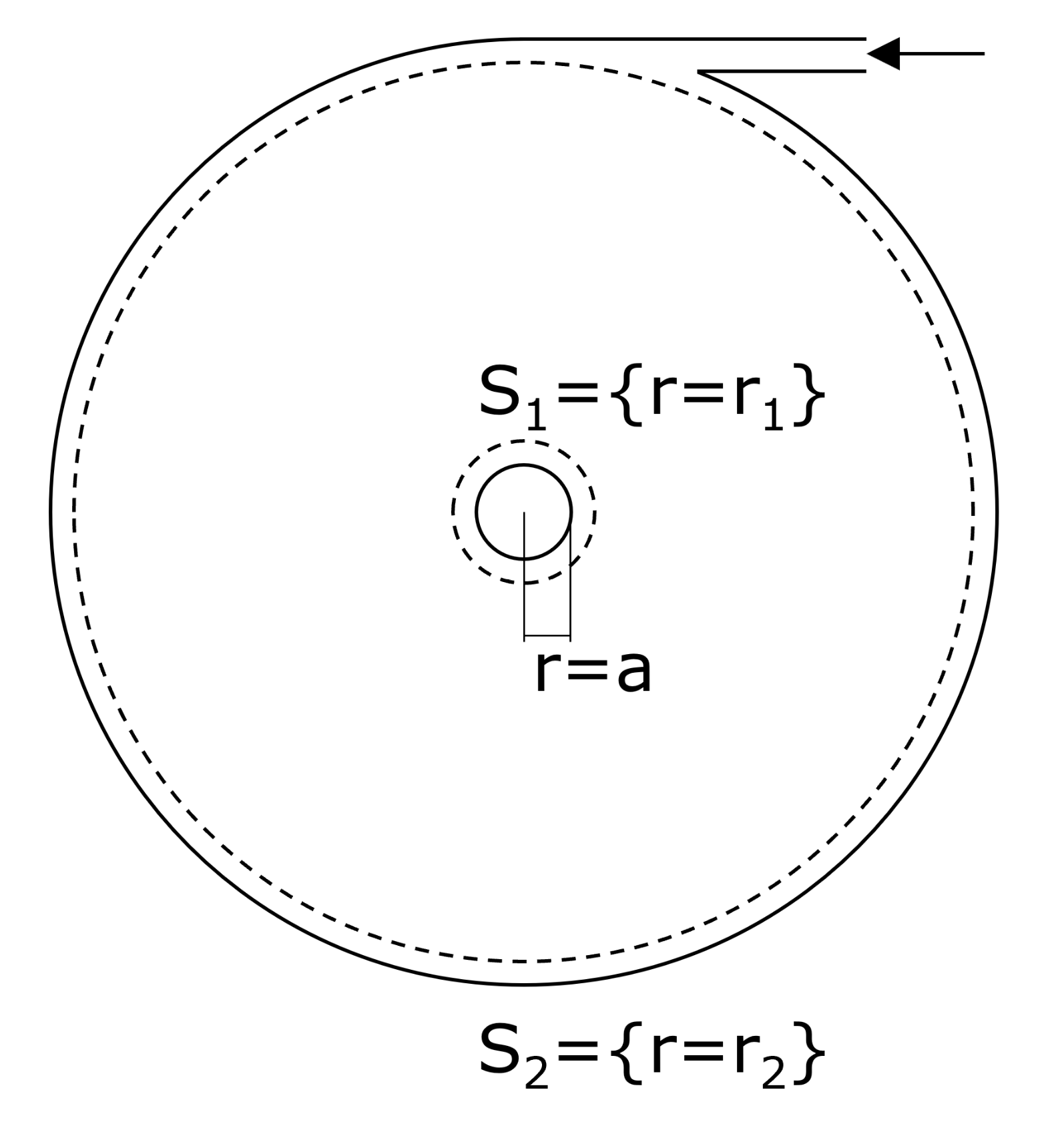}
\caption{Idealised water tank system.
Water flows in through the inlet (dipicted by the black arrow) and out through the drain located in the region $r<a$.
The outer and inner boundaries respectively are taken as the edge of the tank $r=r_2$ and a surface close to the drain $r=r_1$.
Dependence in $\theta$ of the system is contained to a thin layer (depicted as the larger dashed circle) close to $r=r_2$.} \label{fig:circ_setup}
\end{figure}

The set-up we consider is motivated by the wave tank experiment presented in Chapters \ref{chap:super} and \ref{chap:qnm}, with a modified outer boundary to simplify the analysis. 
Adopting cylindrical polar coordinates $\mathbf{x}=(r,\theta,z)$, we consider a drain of radius $a$ located at $z=0$ and centred on the origin. 
The boundary of the system is \mbox{$\delta V= S_1\cup S_2$} for \mbox{$S_{1,2}=\{r=r_{1,2},\theta,z\}$}, where $r_1\simeq a$ and $r_2\gg a$.
The outlet condition is specified on $r=r_1$ and the inlet condition on $r=r_2$ for a small $\theta$ range.
We assume that any $\theta$ dependence is confined to a small layer close to $r=r_2$ such that the majority of our system is axisymmetric (see Figure~\ref{fig:circ_setup} for a schematic representation). 
Initially, we consider a system in equilibrium with $\mathcal{M}\big|_{S_1}=-\mathcal{M}\big|_{S_2}$, resulting in a stationary background state $(H,\mathbf{V})$. 
We now perturb the background variables,
\begin{equation}
\begin{split}
H & \ \to H + \epsilon h, \\
\mathbf{V} & \ \to \mathbf{V} + \epsilon\mathbf{v},
\end{split}
\end{equation}
where $\mathcal{O}(\epsilon)$ terms describe waves oscillating about the stationary background. The parameter $\epsilon$ is a small expansion parameter which can be understood as the ratio of wave amplitude to the water height.

It is well-known \cite{philips1977dynamics} that linear surface waves propagating on a background flow produce a mass flux $\mathbf{j}$ in the direction of wave propagation. This mass flux is related to the phenomena of Stokes drift \cite{kenyon1969stokes,hasselmann1963conservation}. We can extract the rate of flow of mass through a radial surface due to the waves by evaluating the integral in Eq.~\eqref{massflow},
\begin{equation} \label{MomentumTotal}
\begin{split}
\mathcal{M} = & \ \int^{2\pi}_0\int^{H+\epsilon h}_0 \rho(\mathbf{V}+\epsilon\mathbf{v})\cdot\mathbf{r} dz d\theta \\
= & \ \int^{2\pi}_0 \rho\left[ \int^{H}_0\mathbf{V} dz + \epsilon\int^{H}_0 \mathbf{v} dz + \epsilon h\mathbf{V} + \epsilon^2h\mathbf{v} \right]\cdot\mathbf{r}d\theta,
\end{split}
\end{equation}
where in the last two terms, we have used the fact that the integral from $H$ to \mbox{$H+\epsilon h$} reduces to a multiplication by $\epsilon h$ in the limit of small $\epsilon$ with $\mathbf{V}$ and $\mathbf{v}$ evaluated at \mbox{$z=H$}.
The $\mathcal{O}(1)$ term is the contribution from the stationary background which, by assumption, will vanish when taking contributions from $S_1$ and $S_2$.
The $\mathcal{O}(\epsilon)$ terms are linear in the perturbations and therefore oscillatory, i.e. they represent the sloshing of water back and forth as the wave propagates but contribute no net transport of mass.
The final $\mathcal{O}(\epsilon^2)$ term is quadratic in the perturbation, containing an oscillatory component (with twice the excitation frequency) and a part which is constant in time. 
This non-oscillatory component allows in principle for the transport of mass over large scales if enough time is allowed to pass. 
Thus, we define the rate of flow of mass through a (radial) surface due to the waves,
\begin{equation} \label{massflow1}
\mathcal{M}_\mathrm{waves} = \int^{2\pi}_0 \rho h\mathbf{v}\cdot\mathbf{r}d\theta,
\end{equation}
which is also the wave induced radial momentum per unit length. Furthermore, since the background is (quasi)-stationary and axisymmetric, we may write the perturbations at $z=H$ as,
\begin{equation} \label{VecAnsatz}
\begin{pmatrix}
h(t,r,\theta) \\ \mathbf{v}(t,r,\theta)
\end{pmatrix} = \mathrm{Re}\left[\int^\infty_{-\infty} \sum_m \begin{pmatrix}
h_m(r) \\ \mathbf{v}_m(r)
\end{pmatrix} e^{im\theta-i\omega t} d\omega \right],
\end{equation}
where the $m$ is the azimuthal number and $\omega$ is the frequency of the wave (we select a specific frequency in what follows to avoid having to perform the integration in $\omega$). As shown in Appendix \ref{app:momflow}, Eq.~\eqref{massflow1} becomes,
\begin{equation} \label{Massflow2}
\mathcal{M}_\mathrm{waves} = \pi\rho\sum_m \mathrm{Re}\left[h_m^*\mathbf{v}_m\cdot\mathbf{r}\right],
\end{equation}
where $*$ denotes the complex conjugate of the field. 
Finally, we can evaluate the height change in our system. 
In Eq.~\eqref{mdot}, the $\mathcal{O}(1)$ contributions coming from the two boundaries cancel. 
The $\mathcal{O}(\epsilon)$ terms time integrate to zero and we drop them in the following.
Since the system is closed at $r=r_2$ (ignoring the water inlet which is small) $\mathcal{M}_\mathrm{waves}$ is zero there.
Hence, using Eq.~\eqref{mhdot}, the only contribution to the mean height change comes from the wave induced mass flow at $S_1$,
\begin{equation} \label{height_change}
\langle\dot{H}\rangle = \frac{\pi}{\mathcal{A}}\sum_m \mathrm{Re}\left[h_m^*\mathbf{v}_m\cdot\mathbf{r}\right]\Big|_{r=r_1}.
\end{equation}
The term on the right is independent of time (ignoring oscillations), hence the waves induce a change in the mean water height which grows linearly in time.

\subsection{Shallow water} \label{sec:HchangeShal}

The argument above made no assumption about the depth of the water, hence Eq.~\eqref{height_change} is the general result for an asymmetric flow.
However, we may gain further insight into this result by deriving it from the shallow water equations.

The fluid state variables perturbed to second order are,
\begin{equation}
\begin{split}
H & \ \to H + \epsilon h + \epsilon^2 \eta, \\
\mathbf{V} & \ \to \mathbf{V} + \epsilon\mathbf{v} + \epsilon^2\mathbf{u},
\end{split}
\end{equation}
where the $\mathcal{O}(\epsilon^2)$ terms contain corrections to the background induced by the waves.
The continuity equation at $\mathcal{O}(\epsilon^2)$ is,
\begin{equation} \label{O2cont}
\partial_t\eta + \bm\nabla\cdot\left(h\mathbf{v} + \eta\mathbf{V} + H\mathbf{u}\right) = 0.
\end{equation}
Since we seek a solution for the height change at early times, we may write a Taylor expansion about $t=0$,
\begin{equation}
\begin{split}
\eta = \eta_0 + \dot{\eta}_0t + \mathcal{O}(t^2), \\
\mathbf{u} = \mathbf{u}_0 + \dot{\mathbf{u}}_0t + \mathcal{O}(t^2),
\end{split}
\end{equation}
where zero indicates a quantity is evaluated at $t=0$. We define $t=0$ to be the instant when the backreaction begins (a similar procedure is followed in \cite{balbinot2006hawking}), hence $\eta_0$ and $\mathbf{u}_0$ are both zero. 
In Eq.~\eqref{O2cont}, only the quadratic term $h\mathbf{v}$ has a contribution which is independent of $t$. 
Hence at leading order in the $t$ expansion we have,
\begin{equation}
\dot{\eta}_0 + \bm\nabla\cdot(h\mathbf{v}) = \mathcal{O}(t).
\end{equation}
Integrating over $\mathcal{A}$ and applying the divergence theorem, this becomes,
\begin{equation}
\langle\dot{\eta}_0\rangle = - \frac{1}{\mathcal{A}}\left[\int^{2\pi}_0 h\mathbf{v}\cdot\mathbf{r}~ d\theta\right]^{r=r_2}_{r=r_1} + \mathcal{O}(t),
\end{equation}
where we have used the definition of the spatial average in Eq.~\eqref{DefMean}. 
Using \mbox{$\mathbf{v}(r=r_2)=0$} and the ansatz in Eq.~\eqref{VecAnsatz}, we again recover Eq.~\eqref{height_change} with $\langle\dot{H}\rangle=\langle\dot{\eta}_0\rangle$, which is true since the time independent part of $H$ is zero under the derivative.

This derivation based on a perturbative expansion makes it clear that our formula is only valid at early times when the second order variables in Eq.~\eqref{O2cont} can be neglected. 
After enough time has elapsed, these terms will grow to the same size as $h\mathbf{v}$.
When this happens, $\dot{\eta}$ depends on $\eta$, leading to exponential behaviour at late times.

\subsection{Energy current}

We can gain some intuition for the form of $\mathcal{M}_\mathrm{waves}$ in Eq.~\eqref{massflow1} by considering the equation for energy conservation in shallow water. 
A full discussion of conserved currents is provided in Section~\ref{sec:currents}.
For convenience, we state energy equation again here, which is obtained by contracting the shallow water equations in vector form with $(gh,H\mathbf{v})$, see e.g. \cite{buhler2014waves}:
\begin{equation} \label{energy}
\partial_tE + \bm\nabla\cdot\mathbf{I}=0,
\end{equation}
where the energy and it's associated current are defined respectively as,
\begin{equation}
\begin{split}
E = \ & \frac{1}{2}gh^2 + \frac{1}{2}H\mathbf{v}^2 + h\mathbf{V}\cdot\mathbf{v}, \\
\mathbf{I} = \ & (h\mathbf{V}+H\mathbf{v})(gh+\mathbf{V}\cdot\mathbf{v}).
\end{split}
\end{equation}
If the perturbations are stationary then $\partial_tE=0$ and applying the divergence theorem over the region $\mathcal{A}$ yields,
\begin{equation} \label{RadialEnergy}
\left[ \int^{2\pi}_0 c^2 h\mathbf{v}\cdot\mathbf{r} + gh^2\mathbf{V}\cdot\mathbf{r} + \mathbf{V}\cdot\mathbf{v}(h\mathbf{V}+H\mathbf{v})\cdot\mathbf{r} \right]^{r=r_2}_{r=r_1} = 0,
\end{equation}
which states that the energy current is conserved as the wave propagates radially. The first term in this expression is simply $\mathcal{M}_\mathrm{waves}$ multiplied by a factor $c^2/\rho$. Hence for the special case $\mathbf{V}=0$, the notion of energy and mass transport coincide and the flow of mass is conserved throughout the fluid.

In the general case $\mathbf{V}\neq0$ however, the mass transport term only makes up a part of the energy current. 
Therefore if the energy current points in a specific direction, there is no reason to believe that mass will be transported in the same direction. 
For example, superradiant scattering corresponds to the extraction of energy from the system to infinity but, based on this argument, we do not expect there to be a net transport of mass out of a superradiating system.

\subsection{DBT flow}

There has been no mention thus far of the form of the velocity field.
To keep the argument simple, we will assume an irrotational flow in shallow water with a flat free surface, which means we are dealing with the $1/r$ profiles for $V_r$ and $V_\theta$ in Eq.~\eqref{DBT}.
This flow exhibits a horizon at $r=r_h$ which, as we argued in Section~\ref{sec:SRresults}, we expect to be close to the drain.
Hence, for the outlet boundary of our system we take $r_1\simeq r_h$.
Since this flow is irrotational, the perturbations are described by the scalar function \mbox{$\phi=\sum_m\mathrm{Re}[f_m e^{im\theta-i\omega t}]$}, which is related to the physical quantities of interest by,
\begin{equation} \label{HmVm}
h_m = -i\tilde{\omega}f_m + V_r\partial_rf_m, \qquad \mathbf{e}_r\cdot\mathbf{v}_m = v_{r,m} = \partial_rf_m,
\end{equation}
and we recall that $\tilde{\omega}=\omega-mC/r^2$.
This has useful consequences since we already have an exact solution about the horizon in the form of a Frobenius expansion, see Eq.~\eqref{frob}. Reinserting the amplitude on the horizon, this is,
\begin{equation} \label{frobx}
f_m(x) = A_m^h\left\{1 - \frac{\sigma_h^2-2imB-m^2}{2(1-i\sigma_h)}(x-1) + \mathcal{O}\left((x-1)^2\right)\right\},
\end{equation}
where we have made use of the dimensionless variables defined in Eq.~\eqref{adim} and introduced \mbox{$\sigma_h=\sigma-mB$} (i.e. the dimensionless version of $\tilde{\omega}_h$) to further simplify the notation.
Using Eqs.~\eqref{HmVm} and~\eqref{frobx}, the induced mass flow across the horizon is,
\begin{equation} \label{MomentumHorizon}
\mathcal{M}_\mathrm{waves}\big|_{r=r_h} = -\frac{\pi\rho}{4gc}\sum_m\tilde{\omega}_h^2Q_m\big|A_m^h\big|^2,
\end{equation}
which results in a mean height change,
\begin{equation} \label{HeightChange_Shallow}
\langle\dot{H}\rangle = -\frac{\pi}{4gc\mathcal{A}}\sum_m \tilde{\omega}_h^2Q_m\big|A_m^h\big|^2.
\end{equation} 
The parameter $Q_m$ is defined,
\begin{equation} \label{Qexact}
Q_m(\sigma,B) = \frac{\sigma_h^4-m^4-4mB\sigma}{\sigma_h^2(\sigma_h^2+1)},
\end{equation}
where the writing in terms of dimensionless variables makes it clear that $Q_m$ is itself dimensionless.
This factor determines whether a particular $m$-mode induces a net flow of mass toward the origin ($Q_m>0$) or toward infinity ($Q_m<0$).
Notice also the appearance $\tilde{\omega}_h^2$ in Eq.~\eqref{Qexact} which means that superradiant scattering does not lead to a transport of mass away from the drain, as predicted in the previous section.

\begin{figure}
\includegraphics[width=\linewidth]{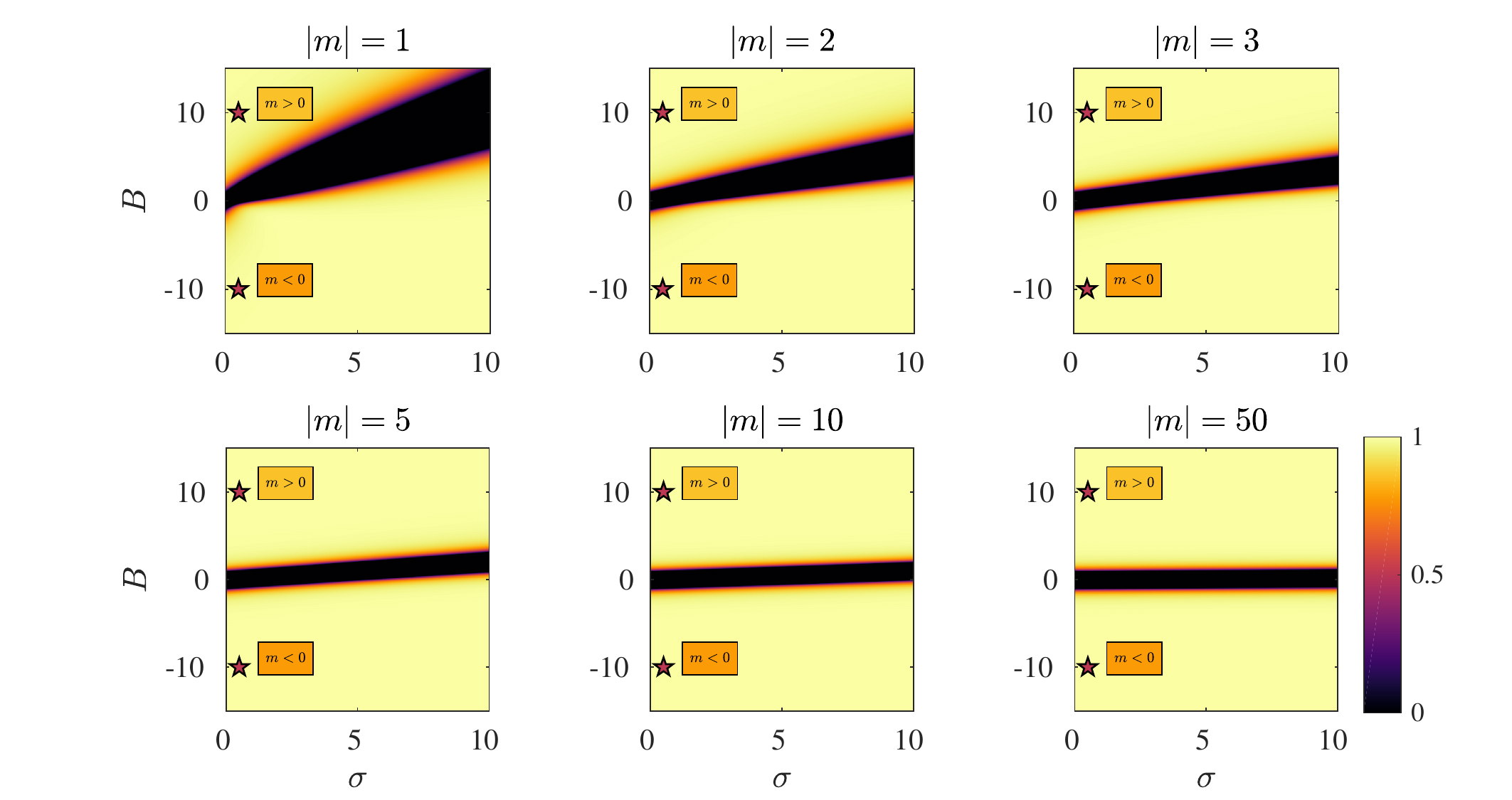}
\caption{The parameter space for $Q_m$ for several $m$ modes. 
$Q_m<0$ in the dark regions. 
Stars represent parameters pertaining to the experiment in the next section, $\sigma\sim1/4$ and $B\sim10$. Since $Q_m$ is invariant under the transformation $m\to-m$, $B\to-B$, the points at $B<0$ represent the $m<0$ mode for $B>0$. Hence, for the regime of our experiments, all $m$-modes act to decrease the water height.} \label{fig:Q_ParamSpace}
\end{figure}

In Figure~\ref{fig:Q_ParamSpace} we display the parameter space for $Q_m(\sigma,B)$.
In most of the parameter space, we see that $Q_m\simeq1$, indicating that the $m$-mode acts to decrease the mean height.
However there are dark bands where $Q_m<0$, meaning that an $m$ mode acts to increase the mean height in this region.
These bands are larger for smaller $|m|$ values.
We do not display the axisymmetric mode ($m=0$), for which $Q_m$ has the particularly simple form \mbox{$Q_0=\sigma^2(\sigma^2+1)^{-1}$}, which is always positive and therefore always removes mass from the system.

To understand the behaviour of $Q_m$, we use the WKB form of the solution in Eq.~\eqref{WKBsol1} to derive an approximate expression for $Q_m$, keeping track in the process of where each term comes from.
First of all, close to the horizon, i.e. $r=r_h(1+\epsilon)$ where $\epsilon$ is a small parameter which we will send to zero, the radial wave vector and eikonal potential in Eqs.~\eqref{kr} and \eqref{GeoPotential} reduce to,
\begin{equation} \label{krVlimrh}
k_r\propto-\frac{\sigma_h}{2}\left(1-\frac{m^2}{\sigma_h^2}\right) + \mathcal{O}(\epsilon), \qquad V_\mathrm{geo}\propto -\sigma_h^2 + (4mB\sigma_h-2m^2)\epsilon + \mathcal{O}(\epsilon^2),
\end{equation}
where the proportionality factor is the relevant dimensionful variable, and we have taken the solution for $k_r$ which does not diverge on the horizon.
Inserting the WKB ansatz of Eq.~\eqref{WKBsol1} into Eq.~\eqref{HmVm}, we find,
\begin{equation}
\begin{split}
\partial_rf_m\big|_{r=r_h} \propto & \ -\frac{1}{2}\left(i\sigma_h-\frac{im^2}{\sigma_h} + 1 + \frac{2mB\sigma_h-m^2}{\sigma_h^2}\right)f_m, \\
h_m(r=r_h) \propto & \ -\frac{1}{2}\left(-i\sigma_h-\frac{im^2}{\sigma_h} + 1 + \frac{2mB\sigma_h-m^2}{\sigma_h^2}\right)f_m.
\end{split}
\end{equation}
In both of these expressions, the first term is what we would expect from the asymptotic expression in Eq.~\eqref{asymp_sols_b} when taking a radial derivative, and the second term is the phase correction to this resulting from the non-vanishing gradient of $V_\mathrm{geo}$ at $r=r_h$ (i.e. the $\mathcal{O}(\epsilon)$ term in Eq.~\eqref{krVlimrh}).
The third term comes from the focussing of the amplitude as a radial mode propagates inwards.
Lastly, the forth term is a result of the adiabatic change in amplitude on an inhomogeneous background.

Using these approximate forms, we again find Eq.~\eqref{MomentumHorizon} for the induced mass flow but now with,
\begin{equation}
Q_m^\mathrm{wkb}(\sigma,B) = 1 - \frac{m^4}{\sigma_h^4} -  \frac{1}{\sigma_h^2}\left(1 + \frac{2mB\sigma_h - m^2}{\sigma_h^2} \right)^2.
\end{equation}
Since this expression is derived in the WKB approximation, the horizon must not be close to a turning point (which means the second term must be small) and the variation in amplitude must be small with respect to the phase (so the last term must also be small).
Both of these conditions are satisfied if $\sigma_h$ is large.
Expanding the exact expression for $Q_m$ in powers of $1/\sigma_h$, we have,
\begin{equation}
Q_m = 1 - \frac{m^4}{\sigma_h^4} - \frac{4mB\sigma}{\sigma_h^4} - \frac{1}{\sigma_h^2} + \mathcal{O}\left(\frac{1}{\sigma_h^6}\right),
\end{equation}
Hence, in the high frequency limit, we may identify the terms $m^4$, $4mB\sigma$ and the 1 on the denominator in Eq.~\eqref{Qexact} as resulting from the gradient of the potential on the horizon, the variation in amplitude on an inhomogeneous background and the focussing of amplitude of a radial mode respectively. 
Whilst each of these terms may decrease the value of $Q_m$, only the first two may cause $Q_m$ to be negative, which results in an outwardly directed flow of mass at the horizon due to an in-going mode.
This is somewhat counter intuitive, so we provide a brief discussion as to why this can be the case.

The $m^4$ term in $Q_m$ comes from the second term in $k_r$ in Eq.~\eqref{krVlimrh}, which causes $k_r$ to become positive for $m>0$ modes in the frequency range \mbox{$\sigma\in[\sigma_-,\sigma_+]$} and for $m<0$ modes in the range \mbox{$\sigma\in[0,\sigma_-]$} where \mbox{$\sigma_\pm=m(B\pm1)$}.
The $4mB\sigma$ term comes from the second term of $V_\mathrm{geo}$ in Eq.~\eqref{krVlimrh} which enters $Q_m$ by taking derivatives of the amplitude.
This term changes the sign of $Q_m$ when the amplitude varies quickly compared to the phase of the mode i.e. out of the WKB regime.
Hence it is not appropriate to think of the mode as a local plane wave, and our notion of in and out-going (determined by the dispersion relation) breaks down. 
In brief, the counter intuitive result that the mode on the horizon can create an outward flow of mass seems to be determined by two factors: firstly, the sign of the radial wavenumber in the eikonal approximation, and secondly the breakdown of the WKB approximation.

Having considered the nature of the height change induced by linear waves in an open DBT system, we are ready to move forward with an experiment to demonstrate this effect.

\section{Detection in the laboratory} \label{sec:ExpPro}

To investigate the wave induced height change, we conducted a series of experiments in the water tank system introduced in Chapter \ref{chap:super}.
Due to modifications with the wave generator, the tank dimensions were $2.65~\mathrm{m}$ in length and $1.38~\mathrm{m}$ in width.
The diameter of the drain hole in the centre was $4~\mathrm{cm}$ and the flow rate is denoted $Q$.
Full PIV measurements were not performed here although we expect the flow field to not differ significantly in its form from that described in Sections~\ref{sec:BgConfig}, \ref{sec:flow_values} and \ref{sec:flow_values2}.

A stationary background was first established by monitoring the system and waiting until $Q$ and the initial water height $H_0$ were constant within our uncertainty.
Waves were then generated from the side of the tank opposite the inlet with amplitude $a$ and frequency $f=\omega/2\pi$ over a time window $\Delta t$, and the change in water height $\Delta H$ was measured. 
A schematic of the apparatus is shown in Figure~\ref{fig:apparatus} of Chapter~\ref{chap:super}.

We performed four experiments to establish (1) consistency over different repeats, (2) the effect of changing frequency, (3) the effect of the wave amplitude and (4) the late time behaviour. 
An absorption beach was placed opposite the wave generator in the last experiment to get rid of large standing waves that we found built up over long periods of stimulation.
Details of the four experiments are summarised in Table \ref{tab:exps}.

\begin{table}
\small
\begin{center}
\begin{tabular}{c|c|c|c|c|c}
Exp. & $Q$~[l/min] & $H_0$~[cm] & $f$~[Hz] & $a$~[mm] & $\Delta t$~[s] \\
\hline
$1$ & $14.0\pm0.4$ & $2.0\pm0.1$ & $4$ & $1.6\pm0.1$ & $120$ \\
\hline
$2$ & $14.0\pm0.4$ & $1.9\pm0.1$ & $2,3,4$ & $2.3\pm0.5$ & $120$ \\
\hline
$3$ & $14.0\pm0.4$ & $1.9\pm0.1$ & $4$ & $0.6~\&~1.7\pm0.5$ & $120$ \\
\hline
$4$ & $29.4\pm0.4$ & $6.5\pm0.1$ & $4$ & $2.1\pm 0.5$ & $600$ \\
\end{tabular}
\end{center}
\caption{Details of the four experiments.}
\label{tab:exps}
\end{table}

\subsection{Method}

To determine $\Delta H$ resulting from the impinging waves, we filmed the free surface from the side of the tank with a high-speed Phantom Miro Lab camera at $24~\mathrm{fps}$.
The free surface was illuminated from above using a Yb-doped laser with mean wavelength $457~\mathrm{nm}$.
This was converted into a thin laser sheet (thickness $\approx 2~\mathrm{mm}$) using a cylindrical lens, and appears as a line of the free surface spanning almost the full length of the tank, which we positioned $17~\mathrm{cm}$ from the drain on the side nearest the camera. 
The line of sight of the camera was at an angle of $\Theta$ to the free surface such that $\cos\Theta=0.91\pm0.01$, which was necessary to avoid shadows in the data caused by waves passing in between the laser-sheet and the camera, which obscured our vision of the free surface. 
The error in $\Theta$ induces an uncertainty on the measured height change, which we include in our error estimates. 
We recorded the height before sending any waves to confirm that the background was steady, and then monitored the water height whilst sending waves for a time $\Delta t$. 
The position of the free surface in each image was identified by finding the pixel of maximum intensity in each column, and interpolating using adjacent points to determine the maximum to a sub pixel accuracy. 

Once the free surface in each image is determined as a function of spatial coordinate $x$, we then take the mean value.
This gave a measurement of the average height across the observed region at each time step. 
We divide this by $\cos\Theta$ to correct for the camera angle, obtaining the variation of the background height over time. 
In all experiments, the initial height was determined to be sufficiently steady (less than $10\%$ of the total change over the experiment) and any slight variation was due to the difficulty in maintaining constant $Q$ over the experiment. We corrected for this (as well as the slightly different initial heights $H_0$) by fitting the curve prior switching on the waves with a straight line and subtracting this from the entire data set, resulting in the height change $\Delta H(t)$. 
This ensured that the measured height change was the result of the waves.

In each experiment, we observed a small amplitude oscillation about the general decrease at the frequency $f$, confirmed by a peak in the Fourier transform of $\Delta H(t)$. 
This corresponds the oscillatory (linear) terms we dropped in Eq.~\eqref{MomentumTotal}. 
Furthermore, the amplitude of this oscillation decreased with increasing $f$ which is expected, since integrating Eq.~\eqref{MomentumTotal} in time brings a factor of $1/f$ in front of the linear terms. 
We also observe a sharp increase in height immediately before the height starts to decrease. 
This is because our wave generator panel is initially fully retracted and must move forward to it's average position, thereby reducing the area of the system and increasing the mean height. 
In accordance with this explanation, a sharper increase was observed for larger wave amplitudes.
 
\begin{figure*}[t!]
    \centering
    \begin{subfigure}[t]{0.5\textwidth}
        \centering
        \includegraphics[height=3in]{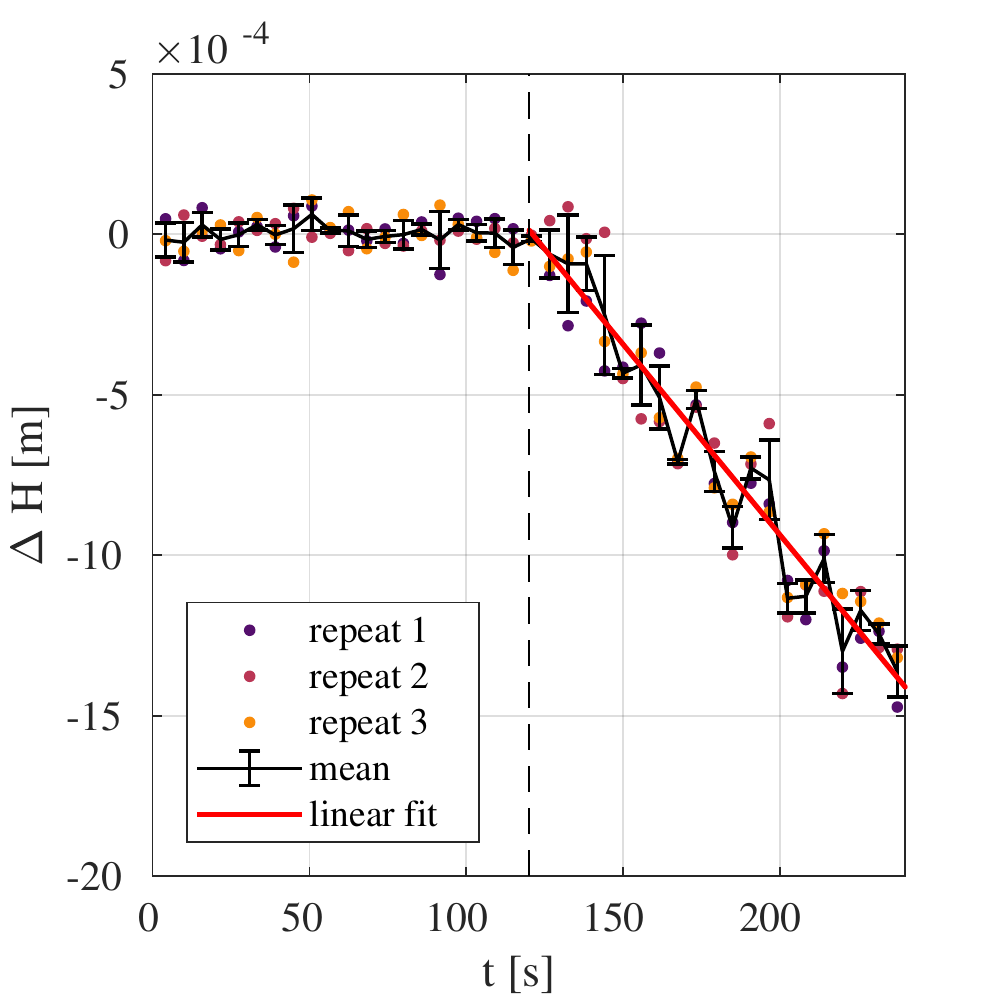}
        \caption{Experiment 1} \label{fig:same}
    \end{subfigure}%
    ~ 
    \begin{subfigure}[t]{0.5\textwidth}
        \centering
        \includegraphics[height=3in]{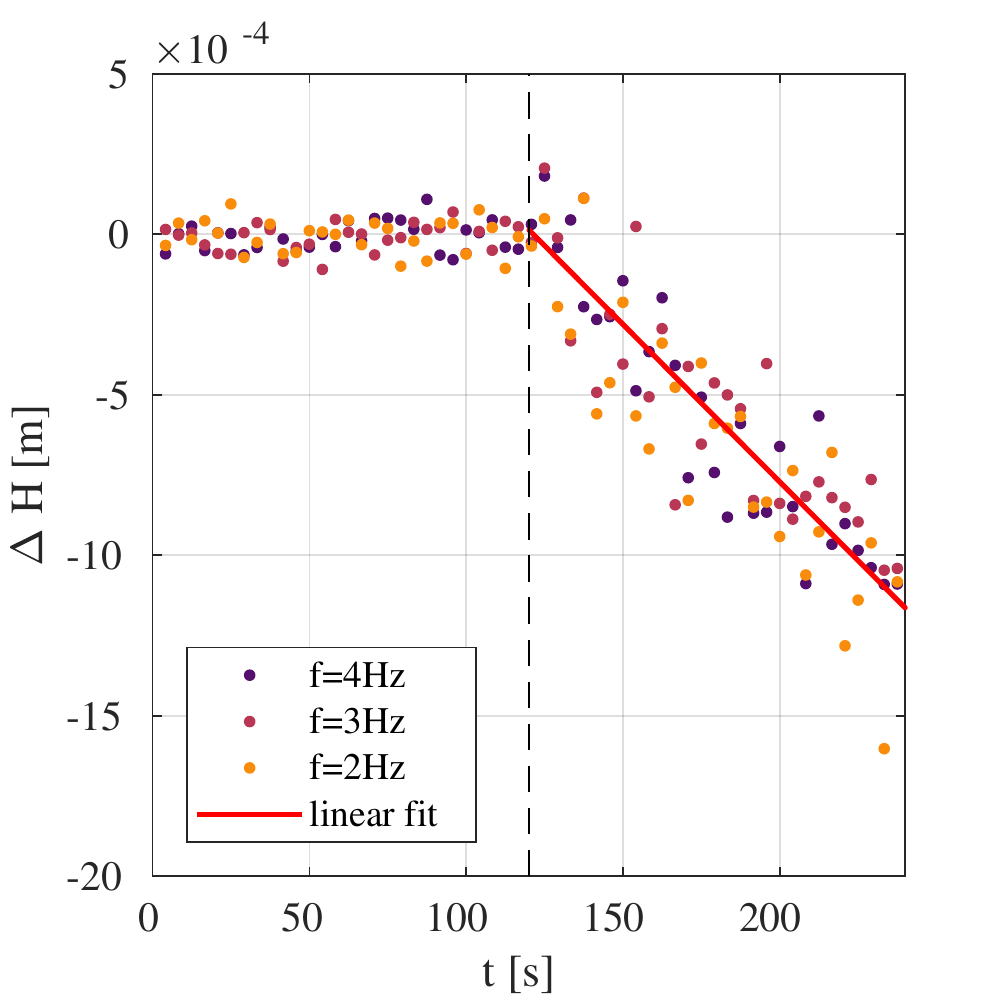}
        \caption{Experiment 2} \label{fig:diff}
    \end{subfigure}
    \caption{Results from experiments 1 (left side) and 2 (right side). The dashed vertical line indicates when the first wave-front passed the vortex. The red line is the best fit line to the mean of the three repeats in each experiment. Experiment 1 shows that the decrease is consistent when parameters remain fixed. Experiment 2 shows that the decrease does not depend significantly on the excitation frequency.}
\end{figure*}

\begin{figure*}[t!]
    \centering
    \begin{subfigure}[t]{0.5\textwidth}
        \centering
        \includegraphics[height=3in]{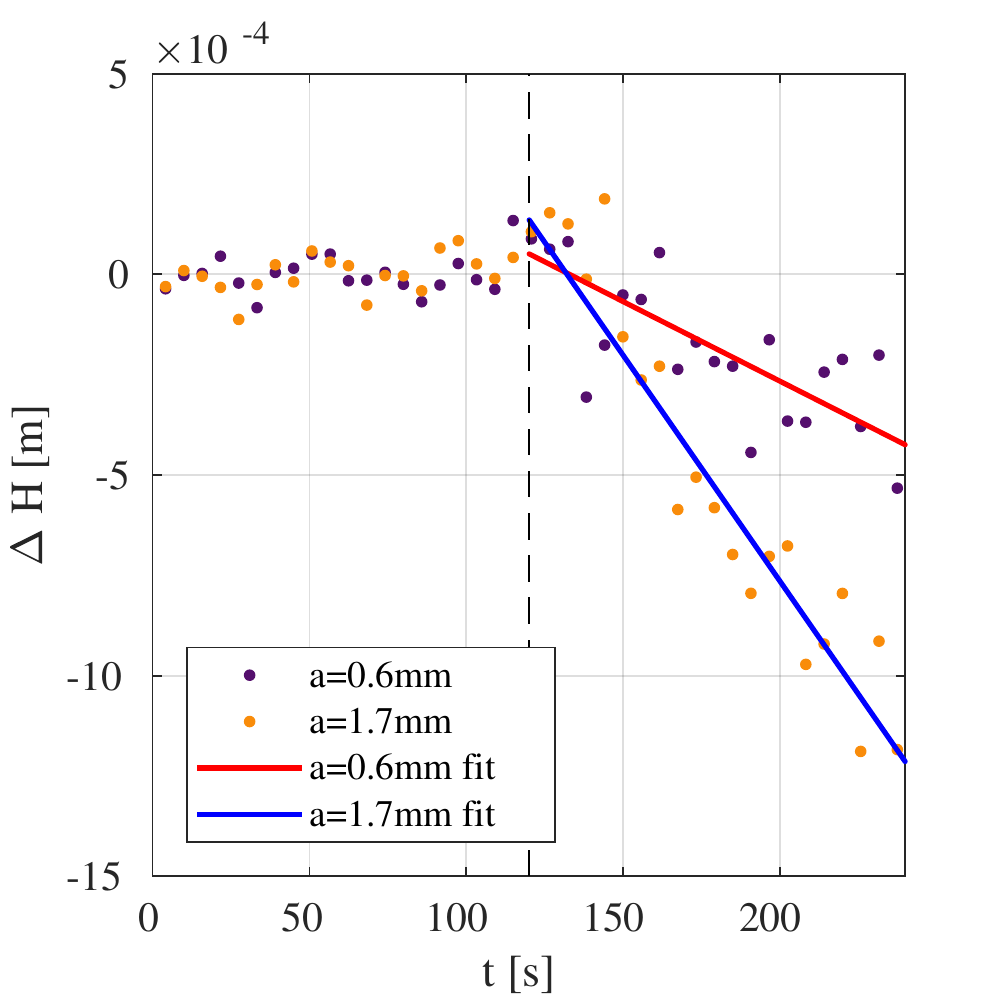}
        \caption{Experiment 3} \label{fig:amps}
    \end{subfigure}%
    ~ 
    \begin{subfigure}[t]{0.5\textwidth}
        \centering
        \includegraphics[height=3in]{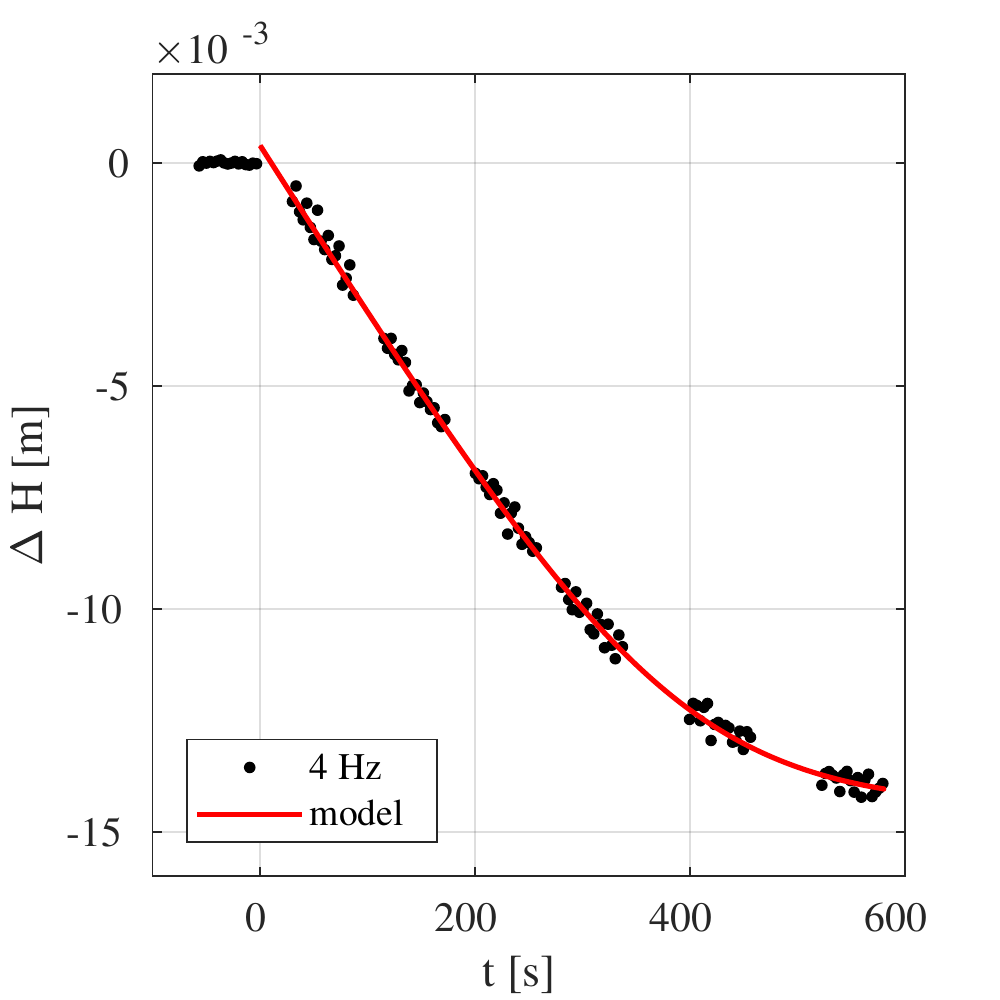}
        \caption{Experiment 4} \label{fig:long}
    \end{subfigure}
    \caption{Left side: the third experiment indicates that the size of the height drop is dependent on the amplitude of the waves. Right side: evolution of the water height under continual stimulation of waves at $4~\mathrm{Hz}$. The results show that the decrease remains linear at early times and a new equilibrium height is approached exponentially at late times. The red line is the best fit with Eq.~\eqref{sech2}.}
\end{figure*}

\subsection{Results} \label{HeightChange_results}

All experiments showed an initial period of linear height decrease. 
The gradient $\dot{H}$ of this decrease was obtained for experiments 1-3 by fitting the data with the function,
\begin{equation}
\Delta H(t) = H_1 + \dot{H}t,
\end{equation}
where $H_1$ accounts for the increase in mean height when the wave generator is moved forward.
Note that really we are measuring $\langle\dot{H}\rangle$ defined earlier, but we drop the angular brackets to simplify the notation.
The fit is computed from a least squares regression, with the error on $\dot{H}$ calculated from the residues of the fit.

In Figure~\ref{fig:same}, we display the result of experiment 1 which indicates that the height change is consistent. 
The errorbars were computed from the mean of three repeats repeats, indicating that we observe consistent behaviour.
The gradient of the best fit line in is \mbox{$\dot{H}=-1.18\pm0.02\times 10^{-2}~\mathrm{mm/s}$}.

In Figure~\ref{fig:diff}, we display the results of experiment 2, with incident waves of frequency  $f=2,3,4~\ \mathrm{Hz}$.
The gradient of the best fit to the mean of the three repeats is \mbox{$\dot{H}=-9.8\pm0.2\times10^{-3}~\mathrm{mm/s}$}.
Since each frequency is spread randomly about the best fit line, we deduce that $\Delta H$ is not sensitive to significant changes in $f$.
This observation is supported by our prediction in Eq.~\eqref{HeightChange_Shallow}, which has weak frequency dependence in the regime $\sigma\ll mB$.

The results of experiment 3 in Figure~\ref{fig:amps} demonstrate the noticeable effect of changing the wave amplitude. 
For the run at small amplitude, the best fit line has a gradient \mbox{$\dot{H}=-4.0\pm0.2\times10^{-3}~\mathrm{mm/s}$} and for the larger amplitude we find $\dot{H}=-1.13\pm0.02\times 10^{-2}~\mathrm{mm/s}$. 
To show that the dependence of $\Delta H$ on amplitude is quadratic, many repeats would need to be performed for each value of $a$ due to the noise in our set-up.
Also, care needs to be taken to ensure that $a$ stays small enough that a perturbative expansion of the fluid equations remains valid.
If $a$ is too large then non-linear corrections enter into the wave equation (e.g. the Boussinesq approximation \cite{boussinesq1872theorie}) and it is unclear how the backreaction occurs.

In Figure~\ref{fig:long}, we display the height change over a longer period of stimulation from the last experiment. 
We observe a constant gradient at early times and a tendency towards a new constant equilibrium which, in line with our argument at the end of Section~\ref{sec:HchangeShal}, we expect to be approached exponentially. 
Since we do not solve the full second order equations, we instead choose a four parameter phenomenological model with amenable analytic properties. 
Such a function is,
\begin{equation} \label{sech2}
\Delta H(t) = H_1 + \frac{\dot{H}}{2}(t+\tau) - \frac{\dot{H}}{\Gamma}\log\left(2\cosh\left[\frac{\Gamma}{2}(t-\tau)\right]\right),
\end{equation}
which has the following asymptotics,
\begin{equation}
\begin{split}
\Delta H(t\to -\infty) = \ & H_1+ \dot{H}t, \\
\Delta H(t\to +\infty) = \ & H_\mathrm{end} - \frac{\dot{H}}{\Gamma}e^{-\Gamma(t-\tau)}, \qquad H_\mathrm{end} = H_1+\dot{H}\tau.
\end{split}
\end{equation}
The extra free parameters of the model are $\tau$, which is approximately the time at which the behaviour switches from linear to exponential, and $\Gamma$, the decay rate at late times. 
This model predicts the new equilibrium height $H_\mathrm{end}$ given above.

The velocity and acceleration of the free surface in this model are,
\begin{equation}
\partial_t (\Delta H) = \frac{\dot{H}}{2}\left(1-\tanh\left[\frac{\Gamma}{2}(t-\tau)\right]\right), \qquad \partial_t^2(\Delta H) = -\frac{\dot{H}\Gamma}{4}\mathrm{sech}^2\left[\frac{\Gamma}{2}(t-\tau)\right].
\end{equation}
Since the $\tanh$ function rapidly tends to $1$ once its argument becomes $\mathcal{O}(1)$, the function in Eq.~\eqref{sech2} is approximately linear until a time \mbox{$t_\mathrm{lin}=\tau-1/\Gamma$}. 
Hence the height close to $t=0$ is well approximated by its value as $t\to-\infty$ provided $\Gamma\tau>1$. 
The maximum acceleration of the free surface is $-\dot{H}\Gamma/4$ which occurs at $t=\tau$, at which time deviations from linearity start to become significant.

Fitting this model to our data in Figure~\ref{fig:long}, we find \mbox{$\dot{H}=-3.89\pm0.05\times 10^{-2}~\mathrm{mm/s}$}, $1/\Gamma = 94\pm4~\mathrm{s}$ and \mbox{$\tau = 383\pm3\mathrm{s}$}. 
The mean height change once a new equilibrium is reached is \mbox{$\Delta H_\mathrm{tot} = -1.4\pm0.1\mathrm{cm}$}.
The fit is performed once the high frequency oscillations have been removed from the data, which is achieved by collecting the data into bins of width $1/f$ and averaging.  The $R^2$ value of the model is $0.9997$.

Comparing the initial slopes across the three experiments we see that $\dot{H}$ depends on $H_0$, supporting our claim that the long term behaviour should be exponential.
This is further corroborated by the late time tail in experiment 4. 
In all experiments we see that at early times the gradient of the height decrease is well approximated as linear, in agreement with our predictions in Eq.~\eqref{HeightChange_Shallow}.
The observation that the early height change is linear (and therefore the gradient in $t$ is constant) justifies our theoretical treatment in Section~\ref{sec:HchangeShal} where we discarded the second order terms on the basis that they should be small initially.

\section{Comparison with previous experiments}


We now compare our theoretical prediction in Eq.~\eqref{HeightChange_Shallow} with the wave measurement data from the superradiance experiment in Chapter~\ref{chap:super}. 
Analysing the data obtained there, we found that in all experiments the height of the water decreased during the time of wave incidence with a gradient of around \mbox{$\dot{H}\sim -2\pm1\times 10^{-2}~\mathrm{mm/s}$}. 
The large error is the result of only recording for 13~s, since during this time the total height change is smaller than the estimated amplitude of the waves.
The flow parameters in Chapter~\ref{chap:super} are similar to those of experiment 4 performed here and correspondingly, we see that the order of magnitude of $\dot{H}$ is in agreement. 

We now estimate the height change using our prediction in Eq.~\eqref{HeightChange_Shallow} with the scattering amplitudes taken from the experiment at \mbox{$f=3.7~\mathrm{Hz}$} of Chapter \ref{chap:super}, for which we have the most repeats. 
To do this, we summed over \mbox{$|m|\leq5$} since higher $m$ modes were not resolvable within the window of observation. 
Using Eq.~\eqref{HeightChange_Shallow}, we predict \mbox{$\dot{H}=-2.3\pm 0.6\times10^{-2}~\mathrm{mm/s}$} which agrees with the measured value within our uncertainty.
The error is estimated using a Monte Carlo method, which involves creating a distribution for each of the parameters in Eq.~\eqref{HeightChange_Shallow}, taking the uncertainty on the parameter as the standard deviation of the distribution, and creating a data set for $\dot{H}$ by sampling randomly from these distributions.
The prediction for $\dot{H}$ above is the mean and standard deviation of this data set.

To improve the agreement between theory and experiment, a theory of wave propagation accounting for dispersion, dissipation, vorticity and gradients in the free surface is required (the former two being important throughout the flow and the latter two arising close to the drain). 
The reason for this is that our predictions depend on the precise form of the solution near the drain, which will be influenced by any of these effects (see \cite{torres2017rays,schutzhold2002gravity,patrick2018QBS,richartz2015rotating} for attempts to account for these phenomena).
However, the estimate provided by the simple irrotational, shallow water theory already gives the correct order of magnitude.
This is a good indication that the proposed mechanism is responsible for our observations.

\subsection{Explanation of PIV particle flux}

\begin{figure}
\includegraphics[width=\linewidth]{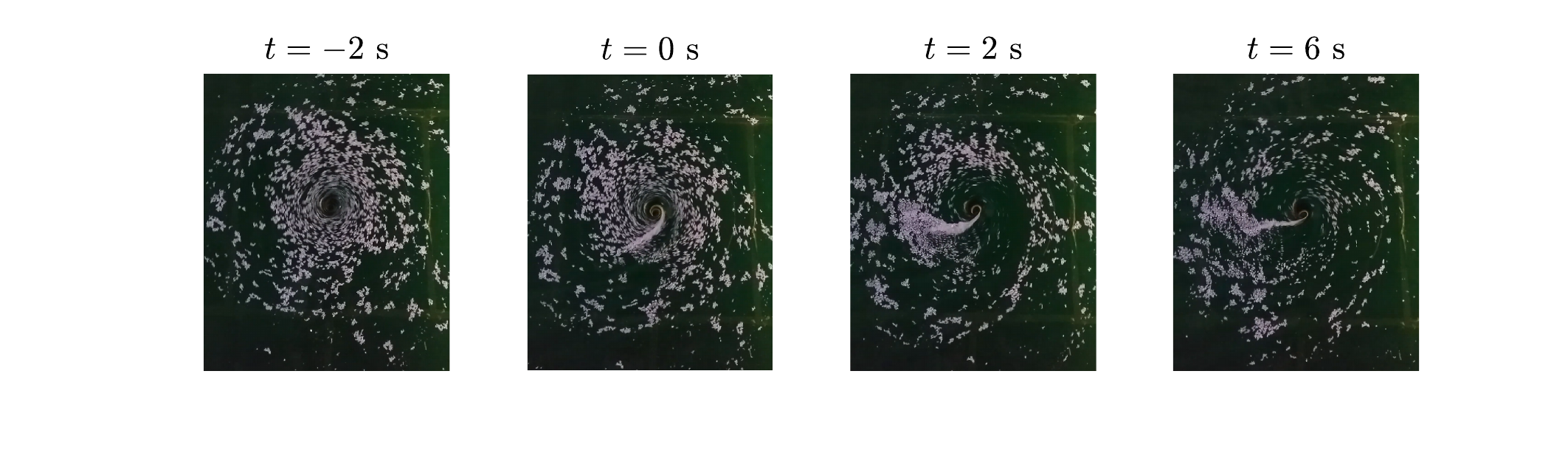}
\centering
\caption{PIV particles drain from the system under stimulation by a plane wave at $f=4~\mathrm{Hz}$.
The flow rotates in the counter-clockwise direction and the plane wave is incident from the bottom of the images.
We define $t=0$ as the moment at which the wave hits the vortex centre.
For $t<0$, particles drain symmetrically and slowly.
For $t>0$, particles drain rapidly through a narrow strip on the left side of the images where the counter-rotating modes are dominant.
This observation is consistent with the form of the wave induced mass flow in Eq.~\eqref{MomentumHorizon}.
The presence of the strip could be explained by the momentum in the $\theta$-direction of the $m<0$ modes, which would counter the momentum of the background and cause the particles to bunch up.
However, we do not study this here.} \label{fig:particles}
\end{figure} 

Another observation during the experiment of Chapter~\ref{chap:super} was the following.
To measure the flow field, the free surface was seeded with paper tracer particles.
Once measurements were complete, the PIV particles took a long time to drain out of the water tank due to the small radial velocity at the water's surface.
The process was speeded up by sending waves, which had the effect of draining particles from the region where the wave was propagating against the flow.
The process is illustrated by a series of photos in Figure~\ref{fig:particles}.
We believe that the theoretical analysis at the start of this chapter provides an explanation for this observation.

To show this, we argue that the $m<0$ components of a wave induce a larger mass flow at the horizon than the $m>0$ components, using WKB waves with the eikonal potential in Eq.~\eqref{GeoPotential} to illustrate our point.
In the WKB limit, we have seen that $Q_m\simeq1$ provided the horizon is not near a turning point of the potential.
Hence the magnitude of mass flow induced by a single $m$-mode is determined by it's amplitude $A_m^h$ at the horizon squared and $\tilde{\omega}_h^2$.
Considering co- and counter-rotating modes with the same $|m|$, clearly $\tilde{\omega}_h^2$ will be larger for $m<0$ than for $m>0$.

\begin{figure*}[t!]
    \centering
    \begin{subfigure}[t]{0.5\textwidth}
        \centering
        \includegraphics[height=3in]{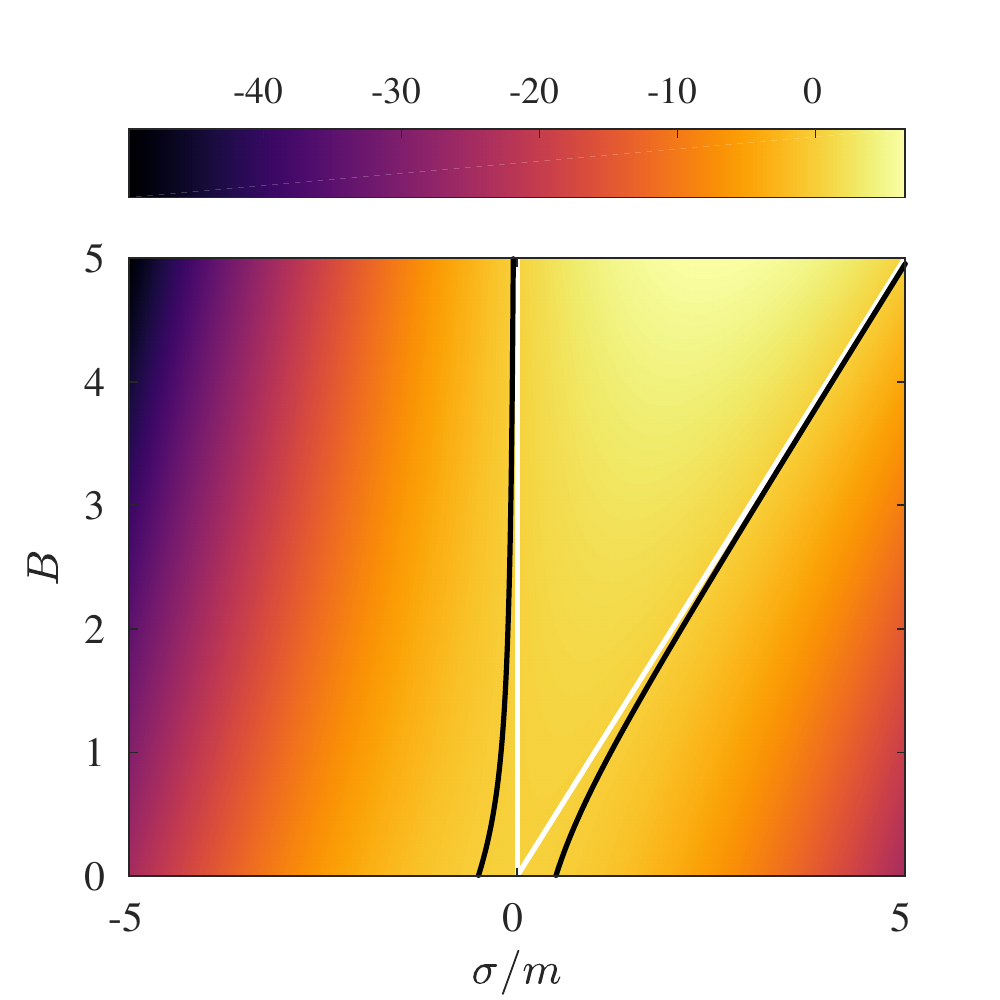}
    \end{subfigure}%
    ~ 
    \begin{subfigure}[t]{0.5\textwidth}
        \centering
        \includegraphics[height=3in]{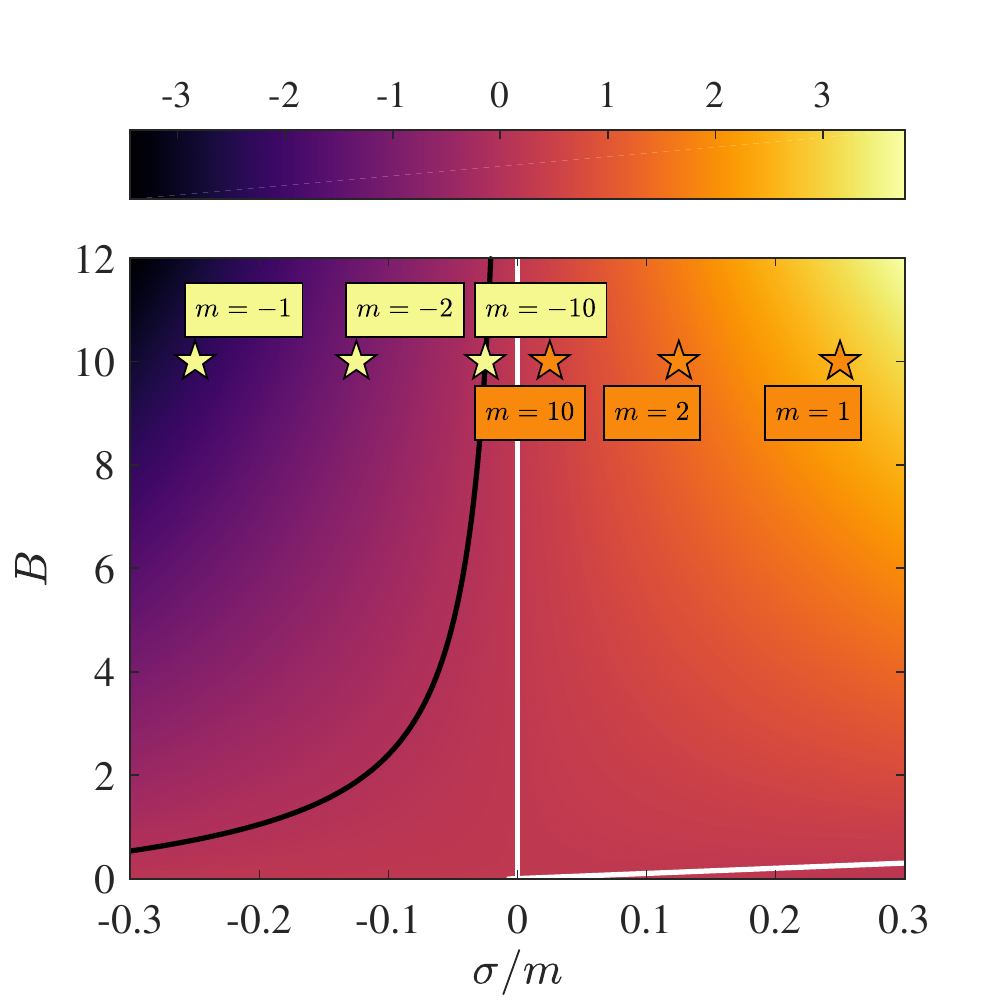}
    \end{subfigure}
    \caption{The region of the parameter space between the black lines is where the eikonal potential has two turning points.
    Outside these lines there are no turning points.    
    The region between the white lines is where superradiant amplification occurs. 
    On the right, we zoom into the region pertaining to the parameters of our experimental set-up, i.e. \mbox{$\sigma\sim1/4$} and \mbox{$B\sim10$}, displaying several $m$-modes.
    } \label{fig:TurningPoints}
\end{figure*}

To compare $A_m^h$ for co- and counter rotating modes, we make use of the WKB transmission coefficient $\mathcal{T}_m$ we derived in Eq.~\eqref{TransWKB}.
If the incident wave has amplitude $A_m^\mathrm{in}$, then the transmission amplitude is given by \mbox{$A_m^h=|\mathcal{T}_m|A_m^\mathrm{in}$}.
Within our approximation, $\mathcal{T}_m$ will be (exponentially) close to $1$ if $V_\mathrm{geo}$ has no turning points, and is otherwise given by Eq.~\eqref{TransWKB}, i.e. expontentially small.
The turning points of the eikonal potential in dimensionless variables are given by,
\begin{equation}
\frac{1}{x^2_\mathrm{tp}} = \frac{1}{B^2+1}\left(\lambda B+\frac{1}{2}\pm\sqrt{\left(\lambda B+\frac{1}{2}\right)^2-(B^2+1)\lambda^2}\right),
\end{equation}
where for brevity we have defined $\lambda=\sigma/m$. 
The term under the square root (i.e. the discriminant) $\mathcal{D}$ can be written in the simplified form,
\begin{equation}
\mathcal{D}(\lambda,B) = \lambda B + \frac{1}{4} - \lambda^2.
\end{equation}
Hence there are three cases,
\begin{enumerate} [noitemsep]
\item $\mathcal{D}>0$: $V_\mathrm{geo}$ has two real turning points,
\item $\mathcal{D}<0$: $V_\mathrm{geo}$ has no real turning points,
\item $\mathcal{D}=0$: $V_\mathrm{geo}$ has one real turning point.
\end{enumerate}
In Figure~\ref{fig:TurningPoints}, we show in which regions of the parameter space $V_\mathrm{geo}$ has turning points by plotting the value of $\mathcal{D}$.
For the purposes of our discussion, we are interested in cases 1 and 2.
Case 3 of course pertains to the light-ring modes we discussed in Chapter~\ref{chap:qnm}, although here it serves as a boundary in parameter space separating the regions of interest.
Solving $\mathcal{D}=0$, this boundary is given by,
\begin{equation}
\lambda_{1,2} = \frac{B}{2}\pm\frac{\sqrt{B^2+1}}{2},
\end{equation}
which are the black lines in Figure~\ref{fig:TurningPoints}. 
In the large $B$ limit, we have $\lambda_1\to B$ and $\lambda_2\to 0$, which are the boundaries of the superradiant regime (superradiance occurs for $0<\lambda<B$).
These are represented by white lines in Figure~\ref{fig:TurningPoints}. 
Thus for large $B$, the majority of the modes in parameter space which scatter off the potential barrier are those which are superradiant.

In the right panel of Figure~\ref{fig:TurningPoints}, we show a zoomed in region of the parameter space where our experiments were performed.
Since modes with large $|m|$ have inherently low amplitude in the plane wave sum for a system of finite size, we deduce that for $m$-modes relevant to our experiment, those with $m>0$ encounter two turning points whilst those with $m<0$ encounter no turning points.
Hence the $m<0$ modes have a larger amplitude on the horizon than those with $m>0$ (in the eikonal approximation).
This statement is corroborated by the small reflection coefficients for $m<0$ in Figures~\ref{fig:ReflB1}, \ref{fig:ReflB5} and \ref{fig:ReflExp}

Taking both $\tilde{\omega}_h^2$ and $|A_m^h|^2$ into account, we conclude that the $m<0$ components contribute more to $\mathcal{M}_\mathrm{waves}$ than $m>0$ modes on the horizon. 
Finally, the $m<0$ modes contribute more in the partial wave sum in the region where the wave propagates against the flow, whereas $m>0$ modes are dominant where the wave moves with the flow.
Hence, there is more mass carried down the drain where the wave moves against the flow, explaining why PIV particles are drained mostly from this region.
It is possible that this effect could be of practical use in clearing free surface contaminants in DBT type systems.
Indeed, in the example of Figure~\ref{fig:particles}, barely any particles drain initially whereas almost all particles have been cleared after about 30 seconds of stimulation by surface waves.


\section{A quantum stepping stone}

We now comment on how our results could be extended to quantum systems which emit Hawking radiation (HR).
Rotation is not typically considered in such systems in order to focus purely on horizon effects, without the complication of having an ergosphere.
This is not a problem for our result in Eq.~\eqref{HeightChange_Shallow}, which can be applied to purely draining systems by setting $C=0$.
Our result, however, assumes a purely in-going mode since we are dealing with a classical system, whereas HR is concerned with quantum tunnelling to the out-going mode \cite{visser2003essential}.
It is feasible that since the in-going mode tends to cause a height decrease, the out-going mode may cause a height increase which for fixed flow rate (at the inlet) corresponds to a decrease in the drain rate.
As the drain is the analogue of the gravitational mass, it is possible that this system could provide an analogue of black hole evaporation.

The final fate of evaporating black holes has implications concerning the information paradox \cite{giddings1995black,hawking2005information,page1993information,giddings2019searching,susskind1992hawking,bekenstein1994entropy} and thus, a clearer picture of the process is desirable.
The main issue is that Hawking's original derivation of HR assumes a non-evolving geometry in which the backreaction is neglected \cite{hawking1974explosions}.
However, since the temperature of the radiation emitted by a black hole is proportional to the inverse of it's mass, the backreaction of HR onto the geometry becomes increasingly important at late times in the black hole's evolution \cite{kodama1980conserved,york1985black}.
Different approaches to the backreaction problem have been considered in the literature \cite{balbinot1984hawking,balbinot1989backreaction,callan1992evanescent,susskind1992hawking,russo1992end,parikh2000hawking,medved2005hawking,banerjee2008quantum,banerjee2008quantum2} with no clear consensus as to what the ultimate fate of evaporating black holes should be.
There are two main problems: (a) for quantum fields living on the geometry, expressions for the energy momentum tensor as general functions of the metric cannot be computed \cite{balbinot1989backreaction}, meaning approximations have to be employed, and (b) for metric perturbations a full quantum theory of gravity is required, which is not currently available.

The uncertainty surrounding these problems in GR has prompted some authors to consider how the backreaction occurs in the analogues \cite{balbinot2005quantum,balbinot2006hawking}, where the small-scale description of the medium is known.
Particularly interesting are the results of \cite{schutzhold2005quantum,schutzhold2007quantum,maia2007quantum}, which show that, in a full quantum mechanical treatment of a BEC, quantum corrections to the classical backreaction equations can play an important role in determining the evolution.
Constraints on the importance of such corrections are necessary if we are to make predictions about the evolution of black holes at late times in the absence of a quantum theory of gravity.

One way to approach this could be to study other examples of backreacting quantum systems, to develop a broader understanding of the problem in general.
This motivates our suggestion to investigate draining quantum fluids with a free surface, which should exhibit similar height change behaviour to that studied here.
An added benefit to such a system would be that a change in height may act as an indicator that the system is emitting HR, although our system needs to be investigated further in order to show this.

\section{Summary}

In this chapter, we have demonstrated that surface waves interacting with an initially stationary vortex will trigger the evolution of the background into a new equilibrium state. 
In particular, we have shown theoretically and experimentally that perturbations to a DBT flow, which is closed at the outer boundary and open at the drain, are able to push mass out of the system, thereby decreasing the water height.

Our theoretical predictions in shallow water suggest that the wave induced mass flow at the horizon can actually be directed radially outward in a small region of the parameter space, which would decrease the drain rate and therefore cause an increase in water height. 
Based on these predictions, it seems possible that perturbations of a particular frequency could be used to control the water height in situations where the flow rate is not under the experimentalist's control.

Our experimental results are also interesting from the analogue gravity perspective.
Due to the flow being externally driven, it was previously unclear whether the backreaction could be observed in analogue gravity simulators.
Our findings show that the backreaction is indeed observable, which indicates that the system does in fact have freedom to re-distribute energy and angular momentum between the incident waves and the analogue black hole.

However, many improvements are required to understand the behaviour of our system in detail.
In particular, it would be desirable to experimentally demonstrate that the height change depends on the wave amplitude squared, thereby validating it as a second order effect.
This could be achieved in our set-up by varying the amplitude of the wave generator for fixed frequency. 
Multiple repeats would need to be performed at each amplitude due to the inherent noise in the system (e.g. 6-15 repeats were performed for the superradiance experiment and 25 repeats for the QNM study).
One difficulty with varying the amplitude is that one must ensure that the maximum amplitude stays sufficiently small that a perturbative expansion remains valid.
If the amplitude is too large, terms at second order creep back up to linear order and the wave equation acquires extra terms, e.g. the leading non-linear corrections in shallow water are given by the Boussinesq approximation \cite{boussinesq1872theorie,jager2006origin,johnson1996two}.
Another improvement would be to extend the theory to include dispersion, vorticity and free surface gradients, all of which are expected to influence the effective field theory close to the drain. 

Penultimately, this demonstration of the backreaction is important since it highlights that one must ensure that any wave effects (e.g. superradiance and QN ringing) are measured on a timescale much shorter than the time it takes for the flow to vary substantially, such that the assumption of a quasi-stationary background is not violated.
For example, in the experiment of Chapter~\ref{chap:super}, the height decrease was around $0.25~\mathrm{mm}$ over the $13.2~\mathrm{s}$ window.
Since the wave amplitude was of the order $2~\mathrm{mm}$, our perturbative analysis seems reasonable, in which case the background changes have negligible effect on linear wave propagation.
The window of observation in experiments of Chapter~\ref{chap:qnm} was about $5.5~s$ for a similar background flow, hence wave propagation there is also unaffected by the backreaction.

Finally, in addition to being the first observation of the classical backreaction in an analogue gravity system, this work represents a natural first step towards experimentally probing the effects of backreaction due to quantum fluctuations.

\chapter{Conclusion} \label{chap:conc}
In this thesis, we have studied a range of effects that occur in a rotating, draining fluid flow called the draining bathtub vortex.
Our initial motivation came from the analogue gravity programme, where this system is widely studied as a toy model of a rotating black hole.
Despite numerous theoretical studies in the literature, an experimental investigation of this system was lacking and to this end, we devised a series of experiments to test how the analogy performs under realistic laboratory conditions.
Furthermore, previous experimental efforts within the analogue gravity community have focussed on the detection of Hawking radiation.
Therefore, our experimental results represent the first successful demonstration of superradiance and quasi-normal ringing around rotating black hole analogues, as well as the first step toward probing the effects of the backreaction in analogue systems.
We briefly review our main results, highlighting the relevance of our findings for both the gravitational and fluid-dynamics communities.

Firstly, our detection of superradiance in the dispersive regime demonstrates the robustness of the effect beyond the regime of the analogy, where it is unclear as to whether an effective field theory for free surface fluctuations even exists.
Specifically, we have demonstrated that superradiance can survive the influence of both dispersion and vorticity, in addition to dissipation and turbulence which are also present in fluids.
This result echoes the earliest success of analogue gravity, which was to demonstrate that the emission of Hawking radiation is not compromised by the effects of high-frequency dispersion.
In the same vein, we also expect that superradiance will persist when subject to high energy modifications to GR.

Similarly, our observation of the light-ring modes in dispersive media demonstrates the generality of quasi-normal ringing.
We found that this phenomenon does not require an external source of excitation (i.e. sending waves) to be stimulated, but instead can be seen in the ``noise'' on the water's surface when the system is in a quasi-stationary state.
Inspired by the idea of black hole spectroscopy, we have used the prevalence of the light-ring modes to propose a new technique for flow measurement that we call \textit{analogue black hole spectroscopy} (ABHS).
We believe the advantage of ABHS over existing methods is its non-invasive nature, compared with e.g. PIV, which requires the presence of tracer particles.
The ABHS scheme has the added benefit that it can be used in reverse to test how fluctuations interact with the background flow, making it an ideal tool for systems such as ours where the effective field theory for the waves is unclear.

In both of these experiments, we observed from our measurements of the velocity field that vortex flows, which form under laboratory conditions, exhibit a column of vorticity concentrated in the vortex core.
This observation is supported by a range of experimental and theoretical studies in the literature.
Although vorticity of the background significantly complicates wave scattering processes by allowing for the excitation of additional degrees of freedom, we have argued that in purely rotating fluids, the effect of vorticity on counter propagating modes can be encoded in a quantity which preserves the effective metric, namely a local effective mass.
We then found that the main effect of vorticity on the resonance spectrum is to introduce extra characteristic modes, which are similar in origin to the bound states of the hydrogen atom.

The results of the QNM experiment and the vorticity study are interesting to consider collectively.
The problem of spectral stability arises naturally in black hole physics when considering how matter distributions surrounding astrophysical black holes can influence the QNM spectrum \cite{barausse2014environment}.
Our experiment shows that the light-ring modes of a dispersive bathtub vortex accurately describe the system's response even when an outer boundary is present, thus demonstrating a point made in \cite{barausse2014environment} that QNMs of the ideal system still appear in the frequency response despite the fact that they are no longer QNMs of the true system.
By contrast, our theoretical study for a purely rotating fluid with vorticity indicates that the characteristic mode spectrum is heavily influenced by modifications to the underlying geometry itself.
This may be relevant for stellar geometries \cite{tikekar1990exact} as well as for black holes in modified theories of gravity, e.g. in certain scalar-tensor theories, perturbations can acquire an effective local mass (much like the one in Eq.~\eqref{waveeqnMass}) due to the coupling of gravity to an additional scalar field \cite{sotiriou2013scalar}.

Finally, we have demonstrated that the presence of waves in our system induces a backreaction which becomes significant on a timescale of roughly $\sim\mathcal{O}(100/f)$ where $f$ is the wave frequency.
Specifically, waves incident on a DBT vortex can lead to the removal of fluid mass from the system, which manifests itself as a change in the mean water level.
This realisation is important for a number of reasons.
Firstly, it suggests the naive expectation that the backreaction can be neglected in analogue experiments is ill-founded, which should be born in mind in all future analogue gravity studies.
Secondly, this experiment represents the first step toward probing the effects of backreaction in quantum analogue systems, which may be of importance in developing our understanding of black hole evaporation.
Lastly, the form of the backreaction specific to the DBT vortex has the potential to find use in practical application.
Our theoretical study of the shallow water equations suggests that wave absorption can lead to a height increase or decrease depending on the frequency of the wave, which suggests that waves could be used to control the water level in circumstances where the flow rate is beyond the control of the experimentalist.
Furthermore, we observed that the momentum flow induced by the wave near the drain leads to a rapid clearance of free surface contaminants, which would otherwise drain from the system extremely slowly in the absence of waves.

\section*{Future prospects for analogue studies}

In the concluding remarks of each chapter, we have discussed how each of these studies could be extended and/or improved.
We provide a condensed list below to serve as a starting point for future investigations:
\begin{enumerate}
\item \textit{Superradiance}. An effective theory beyond that presented in Section~\eqref{wavscat1} is required to explain why the $m=2$ mode is amplified more than $m=1$.
In particular, the effects of dispersion, vorticity and free surface gradients should be accounted for, as well as potentially dissipation and asymmetry of the horizon.
\item \textit{Bound states}. The system considered in Chapter~\ref{chap:vort} should be set up in the laboratory, with the aim of performing a detection (of the kind in Chapter~\ref{chap:qnm}) to detect the bound state resonances.
These resonances could be used in conjunction with the ABHS method to not only measure the parameters of fluid flows with vorticity, but also to test the effective field theory for the perturbations.
\item \textit{Backreaction}. The height change project has opened up several interesting avenues which require further investigation, even at the classical level.
These are:
(a) an experiment should be performed to demonstrate that the water height evolves with the square of the wave amplitude,
(b) a theoretical investigation is required to explain the observed late time behaviour in Figure~\ref{fig:long} and
(c) experimental and theoretical work is required to determine which parameters control the time taken to effectively clear the free surface of contaminants.
\end{enumerate}
In addition to these developments which are specific to our experimental set-up, one of the active areas of research in analogue gravity at present is the study of backreacting quantum systems, which aim to explore how the backreaction in influenced by the atomic nature of a medium.
With a more complete picture of how the backreaction occurs in the classical set-up, extension to the quantum case will be the next natural step for DBT-type systems.

\section*{Outlook}

To summarise, over the course of this thesis we have seen that analogue gravity is an extremely rich field of research, whose remit spans multiple disciplines within the physical sciences.
Given the apparent complexity of nature in many circumstances, it is truly remarkable that two theories describing seemingly disparate areas of physics should share some region of overlap.
However, the real interest of such an analogy lies not in the similarities between systems, but in the differences.
Indeed, in view of our ignorance of the behaviour of spacetime at small scales, any attempt to solve problems of a gravitational and quantum mechanical nature would be at best an educated shot in the dark. 
The analogies however provide us with a guiding light.
By shining this light into new territory, beyond the region where our theories overlap, we are able to make predictions about how we might expect gravitational physics to be modified when GR gives way to a more fundamental theory at small scales.
The prevalence of not only Hawking radiation, but now also superradiance and quasi-normal ringing beyond the analogue regime is a testament to this statement.
We hope that these results will inspire further investigation into this promising field of research.

\appendix



\chapter{Polar wave-vector} \label{app:polar}
For waves propagating on a two dimensional surface, the wave vector can be decomposed in Cartesian coordinates $\mathbf{x}=(x,y)$ exactly as $\mathbf{k}=(k_x,k_y)$. In polar coordinates $\mathbf{x}=(r,\theta)$ however, the decomposition $\mathbf{k}=(k_r,k_\theta)$ is not exact since $k_{\theta}$ diverges as $r\to0$. Consider a wave which is periodic in $\theta$. The wavelength of such a mode is given by $\lambda_{\theta} = 2\pi r/m$ where $m$ is the number of times the wave repeats over the $\theta$ interval, i.e. the azimuthal number of the mode. The corresponding wavenumber is $k_\theta = m/r$, which blows up at the origin. It seems reasonable therefore to suggest that the polar wavevector decomposition in valid only far from the origin.

To see this, consider wave equation in Eq.~\eqref{waveeqn} for the case $\mathbf{V}=0$. Using the decomposition $\phi=\sum_mR_m(r)\exp(im\theta-i\omega t)$, the equation for each mode is,
\begin{equation} \label{bessel}
\partial_r^2R_m + \frac{1}{r}\partial_rR_m + \left(k^2-\frac{m^2}{r^2}\right)R_m = 0,
\end{equation}
where we have used the exact dispersion relation $\omega = \pm ck$ with $k = |\mathbf{k}|$, and the sign dictates the direction of propagation. This is Bessel's equation whose solutions can be written in the form of the Hankel functions of the first and second kinds $H^{(1)}_m(kr)$ and $H^{(2)}_m(kr)$ respectively. These functions asymptote to out-going and in-going plane waves,
\begin{equation}
\begin{split}
H^{(1)}_m(kr\to\infty) & \to \sqrt{\frac{2}{\pi}}e^{-i(m+1/2)\pi/2}\frac{1}{\sqrt{kr}}e^{ikr} = \text{out-going}, \\
H^{(2)}_m(kr\to\infty) & \to \sqrt{\frac{2}{\pi}}e^{i(m+1/2)\pi/2}\frac{1}{\sqrt{kr}}e^{-ikr} = \text{in-going}.
\end{split}
\end{equation}
Note the factor of $1/\sqrt{r}$, which is present in all radial modes, accounts for the focussing of amplitude in 2D as the mode propagates onto a smaller radius. To deduce the validity of the polar decomposition, consider writing $R_m$ using the WKB ansatz in Eq.~\eqref{WKB}. Following the prescription outlined in Section~\ref{sec:wkb} and defining $\partial_rS_0 = k_r$, Bessel's equation in Eq.~\eqref{bessel} at zeroth and first order gives,
\begin{equation}
\begin{split}
\mathcal{O}(1): & \quad k_r = \sqrt{k^2-m^2/r^2} \to \omega = \pm c\sqrt{k_r^2 + m^2/r^2} \\
\mathcal{O}(\epsilon): & \quad \partial_r(rk_rA_0^2) = 0.
\end{split}
\end{equation}
The effective potential $V$ seen by the modes is $V=-k_r^2$, which diverges at the origin for $m\neq0$ (this is the angular momentum barrier). 
The description in terms of propagating WKB modes breaks down at the turning point of $V$ given by $r_\mathrm{tp}=|m|c/\omega$.
For $r<r_\mathrm{tp}$, $k_r$ becomes imaginary which leads to the decay of the Bessel functions near the origin.
This is discussed further in Section~\ref{sec:BesselR0} of Chaper~\ref{chap:vort}.
Thus, the description in terms of propagating modes with wavevector components $k_r$ and $m/r$ is valid in the region $r>r_\mathrm{tp}$.

\begin{figure} 
\centering
\includegraphics[width=0.6\linewidth]{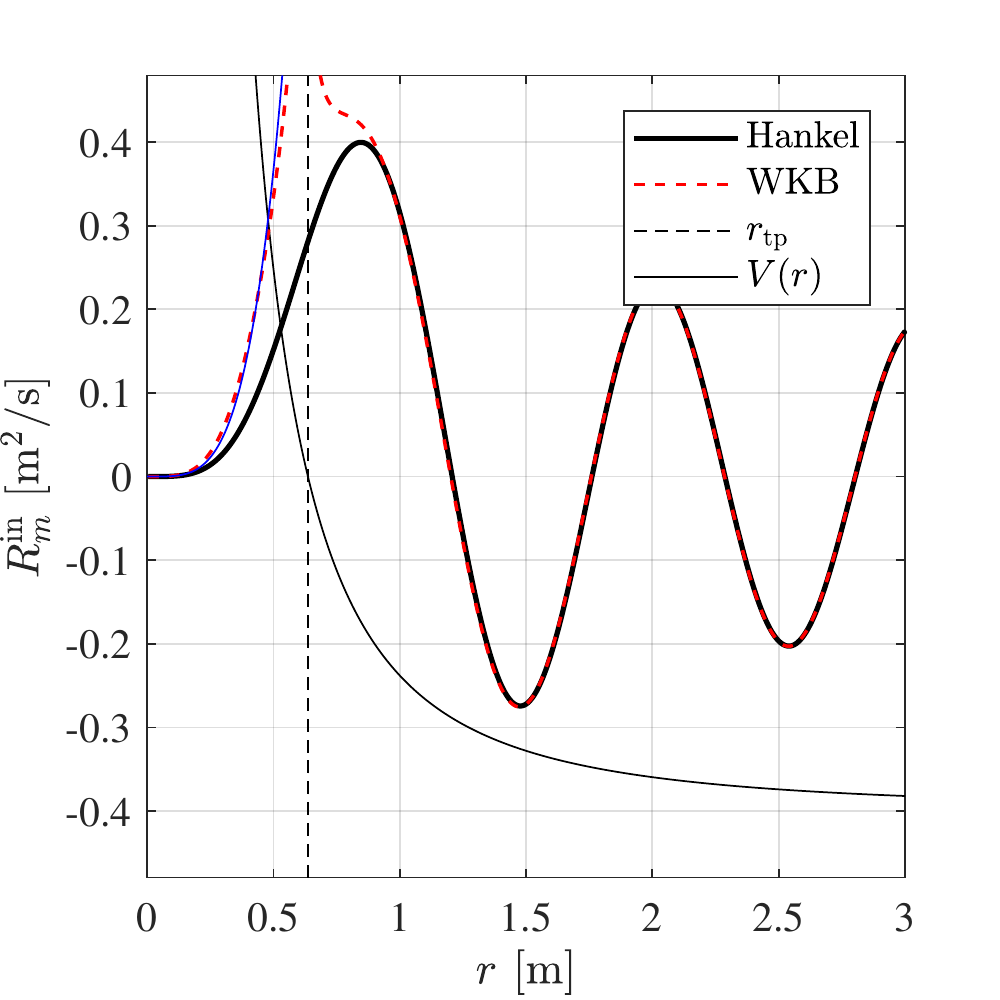}
\caption{Solutions $\psi_m^{in}$ for $k=2\pi$, $m=3$. The agreement between the WKB solution and the exact solution is excellent whilst the solution is oscillatory, i.e. for $r>r_\mathrm{tp}$, where $r_\mathrm{tp}=|m|c/\omega$ is the turning point of the potential barrier $V$. 
The WKB solution is also valid for $r<r_\mathrm{tp}$, and tends to a power law $\propto r^{|m|}$ close to the origin (indicated in blue). 
This is discussed in detail in Section~\ref{sec:BesselR0} of Chapter~\ref{chap:vort}.} \label{fig:WKBhank}
\end{figure}

The integral of $k_r$ can be evaluated exactly and the solutions for $R_m$ are found to be,
\begin{equation}
R_m^{\mathrm{out}} \simeq \frac{1}{\sqrt{k_rr}}e^{ik_rr + im\Theta}, \qquad R_m^{\mathrm{in}} \simeq \frac{1}{\sqrt{k_rr}}e^{-ik_rr - im\Theta},
\end{equation}
where $\Theta(r) = \tan^{-1}(m/k_rr)$ can be understood as the angle formed between the total wavevector $\mathbf{k}$ and the effective radial wavevector $k_r\bm{\mathrm{e}}_r$. These solutions can be matched in the asymptotic limit to the Hankel functions since $k_r(r\to\infty)\to k$ and $\Theta(r\to\infty)\to0$. As shown in Figure~\ref{fig:WKBhank}, the WKB solution is in excellent agreement with the exact solution for $r>r_\mathrm{tp}$. In fact, the solution remains close to the true solution until the angular wavenumber exceeds the radial wavenumber, i.e. $m/r>k_r$. This suggests a more conservative estimate for the region of validity of the polar decomposition is $r>2r_\mathrm{tp}$. 
For $m=0$, the WKB solution is only valid up to near the origin since the factor of $1/\sqrt{r}$ causes divergence, even though $k=k_r$ is exact.

Finally, returning the $t$ and $\theta$ dependence, the WKB approximation for a single $m$ mode is proportional to,
\begin{equation}
\frac{1}{\sqrt{k_rr}}e^{\pm ik_rr + im(\theta\pm\Theta) - i\omega t}.
\end{equation}
From this, we see that far away, we can interpret a single $m$-mode as a plane wave propagating radially with wavenumber $k_r(r)$ and a phase shift $\Theta(r)$ in the azimuthal component which increases toward decreasing $r$.

\chapter{Geodesics} \label{app:geos}
The trajectories of high frequency waves (also called rays) are equivalent in the shallow water regime to the null geodesics of the effective geometry.
In this appendix, we derive the equations governing the geodesics for the DBT flow of Eq.~\eqref{DBT} in the Lagrangian formalism.
This results in a potential for the radial geodesics which is identical to the potential seen by eikonal waves in Eq.~\eqref{GeoPotential}, and it is in this sense that waves of high frequency follow the null geodesics of the underlying metric.
We then show how the trajectories can be obtained in the Hamiltonian formalism, which can be easily extended to include dispersive effects.

\section*{Lagrangian method}

A geodesic is shortest path between two points on a manifold (see e.g. \cite{carroll2019spacetime} for a review) which, for our purposes, can be obtained by minimising the total spacetime distance,
\begin{equation}
S = \int ds = \int \frac{ds}{d\lambda} d\lambda = \int \mathcal{L} d\lambda,
\end{equation}
where the incremental distance $ds$ is given by the line element in Eq.~\eqref{EffMetric} and $\lambda$ parametrises the path along the geodesic.
Requiring that the path be stationary with respect to variations, i.e. $\delta S = 0$, yields the Euler-Lagrange equations,
\begin{equation} \label{EL2}
\partial_\lambda\left(\frac{\partial\mathcal{L}}{\partial \dot{x}^\mu}\right) + \frac{\partial\mathcal{L}}{\partial x^\mu} = 0,
\end{equation}
whose solutions are the geodesics.
Here, the overdot signifies the derivative with respect to $\lambda$.
However, when computing geodesics it is usually more convenient to define the Lagrangian as $\mathcal{L}=\frac{1}{2}(ds/d\lambda)^2$.
Using this definition, the Lagrangian describing the geometry seen by linear perturbations to a shallow fluid in laboratory coordinates $x^\mu=(t,r,\theta)$ is,
\begin{equation} \label{LGRN2}
\mathcal{L} = -\frac{1}{2}\bigg(c^2-V_r^2-V_\theta^2\bigg)\dot{t}^2 - V_r \dot{r}\dot{t} - rV_\theta \dot{\theta}\dot{t} + \frac{1}{2}\dot{r}^2 + \frac{1}{2}r^2\dot{\theta}^2.
\end{equation}
Since $\mathcal{L}$ is independent of $t$ and $\theta$, Eq.~\eqref{EL2} gives,
\begin{subequations}
\begin{align}
\partial_\lambda\left(\frac{\partial\mathcal{L}}{\partial \dot{t}}\right) = 0 & \ \to \ \bigg(c^2-V_r^2-V_\theta^2\bigg)\dot{t} + V_r\dot{r} + rV_\theta \dot{\theta} = \omega, \label{EandLgeo_a} \\
\partial_\lambda\left(\frac{\partial\mathcal{L}}{\partial \dot{\theta}}\right) = 0 & \ \to \ r^2\dot{\theta} - rV_\theta\dot{t} = m, \label{EandLgeo_b}
\end{align}
\end{subequations}
where $\omega$ and $m$ are two conserved quantities related to energy and angular momentum, which we identify with frequency and azimuthal number respectively.

Although we could now use the $r$ component of Eq.~\eqref{EL2} to obtain a second order differential equation for $\ddot{r}$, it is simpler in this case to use the expression for $\mathcal{L}$ to obtain an equation for $\dot{r}$ directly.
To do this, we may rescale $\lambda$ such that their are only three different possibilities for the value of the Lagrangian:
\begin{equation}
\mathcal{L} = 
\begin{cases}
-1/2, \qquad \mathrm{timelike} \\
\ \ \ \ 0, \ \ \qquad \mathrm{null} \\
+1/2, \qquad \mathrm{spacelike} \\
\end{cases},
\end{equation}
where timelike, null and spacelike are synonymous with subluminal, light-like and superluminal respectively.
Since we are interested in null geodesics, we take $\mathcal{L}=0$ and eliminate $\dot{t}$ and $\dot{\theta}$ from Eq.~\eqref{LGRN2} using Eqs.~\eqref{EandLgeo_a} and~\eqref{EandLgeo_b}.
Thus, we arrive at the equation governing the radial geodesics,
\begin{equation} \label{geoEnergy}
\dot{r}^2 + V_\mathrm{geo} = 0,
\end{equation}
where the potential is given by,
\begin{equation} \label{radGeoEqn}
V_\mathrm{geo} = \frac{1}{c^2}\left[-\left(\omega-\frac{mV_\theta}{r}\right)^2 + \left(c^2-V_r^2\right)\frac{m^2}{r^2}\right].
\end{equation}
Note that Eq.~\eqref{geoEnergy} is of the form of an energy equation with $\dot{r}^2$ playing the role of the kinetic energy.
Up to the factor of $1/c^2$, this is precisely the geodesic potential defined in Eq.~\eqref{GeoPotential} of the main text.

Finally, the trajectory of a particle through the effective geometry is determined by specifying $x^\mu(\lambda=0)$ and solving the first order equation in Eq.~\eqref{geoEnergy} followed by Eqs.~\eqref{EandLgeo_a} and~\eqref{EandLgeo_b}.

\section*{Hamiltonian method}

The Hamiltonian formulation is a different way of approaching mechanics whose central object is a system of equations called Hamilton's equations, in place of the Euler-Lagrange equations of Lagrangian mechanics.
Although the two formalism's are completely equivalent to one another, it is often more convenient to use one over the other depending on the problem at hand.
After showing how the shallow water results can be recovered, we will show how Hamilton's equations can be applied to dispersive systems.

We start by writing the Lagrangian in covariant form,
\begin{equation}
\mathcal{L} = \frac{1}{2}g_{\mu\nu}\dot{x}^\mu\dot{x}^\nu.
\end{equation}
To obtain an effective Hamiltonian $\mathcal{H}$ for the system, we first need to define the conjugate momentum to the coordinates $p_\mu=\frac{\partial\mathcal{L}}{\partial x^\mu}=g_{\mu\nu}\dot{x}^\nu$.
In terms of components, these are $p_\mu=(-\omega,k_r,m)$ where we identify $k_r$ with the radial wavenumber.
Then the Hamiltonian is given by,
\begin{equation}
\mathcal{H} = g^{\mu\nu}p_\mu\dot{x}_\nu - \mathcal{L}  = \frac{1}{2}g^{\mu\nu}p_\mu p_\nu.
\end{equation}
Particles that follow null geodesics obey the Hamiltonian constraint $\mathcal{H}=0$, which gives (up to the factor $1/2$) the dispersion relation in Eq.~\eqref{dispersion}.
This equation when written in the form,
\begin{equation}
g^{\mu\nu}\frac{\partial S_0}{\partial x^\mu}\frac{\partial S_0}{\partial x^\nu} = 0,
\end{equation}
is called the Hamilton-Jacobi equation, and we have introduced the Hamilton-Jacobi action $S_0$ through the relation $p_\mu=\frac{\partial S_0}{\partial x^\mu}$.
Note that $S_0$ plays the role the phase of a high frequency wave in the WKB approximation (see Section~\ref{sec:Eik}).

The geodesics of the effective geometry can be obtained through Hamilton's equations,
\begin{equation} \label{HamiltonsEqns}
\dot{p}_\mu=-\frac{\partial\mathcal{H}}{\partial x^\mu}, \quad \dot{x}^\mu=\frac{\partial\mathcal{H}}{\partial p_\mu}.
\end{equation}
However, the notion of an effective geometry breaks down when dispersive effects become important.
Nonetheless, we can still talk about trajectories of high frequency waves by modifying the Hamiltonian to incorporate dispersive effects.
This is a simple task in the Hamiltonian formalism, since $\mathcal{H}$ as defined above is just the dispersion relation up to a factor of $1/2$.
Hence by taking $\mathcal{H}$ as the full dispersion relation for gravity waves in Eq.~\eqref{dispersion2}, we may use Hamilton's equations in Eq.~\eqref{HamiltonsEqns} to find trajectories of dispersive waves.
This is our starting point for Eq.~\eqref{RayHamiltonian} in Chapter~\ref{chap:qnm} and is easily generalised to more complicated dispersion relations, as we do later on in Eq.~\eqref{dispersionFullCap}.


\chapter{Numerical Methods} \label{app:RK4}

In this appendix, we provide the details of the numerical simulations in Section~\ref{sec:ABHSmethod} of Chapter~\ref{chap:qnm} and Section~\ref{sec:BSsims} of Chapter~\ref{chap:vort}, which were performed using a technique called the Method of Lines (MOL).
The basic idea is to reformulate the wave equation, which is a second order partial differential equation (PDE) in space and time, into two PDEs which are first order in time.
The two PDEs can then be solved as a matrix equation on discrete spatial and temporal grids by supplementing the wave equation with (spatial) boundary conditions and (temporal) initial conditions.

We first explain how the reformulation is done for the regular wave equation in Eq.~\eqref{waveeqn}.
Defining the new variable $\pi = D_t\phi$, Eq.~\eqref{waveeqn} becomes,
\begin{equation} \label{vec_waveeqn1}
\partial_t	\begin{pmatrix}
\phi \\ \pi
\end{pmatrix}
= \begin{pmatrix}
-\mathbf{V}\cdot\bm\nabla & \mathbb{I} \\
c^2\nabla^2 & -\mathbf{V}\cdot\bm\nabla
\end{pmatrix} \begin{pmatrix}
\phi \\ \pi
\end{pmatrix},
\end{equation}
where $\mathbb{I}$ is the identity matrix.
Since we consider axisymmetric velocity fields, we consider the problem in polar coordinates $(t,r,\theta)$, and reduce the wave equation to a differential equation in $(t,r)$ for each $m$-mode.
Defining $(\phi,\pi)=\sum_m(y_m,y_m')\exp(im\theta)$, the wave equation assumes the form,
\begin{equation} \label{vec_waveeqn2}
\partial_t	\begin{pmatrix}
y_m \\ y_m'
\end{pmatrix}
= \begin{pmatrix}
-\frac{im}{r}V_\theta-V_r\partial_r & \mathbb{I} \\
c^2\left[\partial_r^2+\frac{1}{r}\partial_r - \frac{m^2}{r^2}\right] & -\frac{im}{r}V_\theta-V_r\partial_r
\end{pmatrix} \begin{pmatrix}
y_m \\ y_m'
\end{pmatrix} ,
\end{equation}
which can be written in the compact form,
\begin{equation} \label{vec_waveeqn3}
\partial_tY_m = \mathcal{D}Y_m.
\end{equation}
This is a vectorial PDE (first order in $t$ and second order in $r$) where $Y_m$ is the vector containing our solution and $\mathcal{D}$ is a matrix containing all the parts of the wave equation except the time derivatives.
Note that here we have opted to contain all the $r$ dependence of the solution within $Y_m$ since the $1/\sqrt{r}$ in Eq.~\eqref{waveansatz}, which simplifies the analytic expressions, only serves to complicate the form of $\mathcal{D}$.
For the regular wave equation in Eq.~\eqref{waveeqn} and the wave equation with the effective local mass in Eq.~\eqref{waveeqnMass}, $Y_m$ and $\mathcal{D}$ are, 
\begin{equation}
Y_m = \begin{pmatrix}
y_m \\ y'_m
\end{pmatrix}, \qquad \mathcal{D} = \begin{pmatrix}
-\frac{im}{r}V_\theta-V_r\partial_r & \mathbb{I} \\
c^2\left[\partial_r^2+\frac{1}{r}\partial_r - \frac{m^2}{r^2} - \zeta^2\right] & -\frac{im}{r}V_\theta-V_r\partial_r
\end{pmatrix}.
\end{equation}
For the former we set $\zeta=0$ since we have an irrotational flow, and for the latter we use $V_r=0$ since we are interested in purely rotating flows. 
In Eqs.~\eqref{EqMot2} and~\eqref{EqMot3}, which describe waves in a rotating fluid with vorticity, the irrotational part of the field $\phi$ is coupled to a rotational degree of freedom which we write as $\bm\xi = \sum_m(a_m,b_m)\exp(im\theta)$, where $a_m$ ($b_m$) are the radial (angular) components respectively.
Eqs.~\eqref{EqMot2} and~\eqref{EqMot3} can then be cast in the form,
\begin{equation}
\begin{split}
Y_m = \begin{pmatrix}
y_m \\ y'_m \\ a_m \\ b_m
\end{pmatrix}, \qquad \mathcal{D} = & \ \begin{pmatrix}
-\frac{im}{r}V_\theta-V_r\partial_r & \mathbb{I} & 0 & 0 \\
c^2\left[\partial_r^2+\frac{1}{r}\partial_r - \frac{m^2}{r^2}\right] & -\frac{im}{r}V_\theta-V_r\partial_r & c^2\left[\partial_r+\frac{1}{r}\right] & \frac{imc^2}{r} \\ \frac{im\zeta}{r} & 0 & -F_1 & \frac{2V_\theta}{r} \\ -\zeta\partial_r & 0 & -\zeta & -F_2
\end{pmatrix} , \\
F_1 = & \ V_r\partial_r+\partial_rV_r+\frac{imV_\theta}{r} , \qquad F_2 = V_r\partial_r+\frac{V_r}{r}+\frac{imV_\theta}{r} .
\end{split}
\end{equation}
We now explain how the discretisation is performed and how the boundary/initial conditions are implemented.

\section*{Spatial discretisation}

To evaluate the derivatives $\partial_r$ and $\partial_r^2$ numerically, we first discretise the radial grid $r=r_1:r_N$ with step size $\Delta r = (r_N-r_1)/(N-1)$. The derivatives can then be approximated using finite difference stencils, which are computed by taking combinations of neighbouring points in such a manner that the error to the relevant order in $\Delta r$ is minimised \cite{gustafsson2007high}. 
The second order centred difference stencils are,
\begin{equation} \label{stencils}
\partial_rf_j = \frac{f_{j+1}-f_{j-1}}{2\Delta r}, \qquad \partial_r^2f_j = \frac{f_{j+1}-2f_j+f_{j-1}}{\Delta r^2} ,
\end{equation}
where $j=1:N$ is the spatial index and $f$ represents the components of $Y_m$. 
The boundary conditions are used to supply the values of the field at the points $j=0$ and $j=N+1$.
In general, since $j=0$ and $j=N+1$ are not in our numerical range, the boundary conditions are used to compute these values in terms of points that are in the range, and the matrix $\mathcal{D}$ is supplemented by a boundary matrix contained these terms.
However, for the simulations we performed the solution is even simpler.

Far away from the vortex at large $r$, we place a hard wall which is sufficiently far away that the mode will not reflect from it during the time scale of our simulation.
This is implement by setting $f_{N+1}=0$ (Dirichlet boundary conditions) for all components of $Y_m$.
In the simulations of Chapter~\ref{chap:vort}, there is an angular momentum barrier in the potential at the origin and hence, in this case we also set $f_0=0$, where the point at $j=0$ is some small distance away from the origin.
In the case that our system contained a horizon, as in Chapter~\ref{chap:qnm}, we used a free boundary condition placed just inside the horizon, which allows the solution to oscillate freely there.
The justification for this is that the interior of the horizon is casually disconnect from the exterior, and thus the value of the field inside cannot effect the solution outside.
Furthermore, in our simulations we are only interested in the physics of the exterior region (specifically near the light-ring).
The free boundary condition is implemented via the use of forward difference stencils at $j=0$.
These are stencils that are not centred on the point where the derivative is approximated, which at second order in $\Delta r$ take the form,
\begin{equation} \label{forward_stenc}
\partial_rf_1 = \frac{-f_3+4f_2-3f_1}{2\Delta r}, \qquad \partial_r^2f_1 = \frac{f_3-2f_2+f_1}{\Delta r^2}.
\end{equation}
Thus, we may write the full matrix approximation of the derivatives.
For a horizon and a hard wall, these are,
\begin{equation} \label{stenc_horz}
\partial_r = \frac{1}{2\Delta r}\begin{pmatrix}
-3 & 4 &  -1    &    &   \\
-1 & 0 &   1    &    &   \\
   &   & \ddots &    &   \\
   &   &  -1    & 0  & 1 \\
   &   &        & -1 & 0 
\end{pmatrix}, \qquad 
\partial_r^2 = \frac{1}{\Delta r^2}\begin{pmatrix}
 1 & -2 &  1     &    &   \\
 1 & -2 &  1     &    &   \\
   &    & \ddots &    &   \\
   &    &   1    & -2 & 1 \\
   &    &        &  1 & -2
\end{pmatrix},
\end{equation} 
and for two hard walls we have,
\begin{equation} \label{stenc_wall}
\partial_r = \frac{1}{2\Delta r}\begin{pmatrix}
 0 & 1 &        &    &   \\
-1 & 0 &   1    &    &   \\
   &   & \ddots &    &   \\
   &   &  -1    & 0  & 1 \\
   &   &        & -1 & 0 
\end{pmatrix}, \qquad 
\partial_r^2 = \frac{1}{\Delta r^2}\begin{pmatrix}
-2 &  1 &        &    &   \\
 1 & -2 &  1     &    &   \\
   &    & \ddots &    &   \\
   &    &   1    & -2 & 1 \\
   &    &        &  1 & -2
\end{pmatrix},
\end{equation}
where empty space is representative of zeros, and the dots indicates that the same sequence is repeated along the diagonal.
Note, that for Eq.~\eqref{stenc_horz}, this means that the approximation for $\partial_r^2$ is the same at the first and the second point, which is a consequence of the second order approximation used.
This could be ameliorated by the use higher order stencils, however, we already found good agreement with our predictions at the presented level of approximation and thus did not pursue such improvements.

\section*{Temporal discretisation}

Once the wave equation has been put in the form of Eq.~\eqref{vec_waveeqn3}, we can solve in time for all $Y_m^k$ where the index denoting each time step is $k=1:M$. 
We first discretise the time vector $t=t_1:t_M$ with spacing $\Delta t=(t_M-t_1)/(M-1)$. 
Next, we initialise the problem at $t=t_1$ with the vector $Y_m^0$.
This contains $y_m^0$, which is a wave packet located far away from $r=r_1$, and it's time derivative $y_m'^0$, which is defined so that the packet propagates toward smaller $r$. 
The form of $y_m'^0$ is computed assuming the wave packet is located in flat space. 
In reality there will be small deviations from flat space meaning that a small wave will also propagate toward larger $r$. 
This will be negligible provided $|\mathbf{V}|\ll c$, and we can further minimise it's influence by placing $r_N$ far enough away that this wave does not have time to reflect and reach the part of the solution we are interested in.
For the simulations of Eqs.~\eqref{EqMot2} and~\eqref{EqMot3}, we set $a_m^0=b_m^0=0$ for all $r$ since the wave packet starts in the irrotational region.

The simplest method to evolve to the next time step is the Euler method,
\begin{equation}
\begin{split}
Y_m^{k+1} = \ & Y_m^{k} + \Delta t~\partial_tY_m^{k} \\
= \ & \left[\mathbb{I}+\mathcal{D}\Delta t\right]Y_m^k,
\end{split}
\end{equation}
which is a rewriting of the definition of the derivative as $\Delta t\to 0$. 
Since $\Delta t\neq 0$, we use a more sophisticated variant of this method called the Fourth-Order-Runge-Kutta algorithm (RK4) \cite{gustafsson2007high}. 
The basis of this method is to approximate the slope four times over the interval $\Delta t$ (once at the start, twice in the middle and once at the end) and use each successive improvement at the next iteration. 
This is simple to implement for us since $\mathcal{D}$ does not depend on $t$.
Let $\gamma_i$ for $i=1:4$ represent the four estimates of the slope.
The RK4 algorithm is,
\begin{equation}
\begin{split}
\gamma_1 = \ & \mathcal{D}Y_m^k \\
\gamma_2 = \ & \mathcal{D}(Y_m^k+\Delta t~\gamma_1/2) \\
\gamma_3 = \ & \mathcal{D}(Y_m^k+\Delta t~\gamma_2/2) \\
\gamma_4 = \ & \mathcal{D}(Y_m^k+\Delta t~\gamma_3) \\
Y_m^{k+1} = \ & Y_m^k + \Delta t\left(\frac{\gamma_1+2\gamma_2+2\gamma_3+\gamma_4}{6}\right).
\end{split}
\end{equation}
Evaluating this for all $k$, we can extract from $Y_m$ the vector $y_m$, which contains an $m$-mode starting far away and scattering off a non-zero velocity field in the centre.

\chapter{Waves in rotational flows} \label{app:clebsch}
In this appendix, we demonstrate that Eq.~\eqref{EqMot1_a} splits in two when the perturbations are defined according to Eq.~\eqref{PhiClebsch}.
The proof is complicated and is proven for perturbations to the density $\rho$ in \cite{visser2004vorticity} using an action principle (see Eq.~(4.2) therein).
We show here that the action for our problem is completely equivalent to theirs with the height field replacing the density in one less spatial dimension.

The action $S^{(3)}$ for a three component incompressible fluid with a free surface $H$ can be written using the Clebsch decomposition in Eq.~\eqref{clebschPert} as, \cite{luke1967variational},
\begin{equation} \label{action1}
S^{(3)} = \iiint\rho\left\{\int^H_0\left(\dot{X}_1+X_2\dot{X}_3 + \frac{1}{2}(\partial_iX_1+X_2\partial_iX_3)^2 + gz\right)dz\right\}d^2\mathbf{x}dt,
\end{equation}
where the overdot denotes the time derivative, $i=x,y,z$ and the squared quantity is to be understood as the scalar product.
To convert this into an effective 2D description at the free surface in the shallow water regime, we assume that the horizontal velocity components are independent of $z$ and $V_z\propto z$. 
Integrating $S^{(3)}$ over $z$, we obtain the effective action in 2D,
\begin{equation} \label{action2}
S^{(2)} = \iiint\rho\left\{H(\dot{X}_1+X_2\dot{X}_3) + \frac{H}{2}(\bm\nabla X_1+X_2\bm\nabla X_3)^2 + \frac{gH^2}{2}\right\}d^2\mathbf{x}dt + \mathcal{O}(H^3),
\end{equation}
where the $z$-dependence in $X_1,X_2,X_3$ and the $z$ piece of the inner product are of order $\mathcal{O}(H^3)$. 
These contributions are neglected from here on and all vector terms are understood to be two dimensional.

The final step is to expand Eq.~\eqref{action2} to quadratic order in the perturbations,
\begin{equation}
S^{(2)} = S^{(2)}_0 + \epsilon S^{(2)}_1 + \epsilon^2 S^{(2)}_2 + ...
\end{equation}
Variation of $S^{(2)}_0$ yields the equations of motion for the background and $S^{(2)}_1$ vanishes under the assumption that the zeroth order equations are satisfied.
The contribution at second order is given by,
\begin{equation}
S^{(2)}_2 = \iiint\rho\left\{ \frac{1}{2}H\mathbf{v}^2 + h\mathbf{V}\cdot\mathbf{v} + h(\dot{\chi}_1+\chi_2\dot{X}_3+X_2\dot{\chi}_3) + H\chi_2\dot{\chi}_3 + \frac{c^2}{2H}h^2  \right\}d^2\mathbf{x}dt
\end{equation}
where we have used $c=\sqrt{gH}$. 
This is completely equivalent to Eq.~(4.2) of \cite{visser2004vorticity} with the replacement $\rho\to H$ (and similarly for the perturbations), which is due to the fact that surface waves in shallow water effectively behave as acoustic perturbations on a 2D plane.
Hence, the analysis found therein also holds for our system.
In particular, the results we quote in Eq.~\eqref{PhiClebsch} of the main text can be found in Eqs.~(4.8) and~(4.12) of \cite{visser2004vorticity}.

\chapter{Wave induced mass flow} \label{app:momflow}
In this appendix, we derive the form of the wave induced mass flow for axisymmetric, stationary modes. 
To do this, we insert the decomposition of Eq.~\eqref{VecAnsatz} into the expression for the mass flow in Eq.~\eqref{massflow1}, which requires us to evaluate the integral of $h\mathbf{v}$ over $\theta$. 
For a perturbation of arbitrary frequency content, we have,
\begin{equation} \label{eq:appD1}
\begin{split}
\int_0^{2\pi}h\mathbf{v}d\theta = & \int_0^{2\pi} \iint \sum_{m,n} \Big\{ \big[ h^\mathrm{R}_m(\omega)\cos(m\theta-\omega t)-h^\mathrm{I}_m(\omega)\sin(m\theta-\omega t)\big] \times \\
& \qquad \qquad \big[\mathbf{v}^\mathrm{R}_n(\omega') \cos(n\theta-\omega' t)-\mathbf{v}^\mathrm{I}_n(\omega')\sin(n\theta-\omega' t)\big]  \Big\} d\omega' d\omega d\theta \\
= & \ \pi\iint \sum_m \Big\{ \mathrm{Re}[h_m^*\mathbf{v}_m]\cos(\omega-\omega')t + \mathrm{Im}[h_m\mathbf{v}_m] \sin(\omega-\omega')t \Big\} d\omega' d\omega,
\end{split}
\end{equation}
where superscript R and I denote the real and imaginary components respectively and in the last line we have evaluated the $\theta$-integral using the relations,
\begin{equation}
\begin{split}
& \int_0^{2\pi}\cos(m\theta-\omega t)\cos(n\theta-\omega' t)d\theta=\pi\delta_{m,n} \cos(\omega-\omega')t, \\
& \int_0^{2\pi}\sin(m\theta-\omega t)\sin(n\theta-\omega' t)d\theta=\pi\delta_{m,n} \cos(\omega-\omega')t, \\ 
& \int_0^{2\pi}\sin(m\theta -\omega t)\cos(n\theta - \omega' t)d\theta=-\pi\delta_{m,n} \sin(\omega-\omega')t.
\end{split}
\end{equation}
It is evident from Eq.~\eqref{eq:appD1} that different frequencies are able to couple to one another to produce oscillatory terms at second order.
The only non-oscillatory term comes from the $\omega=\omega'$ contribution to Eq.~\eqref{eq:appD1}, which when integrated in $t$ leads to secular growth in time. 
The oscillatory terms are accompanied by a prefactor $1/(\omega-\omega')$ after integration and when this factor is small, the oscillatory terms can in principle be large. 
However, if the wave is sharply peaked on a single frequency $\omega$, then the $\omega'\neq\omega$ terms will be inherently lower amplitude and the factor $1/(\omega-\omega')$ will not be sufficient to compensate.
Hence, only the $\omega=\omega'$ term of Eq.~\eqref{eq:appD1} contributes to Eq.\eqref{Massflow2} of the main text.

Note, however, that there is oscillatory behaviour present in the evolution of the mean height measured in the experiments of Section~\ref{HeightChange_results}.
These oscillations are due to the $\mathcal{O}(\epsilon)$ contributions to the momentum in Eq.~\eqref{MomentumTotal} and thus oscillate at frequency $\omega$.
This has been confirmed experimentally by a peak in the Fourier transform of $\Delta H$ at the excitation frequency $f$. 
Even in circumstances where these oscillations are large, the quadratic term at $\omega=\omega'$ is the only one that can grow in time and hence, it will always come to dominate after sufficient time has elapsed.


\addcontentsline{toc}{chapter}{References}
\bibliographystyle{plain}
\bibliography{thesis.bbl}

\end{document}